\documentclass[aip,reprint,amsmath,amssymb,floatfix]{revtex4-2}
\usepackage{float}
\usepackage{txfonts}
\usepackage{graphicx}
\usepackage{bm}
\usepackage{simpler-wick}
\usepackage{simplewick}
\usepackage{dcolumn}

\begin{document}

%Title of paper
%\title{Finite-temperature many-body perturbation theory for anharmonic vibrations: Recursive definitions, algebraic reduction, normal-ordered second-quantized derivation, time-independent diagrammatic rules, linked-diagram theorem, and general-order algorithms}
%\title{Finite-temperature many-body perturbation theory for anharmonic vibrations: Recursions, algebraic reduction, second-quantized reduction, diagrammatic rules, linked-diagram theorem, finite-temperature vibrational self-consistent field, and general-order algorithms}
\title{Finite-temperature many-body perturbation theory for anharmonic vibrations: Recursions, algebraic reduction, second-quantized reduction, diagrammatic rules, linked-diagram theorem, finite-temperature self-consistent field, and general-order algorithm}
\author{Xiuyi Qin}
\affiliation{Department of Chemistry, University of Illinois at Urbana-Champaign, Urbana, Illinois 61801, USA}
\author{So Hirata}\email[Email:~]{sohirata@illinois.edu}
\affiliation{Department of Chemistry, University of Illinois at Urbana-Champaign, Urbana, Illinois 61801, USA}
\date{\today}

\begin{abstract}
A unified theory is presented for finite-temperature many-body perturbation expansions of the anharmonic vibrational contributions to thermodynamic functions, i.e., the free energy, internal energy, and entropy. 
The theory is diagrammatically size-consistent at any order, as ensured by the linked-diagram theorem proved in this study, and thus applicable to molecular gases and solids on an equal footing. 
It is also a basis-set-free formalism, just like its underlying Bose--Einstein theory, capable of summing  anharmonic effects over an infinite number of states analytically. 
It is formulated by the Rayleigh--Schr\"{o}dinger-style recursions, generating sum-over-states formulas for the perturbation series, which unambiguously converges at the finite-temperature 
vibrational full-configuration-interaction limits. 
Two strategies are introduced to reducing these sum-over-states formulas into compact sum-over-modes analytical formulas. 
One is a purely algebraic method that factorizes each many-mode thermal average into a product of one-mode thermal averages, which are then evaluated by the thermal Born--Huang rules. 
Canonical forms of these rules are proposed, dramatically expediting the reduction process. The other is finite-temperature normal-ordered second quantization, which 
is fully developed in this study, including a proof of thermal Wick's theorem and the derivation of a normal-ordered vibrational Hamiltonian at finite temperature. The latter naturally defines 
a finite-temperature extension of size-extensive vibrational self-consistent field theory. These reduced formulas can be represented graphically 
as Feynman diagrams with resolvent lines, which include  anomalous and renormalization diagrams.
Two order-by-order and one general-order algorithms of computing these perturbation corrections are implemented and applied up to the eighth order. The results show no signs of Kohn--Luttinger-type nonconvergence.
\end{abstract}

\maketitle

\section{Introduction}

%\subsection{Anharmonicity} 

Einstein's quantum-mechanical treatment of heat capacity ignited the quantum revolution. 
While the breakdown of the Dulong--Petit law could be explained qualitatively within the harmonic approximation to the potential energy surface (PES) of a crystal, 
quantitative agreement between theory and experiment deteriorates with increasing 
temperature (see, e.g., Fig.\ 6 of Ref.\ \onlinecite{He2012}). 
This is due to anharmonicity in the PES, which tends to lower the calculated vibrational frequencies,  making 
them more accessible to heat and thus elevating the calculated heat capacity to be in better agreement with the observed. 

There are other thermal properties of crystals in which anharmonicity is even more crucial.
Thermal expansion\cite{White1965,Li2015,Salim2016} does not occur in a harmonic crystal because the mean displacement of a harmonic oscillator is zero 
at any quantum number. For the same reason, the volume isotope effect\cite{Salim2016} does not exist in a harmonic crystal, either. 
Since harmonic-oscillator wave functions are stationary states and cannot be scattered by one another, thermal conductivity\cite{Herring1954,Wilson1973,Knauss1974} 
in a harmonic crystal is infinity. Only when anharmonicity is taken into account can the phenomenon known as second sound\cite{Ackerman1968} 
be explained. Furthermore, since a harmonic potential rises forever, a harmonic crystal never melts.\cite{Shukla1971_2}
Needless to say, there are also zero-temperature anharmonic effects in crystals, such as the Fermi and Darling--Dennison resonances
and the breakdown of spectral selection rules (see, e.g., Refs.\ \onlinecite{Sode2013,HirataSode2014,Qin2020}).

%\subsection{Anharmonic theories for crystals} 

Owing to these noticeable thermal anharmonic effects in crystals, mathematical theories\cite{born1988dynamical} for their treatments have been extensively developed  
in solid state physics. Since these theories must be diagrammatically size-consistent,\cite{Hirata2011,HirataARPC2012} they tend to be 
either a mean-field theory or perturbation theory that treats anharmonicity as perturbation.\cite{Hirata_conjecture2014}

The most widely used mean-field theory is the so-called self-consistent-phonon (SCP) method,\cite{Hooton1958,Koehler1966,Choquard1967,Gillis1968,Hermes2013_Dyson} 
which is an effective harmonic approximation whose frequencies include first-order (and occasionally 
second-order) perturbative anharmonic effects.\cite{Shukla1974} 
Thermodynamic functions are then 
evaluated in the framework of the Bose--Einstein theory\cite{march,Fetter1971,mattuck} using these renormalized harmonic frequencies, which implicitly include the second- and all higher-order perturbation corrections of certain (ring-diagram) types. 

A finite-temperature many-body perturbation theory for thermodynamic functions of an anharmonic crystal---crystal phonon perturbation theory---was developed by Maradudin and coworkers,\cite{Maradudin1961,Maradudin1961_2,Flinn1963}
by Cowley,\cite{Cowley1963,Cowley1968} and by Shukla and coworkers\cite{Shukla1971_2,Shukla1974,Shukla1985,Shukla1985_2} (see also Kobashi {\it et al.}\cite{Kobashi1978_1,Kobashi1978_2}).
It is based on the time-dependent diagrammatic perturbation theory\cite{march,Fetter1971,mattuck} as extended to thermodynamics, exploiting the isomorphism of the time-dependent Schr\"{o}dinger and $\beta$-dependent Bloch equations.\cite{bloch,Kohn1960,Luttinger1960,balian,blochbook} 

A derivation of perturbation correction formulas to the free energy proceeds as follows: First, draw all closed linked diagrams of a desired perturbation order. Second, in each diagram, associate a vertex with a force constant, and assign an edge with a zeroth-order finite-temperature Green's function.\cite{Matsubara1955} Third, multiply together 
these algebraic interpretations of the building blocks of each diagram. 
Fourth, integrate and sum this product over all variables and indexes of the Green's functions and force constants to arrive at the diagram's 
algebraic evaluation.  
In the last step, care must be exercised in considering all possible cases of index coincidences, 
as they lead to anomalous-\cite{Kohn1960,Hirata2021} and/or renormalization-diagram\cite{Hirata2021} contributions.  
Shukla and coworkers\cite{Shukla1971_2,Shukla1974,Shukla1985,Shukla1985_2} derived a fourth-order perturbation formula of the free energy using 
this method, correctly accounting for the renormalization diagrams. 

In these perturbation theories, the Hamiltonian is inevitably expressed with force constants in the normal coordinates, and   
the reference wave function takes a harmonic form. Unlike Rayleigh--Schr\"odinger perturbation theory (RSPT), however, 
the perturbation operator is defined according to the Van Hove ordering scheme,\cite{vanhove} 
in which anharmonic force constants of different ranks are assigned different orders. 
Therefore, as the perturbation order is raised, higher-ranked force constants
are introduced to the Hamiltonian, obscuring the meaning of the perturbation order. 
%For example, Shukla and coworkers' fourth-order perturbation theory includes a ring diagram with only one sextet-force-constant vertex,\cite{Shukla1971_2} which would be a first-order contribution in RSPT. 

%\subsection{Anharmonic theories for molecules} 

In molecular applications also, quantitative computational treatments of anharmonicity\cite{Bloino2012} began attracting growing interest 
owing to the underlying electronic structure calculations for PES gaining predictively accuracy.\cite{HEAT2004,HirataYagiCPL2008,Helgaker2008}
Here, the primary goal is the predictions of anharmonic vibrational spectra, including Fermi resonances,\cite{RodriguezGarcia2007}
for which zero-temperature formalisms usually suffice. 

For this purpose, second-order anharmonic vibrational perturbation theory (VPT2) at zero temperature has been developed and refined by many groups.\cite{Amos1991,Martin1995,Kuhler1996,Stanton1998,Ruud2000,Barone2004,Barone2005,Vazquez2005,Vazquez2006,Barone2010,Bloino2012}
It is based on a harmonic reference and formulated with force constants in the normal coordinates. Naturally, its formula for the anharmonic correction to 
the zero-point energy mirrors the second-order correction to the free energy of the crystal phonon perturbation theory in the zero-temperature limit,\cite{Maradudin1961,Maradudin1961_2,Flinn1963,Cowley1963,Cowley1968,Shukla1971_2,Shukla1974,Shukla1985,Shukla1985_2} especially when the Van Hove 
ordering scheme is used in both cases. However, when the theory is applied to finite-temperature problems (thermodynamics and kinetics),\cite{Bloino2012} 
an  {\it ad hoc} adjustment\cite{Truhlar1991,Kuhler1996} of the vibrational partition function by zero-temperature VPT2 results has been proposed and widely used.\cite{Bloino2012}
There are also semiclassical approximations to the anharmonic vibrational free energy (see, e.g., Makri and Miller\cite{Makri2002}). 

For molecules, however, a whole new class of anharmonic vibrational theories has also been developed, 
which may be called modal theory\cite{Bowman1986}
in analogy to molecular orbital theory for electronic structures. The PES used in this class of theories can be expressed by force constants or numerically on a grid. 

Its mean-field approximation is known as vibrational self-consistent-field (VSCF) theory,\cite{Bowman1978,Ratner1986,Bowman1986} which expresses the zero-point vibrational wave function by a single Hartree product of one-mode functions (modals) along the normal coordinates, whose shapes are varied to minimize its energy expectation value. This ansatz leads to an effective one-mode eigenvalue equation with a one-dimensional anharmonic potential that modals must satisfy. 
Typically, the one-mode equations are solved by expanding the modals by a finite number of basis functions not only for the zero-point state
but also for excited states. These state energies are related to
the corresponding eigenvalues, and do not bear a simple relationship with 
one another,  making it difficult to incorporate thermal effects in the theory. This is in contrast with the evenly spaced eigenvalues of a harmonic oscillator, which can be summed over
analytically up to an infinite quantum number in the Bose--Einstein theory and in the perturbation theories based on it.

Starting from a VSCF reference wave function, vibrational configuration-interaction (VCI),\cite{Christoffel1982} vibrational M{\o}ller--Plesset perturbation (VMP),\cite{Norris1996} and vibrational coupled-cluster (VCC)\cite{Christiansen2004} theories were developed to capture the residual one-mode anharmonic effects and anharmonic mode-mode coupling. Their formulations largely follow the corresponding ones for electronic structures\cite{szabo,Shavitt2009} by replacing
molecular orbitals with modals and the electronic Hamiltonian with the vibrational counterpart. 
Each of these hierarchical theories should converge, as a single unambiguous series (especially if the Van Hove ordering scheme is not used) at the exact solution of the Schr\"{o}dinger equation
within a basis set with increasing rank of the hierarchical theory. 

To calculate the partition function and account for the thermal effects, however,
one would have to sum over all states by brute force, which would furthermore be truncated by the finite basis set; unlike the Bose--Einstein theory or the
perturbation theories thereof, a dramatic reduction of these expensive sum-over-states formulas into sum-over-modes expressions 
is unlikely to occur. Consequently, these modal theories are formulated only at 
zero temperature with notable exceptions of finite-temperature VSCF methods,\cite{Njegic2006,Hansen2008,Roy2009} employing
additional {\it ad hoc} approximations, and finite-temperature vibrational full configuration interaction (FCI) as a benchmark method.\cite{Xiuyi2021}
This is despite the fact that molecular vibrational excited states are thermally populated at room temperature.

Furthermore, it was revealed that VSCF is neither diagrammatically size-consistent\cite{Hirata2010,Keceli2011} nor invariant to a unitary transformation
of degenerate modals,\cite{KeceliShio2009} unlike its electronic counterpart, i.e., the Hartree--Fock (HF) theory.\cite{szabo}
The lack of size-consistency does not mean that VSCF is rendered useless for crystals; rather 
it means that VSCF energies and one-mode potentials are dominated by non-closable diagrammatic contributions that vanish in the thermodynamic limit.
As a result, the one-mode potentials are shown, both analytically\cite{Hirata2010} and numerically,\cite{Keceli2011} to become effectively harmonic in this limit, in accordance with Makri's theorem.\cite{Makri1999} These ``harmonic'' potentials are dressed with the effects of even-order force constants of certain (ring-diagram) types (see Fig.\ 1 of Ref.\ \onlinecite{Keceli2011}), whereas the effects 
of cubic and all higher-odd-order force constants are washed out. This implies that VMP and VCC based on a VSCF reference 
also lack diagrammatic size-consistency in the sense that they are saddled with a vast number of vanishing terms for crystals and are exceedingly inefficient, although they should still produce meaningful results. VCI, on the other hand, is non-size-consistent in the sense of not satisfying a linked-diagram theorem and will give meaningless (null-anharmonicity) results in the thermodynamic limit.\cite{Hirata2011,HirataARPC2012} 

%\subsection{Unified anharmonic theories for molecules and crystals}

One of the present authors with coauthors\cite{Hirata2010,Keceli2011,Hermes2012} introduced diagrammatically size-consistent analogs of VSCF by retaining only those terms in the VSCF equations that are nonvanishing in the thermodynamic limit. 
The resulting vastly-streamlined diagrammatic theory, named XVSCF,\cite{Keceli2011,Hermes2012} was identified\cite{private} as variants of SCP (for the precise relationship between XVSCF and SCP,
see Table 1 of Ref.\ \onlinecite{Hermes2013_Dyson}).  They are also closely related to the effective-harmonic-oscillator method of Cao and Voth.\cite{Cao1995}
On this basis, diagrammatically size-consistent analogs of VMP2 and VCC were developed as XVMP2
(Ref.\ \onlinecite{Hermes2013}) and XVCC (Refs.\ \onlinecite{Faucheaux2015,Faucheaux2018}), respectively.
The former, whose applicability to an infinitely extended system has been demonstrated recently,\cite{Qin2020} 
is essentially equivalent to VPT2 for zero-point energy when the harmonic approximation is adopted as the reference. It also corresponds to  
crystal phonon perturbation theory in the zero-temperature limit. On the other hand, XVCC has been identified\cite{Faucheaux2015} as a higher-order generalization of the vibrational coupled-cluster theory of Prasad and coworkers,\cite{Prasad1988,Prasad1994,Banik2008} but differs materially from VCC of Christiansen.\cite{Christiansen2004}

All of these diagrammatically size-consistent vibrational theories\cite{Hermes_review2015,HirataARPC2012} can be formulated expediently by the quantum-field-theoretical methods of normal-ordered second quantization and Wick's theorem,\cite{Shavitt2009}
which have been extended to the vibrational Hamiltonian.\cite{Hirata2014}
When passing from the plain second quantization to the normal-ordered one, the constant part of the Hamiltonian rearranges from the PES value at the origin to the XVSCF zero-point energy expression; the gradients and harmonic force constants are replaced by the XVSCF effective gradients and harmonic force constants
dressed with higher-even-order force constants of certain types; cubic- and higher-order force constants are modified similarly. 
This underscores the central place XVSCF occupies in the framework of anharmonic vibrational theories,\cite{Hirata2011} just as the HF theory does 
in the {\it ab initio} electronic structure theory.\cite{Shavitt2009}

The XVSCF, XVMP, and XVCC theories are now formulated in a unified fashion by the normal-ordered
Hamiltonian and, therefore, in terms of the XVSCF energy expression and effective gradients and force constants. Consequently, a XVMP calculation based 
on any reference will largely recover the XVSCF energy at the first order (taking into account the ring diagrams)
without resorting to an artificial, order-by-order adjustment of the perturbation operator such as the Van Hove ordering scheme. 
It may be said that the anharmonic perturbation theories that have evolved separately in solid state physics and quantum chemistry
are converging at the unified series of XVSCF and XVMP.
Nonetheless, these theories are thus far formulated at zero temperature only.

%\subsection{Finite-temperature anharmonic theories for molecules and crystals} 

In this article, we establish the whole converging series of the finite-temperature many-body perturbation theory for anharmonic vibrations,
which is diagrammatically size-consistent and thus applicable to molecules and crystals on an equal footing. It is a general-order extension of crystal phonon perturbation theory\cite{Maradudin1961,Maradudin1961_2,Flinn1963,Cowley1963,Cowley1968,Shukla1971_2,Shukla1974,Shukla1985,Shukla1985_2}
or a finite-temperature generalization of the XVMP theory.\cite{Hermes2013}
Capitalizing on the recent 
development of the finite-temperature many-body perturbation theory 
for electrons,\cite{Jha2019,Hirata2019,Jha2020,Hirata_2020,Hirata2021} we present the recursive definitions
of thermodynamic functions---free energy, internal energy, and entropy---in the style of the RSPT recursions, which make no assumption or restriction on  
the Hamiltonian or its partitioning, except that its PES is expressed with force constants in 
the normal coordinates and the partitioning does not vary with the perturbation order. The recursions immediately give the sum-over-states
analytical formulas of the perturbation corrections at any arbitrary order, also indicating the presence of anomalous- and renormalization-diagram contributions.\cite{Hirata2021}

The sum-over-states analytical formulas are then reduced to sum-over-modes analytical formulas, which will be useful for actual applications to large molecules and crystals.
These sum-over-states formulas are written in terms of the perturbation corrections to state energies, which are in turn given by 
the degenerate RSPT (Ref.\ \onlinecite{Hirschfelder1974}) as eigenvalues of an effective perturbation operator matrix in each subspace spanned by degenerate reference states. 
Using the same mathematical trick enabling the reduction in electronic finite-temperature many-body perturbation theory,\cite{Hirata2021} i.e., the trace invariance of a cyclic matrix product, we show that the sum-over-states formulas can be written in a closed form involving the Hamiltonian matrix elements in the basis of the Hartree products. These matrix elements 
are in turn evaluated by the Born--Huang rules\cite{born1988dynamical,Keceli2011} elevated to finite temperature, into sums of products of force constants that are 
further simplified into the sum-over-modes analytical formulas. 
We introduce canonical forms of these thermal Born--Huang rules, which facilitate the simplification process. Remarkably, these reduced analytical formulas
can take into account anharmonic effects on infinitely many states, and are not limited by a finite basis set. 

While this algebraic reduction is straightforward and rigorous, % and is furthermore accelerated by the canonical form of the Born--Huang rules, 
it is extremely tedious. We, therefore, present
an alternative method of reduction, which is based on the quantum-field-theoretical tool of normal-ordered second quantization
and Wick's theorem for vibrations at finite temperatures.\cite{march,Nooijen2021} We stipulate normal ordering for an arbitrary number of operators and prove thermal Wick's theorem.
We introduce the finite-temperature normal-ordered form of the vibrational
Hamiltonian, whose constant, gradients, and force constants prove to be the XVSCF counterparts, 
defining a finite-temperature extension of the XVSCF method (see Ref.\ \onlinecite{Hirata2021} for the same mechanics for electronic structures).
This method dramatically expedites the reduction of the sum-over-states formulas into the 
sum-over-modes ones, without compromising the mathematical rigor or straightforward logic of the purely algebraic reduction. 

The application of thermal Wick's theorem consists in repeatedly connecting (contracting)  
a creation and an annihilation  operator belonging to two different normal-ordered products until no  
 operator is left uncontracted.\cite{Shavitt2009} This instantly suggests a graphical representation of the sum of equal-valued full contraction patterns, in which 
a normal-ordered operator is a vertex and a contraction is an edge,
introducing time-independent Feynman diagrammatics.
We document the rules for diagrammatic enumeration and interpretation, which are the most expedient. 
Furthermore, a linked-diagram theorem is proved for finite-temperature anharmonic vibrations, guaranteeing 
size-consistency of the theory at any perturbation order.

The time-independent diagrams introduced here are distinct from the time-dependent ones\cite{mattuck} employed in crystal phonon perturbation theory in the sense that the 
former carry resolvent lines representing denominator factors. They explicitly include anomalous and renormalization diagrams, which lend themselves to instant algebraic interpretations. 
The time-independent diagrammatic logic is nothing but convenient mnemonics for normal-ordered second quantization  and is based on the recursions in contrast to  the time-dependent counterpart 
outlined above, where a user is responsible for taking into account numerous special circumstances that give rise to these counterintuitive diagrams.

We also report three methods of numerically evaluating the perturbation corrections 
to the free energy, internal energy, and entropy as a function of temperature.
The first method implements a general-order algorithm that evaluates the sum-over-states analytical formulas generated  by the recursions, running 
a modified finite-temperature vibrational FCI program.\cite{Xiuyi2021} 
The second uses the $\lambda$-variation method,\cite{Hirata2017,Jha2019} in which some lowest-order perturbation corrections are obtained as finite-difference $\lambda$-derivatives 
of the finite-temperature vibrational FCI results\cite{Xiuyi2021} with a perturbation-scaled Hamiltonian $\hat{H} = \hat{H}^{(0)} + \lambda \hat{V}^{(1)}$. 
The third is based on the sum-over-modes analytical formulas for several lowest perturbation orders. 
They are shown to agree with one another numerically 
exactly insofar as the finite-basis-set effects and numerical errors due to finite-difference approximations are negligible, mutually confirming the correctness
of the formalisms and computer programs. 

The general-order algorithm, furthermore, demonstrates the convergence of the finite-temperature perturbation theory at the 
exact (finite-temperature vibrational FCI) limit as the perturbation order is raised, numerically proving the correctness of the theory in the sense that no diagram is left out.  
In the zero-temperature limit, the finite-temperature theory recovers the zero-temperature counterpart both numerically and analytically insofar as there is no zero-frequency mode. 
For vibrations, the Kohn--Luttinger-type nonconvergence problem\cite{Kohn1960,Hirata_KL2021,Hirata_KL2022} does not exist.

Below, we give the most comprehensive and complete exposition of this unified many-body perturbation theory for anharmonic
vibrations, which is valid for gases of molecules and crystals at zero and nonzero temperatures. 

% =====================================
% Thermodynamics
% =====================================
\section{Thermodynamics of anharmonic vibrations}

Here, we start with thermodynamic functions for an ideal gas of identical, non-rotating, anharmonic molecules, which are defined 
at an exact or mean-field-theoretical level. Electronic and rotational degrees of freedom
are suppressed and the nonadiabatic and vibration-rotation couplings are not considered. It is applicable to crystals.

The exact partition function for vibrations per molecule is given by\cite{Xiuyi2021}
\begin{eqnarray}
\label{eq:Xi}
    \Xi = \sum_{N=0}^\infty \exp(-\beta E_N),
\end{eqnarray}
where $\beta=(k_{\mathrm{B}} T)^{-1}$, $E_N$ is the exact energy of the $N$th vibrational state ($N=0$ being the zero-point state), and the summation runs over all infinitely many states. 
The chemical potential is zero for vibrations,\cite{mcquarrie1975} and, therefore, the grand canonical and canonical ensembles coincide with each other. In this article, 
we elect to use the terminology of the grand canonical ensemble, and call the free energy the grand potential, which is synonymous with the Helmholtz energy adopted by other authors.\cite{Shukla1971_2,Shukla1974,Shukla1985,Shukla1985_2}

The exact grand potential $\Omega$,  internal energy $U$, and  entropy $S$ are derived from $\Xi$ as
\begin{eqnarray}
\label{eq:Omg_exact}
\Omega &=& - \frac{1}{\beta}\ln \Xi,\\
\label{eq:U_exact}
U &=& - \frac{\partial }{\partial \beta} \ln\Xi = \Omega + \beta \frac{\partial\Omega }{\partial \beta},\\
\label{eq:S_exact}
  S &=& \frac{U-\Omega}{T} = \frac{\beta}{T} \frac{\partial \Omega}{\partial \beta}.
\end{eqnarray}
We will hereafter focus on $\Omega$ and $U$ because $S$ can be readily inferred from them.

The energy in Eq.\ (\ref{eq:Xi}) is the eigenvalue solution of the vibrational Schr\"{o}dinger equation,
\begin{eqnarray}
\label{eq:HPsi}
    \hat{H}  \Psi_{N}  = E_N  \Psi_{N},
\end{eqnarray}
in which the pure vibrational Hamiltonian $\hat{H}$ is written as
\begin{eqnarray}
\label{eq:full_vibhamiltonian}
\hat{H} &=& -\frac{1}{2} \sum_{i}^m \frac{\partial^{2}}{\partial Q_{i}^{2}} + V_{\mathrm{ref}}+\sum_{i}^m F_{i} Q_{i}+\frac{1}{2!} \sum_{i, j}^m F_{i j} Q_{i} Q_{j}  \nonumber\\
& &+ \frac{1}{3!} \sum_{i, j, k}^m F_{i j k} Q_{i} Q_{j} Q_{k} + \frac{1}{4!} \sum_{i, j, k, l}^m F_{i j k l} Q_{i} Q_{j} Q_{k} Q_{l} + \ldots, \nonumber \\
\end{eqnarray}
where $Q_i$ is the $i$th normal coordinate, $V_{\mathrm{ref}}$ is the value of the PES at the reference geometry, which 
is typically but not limited to the equilibrium geometry, and $F_i$, $F_{ij}$, $F_{ijk}$, and $F_{ijkl}$ are a first-, second-, third-, and fourth-order force constant, respectively, 
which have the  permutation symmetry, e.g., $F_{ijk} = F_{ikj} = F_{jik} = F_{jki} = F_{kij}= F_{kji}$. 
The summations run over all $m$ vibrational degrees of freedom or all $m$ modes. 

In the formulations and numerical tests discussed below, a quartic force field (QFF)\cite{YagiQFF2004} will be employed, which truncates 
the above Hamiltonian after the fourth-order force constants, but they can be straightforwardly extended to higher-order force fields, although 
a cubic force field  cannot be used because it is ill-conditioned (see page 508 of Ref.\ \onlinecite{Ashcroft}). The coordinates can also be 
generalized to any rectilinear, orthogonal, mass-weighted coordinates without altering the following formulations, 
but we call them normal coordinates in this article for the sake of convenience. 

The zeroth-order Hamiltonian $\hat{H}^{(0)}$ for a mean-field theory takes a harmonic form,
\begin{eqnarray}
\label{eq:H0}
\hat{H}^{(0)} =- \frac{1}{2} \sum_i^m \frac{\partial^{2}}{\partial Q_{i}^{2}} + V_{\mathrm{ref}}+ \frac{1}{2} \sum_i^m \omega_i^2 Q_{i}^2,
\end{eqnarray}
where $\omega_i$ is the harmonic frequency corresponding to the $i$th normal mode. Superscript `(0)' indicates that this 
mean-field theory will furnish the reference (zeroth-order) 
wave functions and energies for the perturbation theory. 
The frequency $\omega_i$ can be the frequency ($\omega_i = F_{ii}^{1/2}$) of the harmonic approximation, but
it can also be a (finite-temperature) XVSCF frequency, in which case the M{\o}ller--Plesset partitioning of the Hamiltonian is adopted. 

In either case, the reference vibrational Schr\"odinger equation has analytical solutions for all states, i.e.,
\begin{eqnarray}
\label{eq:H0Phi0}
    \hat{H}^{(0)}  \Phi_{N}^{(0)}  = E_N^{(0)}  \Phi_{N}^{(0)} ,
\end{eqnarray}
where $\Phi_{N}^{(0)}$ is the wave function for the $N$th state, which takes a Hartree-product form,
\begin{eqnarray}
\label{eq:Phi_Nwavefunc}%9
\Phi_{N}^{(0)} = | n_1 n_2 \dots n_m \rangle = \prod_i^m | n_i\rangle,
\end{eqnarray}
where $| n_i \rangle$ is a one-dimensional harmonic-oscillator wave function along the $i$th normal coordinate with quantum number $n_i$. 
The corresponding energy $E_N^{(0)}$ is given by
\begin{eqnarray}
\label{eq:E_I0} 
E_{N}^{(0)} = V_{\text{ref}} +  \sum_{i}^m \omega_{i} \left(n_{i}+1 / 2\right) .
\end{eqnarray}

Owing to the additive nature of the energy, the infinite summation in the zeroth-order partition function [Eq.\ (\ref{eq:Xi})] can be carried out
analytically, leading to the following expressions for the thermodynamic functions:\cite{Fetter1971,march,mattuck,Xiuyi2021} 
\begin{eqnarray}
\label{eq:Omg_0}
\Omega^{(0)} &=& V_{\mathrm{ref}}-\sum_{i}^m\frac{\omega_i}{2}-\frac{1}{\beta} \sum_{i}^m \ln f_{i} ,  \\
\label{eq:U_0}
U^{(0)} &=&V_{\mathrm{ref}}+ \sum_{i}^m \omega_{i} (f_{i}+1/2)  ,\\
\label{eq:S_0}
S^{(0)} &=&k_{\mathrm{B}} \sum_{i}^m\left\{-f_{i} \ln f_{i}+\left(f_{i}+1\right) \ln \left(f_{i}+1\right)\right\},
\end{eqnarray}
where $f_i$ is the Bose--Einstein distribution function given by
\begin{eqnarray}
\label{eq:BEfunction}
f_i = \frac{1}{\exp(\beta \omega_i) - 1}.
\end{eqnarray}
This Bose--Einstein theory\cite{Fetter1971,march,mattuck} is an infinite-basis or basis-set-free formalism, whose
$T \to \infty$ limit of $S^{(0)}$ is infinity. This is in contrast with any finite-basis-set theory such as VSCF, VCI (including vibrational FCI), VMP, and VCC, which, if extended to a finite temperature, would have its entropy display a spurious plateau in the high-temperature limit.\cite{Xiuyi2021} 

In the subsequent sections, we drop the upper limit ($m$) of the summation index range over modes to reduce clutter.

% =====================================
% Recursion
% =====================================

\section{Rayleigh--Schr\"{o}dinger recursions \label{section:recursion}}

We consider a Rayleigh--Schr\"{o}dinger perturbation expansion of the thermodynamic functions, starting with the Bose--Einstein theory as the reference (zeroth order)
and thus treating anharmonicity of the PES as a perturbation. 
%Being diagrammatically size-consistent (see Sec.\ \ref{section:linkedtheorem}), the formalisms are directly applicable to crystals, although analytically eliminating terms vanishing by translational symmetry may be considered for efficiency.\cite{Hirata2010} 
Quantity $X$ (in this study, $X = \Omega$, $U$, $S$, $E_N$, etc.)\ is expanded\cite{Hirata2017,Hirata2021} in a power series as
\begin{eqnarray}
\label{eq:X_perturbed2}
X(\lambda) &=& X^{(0)} + \lambda X^{(1)} + \lambda^2 X^{(2)}+ \lambda^3 X^{(3)} + \ldots
\end{eqnarray}
or
\begin{eqnarray}
\label{eq:X_perturbed}
X^{(n)}=\left.\frac{1}{n !} \frac{\partial^{n} X(\lambda)}{\partial \lambda^{n}}\right|_{\lambda=0},
\end{eqnarray}
where $X(\lambda)$ is the exact value of $X$, i.e., calculated by the finite-temperature vibrational FCI (Ref.\ \onlinecite{Xiuyi2021}) in the infinite-basis-set limit
for a 
perturbation-scaled Hamiltonian, $\hat{H}=\hat{H}^{(0)} + \lambda \hat{V}^{(1)}$.
The parenthesized superscript indicates the perturbation order. 
%The Hamiltonian $\hat{H}$ is given by Eq.\ (\ref{eq:full_vibhamiltonian}), whereas  the zeroth-order Hamiltonian $\hat{H}^{(0)}$ by Eq.\ (\ref{eq:H0}), corresponding to the Bose--Einstein reference (which is not necessarily synonymous with the harmonic approximation). 

A recursion for $\Omega^{(n)}$ can be derived by differentiating the exact definition of $\Omega$ [Eq.\ (\ref{eq:Omg_exact})] with respect to $\lambda$.
After some nontrivial algebraic manipulations,\cite{Hirata2021} it becomes
\begin{eqnarray}
\label{eq:recursionOmg}
\Omega^{(n)}&=&\left[ E_{N}^{(n)}\right]+\frac{(-\beta)}{2 !} \sum_{i=1}^{n-1}\left(\left[ E_{N}^{(i)} E_{N}^{(n-i)}\right]-\Omega^{(i)} \Omega^{(n-i)}\right)\nonumber \\
& &+\frac{(-\beta)^{2}}{3 !} \sum_{i=1}^{n-2} \sum_{j=1}^{n-i-1}\left(\left[ E_{N}^{(i)} E_{N}^{(j)} E_{N}^{(n-i-j)}\right]-\Omega^{(i)} \Omega^{(j)} \Omega^{(n-i-j)}\right)\nonumber \\
& &+\cdots+\frac{(-\beta)^{n-1}}{n !}\left\{\left[(E_{N}^{(1)})^{n}\right]-(\Omega^{(1)})^{n}\right\},
\end{eqnarray}
where the square bracket denotes a zeroth-order thermal average of quantity $X_N$, i.e.,
\begin{eqnarray}
\label{eq:thermalavg_XN}
    \Big[ X_{N} \Big] \equiv \frac{\sum\limits_{N=0}^\infty X_N \exp(-\beta E_N^{(0)}) }{\sum\limits_{N=0}^\infty \exp(-\beta E_N^{(0)})}.
\end{eqnarray}
The derivation of this recursion is identical to the electronic case\cite{Hirata2021} except that the chemical potential and its perturbation corrections are all set to zero.
The second term and onwards are multiplied by a power of $\beta$ and are responsible for the anomalous-diagram contributions\cite{Kohn1960,Hirata2021} (see below). 
Hereafter, a ``thermal average'' refers to the zeroth-order thermal average defined by Eq.\ (\ref{eq:thermalavg_XN}).

Differentiating Eq.\ (\ref{eq:U_exact}) with respect to $\lambda$, we obtain the recursion for $U^{(n)}$ as
\begin{eqnarray}
\label{eq:recursion_U}
U^{(n)} &=& \left[ {E_N^{(n)}} \right] + (-\beta)\sum_{i=1}^{n}\left( \left[ E_N^{(i)} E_N^{(n-i)}\right] - \Omega^{(i)} U^{(n-i)}\right) \nonumber \\
&&+ \frac{(-\beta)^2}{2!} \sum_{i=1}^{n-1}  \sum_{j=1}^{n-i} \left( \left [ E_N^{(i)}  E_N^{(j)} E_N^{(n - i - j)} \right] -\Omega^{(i)}\Omega^{(j)} U^{(n-i-j)} \right) \nonumber\\ 
&& +\ldots +\frac{(-\beta)^n}{n!} \left\{  \left[ (E_N^{(1)})^n E_N^{(0)} \right] - (\Omega^{(1)})^n U^{(0)} \right\} ,
\end{eqnarray}
which can also be reached by setting all chemical potentials to zero in the electronic counterpart, i.e., Eq.\ (38) of Ref.\ \onlinecite{Hirata2021}.

The recursion for $S^{(n)}$ is obtained as a concatenation of the recursions for $\Omega^{(n)}$ and $U^{(n)}$ from the relation,
\begin{eqnarray}
    S^{(n)} = \frac{U^{(n)} - \Omega^{(n)}}{T}.
\end{eqnarray}

The perturbation corrections to the state energies $\{E_N^{(n)}\}$ are given by degenerate RSPT (Ref.\ \onlinecite{Hirschfelder1974}) because some zeroth-order 
states are almost always degenerate. A degenerate RSPT valid for any Hamiltonian was developed by Hirschfelder and Certain\cite{Hirschfelder1974} and serves as 
the foundation of the present theory.

According to this degenerate RSPT,\cite{Hirschfelder1974,Hirata2021} the $n$th-order corrections $\{E_N^{(n)}\}$  are the eigenvalues of an effective perturbation operator matrix $\bm{E}^{(n)}$.
This matrix is block-diagonal, and each of its nondiagonal blocks is spanned by zeroth-order degenerate states. The $\gamma$th block 
 is defined recursively by\cite{Hirschfelder1974,Hirata2021}
\begin{eqnarray}
\label{eq:E_In_degen}
    \left(\bm{E}_\gamma^{(n)}\right)_{IJ} = \langle \Phi_J^{(0)} | \hat{V}^{(1)} | \Phi_I^{(n-1)}\rangle
\end{eqnarray}
with
\begin{eqnarray}
\label{eq:Phi_In_degen}
\Phi_{I}^{(n)}=\hat{R}\left\{\hat{V}^{(1)} \Phi_{I}^{(n-1)}-\sum_{i=1}^{n-1} \sum_{J \in \gamma} \left( \bm{E}_\gamma^{(i)} \right)_{IJ} \Phi_{J}^{(n-i)}\right\},
\end{eqnarray}
for each $I$, where $I$ and $J$ run over all Hartree-product states in the $\gamma$th block, which share the same zeroth-order energy of $E_\gamma^{(0)}$. The resolvent operator $\hat{R}$ (Ref.\ \onlinecite{Shavitt2009}), when acting within the $\gamma$th degenerate subspace, is defined as
\begin{eqnarray}
\label{eq:R_Nresolvent}
\hat{R} = \sum_{A\notin\gamma} \frac{| \Phi_{A}^{(0)} \rangle\langle \Phi_{A}^{(0)} |}{E_{\gamma}^{(0)}-E_{A}^{(0)}} 
= \sum_{A}^{\text {denom.} \neq 0} \frac{| \Phi_{A}^{(0)} \rangle\langle \Phi_{A}^{(0)} |}{E_{\gamma}^{(0)}-E_{A}^{(0)}} ,
\end{eqnarray}
where $A$ runs over all zeroth-order states that are {\it not} degenerate with the $I$th or $J$th state and are thus outside the $\gamma$th block.
This summation restriction is equivalent to the condition ``denom.$\neq$0,'' which demands that only those summands with nonzero denominators be summed. 
When the rank of the block is one, the above recursion reduces to the familiar, nondegenerate RSPT.\cite{Shavitt2009}

While the elements of $\bm{E}^{(n)}$ are expressed in a closed form according to Eqs.\ (\ref{eq:E_In_degen})--(\ref{eq:R_Nresolvent}),
its eigenvalues in general cannot be. This may give a false impression that the sum-over-states formulas for $\Omega^{(n)}$ or $U^{(n)}$ obtained from Eq.\ (\ref{eq:recursionOmg}) or (\ref{eq:recursion_U})
cannot be simplified any further and hence the finite-temperature perturbation theory has a limited practical utility. This is not the case. 

Since these eigenvalues  are thermal-averaged with an equal weight [see Eq.\ (\ref{eq:thermalavg_XN})], all we need is the sum of the eigenvalues 
over zeroth-order degenerate states rather than their individual values. This sum can furthermore be expressed in a closed form by virtue of the trace invariance
of a matrix or of a cyclic matrix product,\cite{Hirata2021} i.e.,
\begin{eqnarray}
\label{eq:traceinvariance}
\left[ E_N^{(n)} \right] &=& \left[ \mathrm{Tr} \left( \bm{E}^{(n)} \right) \right] , \\
\label{eq:traceinvariance2}
\left[ E_N^{(i)}E_N^{(n-i)} \right] &=& \left[ \mathrm{Tr}\left( \bm{E}^{(i)}\bm{E}^{(n-i)} \right) \right] , \\
\label{eq:traceinvariance3}
\left[ E_N^{(i)}E_N^{(j)}E_N^{(n-i-j)} \right] &=& \left[ \mathrm{Tr}\left( \bm{E}^{(i)}\bm{E}^{(j)}\bm{E}^{(n-i-j)} \right) \right] , 
\end{eqnarray}
and so on, where the left-hand sides are the thermal averages of eigenvalues or their products, while the right-hand sides are
the thermal averages of matrix traces, defined by
\begin{eqnarray}
\label{eq:thermalavg_XN2}
\Big[ \mathrm{Tr}\left(\bm{X} \right) \Big] \equiv \frac{\sum\limits_{\gamma}\mathrm{Tr}\left( \bm{X}_\gamma \right) \exp(-\beta E_\gamma^{(0)}) }{\sum\limits_{N=0}^\infty \exp(-\beta E_N^{(0)})}.
\end{eqnarray}
Here, $\gamma$ runs over all blocks of the block-diagonal matrix $\bm{X}$ with the $\gamma$th block ($\bm{X}_\gamma$) being
spanned by degenerate states with a common zeroth-order energy of $E_\gamma^{(0)}$. 
The matrices in the right-hand sides of Eqs.\ (\ref{eq:traceinvariance})--(\ref{eq:traceinvariance3}) and, therefore, their traces are expressible in closed forms (unlike their eigenvalues), lending them to dramatic reductions into sum-over-modes analytical formulas 
as demonstrated in Secs.\ \ref{section:AGderivation} and \ref{section:2ndquan}. 
Note that the off-diagonal elements of $\bm{E}^{(n)}$ 
enter the final results through  matrix products such as in Eqs.\ (\ref{eq:traceinvariance2}) and (\ref{eq:traceinvariance3}), ultimately giving rise to the renormalization-diagram contributions\cite{Hirata2021} (see Sec.\ \ref{section:thirdSQ}). 

% =====================================
% Algebraic reduction
% =====================================
\section{Algebraic reduction based on thermal Born--Huang rules and canonical forms\label{section:AGderivation}}

In this section, we reduce the sum-over-states analytical formulas for $\Omega^{(n)}$ and $U^{(n)}$ into sum-over-modes analytical formulas
algebraically, i.e., without resorting to a time-dependent diagrammatic logic involving integration of zeroth-order $\beta$-dependent Green's functions. 
While the reduction is extremely tedious, it uses only the most basic algebraic manipulations, placing the theory  on a firm mathematical footing.

We take the first- and second-order corrections in a QFF as examples. The reduction strategy is general and applicable to any perturbation order or force-constant rank, in principle, although going beyond
the second order would be impractical. 
It relies on a finite-temperature extension of the Born--Huang rules,\cite{born1988dynamical,Keceli2011} which is fully developed 
in this study and whose derivation is given in Appendix  \ref{Appendix:BHrules}. In Appendix \ref{appendix:BHfactorization}, we introduce canonical forms of these thermal Born--Huang rules, 
which facilitate the consolidation of numerous terms with complex summation index restrictions into the most streamlined final expressions without such restrictions. In what follows, we illustrate the process of this reduction for several representative terms of $\Omega^{(1)}$, $U^{(1)}$, and $\Omega^{(2)}$, relegating 
lengthier reduction processes of some terms in Appendixes \ref{appendix:algebraicreduction} and \ref{appendix:algebraicreduction2}.

\subsection{Perturbation operator and scaled force constants}

Let us write the perturbation operator $\hat{V}^{(1)} =  \hat{H}-\hat{H}^{(0)}$ 
in the $n$-mode representation\cite{YagiNMR2007} for future convenience. For a QFF, it reads
\begin{eqnarray}
\label{eq:sorted_V}
\hat{V}^{(1)} &=& \hat{V}_1 + \hat{V}_2 + \hat{V}_3 + \hat{V}_4
\end{eqnarray}
with
\begin{eqnarray}
\hat{V}_1 &=& \sum_{i}F_i Q_i, \label{eq:sorted_V1} \\
\hat{V}_2 &=& \sum_i \frac{1}{2} \bar{F}_{ii} Q_i^2 +\sum_{i,j}^{i \neq j} \frac{1}{2}\bar{F}_{ij} Q_iQ_j, \label{eq:sorted_V2}\\
\hat{V}_3 &=& \sum_i \frac{1}{6} F_{iii} Q_i^3 + \sum_{i,j}^{i \neq j} \frac{1}{2} F_{i j j} Q_{i} Q_{j}^2  +\sum_{i,j,k}^{\substack{i\neq j, i \neq k \\ j \neq k}}\frac{1}{6} F_{i j k} Q_{i} Q_{j} Q_{k}, \nonumber\\
\label{eq:sorted_V3} \\
\hat{V}_4 &=& \sum_i \frac{1}{24} F_{i i i i} Q_{i}^{4} +\sum_{i,j}^{i \neq j}\left(  \frac{1}{8} F_{i i j j} Q_{i}^{2} Q_{j}^{2}  +\frac{1}{6} F_{i jj j} Q_{i} Q_{j}^{3}\right)\nonumber\\
&&+\sum_{i,j,k}^{\substack{i\neq j, i \neq k \\ j \neq k}} \frac{1}{4} F_{i j k k } Q_{i} Q_{j} Q_{k}^{2}+\sum_{i,j,k,l}^{\substack{i\neq j , i \neq k \\ i \neq l , j \neq k \\ j \neq l , k \neq l}} \frac{1}{24} F_{i j k l} Q_{i} Q_{j} Q_{k} Q_{l}, \label{eq:sorted_V4}
\end{eqnarray}
and
\begin{eqnarray}
    \bar{F}_{ij} = F_{ij} -\delta_{ij}\omega_i^2,
\end{eqnarray}
where we explicitly showed the summation restrictions for coincident indexes. In an algebraic reduction process, these restrictions are gradually lifted as
multiple terms are consolidated, which is the most arduous aspect of the whole process.\cite{Hirata_2020} 

We also define, again for future use, scaled force constants as
\begin{eqnarray}
      \tilde{F}_{i} &=& \frac{F_{i}}{(2\omega_i)^{1/2}}, \label{eq:F1tilde} \\ 
      \tilde{F}_{ij} &=& \frac{F_{ij}}{(2\omega_i)^{1/2}(2\omega_j)^{1/2}}, \label{eq:F2tilde}\\ 
      \tilde{F}_{ijk} &=& \frac{F_{ijk}}{(2\omega_i)^{1/2}(2\omega_j)^{1/2}(2\omega_k)^{1/2}}, \label{eq:F3tilde}\\
    \tilde{F}_{ijkl} &=& \frac{F_{ijkl}}{(2\omega_i)^{1/2}(2\omega_j)^{1/2}(2\omega_k)^{1/2}(2\omega_l)^{1/2}} \label{eq:F4tilde}
\end{eqnarray}
with
\begin{eqnarray}
\label{eq:Ftildebar}
    \tilde{\bar{F}}_{ij} = \tilde{F}_{ij} - \delta_{ij}\frac{\omega_i}{2}.
\end{eqnarray}
Neither $\bar{F}_{ij}$ nor $\tilde{\bar{F}}_{ij}$ should be assumed zero. 

\subsection{Inner projection operator\label{section:projector}} 

When evaluating traces of matrix products such as Eqs.\ (\ref{eq:traceinvariance2}) and (\ref{eq:traceinvariance3}), we need to consider off-diagonal
elements of $\bm{E}^{(i)}$ within each degenerate subspace. The closed-form expression of an off-diagonal element varies depending on the precise relationship between the two degenerate Hartree-product basis functions involved. 
Therefore, we classify, for a given Hartree-product state (say the $Z$th state), its degenerate states according to the number of constituent modals that are different
from it. 
We thus rewrite, e.g., Eq.\ (\ref{eq:traceinvariance2}) as
\begin{eqnarray}
\label{eq:traceinvariance2new}
\left[ E_N^{(i)}E_N^{(n-i)} \right] &=& \left[ \mathrm{Tr}\left( \bm{E}^{(i)}\hat{P}_0\bm{E}^{(n-i)} \right) \right]  
+ \left[ \mathrm{Tr}\left( \bm{E}^{(i)}\hat{P}_1\bm{E}^{(n-i)} \right) \right] 
\nonumber\\
&& + \left[ \mathrm{Tr}\left( \bm{E}^{(i)}\hat{P}_2\bm{E}^{(n-i)} \right) \right]  
+ \left[ \mathrm{Tr}\left( \bm{E}^{(i)}\hat{P}_3\bm{E}^{(n-i)} \right) \right] 
\nonumber\\
&& + \left[ \mathrm{Tr}\left( \bm{E}^{(i)}\hat{P}_4\bm{E}^{(n-i)} \right) \right]  
+ \dots,
\end{eqnarray}
where $\hat{P}_n$ is an inner projection operator onto all $n$-modal-difference Hartree products within each degenerate subspace (we use matrices and operators interchangeably). 
When acting on a state in the $\gamma$th degenerate subspace, they are defined as
 \begin{eqnarray}
 \label{eq:P}
 \hat{P}_0 &=&  | \Phi^{(0)}_Z \rangle \langle \Phi_Z^{(0)} |, \\
 \hat{P}_1 &=& \sum_{S \in \gamma} | \Phi^{(0)}_S \rangle \langle \Phi_S^{(0)} | = \sum_{S}^{{\text{denom.}=0 }} | \Phi^{(0)}_S \rangle \langle \Phi_S^{(0)} |, \label{eq:PS}\\
 \hat{P}_2 &=& \sum_{D \in \gamma} | \Phi^{(0)}_D \rangle \langle \Phi_D^{(0)} | = \sum_{D}^{{\text{denom.}=0  }} | \Phi^{(0)}_D \rangle \langle \Phi_D^{(0)} |,\label{eq:PD} \\
 \hat{P}_3 &=& \sum_{T \in \gamma} | \Phi^{(0)}_T \rangle \langle \Phi_T^{(0)} | = \sum_{T}^{{\text{denom.}=0 }} | \Phi^{(0)}_T \rangle \langle \Phi_T^{(0)} |, \label{eq:PT}\\
 \hat{P}_4 &=& \sum_{Q \in \gamma} | \Phi^{(0)}_Q \rangle \langle \Phi_Q^{(0)} | = \sum_{Q}^{{\text{denom.}=0 }} | \Phi^{(0)}_Q \rangle \langle \Phi_Q^{(0)} |,\label{eq:PQ} 
 \end{eqnarray}
and so on, where $\Phi_Z^{(0)} = | n_1 \dots n_i \dots n_m\rangle$ is the $Z$th (reference) Hartree product, $S$, $D$, $T$, and $Q$ stand for, respectively,
one-, two-, three-, and four-modal-difference Hartree products relative to the reference ($Z$th) state. 
The summation index restriction, e.g., $S \in \gamma$, can be replaced by an equivalent one, ``demon.=0,'' meaning that the fictitious denominator $E_{\gamma}^{(0)}-E_{S}^{(0)}$ is zero.
We furthermore stipulate that $\hat{P}_n$ erases all 
off-diagonal matrix elements belonging to two separate degenerate subspaces, i.e.,
 \begin{eqnarray}
 \label{eq:projector}
 \langle \Phi_I^{(0)} | \hat{P}_n | \Phi_J^{(0)} \rangle = 0,
 \end{eqnarray}
when $I $ and $J $ belong to two different (degenerate) blocks. 
The total inner projection operator $\hat{P}$ is given by
 \begin{eqnarray}
 \label{eq:totalprojector}
 \hat{P} = \hat{P}_0 + \hat{P}_1 + \hat{P}_2+ \hat{P}_3 + \hat{P}_4 + \dots.
 \end{eqnarray}

More precisely, $S$ in Eq.\ (\ref{eq:PS}) is specified by its occupation number vector as 
\begin{eqnarray}
\label{eq:S}
\left\{ \Phi_S^{(0)} \right \} &=& \left\{ \Phi_{S(+1)}^{(0)} \right \} \cup \left\{ \Phi_{S(-1)}^{(0)} \right \}
\end{eqnarray}
with
\begin{eqnarray}
\left\{ \Phi_{S(+1)}^{(0)} \right \} &=& \Big\{ |n_1 \dots (n_i+1) \dots n_m \rangle \Big\}, \\
\left\{ \Phi_{S(-1)}^{(0)} \right \} &=& \Big\{ |n_1 \dots (n_i-1) \dots n_m \rangle \Big\}.
\end{eqnarray} 
Since these one-modal-difference states must be degenerate with the reference ($Z$th) state, $\omega_i = 0$. The component inner projectors are given by
\begin{eqnarray}
 \hat{P}_1^{(+1)} &=& \sum_{S(+1)}^{{\text{denom.}=0 }} | \Phi^{(0)}_{S(+1)} \rangle \langle \Phi_{S(+1)}^{(0)} |, \label{eq:PS1}\\
 \hat{P}_1^{(-1)} &=& \sum_{S(-1)}^{{\text{denom.}=0 }} | \Phi^{(0)}_{S(-1)} \rangle \langle \Phi_{S(-1)}^{(0)} |, \label{eq:PSminus1}
 \end{eqnarray}
which satisfy $\hat{P}_1 = \hat{P}_1^{(+1)} +  \hat{P}_1^{(-1)}$.

Likewise, $\{D\}$ in Eq.\ (\ref{eq:PD}) consists of three subsets:
\begin{eqnarray}
\label{eq:D}
\left\{ \Phi_D^{(0)} \right \} &=&  \left\{ \Phi_{D(+2)}^{(0)} \right \} \cup  \left\{ \Phi_{D(\pm0)}^{(0)} \right \} \cup \left\{ \Phi_{D(-2)}^{(0)} \right \}
\end{eqnarray}
with
\begin{eqnarray}
 \left\{ \Phi_{D(+2)}^{(0)} \right \} &=& \Big\{ |n_1 \dots (n_i+2) \dots n_m \rangle  \Big\} \nonumber\\
&& \cup\, \Big\{|n_1 \dots (n_i+1) \dots (n_j +1) \dots n_m \rangle  \Big\}, \label{eq:D2} \\
\left\{ \Phi_{D(\pm0)}^{(0)} \right \}&=&  \Big\{  |n_1 \dots (n_i+1) \dots (n_j -1) \dots n_m \rangle  \Big\}, \label{eq:D0} \\
\left\{ \Phi_{D(-2)}^{(0)} \right \} &=&  \Big\{  |n_1 \dots (n_i-2) \dots n_m \rangle  \Big\} \nonumber\\
&& \cup\, \Big\{  |n_1 \dots (n_i-1) \dots (n_j -1) \dots n_m \rangle  \Big\}. \label{eq:Dminus2} 
\end{eqnarray} 
Each set defined by Eq.\ (\ref{eq:D2}), (\ref{eq:D0}), or (\ref{eq:Dminus2}) implies the condition $-2\omega_i = 0$ or $-\omega_i - \omega_j = 0$, $-\omega_i + \omega_j = 0$, 
 $2 \omega_i = 0$ or $\omega_i + \omega_j = 0$, respectively, because the states in these sets are degenerate with the reference ($Z$th) state. 
We then define the component inner projectors as
\begin{eqnarray}
 \hat{P}_2^{(+2)} &=& \sum_{D(+2)}^{{\text{denom.}=0 }} | \Phi^{(0)}_{D(+2)} \rangle \langle \Phi_{D(+2)}^{(0)} |, \label{eq:PD2}\\
 \hat{P}_2^{(\pm0)} &=& \sum_{D(\pm0)}^{{\text{denom.}=0 }} | \Phi^{(0)}_{D(\pm0)} \rangle \langle \Phi_{D(\pm0)}^{(0)} |, \label{eq:PD0}\\
 \hat{P}_2^{(-2)} &=& \sum_{D(-2)}^{{\text{denom.}=0 }} | \Phi^{(0)}_{D(-2)} \rangle \langle \Phi_{D(-2)}^{(0)} |, \label{eq:PDminus2}
 \end{eqnarray}
where $\hat{P}_2 = \hat{P}_2^{(+2)} +  \hat{P}_2^{(\pm0)} + \hat{P}_2^{(-2)}$.

$T$ in Eq.\ (\ref{eq:PT}) stands for a Hartree product that differs by three in quantum numbers from the reference ($Z$th) state but has
the same zeroth-order energy. Therefore, $\hat{P}_3= \hat{P}_3^{(+3)} + \hat{P}_3^{(+1)} + \hat{P}_3^{(-1)} + \hat{P}_3^{(-3)}$ with
\begin{eqnarray}
 \hat{P}_3^{(+3)} &=& \sum_{T(+3)}^{{\text{denom.}=0 }} | \Phi^{(0)}_{T(+3)} \rangle \langle \Phi_{T(+3)}^{(0)} |, \label{eq:PT3}\\
 \hat{P}_3^{(+1)} &=& \sum_{T(+1)}^{{\text{denom.}=0 }} | \Phi^{(0)}_{T(+1)} \rangle \langle \Phi_{T(+1)}^{(0)} |, \label{eq:PT1}\\
 \hat{P}_3^{(-1)} &=& \sum_{T(-1)}^{{\text{denom.}=0 }} | \Phi^{(0)}_{T(-1)} \rangle \langle \Phi_{T(-1)}^{(0)} |, \label{eq:PTminus1}\\
 \hat{P}_3^{(-3)} &=& \sum_{T(-3)}^{{\text{denom.}=0 }} | \Phi^{(0)}_{T(-3)} \rangle \langle \Phi_{T(-3)}^{(0)} |, \label{eq:PTminus3}
 \end{eqnarray}
 where
\begin{eqnarray}
\label{eq:T}
 \left\{ \Phi_{T(+3)}^{(0)} \right \} &=& \Big\{ |n_1 \dots (n_i+3) \dots n_m \rangle  \Big\} \nonumber\\
&& \cup\, \Big\{|n_1 \dots (n_i+2) \dots (n_j +1) \dots n_m \rangle  \Big\} \nonumber\\
&& \cup\, \Big\{|n_1 \dots (n_i+1) \dots (n_j +1) \dots (n_k + 1) \dots n_m \rangle  \Big\} , \label{eq:T3} \nonumber\\ \\
\left\{ \Phi_{T(+1)}^{(0)} \right \}&=&  \Big\{  |n_1 \dots (n_i+2) \dots (n_j -1) \dots n_m \rangle  \Big\} \nonumber\\ 
&& \cup\, \Big\{|n_1 \dots (n_i+1) \dots (n_j +1) \dots (n_k - 1) \dots n_m \rangle  \Big\} , \label{eq:T1}   \nonumber\\
\end{eqnarray} 
and so on. Each line in the above implies $-3\omega_i = 0$, $-2\omega_i - \omega_j= 0$, $-\omega_i - \omega_j - \omega_k=0$, $-2\omega_i + \omega_j = 0$, and $-\omega_i - \omega_j + \omega_k = 0$, respectively. 

$Q$ in Eq.\ (\ref{eq:PQ}) is defined completely analogously. We have $\hat{P}_4 = \hat{P}_4^{(+4)} + \hat{P}_4^{(+2)} + \hat{P}_4^{(\pm0)} + \hat{P}_4^{(-2)} + \hat{P}_4^{(-4)}$.

 Let us also introduce a breakdown of the resolvent operator $\hat{R}$ into the singles, doubles, etc.\ components, i.e., $\hat{R} = \hat{R}_1 + \hat{R}_2 + \dots$.
 When acting on a state in the $\gamma$th degenerate subspace, they are given by
\begin{eqnarray}
\label{eq:R}
\hat{R}_1 &=& \sum_{S\notin\gamma}^{} \frac{| \Phi_{S}^{(0)} \rangle\langle \Phi_{S}^{(0)} |}{E_{\gamma}^{(0)}-E_{S}^{(0)}} 
= \sum_{S}^{{\text {denom.} \neq 0 }} \frac{| \Phi_{S}^{(0)} \rangle\langle \Phi_{S}^{(0)} |}{E_{\gamma}^{(0)}-E_{S}^{(0)}} ,\\
\label{eq:R2}
\hat{R}_2 &=& \sum_{D\notin\gamma}^{} \frac{| \Phi_{D}^{(0)} \rangle\langle \Phi_{D}^{(0)} |}{E_{\gamma}^{(0)}-E_{D}^{(0)}} 
= \sum_{D}^{{\text {denom.} \neq 0 }} \frac{| \Phi_{D}^{(0)} \rangle\langle \Phi_{D}^{(0)} |}{E_{\gamma}^{(0)}-E_{D}^{(0)}} ,
\end{eqnarray}
and so on, where $S$, $D$, $T$, and $Q$ have the same meanings as before, and 
the summation index restriction, e.g., $S \notin \gamma$, can be actuated by ``denom.$\neq$0,'' demanding that the denominator $E_{\gamma}^{(0)}-E_{S}^{(0)}$ be nonzero.

We further subdivide $\hat{R}_n$ according to the number of modal differences in the degenerate Hartree products involved. For instance,
\begin{eqnarray}
\hat{R}_2 = \hat{R}_2^{(+2)} + \hat{R}_2^{(\pm0)} + \hat{R}_2^{(-2)} \label{eq:R2breakdown}
\end{eqnarray}
with
\begin{eqnarray}
\label{eq:R2breakdown2}
\hat{R}_2^{(+2)} &=&  \sum_{D(+2)}^{{\text {denom.} \neq 0 }} \frac{| \Phi_{D(+2)}^{(0)} \rangle\langle \Phi_{D(+2)}^{(0)} |}{E_{\gamma}^{(0)}-E_{D(+2)}^{(0)}} ,  \\
\hat{R}_2^{(\pm0)} &=&  \sum_{D(\pm0)}^{{\text {denom.} \neq 0 }} \frac{| \Phi_{D(\pm0)}^{(0)} \rangle\langle \Phi_{D(\pm0)}^{(0)} |}{E_{\gamma}^{(0)}-E_{D(\pm0)}^{(0)}} ,  \\
\hat{R}_2^{(-2)} &=&  \sum_{D(-2)}^{{\text {denom.} \neq 0 }} \frac{| \Phi_{D(-2)}^{(0)} \rangle\langle \Phi_{D(-2)}^{(0)} |}{E_{\gamma}^{(0)}-E_{D(-2)}^{(0)}} ,  
\end{eqnarray}
where $\Phi_{D(+2)}^{(0)}$, $\Phi_{D(\pm0)}^{(0)}$, and $\Phi_{D(-2)}^{(0)}$ are given by Eqs.\ (\ref{eq:D2}), (\ref{eq:D0}), and (\ref{eq:Dminus2}), respectively.

% =====================================
% Omega(1)
% =====================================
\subsection{First-order correction to the grand potential \label{section:firstorderFTPT}}

Here, we illustrate the algebraic reduction of the sum-over-states formula of $\Omega^{(1)}$ in a QFF. 
Substituting the trace invariance [Eq.\ (\ref{eq:traceinvariance})] and the degenerate RSPT energy formula [Eq.\ (\ref{eq:E_In_degen})] into the recursion [Eq.\ (\ref{eq:recursionOmg})], we immediately arrive at the sum-over-states analytical formula of $\Omega^{(1)}$ as
\begin{eqnarray}
\label{eq:Omg1_definition}
    \Omega^{(1)} = \left[{E_N^{(1)}}\right] = \left[\mathrm{Tr}\left(\bm{E}^{(1)}\right) \right]= \left[ \langle N | \hat{V}^{(1)} | N \rangle \right],
\end{eqnarray}
where ``$[\dots]$'' refers to a zeroth-order thermal average [Eq.\ (\ref{eq:thermalavg_XN})] and $N$ stands for $\Phi^{(0)}_N$ as defined by Eq.\ (\ref{eq:Phi_Nwavefunc}). 
Furthermore, using $\hat{V}^{(1)}$ given in Eq.\ (\ref{eq:sorted_V}), we can expand the right-hand side of the above equation as
\begin{widetext}
\begin{eqnarray}
\label{eq:NVNall}
\left[ \langle N | \hat{V}^{(1)}| N \rangle \right]  &=& \sum_{i}\left(F_i \Big[ \langle N | Q_i | N \rangle \Big] + \frac{1}{2} \bar{F}_{ii} \left[ \langle N |Q_i^2| N \rangle \right] 
%\right. \nonumber\\&& \left.
+ \frac{1}{6} F_{i i i} \left[ \langle N |Q_{i}^{3}| N \rangle \right]+\frac{1}{24} F_{i i i i} \left[ \langle N |Q_{i}^{4}| N \rangle \right] \right) \nonumber\\
&&+\sum_{i,j}^{i \neq j}\left(\frac{1}{2}\bar{F}_{ij} \left[ \langle N |Q_iQ_j| N \rangle \right]+ \frac{1}{2} F_{i j j} \left[ \langle N |Q_{i} Q_{j}^2| N \rangle \right] 
%\right. \nonumber\\&& \left.
+\frac{1}{8} F_{i i j j} \left[ \langle N |Q_{i}^{2} Q_{j}^{2}| N \rangle \right] +\frac{1}{6} F_{i jj j} \left[ \langle N |Q_{i} Q_{j}^{3}| N \rangle \right]\right)\nonumber\\
&&+\sum_{i,j,k}^{\substack{i\neq j, i \neq k \\ j \neq k}}\left(\frac{1}{6} F_{i j k} \left[ \langle N |Q_{i} Q_{j} Q_{k}| N \rangle \right]
 %\right. \nonumber\\&& \left.
 +\frac{1}{4} F_{i j k k } \left[ \langle N |Q_{i} Q_{j} Q_{k}^{2}| N \rangle \right]\right) 
 %\nonumber\\&&
 +\sum_{i,j,k,l}^{\substack{i\neq j , i \neq k \\ i \neq l , j \neq k \\ j \neq l , k \neq l}} \frac{1}{24} F_{i j k l} \left[ \langle N |Q_{i} Q_{j} Q_{k} Q_{l}| N \rangle \right],
\end{eqnarray}
\end{widetext}
where we have used the distributive property of a thermal average.

Each term can be factored. The first term, which is a many-mode thermal average, is written as a product of one-mode thermal averages as follows:
\begin{eqnarray}
\label{eq:[FiQi]}
&& \sum_i F_i \Big[\langle N | Q_i | N \rangle\Big] \nonumber\\
&&= \sum\limits_i F_i \frac{\sum \limits_{N=0}^\infty \langle N | Q_i | N \rangle \exp(-\beta E_N^{(0)})}{\sum \limits_{N=0}^\infty \exp(-\beta E_N^{(0)})} \nonumber\\
 &&=  \sum\limits_i F_i \,\frac{ \sum\limits_{n_i=0}^{\infty}  \langle n_i |  Q_i | n_i \rangle \exp\{-\beta  (n_i+1/2)\omega_i \}}{ \sum\limits_{n_i=0}^{\infty} \exp\{-\beta (n_i+1/2) \omega_i \}} \nonumber \\
    \label{eq:[FiQi]6}
&& \equiv    \sum_i  F_{i} \Big[ \langle Q_i \rangle_{0} \Big] = 0.
\end{eqnarray}
In the last line, ``[\dots]'' now refers to a zeroth-order thermal average over a single mode, i.e.,
\begin{eqnarray}
\label{eq:fQnotation1}
     \left[ \langle \hat{F}(Q_i) \rangle_{0}\right] &\equiv& \frac{ \sum \limits_{n_i=0}^\infty \langle n_i | \hat{F}(Q_i) | n_i  \rangle \exp\{-\beta (n_i+1/2)\omega_i \}}{ \sum\limits_{n_i=0}^{\infty} \exp\{-\beta (n_i+1/2)\omega_i \}}, 
\end{eqnarray}
which can be analytically evaluated for a variety of operators made of $Q_i$, $\hat{F}(Q_i)$. 
They are tabulated as the thermal Born--Huang rules\cite{born1988dynamical,Keceli2011} 
in Table \ref{table:thermalavg_EN1}, whose derivation is found in Appendix \ref{Appendix:BHrules}. 
Using the second column of this table, we find this thermal average [Eq.\ (\ref{eq:[FiQi]})] to be zero. 

% =====================================
%Table I
% =====================================
\begin{table}
\caption{\label{table:thermalavg_EN1} The thermal Born--Huang rules for the integrals of the type Eq.\ (\ref{eq:fQnotation1}), where $Q_i$ is the $i$th normal coordinate with frequency $\omega_i$ and Bose--Einstein distribution function $f_i$ [Eq.\ (\ref{eq:BEfunction})]. All other cases are either zero for a QFF or unnecessary for our objectives. See Appendix \ref{Appendix:BHrules} for the derivation.}
\begin{ruledtabular}
\begin{tabular}{ccccc}
 $[ \langle{\partial^2}/{\partial Q_i^2} \rangle_{0} ]$ & $[ \langle Q_i \rangle_{0}]$ &  $[\langle Q_i^2 \rangle_{0}]$  &  $[\langle Q_i^3 \rangle_{0}]$ &  $[\langle Q_i^4 \rangle_{0}]$ \\
 \hline
 $-{\omega_i (f_i+1/2)}$ & 0 & ${(f_i+1/2)}/{\omega_i}$ & 0 & ${3(f_i+1/2)^2}/{\omega_i^{2}}$ \\
\end{tabular}
\end{ruledtabular} 
\end{table}

The second term in Eq.\ (\ref{eq:NVNall}) can be evaluated similarly as
\begin{eqnarray}
\frac{1}{2}\sum_i \bar{F}_{ii} \left [\langle N | Q_i^2 | N \rangle \right] &=&  \frac{1}{2} \sum_i \bar{F}_{ii} \left[ \langle Q_i^2 \rangle_{0} \right] \nonumber\\
&=& \frac{1}{2} \sum_i \bar{F}_{ii} \frac{ f_i  +1/2}{\omega_i} \nonumber\\
&=& \sum_i \tilde{\bar{F}}_{ii} {(f_i+1/2)},
\end{eqnarray}
where the third column of Table \ref{table:thermalavg_EN1} was consulted with in the second equality and $\tilde{\bar{F}}_{ii}$ is defined by Eq.\ (\ref{eq:Ftildebar}).
This term is nonzero. 

Repeating this process for the remaining terms of Eq.\ (\ref{eq:NVNall}), we find that there are only two other terms that are nonvanishing. The first of them is
\begin{eqnarray}
\label{eq:FiiiiAG}
\frac{1}{24} \sum_{i} F_{iiii} \left[\langle N | Q_i^4 | N \rangle\right] 
&=& \frac{1}{24} \sum_{i} F_{iiii} \left[ \langle{Q_i^4}\rangle_{0} \right] \nonumber\\
&=& \frac{1}{2} \sum_i \tilde{F}_{iiii}{(f_i+1/2)^2} ,
\end{eqnarray}
where the last column of Table \ref{table:thermalavg_EN1} was used and $\tilde{F}_{iiii}$ is given by Eq.\ (\ref{eq:F4tilde}). The other nonvanishing term is
\begin{eqnarray}
\label{eq:FiijjAG}
&&\frac{1}{8}\sum_{i,j}^{i\neq j}F_{iijj} \left[\langle N | Q_i^2 Q_j^2 | N \rangle \right] 
\nonumber\\
&&= \frac{1}{8}\sum_{i,j}^{i\neq j}F_{iijj} \frac{\sum\limits_{n_i=0}^{\infty} \langle n_i |     Q_i^2  | n_i \rangle \exp\{-\beta (n_i+1/2) \omega_i \}}{ \sum \limits_{n_i=0}^{\infty} \exp\{-\beta (n_i+1/2)\omega_i \}} 
\nonumber\\&&\,\,\,\,\,\times\,
\frac{\sum\limits_{n_j=0}^{\infty} \langle n_j |     Q_j^2 | n_j \rangle \exp\{-\beta  (n_j+1/2)\omega_j\}}{ \sum \limits_{n_j=0}^{\infty} \exp\{-\beta(n_j+1/2) \omega_j \}} \nonumber\\
%&=& \frac{1}{8}\sum_{i\neq j}F_{iijj} [ \langle n_i | Q_i^2 | n_i \rangle \langle n_j | Q_j^2 | n_j \rangle ] \nonumber\\
%& =&\frac{1}{8}\sum_{i\neq j}F_{iijj} [ \langle Q_i^2 \rangle_{0}  \langle Q_j^2 \rangle_{0} ] \nonumber\\
&&= \frac{1}{8}\sum_{i,j}^{i\neq j}F_{iijj} \left[ \langle Q_i^2 \rangle_{0} \right]\left[ \langle Q_j^2 \rangle_{0} \right] \nonumber\\
&&=\frac{1}{8}\sum_{i,j}^{i\neq j}F_{iijj}\frac{f_i+1/2}{\omega_i}\frac{f_j+1/2}{\omega_j} \nonumber\\
&&= \frac{1}{2}\sum_{i,j}^{i\neq j}\tilde{F}_{iijj} (f_i+1/2)(f_j+1/2).
\end{eqnarray}
Summing the above two equations, we obtain
\begin{eqnarray}
&& \frac{1}{24} \sum_{i} F_{iiii} \left[\langle N | Q_i^4 | N \rangle\right] 
+ \frac{1}{8}\sum_{i,j}^{i\neq j}F_{iijj} \left[\langle N | Q_i^2 Q_j^2 | N \rangle \right] 
\nonumber\\
&& =  \frac{1}{2} \sum_i \tilde{F}_{iiii}{(f_i+1/2)^2} + \frac{1}{2}\sum_{i,j}^{i\neq j}\tilde{F}_{iijj} (f_i+1/2)(f_j+1/2)
\nonumber\\
&& =   \frac{1}{2}\sum_{i,j} \tilde{F}_{iijj} (f_i+1/2)(f_j+1/2).
\end{eqnarray}
This is the simplest example of the lifting of summation index restrictions as multiple terms are consolidated. 
For higher-order perturbation contributions, reorganizing terms to lift summation index restrictions is an extremely 
arduous task. We introduce canonical forms of the thermal Born--Huang rules to facilitate this process, which are 
summarized in Table \ref{table:canonical} and justified in Appendix \ref{appendix:BHfactorization}.
For these two terms, canonical form (1) of Table \ref{table:canonical} is salient.
%\begin{eqnarray}
%\left [ \langle{Q_i^4}\rangle_{0} \right] = 3 \left[ \langle Q_i^2 \rangle_{0} \right]\left[ \langle Q_i^2 \rangle_{0} \right].
%\end{eqnarray}
Substitution of this into Eq.\ (\ref{eq:FiiiiAG}) early will expedite the consolidation process. 

That all the other terms in Eq.\ (\ref{eq:NVNall}) vanish can be inferred from Table \ref{table:thermalavg_EN1}.

%============================================
% Table II
%============================================
\begin{table*}
\caption{\label{table:canonical}
Canonical forms of the thermal Born--Huang rules. See Eq.\ (\ref{eq:fQnotation2}) for the definition of the thermal averages and Eq.\ (\ref{eq:symmetry}) for their permutation symmetry.}
\begin{ruledtabular}
\begin{tabular}{rlcl}
& Original form &&  Canonical form \\ \hline
(1) & $[ \langle Q_i^4 \rangle_{0} ]$ & =& $  3 [ \langle Q_i^2 \rangle_{0} ][ \langle Q_{i}^2 \rangle_{0} ]$ \\
(2) & $[ \langle Q_i^2 \rangle_{0} \langle Q_i^2 \rangle_{0}] $ &=& $ [ \langle Q_i^2 \rangle_{0} ] [ \langle Q_{i}^2 \rangle_{0} ]  + 4[ \langle Q_i \rangle_{1} \langle Q_i \rangle_{-1}][ \langle Q_i \rangle_{-1} \langle Q_i \rangle_{1}]$ \\
(3) & $[\langle Q_i^2 \rangle_{2} \langle Q_i^2 \rangle_{-2} ] $ &=& $2  [ \langle Q_i \rangle_{1} \langle Q_i \rangle_{-1}][ \langle Q_i \rangle_{1} \langle Q_i \rangle_{-1}]$ \\
(4) & $[\langle Q_i^2 \rangle_{-2} \langle Q_i^2 \rangle_{2} ] $ &=& $2  [ \langle Q_i \rangle_{-1} \langle Q_i \rangle_{1}][ \langle Q_i \rangle_{-1} \langle Q_i \rangle_{1}]$ \\
 (5) & $[\langle Q_i^2 \rangle_{0} \langle {\partial^2}/{\partial Q_i^2} \rangle_{0}]$ &=& $  [ \langle Q_i^2 \rangle_{0} ] [ \langle {\partial^2}/{\partial Q_i^2}  \rangle_{0} ]  - 4\omega_i^2[ \langle Q_i \rangle_{1} \langle Q_i \rangle_{-1}][ \langle Q_i \rangle_{-1} \langle Q_i \rangle_{1}]$\\
 (6) & $[\langle Q_i^2 \rangle_{2} \langle {\partial^2}/{\partial Q_i^2} \rangle_{-2}]$ &=& $2 \omega_i^2 [ \langle Q_i \rangle_{1} \langle Q_i \rangle_{-1}][ \langle Q_i \rangle_{1} \langle Q_i \rangle_{-1}]$\\
 (7) & $[\langle Q_i^2 \rangle_{-2} \langle {\partial^2}/{\partial Q_i^2} \rangle_{2}]$ &=&  $2 \omega_i^2 [ \langle Q_i \rangle_{-1} \langle Q_i \rangle_{1}][ \langle Q_i \rangle_{-1} \langle Q_i \rangle_{1}]$\\
(8) & $[ \langle Q_i \rangle_{1} \langle Q_i^3 \rangle_{-1} ]$ & = & $3 [ \langle Q_i \rangle_{1} \langle Q_i \rangle_{-1}] [ \langle Q_i^2 \rangle_{0} ]$\\
 (9) & $ [ \langle Q_i \rangle_{-1} \langle Q_i^3 \rangle_{1} ]$ & = & $3   [ \langle Q_i \rangle_{-1} \langle Q_i \rangle_{1}] [ \langle Q_i^2 \rangle_{0} ] $\\
 (10) & $[ \langle Q_i^4 \rangle_{0} \langle Q_i^2 \rangle_{0} ]$ &=&  $24 [ \langle Q_i \rangle_{1} \langle Q_i \rangle_{-1}][ \langle Q_i \rangle_{-1} \langle Q_i \rangle_{1}] [ \langle Q_i^2 \rangle_{0} ] + 3   [ \langle Q_i^2 \rangle_{0} ][ \langle Q_i^2 \rangle_{0} ] [ \langle Q_i^2 \rangle_{0} ] $\\
 (11) & $[ \langle Q_i^4 \rangle_{2} \langle Q_i^2 \rangle_{-2} ]$ &=& $ 12 [ \langle Q_i \rangle_{1} \langle Q_i \rangle_{-1}][ \langle Q_i \rangle_{1} \langle Q_i \rangle_{-1}] [ \langle Q_i^2 \rangle_{0} ] $\\
 (12) & $ [ \langle Q_i^4 \rangle_{-2} \langle Q_i^2 \rangle_{2} ]$ &=& $ 12 [ \langle Q_i \rangle_{-1} \langle Q_i \rangle_{1}][ \langle Q_i \rangle_{-1} \langle Q_i \rangle_{1}] [ \langle Q_i^2 \rangle_{0} ] $\\
 (13) & $[\langle Q_i^4 \rangle_{0} \langle {\partial^2}/{\partial Q_i^2} \rangle_{0}]$ &=&  $24 [ \langle Q_i \rangle_{1} \langle Q_i \rangle_{-1}][ \langle Q_i \rangle_{-1} \langle Q_i \rangle_{1}] [ \langle {\partial^2}/{\partial Q_i^2}  \rangle_{0} ]  + 3  [ \langle Q_i^2 \rangle_{0} ][ \langle Q_i^2 \rangle_{0} ] [ \langle {\partial^2}/{\partial Q_i^2} \rangle_{0} ]$\\
 (14) & $[ \langle Q_i^4 \rangle_{2} \langle  {\partial^2}/{\partial Q_i^2} \rangle_{-2} ]$ &=& $ -12 [ \langle Q_i \rangle_{1} \langle Q_i \rangle_{-1}][ \langle Q_i \rangle_{1} \langle Q_i \rangle_{-1}] [ \langle {\partial^2}/{\partial Q_i^2}  \rangle_{0} ]$ \\
 (15) & $[ \langle Q_i^4 \rangle_{-2} \langle {\partial^2}/{\partial Q_i^2} \rangle_{2} ]$ & = & $ -12 [ \langle Q_i \rangle_{-1} \langle Q_i \rangle_{1}][ \langle Q_i \rangle_{-1} \langle Q_i \rangle_{1}] [ \langle {\partial^2}/{\partial Q_i^2}  \rangle_{0} ]$ \\
 (16) & $[ \langle Q_i^3 \rangle_{1} \langle Q_i^3 \rangle_{-1}]  $ &=& $ 9   [ \langle Q_i \rangle_{1} \langle Q_i \rangle_{-1}] [ \langle Q_i^2 \rangle_{0} ] [ \langle Q_i^2 \rangle_{0} ]  +  18  [ \langle Q_i \rangle_{1} \langle Q_i \rangle_{-1}] [ \langle Q_i \rangle_{1} \langle Q_i \rangle_{-1}] [ \langle Q_i \rangle_{-1} \langle Q_i \rangle_{1}] $ \\
(17) & $[ \langle Q_i^3 \rangle_{3} \langle Q_i^3 \rangle_{-3} ] $ & =&$ 6 [ \langle Q_i \rangle_{1} \langle Q_i \rangle_{-1}][ \langle Q_i \rangle_{1} \langle Q_i \rangle_{-1}] [ \langle Q_i \rangle_{1} \langle Q_i \rangle_{-1}]$ \\
(18) & $[ \langle Q_i^3 \rangle_{-1} \langle Q_i^3 \rangle_{1} ] $ &=& $ 9   [ \langle Q_i \rangle_{-1} \langle Q_i \rangle_{1}] [ \langle Q_i^2 \rangle_{0} ] [ \langle Q_i^2 \rangle_{0} ] +  18   [ \langle Q_i \rangle_{-1} \langle Q_i \rangle_{1}] [ \langle Q_i \rangle_{-1} \langle Q_i \rangle_{1}] [ \langle Q_i \rangle_{1} \langle Q_i \rangle_{-1}]$ \\
(19) & $[ \langle Q_i^3 \rangle_{-3} \langle Q_i^3 \rangle_{3} ] $ &=& $6  [ \langle Q_i \rangle_{-1} \langle Q_i \rangle_{1}][ \langle Q_i \rangle_{-1} \langle Q_i \rangle_{1}] [ \langle Q_i \rangle_{-1} \langle Q_i \rangle_{1}]$ \\
(20) & $[ \langle Q_i^4 \rangle_{0} \langle Q_i^4 \rangle_{0}] $ &=& $  144  [ \langle Q_i \rangle_{1} \langle Q_i \rangle_{-1}] [ \langle Q_i \rangle_{-1} \langle Q_i \rangle_{1}][ \langle Q_i^2 \rangle_{0} ] [ \langle Q_i^2 \rangle_{0} ] 
 + 144  [ \langle Q_i \rangle_{1} \langle Q_i \rangle_{-1}] [ \langle Q_i \rangle_{-1} \langle Q_i \rangle_{1}][ \langle Q_i \rangle_{1} \langle Q_i \rangle_{-1}] [ \langle Q_i \rangle_{-1} \langle Q_i \rangle_{1}]$ \\
 & & &  $+9[ \langle Q_i^2 \rangle_{0} ] [ \langle Q_i^2 \rangle_{0} ][ \langle Q_i^2 \rangle_{0} ] [ \langle Q_i^2 \rangle_{0} ]$ \\
 (21) & $[\langle Q_i^4 \rangle_{2} \langle Q_i^4 \rangle_{-2} ]$ & = & $72  [ \langle Q_i \rangle_{1} \langle Q_i \rangle_{-1}] [ \langle Q_i \rangle_{1} \langle Q_i \rangle_{-1}][ \langle Q_i^2 \rangle_{0} ] [ \langle Q_i^2 \rangle_{0} ] +  96   [ \langle Q_i \rangle_{1} \langle Q_i \rangle_{-1}] [ \langle Q_i \rangle_{1} \langle Q_i \rangle_{-1}][ \langle Q_i \rangle_{1} \langle Q_i \rangle_{-1}] [ \langle Q_i \rangle_{-1} \langle Q_i \rangle_{1}] $\\
 (22) & $[\langle Q_i^4 \rangle_{4} \langle Q_i^4 \rangle_{-4} ]$ &=&  $24 [ \langle Q_i \rangle_{1} \langle Q_i \rangle_{-1}][ \langle Q_i \rangle_{1} \langle Q_i \rangle_{-1}] [ \langle Q_i \rangle_{1} \langle Q_i \rangle_{-1}][ \langle Q_i \rangle_{1} \langle Q_i \rangle_{-1}] $\\
 (23) & $ [\langle Q_i^4 \rangle_{-2} \langle Q_i^4 \rangle_{2} ]$  &=& $72   [ \langle Q_i \rangle_{-1} \langle Q_i \rangle_{1}] [ \langle Q_i \rangle_{-1} \langle Q_i \rangle_{1}][ \langle Q_i^2 \rangle_{0} ] [ \langle Q_i^2 \rangle_{0} ]  + 96  [ \langle Q_i \rangle_{-1} \langle Q_i \rangle_{1}] [ \langle Q_i \rangle_{-1} \langle Q_i \rangle_{1}][ \langle Q_i \rangle_{-1} \langle Q_i \rangle_{1}] [ \langle Q_i \rangle_{1} \langle Q_i \rangle_{-1}]$\\
 (24) & $[\langle Q_i^4 \rangle_{-4} \langle Q_i^4 \rangle_{4} ]$ &=& $24  [ \langle Q_i \rangle_{-1} \langle Q_i \rangle_{1}][ \langle Q_i \rangle_{-1} \langle Q_i \rangle_{1}] [ \langle Q_i \rangle_{-1} \langle Q_i \rangle_{1}][ \langle Q_i \rangle_{-1} \langle Q_i \rangle_{1}] $
\end{tabular}
\end{ruledtabular} 
\end{table*}

Collecting all nonvanishing terms, we obtain the final sum-over-modes analytical formula for $\Omega^{(1)}$ as
\begin{eqnarray}
\label{eq:Omg1_Ag}
\Omega^{(1)} &=& \sum_i \tilde{\bar{F}}_{ii} {(f_i+1/2)}
\nonumber\\&& + \frac{1}{2}\sum_{i,j}\tilde{F}_{iijj} (f_i+1/2)(f_j+1/2)\\
&\equiv& \left[{E_N^{(1)}}\right]_L, \label{eq:linked}
\end{eqnarray}
which is linked. Algebraically, an unlinked term
is a simple product (as opposed to a contraction with at least one common summation index) of two or more extensive 
quantities and, therefore, increases superlinearly with system size.\cite{Hirata2011}  
Neither term in Eq.\ (\ref{eq:Omg1_Ag}) can be written as such and, therefore,
$\Omega^{(1)}$ is linked and size-consistent. This is indicated by subscript $L$ in Eq.\ (\ref{eq:linked}).

% =====================================
% U(1)
% =====================================
\subsection{First-order correction to the internal energy} 

From the recursion for $U^{(n)}$ [Eq.\ (\ref{eq:recursion_U})], we immediately obtain the sum-over-states formula for $U^{(1)}$ as
\begin{eqnarray}
\label{eq:U1recursion}
 U^{(1)} =\left[E_N^{(1)}\right] - \beta \left( \left[E_N^{(1)} E_N^{(0)}\right] - \Omega^{(1)} U^{(0)}\right) .
\end{eqnarray}
We have already reduced the first term of the right-hand side in Sec.\ \ref{section:firstorderFTPT}. 

Using Eq.\ (\ref{eq:traceinvariance2new}), we can rewrite the second term as
\begin{eqnarray}
\label{eq:EN1EN0AG}
\left[E_N^{(1)} E_N^{(0)}\right] &=& \left[ \mathrm{Tr}\left( \bm{E}^{(1)}\hat{P}_0\bm{E}^{(0)} \right) \right]  
+ \left[ \mathrm{Tr}\left( \bm{E}^{(1)}\hat{P}_1\bm{E}^{(0)} \right) \right] 
\nonumber\\
&& + \left[ \mathrm{Tr}\left( \bm{E}^{(1)}\hat{P}_2\bm{E}^{(0)} \right) \right]  
+ \left[ \mathrm{Tr}\left( \bm{E}^{(1)}\hat{P}_3\bm{E}^{(0)} \right) \right] 
\nonumber\\
&& + \left[ \mathrm{Tr}\left( \bm{E}^{(1)}\hat{P}_4\bm{E}^{(0)} \right) \right]  \label{eq:U1_2} \nonumber\\
&=& \left[ \mathrm{Tr}\left( \bm{E}^{(1)}\hat{P}_0\bm{E}^{(0)} \right) \right]  \nonumber\\
&=&\left [\langle N | \hat{V}^{(1)} \hat{P}_0 \hat{H}^{(0)} | N \rangle\right] \nonumber\\
&=&\left [\langle N | \hat{V}^{(1)} | N \rangle \langle N | \hat{H}^{(0)} | N \rangle\right] ,
\end{eqnarray}
where we assumed a QFF in the first equality and used the fact that $\bm{E}^{(0)}$ is diagonal
and, therefore, $\bm{E}^{(1)}\hat{P}_i\bm{E}^{(0)} = \bm{0}$ for $i \geq 1$. 

Recalling the definition of $\hat{H}^{(0)}$ [Eq.\ (\ref{eq:H0})] and keeping only those terms in 
$\hat{V}^{(1)}$ that give rise to nonvanishing results (see below for a justification), we can expand the last expression as
\begin{widetext}
\begin{eqnarray}
\label{eq:E1E0Z}
\left[\langle N | \hat{V}^{(1)} | N \rangle \langle N | \hat{H}^{(0)} | N \rangle\right] 
&=& \left[ \langle N | \,\frac{1}{2} \sum_i \bar{F}_{ii} Q_i^2 + \frac{1}{8}\sum_{i,j}^{i\neq j} F_{iijj} Q_i^2 Q_j^2 + \frac{1}{24}\sum_i F_{iiii} Q_i^4 |N \rangle \langle N |  -\frac{1}{2} \sum_k\frac{\partial^2}{\partial Q_k^2} + V_{\mathrm{ref}}+ \frac{1}{2}  \sum_k\omega_k^2 Q_k^2 | N \rangle \right] \nonumber\\
&=& -\frac{1}{4} \sum_{i,k} \bar{F}_{ii} \left[ \langle N | Q_i^2 | N \rangle \langle N| \frac{\partial^2}{\partial Q_k^2} |N \rangle \right]
+ \frac{1}{4} \sum_{i,k} \bar{F}_{ii}\omega_k^2 \left[ \langle N | Q_i^2 | N \rangle \langle N| Q_k^2 |N \rangle \right]  \nonumber\\
&& + \left[ \langle N | \, \frac{1}{8}\sum_{i,j}^{i\neq j} F_{iijj} Q_i^2 Q_j^2 + \frac{1}{24}\sum_i F_{iiii} Q_i^4 |N \rangle \langle N |  -\frac{1}{2} \sum_k\frac{\partial^2}{\partial Q_k^2} +  \frac{1}{2}  \sum_k\omega_k^2 Q_k^2 | N \rangle \right] + \Omega^{(1)} V_\mathrm{ref},
\end{eqnarray}
where we used $\left[ \langle N | \hat{V}^{(1)} | N \rangle \langle N | V_\mathrm{ref} | N \rangle\right]  = \Omega^{(1)} V_\mathrm{ref}$ 
according to Eq.\ (\ref{eq:Omg1_definition}).

Let us first reduce the first term in the right-hand side.
We need to consider two cases, $i\neq k$ and $i=k$, separately. For the $i\neq k$ case, we have
\begin{eqnarray}
\label{eq:E1E0Z1a}
&&-\frac{1}{4}\sum_{i,k}^{i\neq k}\bar{F}_{ii} \left[\langle N | Q_i^2 | N \rangle \langle N | \frac{\partial^2}{\partial Q_k^2} | N \rangle \right] 
\nonumber\\&&
= - \frac{1}{4}\sum_{i,k}^{i\neq k}\bar{F}_{ii} \frac{\sum\limits_{n_i=0}^{\infty} \langle n_i |     Q_i^2  | n_i \rangle \exp\{-\beta  (n_i+1/2)\omega_i\}}{ \sum \limits_{n_i=0}^{\infty} \exp\{-\beta  (n_i+1/2)\omega_i\}} 
%\nonumber\\&&\,\,\,\,\,\times\,
\frac{\sum\limits_{n_k=0}^{\infty} \langle n_k |     {\partial^2}/{\partial Q_k^2} | n_k \rangle \exp\{-\beta (n_k+1/2)\omega_k \}}{ \sum \limits_{n_k=0}^{\infty} \exp\{-\beta  (n_k+1/2)\omega_k\}} 
\nonumber\\
&&= - \frac{1}{4}\sum_{i,k}^{i\neq k}\bar{F}_{ii}\left[ \langle Q_i^2 \rangle_{0} \right]\left[ \left\langle \frac{\partial^2}{\partial Q_k^2} \right\rangle_{0} \right] 
%\nonumber\\
=- \frac{1}{4}\sum_{i,k}^{i\neq k}\bar{F}_{ii}\frac{f_i+1/2}{\omega_i}\left\{-{\omega_k(f_k+1/2)}\right\} 
%\nonumber\\&&
= \frac{1}{2}\sum_{i,k}^{i\neq k}\tilde{\bar{F}}_{ii} \omega_k (f_i+1/2)(f_k+1/2),
\end{eqnarray}
where we used the first and third columns of Table \ref{table:thermalavg_EN1} in the penultimate equality.

For the $i=k$ case, we must instead  write
\begin{eqnarray}
\label{eq:E1E0Z1b}
-\frac{1}{4}\sum_{i,k}^{i=k}\bar{F}_{ii} \left[\langle N | Q_i^2 | N \rangle \langle N | \frac{\partial^2}{\partial Q_k^2} | N \rangle \right] 
%\nonumber\\
&=& - \frac{1}{4}\sum_{i}^{}\bar{F}_{ii} \frac{\sum\limits_{n_i=0}^{\infty} \langle n_i |     Q_i^2  | n_i \rangle \langle n_i |     {\partial^2}/{\partial Q_i^2} | n_i \rangle  \exp\{-\beta  (n_i+1/2)\omega_i\}}{ \sum \limits_{n_i=0}^{\infty} \exp\{-\beta (n_i+1/2)\omega_i \}}  \nonumber\\
&=& - \frac{1}{4}\sum_{i}^{}\bar{F}_{ii}\left[ \langle Q_i^2 \rangle_{0}  \left\langle \frac{\partial^2}{\partial Q_i^2} \right\rangle_{0} \right] 
%\nonumber \\
%&&= - \frac{1}{4}\sum_{i,j}^{i= j}\bar{F}_{ii}\left[ \langle Q_i^2 \rangle_{0} \right]\left[ \left\langle \frac{\partial^2}{\partial Q_j^2}\right \rangle_{0} \right] \label{E1E0Z1b3} \\
= - \frac{1}{4}\sum_{i}^{}\bar{F}_{ii}\left(-\frac{8f_i^2+8f_i+1}{4}\right) 
\nonumber \\
&=& \frac{1}{4}\sum_{i}^{}\bar{F}_{ii}\left\{(f_i+1/2)^2 + f_i (f_i +1)\right\} 
%\nonumber \\&&
= \frac{1}{2}\sum_{i}^{}\tilde{\bar{F}}_{ii} \omega_i (f_i+1/2)^2 + \frac{1}{2}\sum_{i}^{}\tilde{\bar{F}}_{ii} \omega_i f_i (f_i+1),
\end{eqnarray}
where the one-mode thermal average $[\langle\dots\rangle\langle\dots\rangle]$ appearing above is defined by
\begin{eqnarray}
\label{eq:fQnotation2}
  \Big[   \langle \hat{F}(Q_i) \rangle_{\delta_i} \langle \hat{G}(Q_i) \rangle_{-\delta_i}\Big] \equiv \frac{ \sum \limits_{n_i} \langle n_i | \hat{F}(Q_i) | n_i - \delta_i \rangle
   \langle n_i - \delta_i | \hat{G}(Q_i) | n_i \rangle \exp\{-\beta (n_i+1/2)\omega_i \}}{ \sum\limits_{n_i} \exp\{-\beta (n_i+1/2)\omega_i \}}, 
\end{eqnarray}
\end{widetext} 
where $\hat{F}(Q_i)$ and $\hat{G}(Q_i)$ are operators made of $Q_i$, and $n_i$ runs over all appropriate integers. The values of the thermal averages of this type are summarized in Table \ref{table:thermalavg_EN2},
whose derivation is given in Appendix \ref{Appendix:BHrules}. The third equality of Eq.\ (\ref{eq:E1E0Z1b}) made use of this table.

% =====================================
%Table III
% =====================================
\begin{table*}
\caption{\label{table:thermalavg_EN2} The thermal Born--Huang rules for the integrals of the type Eq.\ (\ref{eq:fQnotation2}), where $Q_i$ is the $i$th normal coordinate with frequency $\omega_i$ and Bose--Einstein distribution function $f_i$ [Eq.\ (\ref{eq:BEfunction})]. All other cases are either zero for a QFF or unnecessary for our objective. See Appendix \ref{Appendix:BHrules} for the derivation.}
\begin{ruledtabular}
   \begin{tabular}{ccccc}
 $ \delta_i   $ 
 & $[ \langle Q_i \rangle_{\delta_i}  \langle Q_i  \rangle_{-\delta_i} ]$ 
 & $[  \langle Q_i^2  \rangle_{\delta_i} \langle Q_i^2 \rangle_{-\delta_i} ]$ 
& $[\langle Q_i^2 \rangle_{\delta_i} \langle {\partial^2}/{\partial Q_i^2} \rangle_{-\delta_i}]$ 
& $[\langle Q_i        \rangle_{\delta_i}\langle Q_i^3 \rangle_{-\delta_i}]$\\
  \hline
0  & 0                         & $\frac{8 f_i^2 + 8 f_i + 1}{4 \omega_i^{2}}$ & $- \frac{8 f_i^2 + 8 f_i + 1}{4}$ & 0 \\
1  & $\frac{f_i}{2 \omega_i}$  & 0                                            & 0 & $\frac{6f_i^{2} +  3f_i}{4 \omega_i^{2}}$ \\
2  & 0                         & $\frac{f_i^{2}}{2 \omega_i^{2}}$             & $\frac{f_i^2}{2}$ & 0 \\
%3  & 0                         & 0                                            & 0 & 0 \\
%4  & 0                         & 0                                            & 0&0\\
$-1$ &$\frac{f_i+1}{2 \omega_i}$ & 0                                      & 0 & $\frac{6 f_i^{2} + 9 f_i + 3}{4 \omega_i^{2}}$ \\      
$-2$ & 0                         & $\frac{( f_i + 1)^2}{2 \omega_i^{2}}$        & $\frac{(f_i+1)^2}{2}$ & 0 \\
%$-3$ & 0                         & 0              & 0 & 0 \\                                   
%$-4$ & 0                         & 0             & 0 & 0 \\
\hline                                    
 $ \delta_i   $ 
 & $[ \langle Q_i^4      \rangle_{\delta_i} \langle Q_i^2 \rangle_{-\delta_i} ]$
 & $[\langle Q_i^4 \rangle_{\delta_i} \langle {\partial^2}/{\partial Q_i^2} \rangle_{-\delta_i}]$ 
 & $[   \langle Q_i^3 \rangle_{\delta_i} \langle Q_i^3 \rangle_{-\delta_i}  ]$ 
 & $[  \langle Q_i^4  \rangle_{\delta_i} \langle Q_i^4 \rangle_{-\delta_i} ]$ \\ 
 \hline
0      & $\frac{ 72 f_i^{3} + 108 f_i^{2} + 42 f_i + 3}{8 \omega_i^3}$ & -$\frac{ 72 f_i^{3} + 108 f_i^{2} + 42 f_i + 3}{8 \omega_i}$ & 0                                                     & $\frac{864f_i^{4}+1728f_i^{3}+1080f_i^{2}+216f_i+9}{16\omega_i^{4}}$ \\
1      & 0 & 0 & $\frac{54f_i^{3}+54f_i^{2}+9f_i}{8 \omega_i^{3}}$     & 0                                                                   \\
2      & $ \frac{6 f_i^{3} +  3f_i^{2}}{2 \omega_i^{3}}$ & $\frac{6 f_i^{3} +  3f_i^{2}}{2 \omega_i}$ & 0                                                     & $\frac{48 f_i^{4} + 48 f_i^{3} + 9 f_i^{2}}{2 \omega_i^{4}}$        \\
3      & 0 & 0     & $\frac{3 f_i^{3} }{4 \omega_i^{3}}$          & 0                                                                         \\
4      & 0 & 0 & 0                                                     & $\frac{3 f_i^{4} }{2 \omega_i^{4}}$                                  \\
$-1$ & 0 &0  & $\frac{54f_i^{3}+108f_i^{2}+63f_i+9}{8\omega_i^{3}}$ & 0                      \\                                                        
$-2$ & $\frac{6 f_i^{3} + 15 f_i^{2} + 12 f_i + 3}{2 \omega_i^{3}}$ & $\frac{6f_i^{3} + 15 f_i^{2} + 12 f_i + 3}{2 \omega_i}$ & 0                                                     & $\frac{48f_i^{4}+144f_i^{3}+153f_i^{2}+66f_i+9}{2\omega_i^{4}}$     \\
$-3$ & 0 & 0  & $\frac{3(f_i+1)^3}{4 \omega_i^{3}}$                   & 0 \\
$-4$ & 0 & 0 & 0                                                     & $\frac{3(f_i + 1)^4}{2 \omega_i^{4}}$                                \\ 
\end{tabular}
\end{ruledtabular} 
\end{table*}

Summing together the two cases [Eqs.\ (\ref{eq:E1E0Z1a}) and (\ref{eq:E1E0Z1b})], we observe that the summation index restrictions are lifted, leading to
the following result without any such restriction:
\begin{eqnarray}
\label{eq:E1E0Z1c}
&&-\frac{1}{4}\sum_{i,k}\bar{F}_{ii} \left[\langle N | Q_i^2 | N \rangle \langle N | \frac{\partial^2}{\partial Q_k^2} | N \rangle \right] \nonumber\\
%&&= -\frac{1}{4}\sum_{i,k}^{i\neq k}\bar{F}_{ii} \left[\langle N | Q_i^2 | N \rangle \langle N | \frac{\partial^2}{\partial Q_k^2} | N \rangle \right] 
%\nonumber\\&&
%-\frac{1}{4}\sum_{i,k}^{i=k}\bar{F}_{ii} \left[\langle N | Q_i^2 | N \rangle \langle N | \frac{\partial^2}{\partial Q_k^2} | N \rangle \right] \nonumber\\
%&&= \frac{1}{2}\sum_{i,k}^{i\neq k}\tilde{\bar{F}}_{ii} \omega_k (f_i+1/2)(f_k+1/2) +\frac{1}{2}\sum_{i}^{}\tilde{\bar{F}}_{ii} \omega_i (f_i+1/2)^2 \nonumber\\
%&&+ \frac{1}{2}\sum_{i}^{}\tilde{\bar{F}}_{ii} \omega_i f_i (f_i+1) \nonumber\\
&&= \frac{1}{2}\sum_{i,k}^{}\tilde{\bar{F}}_{ii} \omega_k (f_i+1/2)(f_k+1/2) + \frac{1}{2}\sum_{i}^{}\tilde{\bar{F}}_{ii} \omega_i f_i (f_i +1).
\nonumber\\
\end{eqnarray}
The first term in the right-hand side is identified as unlinked because it is a simple product of 
two extensive quantities, $\sum_i\tilde{\bar{F}}_{ii}(f_i+1/2)$ and $\sum_k \omega_k (f_k+1/2)$, and, therefore, it increases quadratically with size. It is non-size-consistent and
expected to be canceled out somehow to restore size-consistency for a whole perturbation correction (see below and also Sec.\ \ref{section:linkedtheorem} for systematic
cancellation at any order). 
The second term of Eq.\ (\ref{eq:E1E0Z1c}) is linked.

The lifting of the summation index restriction is facilitated by canonical form (5) of Table \ref{table:canonical} (see Appendix \ref{appendix:BHfactorization} for derivation), 
%\begin{eqnarray}
%\left[ \langle Q_i^2 \rangle_{0}  \left\langle \frac{\partial^2}{\partial Q_i^2} \right\rangle_{0} \right] &=& 
%\left[ \langle Q_i^2 \rangle_{0} \right]\left[ \left\langle \frac{\partial^2}{\partial Q_i^2} \right\rangle_{0} \right]
%\nonumber\\&&
%- 4 \omega_i^2 \Big[ \langle Q_i \rangle_{1}  \langle Q_i \rangle_{-1} \Big] \Big[ \langle Q_i \rangle_{-1}  \langle Q_i \rangle_{1} \Big],
%\nonumber\\
%\end{eqnarray}
which anticipates the correct breakdown of the factor $(8f_i^2 + 8 f_i + 1)/4 = (f_i+1/2)^2 + f_i (f_i + 1)$ in the penultimate equality of Eq.\ (\ref{eq:E1E0Z1b}). 
We shall illustrate more directly the utility of the canonical forms starting with the next paragraph.

For the second term in the right-hand side of Eq.\ (\ref{eq:E1E0Z}), we again need to consider two cases ($i\neq k$ and $i=k$) separately.
\begin{widetext}
\begin{eqnarray}
\label{eq:E1E0Z1d}
&&\frac{1}{4}\sum_{i,k}^{}\bar{F}_{ii} \omega_j^2 \left[\langle N | Q_i^2 | N \rangle \langle N | Q_k^2 | N \rangle \right] 
\nonumber\\&&
=\frac{1}{4}\sum_{i,k}^{i\neq k}\bar{F}_{ii} \omega_k^2 \left[\langle N | Q_i^2 | N \rangle \langle N | Q_k^2 | N \rangle \right] 
%\nonumber\\&&
+\frac{1}{4}\sum_{i,k}^{i=k}\bar{F}_{ii} \omega_k^2 \left[\langle N | Q_i^2 | N \rangle \langle N | Q_k^2 | N \rangle \right] 
\nonumber\\
&&= \frac{1}{4}\sum_{i,k}^{i\neq k}\bar{F}_{ii} \omega_k^2  \left[ \langle Q_i^2 \rangle_{0} \right]\left[ \langle  Q_k^2 \rangle_{0} \right] 
+ \frac{1}{4}\sum_{i}^{}\bar{F}_{ii}\omega_i^2  \left[ \langle Q_i^2 \rangle_{0} \langle  Q_i^2 \rangle_{0} \right] \nonumber\\
&&= \frac{1}{4}\sum_{i,k}^{i\neq k}\bar{F}_{ii} \omega_k^2  \left[ \langle Q_i^2 \rangle_{0} \right]\left[ \langle  Q_k^2 \rangle_{0} \right] 
+ \frac{1}{4}\sum_{i}^{}\bar{F}_{ii}\omega_i^2  \left[ \langle Q_i^2 \rangle_{0}\right]\left[ \langle  Q_i^2 \rangle_{0} \right] 
+ \sum_{i}^{}\bar{F}_{ii}\omega_i^2  \Big[ \langle Q_i \rangle_{1}\langle Q_i \rangle_{-1}\Big]\Big[ \langle Q_i \rangle_{-1}\langle Q_i \rangle_{1}\Big] \nonumber\\
&&= \frac{1}{4}\sum_{i,k}^{}\bar{F}_{ii} \omega_k^2  \left[ \langle Q_i^2 \rangle_{0} \right]\left[ \langle  Q_k^2 \rangle_{0} \right] 
+ \sum_{i}^{}\bar{F}_{ii}\omega_i^2  \Big[ \langle Q_i \rangle_{1}\langle Q_i \rangle_{-1}\Big]\Big[ \langle Q_i \rangle_{-1}\langle Q_i \rangle_{1}\Big] \nonumber\\
&&= \frac{1}{4}\sum_{i,k} \bar{F}_{ii}\omega_k^2 \frac{f_i+1/2}{\omega_i}\frac{f_k+1/2}{\omega_k} 
+ \sum_{i}^{}\bar{F}_{ii}\omega_i^2 \frac{f_i (f_i+1)}{(2\omega_i)^2} 
\nonumber\\
&&= \frac{1}{2}\sum_{i,k}^{}\tilde{\bar{F}}_{ii} \omega_k (f_i+1/2)(f_k+1/2) + \frac{1}{2}\sum_{i}^{}\tilde{\bar{F}}_{ii} \omega_i f_i(f_i+1) , \nonumber\\
\end{eqnarray}
\end{widetext}
where canonical form (2) of Table \ref{table:canonical} was used in the third equality, which lifts instantly the summation index restriction ($i\neq k$) introduced in the beginning.
The thermal Born--Huang rules (Tables \ref{table:thermalavg_EN1} and \ref{table:thermalavg_EN2}) were consulted with in the penultimate equality.  

It is unsurprising that the first and second terms of Eq.\ (\ref{eq:E1E0Z})
are reduced to the same formula [Eqs.\ (\ref{eq:E1E0Z1c}) and (\ref{eq:E1E0Z1d})]. This is due to the fact that these two terms originate, respectively, from the kinetic- and potential-energy operators of the same harmonic oscillator, 
whose corresponding energies obey the virial theorem and hence have the equal value.

The first term in the right-hand side of Eq.\ (\ref{eq:E1E0Z1d}) is unliked and will be canceled out later (see below), 
whereas the second term is linked and will persist.

The third term in the right-hand side of Eq.\ (\ref{eq:E1E0Z}) can be expanded as
\begin{widetext}
\begin{eqnarray}
\label{eq:E1E0Z2}
&& \left[ \langle N | \, \frac{1}{8}\sum_{i,j}^{i\neq j} F_{iijj} Q_i^2 Q_j^2 + \frac{1}{24}\sum_i F_{iiii} Q_i^4 |N \rangle \langle N |  -\frac{1}{2} \sum_k\frac{\partial^2}{\partial Q_k^2} +  \frac{1}{2}  \sum_k\omega_k^2 Q_k^2 | N \rangle \right] \nonumber\\
&&=\left[ \langle N | \, \frac{1}{8}\sum_{i,j}^{i\neq j} F_{iijj} Q_i^2 Q_j^2 + \frac{1}{24}\sum_i F_{iiii} Q_i^4 |N \rangle \langle N |  -\frac{1}{2} \sum_k\frac{\partial^2}{\partial Q_k^2}  | N \rangle \right] 
%\nonumber\\&& 
+ \left[ \langle N | \, \frac{1}{8}\sum_{i,j}^{i\neq j} F_{iijj} Q_i^2 Q_j^2 + \frac{1}{24}\sum_i F_{iiii} Q_i^4 |N \rangle \langle N |   \frac{1}{2}  \sum_k\omega_k^2 Q_k^2 | N \rangle \right]. \nonumber\\
\end{eqnarray}
Invoking the same argument given above based on the virial theorem, we expect that the two terms in the right-hand side lead to the same reduced formula. 
Therefore, focusing only on the first term, we can further expand it as
\begin{eqnarray}
\label{eq:E1E0Z2x}
&& \left[ \langle N | \, \frac{1}{8}\sum_{i,j}^{i\neq j} F_{iijj} Q_i^2 Q_j^2  |N \rangle \langle N |   -\frac{1}{2} \sum_k\frac{\partial^2}{\partial Q_k^2} | N \rangle \right] 
%\nonumber\\&& 
+ \left[ \langle N | \,  \frac{1}{24}\sum_i F_{iiii} Q_i^4 |N \rangle \langle N |   -\frac{1}{2} \sum_k\frac{\partial^2}{\partial Q_k^2} | N \rangle \right] \nonumber\\
&& = -\frac{1}{16}\sum_{i,j,k}^{\substack{i\neq j, i\neq k \\ j \neq k}} F_{iijj}  \left[ \langle Q_i^2 \rangle_0 \right] \left[ \langle Q_j^2 \rangle_0 \right]  \left[ \left\langle \frac{\partial^2}{\partial Q_k^2} \right\rangle_0 \right] 
%\nonumber\\&& 
-\frac{1}{16}\sum_{i,j,k}^{\substack{i\neq j, i\neq k \\ j = k}}F_{iijj}  \left[ \langle Q_i^2 \rangle_0 \right] \left[ \langle Q_j^2 \rangle_0 \left\langle \frac{\partial^2}{\partial Q_j^2} \right\rangle_0 \right] 
%\nonumber\\&& 
-\frac{1}{16}\sum_{i,j,k}^{\substack{i\neq j, i = k \\ j \neq k}}F_{iijj}  \left[ \langle Q_j^2 \rangle_0 \right] \left[ \langle Q_i^2 \rangle_0 \left\langle \frac{\partial^2}{\partial Q_i^2} \right\rangle_0 \right] \nonumber\\
&& -\frac{1}{48}\sum_{i,j,k}^{\substack{i = j, i \neq k \\ j \neq k}} F_{iiii} \left[ \langle Q_i^4 \rangle_0 \right]  \left[ \left\langle \frac{\partial^2}{\partial Q_k^2} \right\rangle_0 \right] 
%\nonumber\\&& 
-\frac{1}{48}\sum_{i,j,k}^{\substack{i =  j, i = k \\ j = k}} F_{iiii}  \left[ \langle Q_i^4 \rangle_0 \left\langle \frac{\partial^2}{\partial Q_i^2} \right\rangle_0 \right] \nonumber\\
&& = -\frac{1}{16}\sum_{i,j,k}^{\substack{i\neq j, i\neq k \\ j \neq k}} F_{iijj}  \left[ \langle Q_i^2 \rangle_0 \right] \left[ \langle Q_j^2 \rangle_0 \right]  \left[ \left\langle \frac{\partial^2}{\partial Q_k^2} \right\rangle_0 \right] 
%\nonumber\\&& 
-\frac{1}{16}\sum_{i,j,k}^{\substack{i\neq j, i\neq k \\ j = k}} F_{iijj}  \left[ \langle Q_i^2 \rangle_0 \right] \left[ \langle Q_j^2 \rangle_0 \right]  \left[ \left\langle \frac{\partial^2}{\partial Q_k^2} \right\rangle_0 \right] %\nonumber\\&& 
-\frac{1}{16}\sum_{i,j,k}^{\substack{i\neq j, i = k \\ j \neq k}} F_{iijj}  \left[ \langle Q_i^2 \rangle_0 \right] \left[ \langle Q_j^2 \rangle_0 \right]  \left[ \left\langle \frac{\partial^2}{\partial Q_k^2} \right\rangle_0 \right] 
\nonumber\\&& 
+\frac{1}{2}\sum_{i,j}^{{i\neq j}} F_{iijj}\omega_j^2  \left[ \langle Q_i^2 \rangle_0 \right]  \left[ \langle Q_j \rangle_1\langle Q_j \rangle_{-1}  \right]\left[  \langle Q_j \rangle_{-1}\langle Q_j \rangle_1 \right]  
%\nonumber\\&& 
-\frac{1}{16}\sum_{i,j,k}^{\substack{i =  j, i \neq k \\ j \neq k}} F_{iiii} \left[ \langle Q_i^2 \rangle_0 \right]\left[ \langle Q_i^2 \rangle_0 \right]   \left[ \left\langle \frac{\partial^2}{\partial Q_k^2} \right\rangle_0 \right] 
\nonumber\\ && 
-\frac{1}{2}\sum_{i,j}^{i=j} F_{iiii}  \Big[ \langle Q_i\rangle_1\langle Q_i \rangle_{-1}  \Big]\Big[  \langle Q_i \rangle_{-1}\langle Q_i \rangle_1 \Big]\left[ \left\langle \frac{\partial^2}{\partial Q_i^2} \right\rangle_0 \right] 
%\nonumber\\&&
-\frac{1}{16}\sum_{i,j,k}^{\substack{i =  j, i = k \\ j = k}} F_{iiii}  \left[ \langle Q_i^2 \rangle_0 \right]\left[ \langle Q_i^2 \rangle_0 \right]  \left[\left\langle \frac{\partial^2}{\partial Q_i^2} \right\rangle_0 \right] 
\nonumber\\&& 
= -\frac{1}{16}\sum_{i,j,k} F_{iijj}  \left[ \langle Q_i^2 \rangle_0 \right] \left[ \langle Q_j^2 \rangle_0 \right]  \left[ \left\langle \frac{\partial^2}{\partial Q_k^2} \right\rangle_0 \right] 
%\nonumber\\&& 
+ \frac{1}{2} \sum_{i,j} F_{iijj}\omega_j^2  \left[ \langle Q_i^2 \rangle_0 \right]  \left[ \langle Q_j \rangle_1\langle Q_j \rangle_{-1}  \right]\left[  \langle Q_j \rangle_{-1}\langle Q_j \rangle_1 \right]  \nonumber\\
&& = \frac{1}{4}\sum_{i,j,k} \tilde{F}_{iijj} \omega_k (f_i+1/2)(f_j+1/2)(f_k+ 1/2) 
%\nonumber\\&& 
+ \frac{1}{2} \sum_{i,j} \tilde{F}_{iijj} \omega_j (f_i+1/2) f_j (f_j+1),
\end{eqnarray}
\end{widetext} 
where we used canonical forms (1), (5), and (13) of Table \ref{table:canonical} in the second equality, which helped lift summation index restrictions in 
the penultimate equality. We also invoked $\omega_i^2 [\langle Q_i^2\rangle_0] = -[\langle \partial^2 / \partial Q_i^2 \rangle_0]$ (see Table \ref{table:thermalavg_EN1}) as well as 
\begin{eqnarray}
\label{eq:sums}
\sum_{i,j,k} X_{ijk} &=& \sum_{i,j,k}^{\substack{i\neq j, i\neq k \\ j \neq k}} X_{ijk} +   \sum_{i,j,k}^{\substack{i\neq j, i =  k \\ j \neq k}} X_{iji} +  \sum_{i,j,k}^{\substack{i\neq j, i \neq  k \\ j = k}} X_{ijj} +  \sum_{i,j,k}^{\substack{i =  j, i \neq k \\ j \neq k}} X_{iik} \nonumber\\
&&+  \sum_{i,j,k}^{\substack{i =  j, i = k \\ j = k}} X_{iii} ,
\end{eqnarray}
where $X_{ijk}$ is a general summand (no index permutation symmetry is assumed).
The last equality of Eq.\ (\ref{eq:E1E0Z2x}) follows from the thermal Born--Huang rules (Tables \ref{table:thermalavg_EN1} and \ref{table:thermalavg_EN2}).

An inspection into Tables \ref{table:thermalavg_EN1} and  \ref{table:thermalavg_EN2} indicates that 
the terms in $\hat{V}^{(1)}$ appearing in Eq.\ (\ref{eq:E1E0Z}) exhaust all cases of nonvanishing contributions. 
Remembering that the second term of Eq.\ (\ref{eq:E1E0Z2}) should lead to the same reduced expression as Eq.\ (\ref{eq:E1E0Z2x}), we gather
all the terms and find
\begin{eqnarray}
\left[ E_N^{(1)}E_N^{(0)} \right] &=&  \sum_{i,k}  \tilde{\bar{F}}_{ii} \omega_k (f_i+1/2) (f_k+1/2) \nonumber\\
&& + \sum_{i} \tilde{\bar{F}}_{ii} \omega_i f_i(f_i+1) \nonumber\\
&& + \frac{1}{2} \sum_{i,j,k} \tilde{F}_{iijj}\omega_k (f_i+1/2)  (f_j+1/2)(f_k+1/2) \nonumber\\
&& + \sum_{i,j} \tilde{F}_{iijj}\omega_j (f_i+1/2) f_j (f_j+1)  \nonumber\\
&& + \Omega^{(1)}V_\mathrm{ref} \nonumber\\
&=& \sum_{i} \tilde{\bar{F}}_{ii} \omega_i f_i(f_i+1) 
\nonumber\\
&& +  \sum_{i,j} \tilde{F}_{iijj}\omega_j (f_i+1/2) f_j (f_j+1)  + \Omega^{(1)}U^{(0)} \nonumber\\
%&\equiv& \left[ E_N^{(1)}E_N^{(0)} \right]_L + \left[ E_N^{(1)} \right]_L \left[ E_N^{(0)} \right]_L \nonumber\\
&\equiv& \left[ E_N^{(1)}E_N^{(0)} \right]_L + \Omega^{(1)}U^{(0)},
\end{eqnarray}
where the reduced analytical formulas for $\Omega^{(1)}$ [Eq.\ (\ref{eq:Omg1_Ag})] and $U^{(0)}$ [Eq.\ (\ref{eq:U_0})] were used. 
The unlinked terms add up to become $\Omega^{(1)}U^{(0)}$ (which is quadratic with system size and non-size-consistent).
The rest of the terms consist in the linked part of $[ E_N^{(1)}E_N^{(0)}]$, which is indicated by subscript $L$.

Going back to Eq.\ (\ref{eq:U1recursion}), we obtain the sum-over-modes analytical formula of $U^{(1)}$ as
\begin{eqnarray}
\label{eq:U1AG}
U^{(1)} &=& \left[E_N^{(1)}\right] - \beta \left( \left[E_N^{(1)} E_N^{(0)}\right] - \Omega^{(1)} U^{(0)} \right) \nonumber\\
&=& \sum_i \tilde{\bar{F}}_{ii} {(f_i+1/2)}{}+ \frac{1}{2}\sum_{i,j}\tilde{F}_{iijj} (f_i+1/2)(f_j+1/2) \nonumber\\
&&- {\beta}\sum_i\tilde{\bar{F}}_{ii}\omega_i f_i (f_i+1) \nonumber\\
  & & - {\beta} \sum_{i,j} \tilde{F}_{iijj}\omega_j (f_i+1/2) f_j (f_j+1)\\
  &\equiv& \left[E_N^{(1)}\right]_L - \beta \left[E_N^{(1)} E_N^{(0)}\right]_L, \label{eq:U1linked}
\end{eqnarray}
where the exact cancellation of unlinked terms takes place, leaving only linked contributions. 
Therefore, $U^{(1)}$ is diagrammatically size-consistent and written symbolically as Eq.\ (\ref{eq:U1linked}).

An alternative derivation of $U^{(1)}$ relies on Eq.\ (\ref{eq:U_exact}). Identifying the $n$th $\lambda$-derivative as the $n$th-order 
perturbation correction [Eq.\ (\ref{eq:X_perturbed})], we immediately obtain the same sum-over-modes formula as
\begin{eqnarray}
\label{eq:U1derivative}
    U^{(1)} &=& \Omega^{(1)} + \beta \frac{\partial \Omega^{(1)}}{\partial \beta} \nonumber\\
     &=& \sum_i \tilde{\bar{F}}_{ii} {(f_i+1/2)}{}
      + \frac{1}{2}\sum_{i,j}\tilde{F}_{iijj} (f_i+1/2)(f_j+1/2)\nonumber\\
     &&- {\beta}\sum_i\tilde{\bar{F}}_{ii}\omega_i f_i (f_i+1) \nonumber\\
  & & - {\beta} \sum_{i,j}  \tilde{F}_{iijj}\omega_j (f_i+1/2) f_j (f_j+1),
           % &=& \Omega^{(1)} + \beta\frac{\partial \Omega^{(1)}}{\partial f_i} \frac{\partial f_i}{\partial \beta} + \beta\frac{\partial \Omega^{(1)}}{\partial f_j} \frac{\partial f_j}{\partial \beta},
\end{eqnarray}
where we used Eq.\ (\ref{eq:Omg1_Ag}) for $\Omega^{(1)}$ and
\begin{eqnarray}
    \frac{\partial f_i}{\partial \beta} = - \omega_i f_i(f_i+1). \label{eq:f_deriv}
\end{eqnarray}
%which comes from Eq.\ (\ref{eq:BEfunction}).

% =====================================
% Omega(2)
% =====================================
\subsection{Second-order correction to the grand potential\label{section:secondorderFTPT}}

From the recursion of Eq.\ (\ref{eq:recursionOmg}), the sum-over-states analytical formula
for $\Omega^{(2)}$ is obtained as
\begin{eqnarray}
\label{eq:E2PT_recursion}
    \Omega^{(2)} = \left [E^{(2)}_N\right] - \frac{\beta}{2} \left( \left[{E^{(1)}_N E^{(1)}_N} \right] - \Omega^{(1)} \Omega^{(1)}  \right).
\end{eqnarray}
Each term can then be reduced to a sum-over-modes analytical formula
by evaluating the thermal averages using the thermal Born--Huang rules (Tables \ref{table:thermalavg_EN1} and \ref{table:thermalavg_EN2}) 
and then by consolidating the resulting terms with the aid of the canonical forms (Table \ref{table:canonical}).
After tedious, but straightforward algebraic manipulations described in Appendixes \ref{appendix:algebraicreduction} and \ref{appendix:algebraicreduction2}, 
we can formally write
\begin{eqnarray}
 \Omega^{(2)} &=& \left [E^{(2)}_N\right] _L - \frac{\beta}{2}\left( \left[{E^{(1)}_N E^{(1)}_N} \right]_L + \left[ E^{(1)}_N \right]_L\left[ E^{(1)}_N \right]_L - \Omega^{(1)} \Omega^{(1)}  \right)
 \nonumber\\&=& \left [E^{(2)}_N\right] _L- \frac{\beta}{2} \left[{E^{(1)}_N E^{(1)}_N} \right]_L. \label{eq:linked2}
\end{eqnarray}
The exact cancellation of the unlinked terms 
takes place and leaves
only the linked, and thus size-consistent contributions. 
In a QFF, according to Eqs.\ (\ref{eq:E2linked}) and (\ref{eq:E1E1linked}), the final formula reads
\begin{widetext}
\begin{eqnarray}
\label{eq:Omg2_AG_final}
\Omega^{(2)}& =& 
\sum_{i}^{\mathrm{denom.} \neq 0} \frac{\tilde{F}_{i} \tilde{F}_{i}}{\omega_{i}}f_i 
+ \sum_{i}^{\mathrm{denom.} \neq 0} \frac{\tilde{F}_{i} \tilde{F}_{i}}{-\omega_{i}}(f_i+1)  
+ 2 \sum_{i, j}^{\mathrm{denom.} \neq 0} \frac{\tilde{F}_{i} \tilde{F}_{i j j}}{\omega_{i}}f_i(f_j+1/2) 
+ 2 \sum_{i, j}^{\mathrm{denom.} \neq 0} \frac{\tilde{F}_{i} \tilde{F}_{i j j}}{-\omega_{i}}(f_i+1)(f_j+1/2) \nonumber\\
& & + \sum_{i, j, k}^{\mathrm{denom.} \neq 0} \frac{\tilde{F}_{i j j} \tilde{F}_{i k k}}{\omega_{i}}f_i(f_j+1/2)(f_k+1/2)  
+ \sum_{i, j, k}^{\mathrm{denom.} \neq 0} \frac{\tilde{F}_{i j j} \tilde{F}_{i k k}}{-\omega_{i}}(f_i+1)(f_j+1/2)(f_k+1/2) 
+\frac{1}{2} \sum_{i, j}^{\mathrm{denom.} \neq 0} \frac{\tilde{\bar{F}}_{i j} \tilde{\bar{F}}_{i j}}{\omega_{i}+\omega_{j}}f_if_j \nonumber\\
& &+ \frac{1}{2} \sum_{i, j}^{\mathrm{denom.} \neq 0} \frac{\tilde{\bar{F}}_{i j} \tilde{\bar{F}}_{i j}}{-\omega_{i}-\omega_{j}}(f_i+1)(f_j+1)
+  \sum_{i, j, k}^{\mathrm{denom.} \neq 0} \frac{\tilde{\bar{F}}_{i j} \tilde{F}_{i j k k}}{\omega_{i}+\omega_{j}} f_if_j(f_k+1/2) 
+  \sum_{i, j, k}^{\mathrm{denom.} \neq 0} \frac{\tilde{\bar{F}}_{i j} \tilde{F}_{i j k k}}{-\omega_{i}-\omega_{j}} (f_i+1)(f_j+1)(f_k+1/2)\nonumber\\
& & + \frac{1}{2} \sum_{i ,j, k, l}^{\mathrm{denom.} \neq 0} \frac{\tilde{F}_{i j k k} \tilde{F}_{i j l l}}{\omega_{i}+\omega_{j}}f_if_j(f_k+1/2)(f_l+1/2)
+ \frac{1}{2} \sum_{i ,j, k, l}^{\mathrm{denom.} \neq 0} \frac{\tilde{F}_{i j k k} \tilde{F}_{i j l l}}{-\omega_{i}-\omega_{j}}(f_i+1)(f_j+1)(f_k+1/2)(f_l+1/2)\nonumber\\
& &+  \sum_{i , j}^{\mathrm{denom.} \neq 0} \frac{\tilde{\bar{F}}_{i j} \tilde{\bar{F}}_{i j}}{\omega_i-\omega_j}{f_i(f_j+1)} 
+2 \sum_{i, j, k}^{\mathrm{denom.} \neq 0}  \frac{\tilde{\bar{F}}_{i j} \tilde{F}_{i j k k}}{\omega_i-\omega_j} {f_i(f_j+1)(f_k+1/2)}
+ \sum_{i , j, k, l}^{\mathrm{denom.} \neq 0} \frac{\tilde{F}_{i j k k} \tilde{F}_{i j l l}}{\omega_i-\omega_j}{f_i(f_j+1)}(f_k+1/2)(f_l+1/2) \nonumber\\
& &+ \frac{1}{6} \sum_{i,j,k}^{\mathrm{denom.\neq 0}}  \frac{\tilde{F}_{ijk} \tilde{F}_{ijk} } {\omega_i+\omega_j+\omega_k} f_if_jf_k 
+ \frac{1}{6} \sum_{i,j,k}^{\mathrm{denom.\neq 0}}  \frac{\tilde{F}_{ijk} \tilde{F}_{ijk} } {-\omega_i-\omega_j-\omega_k} (f_i+1)(f_j+1)(f_k+1)
+ \frac{1}{2} \sum_{i,j,k}^{\mathrm{denom.\neq 0}}  \frac{\tilde{F}_{ijk} \tilde{F}_{ijk} }{\omega_i+\omega_j-\omega_k} f_if_j(f_k+1) \nonumber\\
& &+ \frac{1}{2} \sum_{i,j,k}^{\mathrm{denom.\neq 0}}  \frac{\tilde{F}_{ijk} \tilde{F}_{ijk} }{-\omega_i-\omega_j+\omega_k} (f_i+1)(f_j+1)f_k
+\frac{1}{24} \sum_{i,j,k,l}^{\mathrm{denom.}\neq 0}  \frac{\tilde{F}_{ijkl} \tilde{F}_{ijkl} }{{\omega_i+\omega_j+\omega_k+\omega_l}} {f_if_jf_kf_l} \nonumber\\
& &+\frac{1}{24} \sum_{i,j,k,l}^{\mathrm{denom.}\neq 0}  \frac{\tilde{F}_{ijkl} \tilde{F}_{ijkl} }{{-\omega_i-\omega_j-\omega_k-\omega_l}} (f_i+1)(f_j+1)(f_k+1)(f_l+1)  
+ \frac{1}{6} \sum_{i,j,k,l}^{\mathrm{denom.}\neq 0}  \frac{\tilde{F}_{ijkl} \tilde{F}_{ijkl} }{\omega_i+\omega_j+\omega_k-\omega_l} f_if_jf_k(f_l+1) \nonumber\\
& & + \frac{1}{6} \sum_{i,j,k,l}^{\mathrm{denom.}\neq 0}  \frac{\tilde{F}_{ijkl} \tilde{F}_{ijkl} }{-\omega_i-\omega_j-\omega_k+\omega_l} (f_i+1)(f_j+1)(f_k+1)f_l 
+  \frac{1}{4} \sum_{i,j,k,l}^{\mathrm{denom.}\neq 0} \frac{\tilde{F}_{ijkl} \tilde{F}_{ijkl} }{\omega_i+\omega_j-\omega_k-\omega_l} {f_if_j(f_k+1)(f_l+1)}{} \nonumber\\
& & - \frac{\beta}{2}\sum_{i}^{\mathrm{denom.} = 0} {\tilde{F}_{i} \tilde{F}_{i}}{f_i} 
- \frac{\beta}{2}  \sum_{i}^{\mathrm{denom.} = 0} {\tilde{F}_{i} \tilde{F}_{i}}{(f_i+1)} 
- {\beta} \sum_{i, j}^{\mathrm{denom.} = 0} {\tilde{F}_{i} \tilde{F}_{i j j}}{f_i}(f_j+1/2) 
- {\beta} \sum_{i, j}^{\mathrm{denom.} = 0} {\tilde{F}_{i} \tilde{F}_{i j j}}{(f_i+1)}(f_j+1/2) \nonumber\\
& & - \frac{\beta}{2} \sum_{i, j, k}^{\mathrm{denom.} = 0} {\tilde{F}_{i j j} \tilde{F}_{i k k}}f_i(f_j+1/2)(f_k+1/2)
 - \frac{\beta}{2}  \sum_{i, j, k}^{\mathrm{denom.} = 0} {\tilde{F}_{i j j} \tilde{F}_{i k k}}(f_i+1)(f_j+1/2)(f_k+1/2) 
- \frac{\beta}{4} \sum_{i, j}^{\mathrm{denom.} = 0} \tilde{\bar{F}}_{i j} \tilde{\bar{F}}_{i j} f_if_j \nonumber\\ 
& &- \frac{\beta}{4}  \sum_{i, j}^{\mathrm{denom.} = 0} \tilde{\bar{F}}_{i j} \tilde{\bar{F}}_{i j} (f_i+1)(f_j+1) 
- \frac{\beta}{2}  \sum_{i, j, k}^{\mathrm{denom.} = 0} {\tilde{\bar{F}}_{i j} \tilde{F}_{i j k k}} {f_i f_j}(f_k+1/2)   
- \frac{\beta}{2}   \sum_{i, j, k}^{\mathrm{denom.} = 0} {\tilde{\bar{F}}_{i j} \tilde{F}_{i j k k}} {(f_i+1)(f_j+1)}(f_k+1/2) \nonumber\\
& & - \frac{\beta}{4}  \sum_{i ,j, k, l}^{\mathrm{denom.} = 0} {\tilde{F}_{i j k k} \tilde{F}_{i j l l}}f_if_j(f_k+1/2)(f_l+1/2)
- \frac{\beta}{4} \sum_{i ,j, k, l}^{\mathrm{denom.} = 0} {\tilde{F}_{i j k k} \tilde{F}_{i j l l}}(f_i+1)(f_j+1)(f_k+1/2)(f_l+1/2)\nonumber\\
& & - \frac{\beta}{2} \sum_{i , j}^{\mathrm{denom.} = 0} \tilde{\bar{F}}_{i j} \tilde{\bar{F}}_{i j}{f_i(f_j+1)} 
- {\beta} \sum_{i , j, k}^{\mathrm{denom.} = 0} {\tilde{\bar{F}}_{i j} \tilde{F}_{i j k k}} {f_i(f_j+1)}{(f_k+1/2) } 
- \frac{\beta}{2} \sum_{i, j, k, l}^{\mathrm{denom.} = 0} {\tilde{F}_{i j k k} \tilde{F}_{i j l l}}{f_i(f_j+1)}(f_k+1/2)(f_l+1/2) 
\nonumber\\& & 
- \frac{\beta}{12} \sum_{i,j,k}^{\mathrm{denom.= 0}}  \tilde{F}_{ijk} \tilde{F}_{ijk}  {f_if_jf_k}{}- \frac{\beta}{12} \sum_{i,j,k}^{\mathrm{denom.= 0}}  \tilde{F}_{ijk} \tilde{F}_{ijk}  {(f_i+1)(f_j+1)(f_k+1)} 
 - \frac{\beta}{4} \sum_{i,j,k}^{\mathrm{denom.= 0}}  \tilde{F}_{ijk} \tilde{F}_{ijk}  {f_if_j(f_k+1)}
\nonumber\\& &
- \frac{\beta}{4} \sum_{i,j,k}^{\mathrm{denom.= 0}}  \tilde{F}_{ijk} \tilde{F}_{ijk}  {(f_i+1)(f_j+1)f_k} 
- \frac{\beta}{48} \sum_{i,j,k,l}^{\mathrm{denom.}= 0}  {\tilde{F}_{ijkl} \tilde{F}_{ijkl} } f_if_jf_kf_l 
\nonumber\\& &
- \frac{\beta}{48} \sum_{i,j,k,l}^{\mathrm{denom.}= 0}  {\tilde{F}_{ijkl} \tilde{F}_{ijkl} }(f_i+1)(f_j+1)(f_k+1)(f_l+1)  {} 
 - \frac{\beta}{12} \sum_{i,j,k,l}^{\mathrm{denom.}= 0}  {\tilde{F}_{ijkl} \tilde{F}_{ijkl} }{} f_if_jf_k(f_l+1) 
\nonumber\\& & - \frac{\beta}{12} \sum_{i,j,k,l}^{\mathrm{denom.}= 0}  {\tilde{F}_{ijkl} \tilde{F}_{ijkl} } (f_i+1)(f_j+1)(f_k+1)f_l{}
 - \frac{\beta}{8} \sum_{i,j,k,l}^{\mathrm{denom.}= 0}  {\tilde{F}_{ijkl} \tilde{F}_{ijkl} } f_if_j(f_k+1)(f_l+1) ,
\end{eqnarray}
\end{widetext}
where the corresponding terms in $[E^{(2)}_N]_L$ and $[E^{(1)}_NE^{(1)}_N]_L$ are listed in the same order, so that the fictitious denominators in the latter (which are required to be zero) 
can be inferred from the existing denominators in the former (which must be nonzero). 
The terms with no denominator (or those with a fictitious denominator) are divided by $k_\text{B}T$ (or multiplied by $\beta$) to restore
 the correct dimension of energy. Diagrammatically, they correspond to the so-called anomalous diagrams, of which one or more resolvent lines are erased (see below). 
 
% =====================================
% U(2)
% =====================================
\subsection{Second-order correction to the internal energy}

From Eq.\ (\ref{eq:recursion_U}), we obtain the sum-over-states analytical formula of $U^{(2)}$ as
\begin{eqnarray}
\label{eq:U2_SOS}
U^{(2)} &=& \Big[E_N^{(2)}\Big] - \beta \left(  \Big[E_N^{(1)} E_N^{(1)} \Big] - \Omega^{(1)} U^{(1)} \right) 
\nonumber\\&& - \beta \left( \Big[E_N^{(2)} E_N^{(0)}\Big] - \Omega^{(2)} U^{(0)} \right) \nonumber\\
& &+ \frac{\beta^2}{2} \left( \Big[E_N^{(1)} E_N^{(1)} E_N^{(0)}\Big] -\Omega^{(1)} \Omega^{(1)} U^{(0)} \right).
\end{eqnarray}
The first three thermal averages can be algebraically transformed to sum-over-modes analytical formulas in much the same way $U^{(1)}$ and $\Omega^{(2)}$ were reduced 
using Tables \ref{table:thermalavg_EN1}--\ref{table:thermalavg_EN2}. However, the last thermal average is taken over the product of three matrices, i.e.,
\begin{eqnarray}
\Big[E_N^{(1)} E_N^{(1)} E_N^{(0)}\Big] = \Big[ \text{Tr} \left( \bm{E}^{(1)} \hat{P} \bm{E}^{(1)} \hat{P} \bm{E}^{(0)} \right) \Big],
\end{eqnarray} 
and in order to process this using the same strategy, we will need to extend both the thermal Born--Huang rules and their canonical forms to ternary products. 
This poses a severe limitation of the algebraic reduction method, which is shared by the electronic case,\cite{Hirata2019,Hirata_2020} where new Boltzmann sum rules need 
to be discovered every time the perturbation order is raised. 

An alternative algebraic reduction starts from
\begin{eqnarray}
U^{(2)} &=& \Omega^{(2)} + \beta \frac{\partial \Omega^{(2)}}{\partial \beta} \label{eq:alternativeU2}
\end{eqnarray}
with Eq.\ (\ref{eq:f_deriv}). We can thus arrive at the reduced formula of $U^{(2)}$ by a mechanical application of Eq.\ (\ref{eq:f_deriv}) to each term of 
the $\Omega^{(2)}$ formula [Eq.\ (\ref{eq:Omg2_AG_final})], which will not be reproduced here. 

It may be tempting to think that the cancellation of unlinked terms completes within each pair of parentheses in Eq.\ (\ref{eq:U2_SOS}), but this is not the case. 
Cancellations occur, instead, across all terms except for the first term. Specifically, substituting
\begin{eqnarray}
\Big[E_N^{(2)} \Big] &=& \Big[E_N^{(2)}\Big]_L, \\
\Big[E_N^{(1)} E_N^{(1)} \Big] &=& \Big[E_N^{(1)} E_N^{(1)} \Big]_L + \Big[E_N^{(1)}\Big]_L \Big[ E_N^{(1)} \Big]_L, \\
\Big[E_N^{(2)} E_N^{(0)} \Big] &=& \Big[E_N^{(2)} E_N^{(0)} \Big]_L + \Big[E_N^{(2)}\Big]_L \Big[ E_N^{(0)} \Big]_L, \\
\Big[E_N^{(1)} E_N^{(1)} E_N^{(0)} \Big] &=& \Big[E_N^{(1)} E_N^{(1)} E_N^{(0)} \Big]_L + \Big[E_N^{(1)} E_N^{(1)} \Big]_L \Big[E_N^{(0)}\Big]_L 
\nonumber\\&&
+ 2 \Big[E_N^{(1)} E_N^{(0)} \Big]_L \Big[E_N^{(1)}\Big]_L  + \Big[E_N^{(1)}\Big]_L \Big[ E_N^{(1)} \Big]_L \Big[E_N^{(0)}\Big]_L , \nonumber\\
\end{eqnarray}
into Eq.\ (\ref{eq:U2_SOS}), and also using 
\begin{eqnarray}
U^{(0)} &=&  \Big[E_N^{(0)} \Big]_L, \\
\Omega^{(1)} &=&  \Big[E_N^{(1)} \Big]_L, \\
U^{(1)} &=&  \Big[E_N^{(1)} \Big]_L - \beta \Big[ E_N^{(1)} E_N^{(0)} \Big]_L, \\
\Omega^{(2)} &=&  \Big[E_N^{(2)} \Big]_L - \frac{\beta}{2} \Big[ E_N^{(1)} E_N^{(1)} \Big]_L,
\end{eqnarray}
we get
\begin{eqnarray}
\label{eq:U2_SOS_linked}
U^{(2)} &=& \Big[E_N^{(2)}\Big] - \beta  \Big[E_N^{(1)} E_N^{(1)} \Big]_L 
- \beta  \Big[E_N^{(2)} E_N^{(0)}\Big]_L 
\nonumber\\& &
+ \frac{\beta^2}{2} \Big[E_N^{(1)} E_N^{(1)} E_N^{(0)}\Big]_L,
\end{eqnarray}
proving the size-consistency of $U^{(2)}$. See Sec.\ \ref{section:linkedtheorem} for a linked-diagram theorem, showing that the cancellation occurs systematically 
at any perturbation order.

% =====================================
% Normal-ordered SQ
% =====================================

\section{Normal-ordered second-quantized reduction\label{section:2ndquan}}

In this section, we derive the reduced analytical formulas of $\Omega^{(n)}$ and $U^{(n)}$ in a QFF using normal-ordered second quantization at finite temperature, starting from the same sum-over-states formulas 
obtained from the recursions of Sec.\ \ref{section:recursion}. This may be contrasted with the purely algebraic reduction in Sec.\ \ref{section:AGderivation}.
The reader is referred to Appendix \ref{appendix:normalorderedSQ} for the rules of normal ordering and their rigorous derivations including a proof of
thermal Wick's theorem.

\subsection{First-order correction to the grand potential \label{section:firstSQ}}

From the recursion [Eq.\ (\ref{eq:recursionOmg})] and the degenerate RSPT [Eq.\ (\ref{eq:E_In_degen})], we write
\begin{eqnarray}
\label{eq:Omg1_SQ}
\Omega^{(1)} &=& \Big[ E_N^{(1)} \Big] \nonumber\\
&=& \Big[\langle N | \hat{V}^{(1)} | N \rangle\Big] \nonumber \\
&=& E_{\text{XVSCF}}(T)  -  V_{\text{ref}}  - \sum_i {\omega}_i \left(f_i + {1}/{2}\right),
\end{eqnarray}
where we used the rule of normal-ordered second quantization that a thermal average of normal-ordered operators is zero; hence,
the only surviving contribution comes from the constant part of the finite-temperature normal-ordered form of $\hat{V}^{(1)}$ given in Eq.\ (\ref{eq:VSQ}). 

Recalling the finite-temperature XVSCF energy expression [Eq.\ (\ref{eq:EXVSCFT})] and the definition of $\tilde{\bar{F}}_{ij}$ [Eq.\ (\ref{eq:Ftildebar})],
we see that the above reduced formula of $\Omega^{(1)}$ is the same as Eq.\ (\ref{eq:Omg1_Ag}). Not only is the same final formula 
obtained instantly by normal ordering (because much of the reduction efforts are prepaid), 
but also the final expression is more informative about the connection between  the first-order many-body perturbation
theory and XVSCF both at finite temperature. The latter point is more fully discussed in Sec.\ \ref{section:finiteT_XVSCF}. 

\subsection{First-order correction to the internal energy \label{section:firstUSQ}}

The sum-over-states formula of $U^{(1)}$ is given by Eq.\ (\ref{eq:U1recursion}). The first term has already been simplified in Eq.\ (\ref{eq:Omg1_SQ}), 
and the third term does not need any action. Only the second term needs to be reduced by normal ordering:
\begin{widetext}
\begin{eqnarray}
\label{eq:E1E0SQ}
\Big[ E_N^{(1)} E_N^{(0)} \Big] &=& \Big[ \langle N | \hat{V}^{(1)} \hat{P} \hat{H}^{(0)} | N \rangle \Big]  \nonumber\\
&=& \Big[ \langle N | \Big( E_{\mathrm{XVSCF}}(T) - V_{\text{ref}}- \sum_i \omega_i (f_i+1/2)  \Big) \, \hat{P}_0 \, \Big( V_{\text{ref}} + \sum_i \omega_i (f_i + 1/2) \Big)| N \rangle \Big] 
+ \Big[ \sum_{i,j,k} \langle N | W_{ij} \{\hat{a}_i \hat{a}_j^\dagger \} \hat{P}_2 \omega_k \{ \hat{a}_k \hat{a}_k^\dagger \} |N \rangle \Big] \nonumber\\
&=& \Omega^{(1)} U^{(0)} + \Big[ \sum_{p,q}^{\text{denom.}=0} \sum_{i,j,k}  \langle N | W_{ij} \{
\contraction[1ex]{}{\hat{a}}{_i \hat{a}_j^{\dagger} \} 
\{ \hat{a}_i }{\hat{a}}
\hat{a}_i 
\contraction[0.5ex]{}{\hat{a}}{_i^{\dagger} \} \{}{\hat{a}}
\hat{a}_j^{\dagger} \} 
\{ \bcontraction[1ex]{}{\hat{a}}{_p \hat{a}_q^{\dagger} \} | N \rangle \langle N | \{\hat{a}_p}{\hat{a}}
\hat{a}_p 
\bcontraction[0.5ex]{}{\hat{a}}{_t^{\dagger} \} | N \rangle \langle N | \{}{\hat{a}}
\hat{a}_q^{\dagger} \} | N \rangle \langle N | \{
\contraction[1ex]{}{\hat{a}}{_q \hat{a}_p^{\dagger} \} \omega_k \{\hat{a}_k}{\hat{a}}
\hat{a}_q
\contraction[0.5ex]{}{\hat{a}}{_p^{\dagger} \} \omega_k \{}{\hat{a}}
\hat{a}_p^{\dagger} \} \omega_k \{\hat{a}_k \hat{a}_k^{\dagger} \} | N \rangle \Big] 
\nonumber\\& = & 
\Omega^{(1)}U^{(0)} + \sum_{i} W_{ii}  \omega_i f_i (f_i+1),
\end{eqnarray}
\end{widetext}
where Eqs.\ (\ref{eq:VSQ}) and (\ref{eq:H0definitionSQ}) were substituted for $\hat{V}^{(1)}$ and $\hat{H}^{(0)}$, respectively. Only the constant part and $\sum_{i,j} W_{ij}\{\hat{a}_i \hat{a}_j^\dagger\}$ term
in $\hat{V}^{(1)}$ can lead to nonzero full contractions because $\hat{H}^{(0)}$ also has a constant part and $\sum_k \omega_k \{ \hat{a}_k \hat{a}_k^\dagger\}$ term.  
In the penultimate equality, the $\Omega^{(1)}$ and $U^{(0)}$ expressions in Eqs.\ (\ref{eq:Omg1_SQ}) and (\ref{eq:U_0}) were used. In the last two equalities, 
the Wick contraction rules for a projection operator discussed in Appendix \ref{appendix:PR} were invoked.
The ``denom.=0'' restriction in this case demands $\omega_p - \omega_q = \omega_i - \omega_i = 0$, which is automatically fulfilled, and is, therefore, removed in the final expression. 

Substituting the above and Eq.\ (\ref{eq:Omg1_SQ}) into the sum-over-states formula of $U^{(1)}$ [Eq.\ (\ref{eq:U1recursion})], we obtain
\begin{eqnarray}
U^{(1)} %&=& \Big[ E_N^{(1)} \Big] - \beta \left( \Big[ E_N^{(1)} E_N^{(0)} \Big]  -\Omega^{(1)}U^{(0)}\right) \nonumber\\
&=& E_{\text{XVSCF}}(T)  - V_{\text{ref}} - \sum_i {\omega}_i \left(f_i + {1}/{2}\right) 
\nonumber\\&& 
- \beta \sum_{i} W_{ii}  \omega_i f_i (f_i+1),
\end{eqnarray}
which is equal to Eq.\ (\ref{eq:U1AG}), when the definitions of $E_{\text{XVSCF}}(T)$ [Eq.\ (\ref{eq:EXVSCFT})] and $W_{ij}$ [Eq.\ (\ref{eq:Wij_appendix})] truncated after quartic force constants are substituted.
Not only is the reduction process far more expedient than the algebraic reduction in Sec.\ \ref{section:AGderivation}, but it also leads directly to the final formulas that are written 
in a more compact form given in terms of the XVSCF energy and dressed force constants. 
 This is reminiscent of the electronic case,\cite{Shavitt2009,Hirata2021} where the most streamlined analytical formulas written with
the (finite-temperature) HF quantities are obtained directly and expediently by the (finite-temperature) normal ordering.

\subsection{Second-order correction to the grand potential \label{section:secondSQ}}

Let us first break down the sum-over-states formula of $\Omega^{(2)}$ [Eq.(\ref{eq:E2PT_recursion})] as follows:
\begin{eqnarray}
\label{eq:Omg_SQ}
    \Omega^{(2)} &=& \Big[ E_N^{(2)} \Big]  - \frac{\beta}{2} \left( \Big[ E_N^{(1)} E_N^{(1)} \Big] - \Omega^{(1)} \Omega^{(1)} \right)  \nonumber\\
    & =& \Big[ \langle N | \hat{V}^{(1)} \hat{R} \hat{V}^{(1)} | N \rangle \Big] -  \frac{\beta}{2} \left ( \Big[ \langle N | \hat{V}^{(1)} \hat{P} \hat{V}^{(1)} | N \rangle \Big]  - \Omega^{(1)} \Omega^{(1)} \right) \nonumber\\
    &=& \Big[ \langle N | \hat{V}^{(1)} \hat{R}_1 \hat{V}^{(1)} | N \rangle \Big] -  \frac{\beta}{2}\, \Big[ \langle N | \hat{V}^{(1)} \hat{P}_1 \hat{V}^{(1)} | N \rangle \Big] \nonumber\\
    && + \Big[ \langle N | \hat{V}^{(1)} \hat{R}_2 \hat{V}^{(1)} | N \rangle \Big] -  \frac{\beta}{2}\, \Big[ \langle N | \hat{V}^{(1)} \hat{P}_2 \hat{V}^{(1)} | N \rangle \Big] \nonumber\\
    && + \Big[ \langle N | \hat{V}^{(1)} \hat{R}_3 \hat{V}^{(1)} | N \rangle \Big] -  \frac{\beta}{2}\, \Big[ \langle N | \hat{V}^{(1)} \hat{P}_3 \hat{V}^{(1)} | N \rangle \Big] \nonumber\\
    && + \Big[ \langle N | \hat{V}^{(1)} \hat{R}_4 \hat{V}^{(1)} | N \rangle \Big] -  \frac{\beta}{2}\, \Big[ \langle N | \hat{V}^{(1)} \hat{P}_4 \hat{V}^{(1)} | N \rangle \Big] \nonumber\\
    && -  \frac{\beta}{2} \left ( \Big[ \langle N | \hat{V}^{(1)} \hat{P}_0 \hat{V}^{(1)} | N \rangle \Big]  - \Omega^{(1)} \Omega^{(1)} \right),
\end{eqnarray}
in the case of a QFF. See Sec.\ \ref{section:projector} for the definitions of $\hat{P}_n$ and $\hat{R}_n$. Closely related terms are placed in the same line. 

Each thermal average in the right-hand side can be reduced by a mechanical application of the rules of normal-ordered second quantization (see Appendix \ref{appendix:normalorderedSQ}). 
For instance, the first term of Eq.\ (\ref{eq:Omg_SQ}) is evaluated as
\begin{eqnarray}
\label{eq:EN2_SQ_S}
&& \Big[ \langle N | \hat{V}^{(1)} \hat{R}_1 \hat{V}^{(1)} | N \rangle\Big] \nonumber\\
%&=&[ \langle N | \hat{V} \hat{R}_N \hat{V} | N \rangle ]_{\mathrm{S}} \nonumber\\ 
&&=  {\Big[ \sum_p^{\text{denom.}\neq0} \sum_{i,a} \frac{ \langle N | W_i \{ 
\contraction[0.5ex]{}{\hat{a}}{_i^{\dagger}\}  \{}{\hat{a}}
\hat{a}_i^{\dagger}\}  \{
\bcontraction[0.5ex]{}{\hat{a}}{_p\} | N \rangle \langle N| \{}{\hat{a}}
\hat{a}_p\} | N \rangle\langle N| \{
\contraction[0.5ex]{}{\hat{a}}{_p^{\dagger}\} \{}{\hat{a}}
\hat{a}_p^{\dagger}\} \{\hat{a}_a \}W_a | N \rangle }{\omega_p} \Big]} \nonumber\\
& & +   {\Big[ \sum_p^{\text{denom.}\neq0} \sum_{i,a}  \frac{ \langle N |W_i\{
\contraction[0.5ex]{}{\hat{a}}{_i\}  \{}{\hat{a}}
\hat{a}_i\}  \{
\bcontraction[0.5ex]{}{\hat{a}}{_p^{\dagger}\} | N \rangle \langle N| \{}{\hat{a}}
\hat{a}_p^{\dagger}\} | N \rangle \langle N| \{
\contraction[0.5ex]{}{\hat{a}}{_p\} \{}{\hat{a}}
\hat{a}_p\} \{\hat{a}_a^{\dagger} \}W_a| N \rangle}{-\omega_p } \Big]} \nonumber\\
&& =  \sum_i^{\mathrm{denom.} \neq 0}  \frac{W_i W_i}{\omega_i}f_i + \sum_i^{\mathrm{denom.} \neq 0}\frac{ W_i W_i }{-\omega_i}(f_i+1) ,
%&=& - \sum_i^{\mathrm{denom.} \neq 0} \frac{W_i W_i}{\omega_i}\nonumber\\
%&=& [ E_N^{(2)} ]_{\mathrm{S}} \{\text{Eq.}(\ref{eq:EN2_AG_S})\} \equiv  \{[ E_N^{(2)} ]_{\mathrm{S}}\}_{{L}}
\end{eqnarray}
where the rules pertaining the resolvent operator described in Appendix \ref{appendix:PR} were used. 
Recalling the definition of $W_i$ in Eq.\ (\ref{eq:Wi_appendix}), we see that the above two terms encompass the first six terms (or all terms sharing 
the same denominators) of Eq.\ (\ref{eq:Omg2_AG_final}). It again attests 
to the better organization of the final expressions automatically achieved by normal ordering.

The second term of Eq.\ (\ref{eq:Omg_SQ}) is reduced similarly:
\begin{eqnarray}
\label{eq:EN2_SQ_S_P}
&& -\frac{\beta}{2} \Big[ \langle N | \hat{V}^{(1)} \hat{P}_1 \hat{V}^{(1)} | N \rangle\Big] \nonumber\\
%&=&[ \langle N | \hat{V} \hat{R}_N \hat{V} | N \rangle ]_{\mathrm{S}} \nonumber\\ 
&&=  -\frac{\beta}{2}{\Big[ \sum_p^{\text{denom.}=0} \sum_{i,a} \langle N | W_i \{ 
\contraction[1ex]{}{\hat{a}}{_i^{\dagger}\}  \{}{\hat{a}}
\hat{a}_i^{\dagger}\}  \{
\bcontraction[1ex]{}{\hat{a}}{_p\} | N \rangle \langle N| \{}{\hat{a}}
\hat{a}_p\} | N \rangle  \langle N| \{
\contraction[1ex]{}{\hat{a}}{_p^{\dagger}\} \{}{\hat{a}}
\hat{a}_p^{\dagger}\} \{\hat{a}_a \}W_a | N \rangle\Big]} \nonumber\\
& &   -\frac{\beta}{2}{\Big[ \sum_p^{\text{denom.}=0} \sum_{i,a}  \langle N |W_i\{
\contraction[1ex]{}{\hat{a}}{_i\}  \{}{\hat{a}}
\hat{a}_i\}  \{
\bcontraction[1ex]{}{\hat{a}}{_p^{\dagger}\} | N \rangle \langle N| \{}{\hat{a}}
\hat{a}_p^{\dagger}\} | N \rangle \langle N| \{
\contraction[1ex]{}{\hat{a}}{_p\} \{}{\hat{a}}
\hat{a}_p\} \{\hat{a}_a^{\dagger} \}W_a| N \rangle\Big]} \nonumber\\
&& =  -\frac{\beta}{2}\sum_i^{\mathrm{denom.}= 0}  {W_i W_i}f_i  -\frac{\beta}{2} \sum_i^{\mathrm{denom.} = 0}{ W_i W_i }(f_i+1) ,
%&=& - \sum_i^{\mathrm{denom.} \neq 0} \frac{W_i W_i}{\omega_i}\nonumber\\
%&=& [ E_N^{(2)} ]_{\mathrm{S}} \{\text{Eq.}(\ref{eq:EN2_AG_S})\} \equiv  \{[ E_N^{(2)} ]_{\mathrm{S}}\}_{{L}}
\end{eqnarray}
where ``denom.=0'' means that the fictitious denominators gleaned from the corresponding terms of Eq.\ (\ref{eq:EN2_SQ_S}) are zero, i.e., 
$\omega_i = 0$. They account for the first six denominatorless terms of Eq.\ (\ref{eq:Omg2_AG_final}).

It is clear that the reduced analytical formula for the second term in each line of Eq.\ (\ref{eq:Omg_SQ}) can
be readily inferred from the formula for the first term in the same line. Specifically the former is obtained by removing all denominators and then 
replacing ``denom.$\neq$0'' by ``denom.=0.'' Hereafter, we shall, therefore, focus on the first term of each line in Eq.\ (\ref{eq:Omg_SQ}). 

The third term of Eq.\ (\ref{eq:Omg_SQ}) is transformed as
\begin{widetext}
\begin{eqnarray}
\label{eq:EN2_SQ_D}
 \Big[ \langle N | \hat{V}^{(1)} \hat{R}_2 \hat{V}^{(1)} | N \rangle \Big] 
&=&  \frac{1}{2!2!2!} {\left[ \sum_{p,q}^{\text{denom.}\neq0} \sum_{i,j,a,b} \frac{ \langle N | W_{ij} \{ %1
\contraction[1ex]{}{\hat{a}}{_i^{\dagger} \hat{a}_j^{\dagger}\} \{\hat{a}_t }{\hat{a}}
\hat{a}_i^{\dagger}
\contraction[0.5ex]{}{\hat{a}}{_j^{\dagger}\} \{}{\hat{a}}
\hat{a}_j^{\dagger}\} \{
\bcontraction[1ex]{}{\hat{a}}{_p \hat{a}_u\} | N \rangle \langle N| \{\hat{a}_q}{\hat{a}}
\hat{a}_p 
\bcontraction[0.5ex]{}{\hat{a}}{_q\} | N \rangle \langle N| \{}{\hat{a}}
\hat{a}_q\} | N \rangle \langle N| \{
\contraction[1ex]{}{\hat{a}}{_q\hat{a}_t\} \{ \hat{a}_a }{\hat{a}}
\hat{a}_q^{\dagger}
\contraction[0.5ex]{}{\hat{a}}{_p\} \{}{\hat{a}}
\hat{a}_p^{\dagger}\} \{ 
\hat{a}_a \hat{a}_b \}W_{ab}| N \rangle }{\omega_p + \omega_q} \right]} \times 4
\nonumber \\& & 
+  \frac{1}{2!2!2!}  { \left[ \sum_{p,q}^{\text{denom.}\neq0} \sum_{i,j,a,b} \frac{ \langle N | W_{ij} \{ %2
\contraction[1ex]{}{\hat{a}}{_i \hat{a}_j\} \{\hat{a}_t }{\hat{a}}
\hat{a}_i
\contraction[0.5ex]{}{\hat{a}}{_j\} \{}{\hat{a}}
\hat{a}_j\} \{
\bcontraction[1ex]{}{\hat{a}}{_p^{\dagger} \hat{a}_q^{\dagger}\} | N \rangle  \langle N| \{\hat{a}_q}{\hat{a}}
\hat{a}_p^{\dagger} 
\bcontraction[0.5ex]{}{\hat{a}}{_q^{\dagger}\} | N \rangle \langle N| \{}{\hat{a}}
\hat{a}_q^{\dagger}\} | N \rangle  \langle N| \{
\contraction[1ex]{}{\hat{a}}{_q\hat{a}_p\} \{ \hat{a}_a^{\dagger} }{\hat{a}}
\hat{a}_q
\contraction[0.5ex]{}{\hat{a}}{_t\} \{}{\hat{a}}
\hat{a}_p\} \{ 
\hat{a}_a^{\dagger} \hat{a}_b^{\dagger} \} W_{ab}| N \rangle}{-\omega_p-\omega_q} \right] } \times 4 \nonumber \\
& & +   {\left[ \sum_{p,q}^{\text{denom.}\neq0} \sum_{i,j,a,b} \frac{ \langle N | W_{ij} \{%3
\contraction[1ex]{}{\hat{a}}{_i \hat{a}_j\} \{\hat{a}_t }{\hat{a}}
\hat{a}_i
\contraction[0.5ex]{}{\hat{a}}{_j^{\dagger}\} \{}{\hat{a}}
\hat{a}_j^{\dagger}\} \{
\bcontraction[1ex]{}{\hat{a}}{_p \hat{a}_q^{\dagger}\} | N \rangle  \langle N| \{\hat{a}_q}{\hat{a}}
\hat{a}_p
\bcontraction[0.5ex]{}{\hat{a}}{_q^{\dagger}\} | N \rangle  \langle N| \{}{\hat{a}}
\hat{a}_q^{\dagger}\} | N \rangle  \langle N| \{
\contraction[1ex]{}{\hat{a}}{_q\hat{a}_p\} \{ \hat{a}_a }{\hat{a}}
\hat{a}_q
\contraction[0.5ex]{}{\hat{a}}{_p\} \{}{\hat{a}}
\hat{a}_p^{\dagger}\} \{ 
\hat{a}_a \hat{a}_b^{\dagger} \} W_{ab} | N \rangle }{\omega_p - \omega_q}  \right]}\nonumber\\
&=&  \frac{1}{2}\sum_{i, j}^{\mathrm{denom.} \neq 0} \frac{W_{ij} W_{ij} }{\omega_i+\omega_j}f_if_j +  \frac{1}{2}\sum_{i, j}^{\mathrm{denom.} \neq 0} \frac{W_{ij} W_{ij}}{-\omega_i-\omega_j} (f_i+1)(f_j+1) + \sum_{i, j}^{\mathrm{denom.} \neq 0} \frac{W_{ij} W_{ij} }{-\omega_i + \omega_j}(f_i+1) f_j ,
%&=& - \frac{1}{2!}\sum_{i, j}^{\mathrm{denom.} \neq 0} W_{ij} W_{ij}\frac{f_i + f_j + 1}{\omega_i+\omega_j} + \sum_{i, j}^{\mathrm{denom.} \neq 0} W_{ij} W_{ij}\frac{f_j(f_i+1)}{\omega_i - \omega_j} \nonumber\\
%&=& [ E_N^{(2)} ]_{\mathrm{D}} \{\text{Eq.}(\ref{eq:EN2_AG_D})\} \equiv  \{[ E_N^{(2)} ]_{\mathrm{D}}\}_{{L}}
\end{eqnarray}
where ``$\times 4$'' means that there are four distinct, equal-valued full contractions, given the permutation symmetry of $W_{ij}$ (see Appendix \ref{appendix:PR} for the subtlety arising from the $p\neq q$ versus $p=q$ cases). 
They encompass the seventh through fifteenth terms (sharing the same denominators) of Eq.\ (\ref{eq:Omg2_AG_final}).
The fourth term of Eq.\ (\ref{eq:Omg_SQ}) can be reduced analogously, leading to a similar formula in which all denominators are 
stripped and every  ``denom.$\neq$0'' restriction is replaced by ``denom.=0,'' whose fictitious denominators are 
the ones that have been removed.

The fifth term of Eq.\ (\ref{eq:Omg_SQ}) is evaluated as 
\begin{eqnarray}
\label{eq:EN2_SQ_T}
\Big[ \langle N | \hat{V}^{(1)} \hat{R}_3 \hat{V}^{(1)} | N \rangle \Big]
& = & \frac{1}{3!3!3!}  \left[ \sum_{p,q,r}^{\text{denom.}\neq0} \sum_{\substack {i,j,k \\a,b,c}} \frac{  \langle N | W_{ijk} 
\{\contraction[1.5ex]{}{\hat{a}}{{}_{i}^{\dagger}\hat{a}_{j}^{\dagger}\hat{a}_{k}^{\dagger}\}\{\hat{a}_p\hat{a}_q}{\hat{a}} 
\hat{a}_{i}^{\dagger} 
\contraction[1ex]{}{\hat{a}}{{}_{j}^{\dagger}\hat{a}_{k}^{\dagger} \}\{ \hat{a}_p}{\hat{a}} 
\hat{a}_{j}^{\dagger} 
\contraction[0.5ex]{}{\hat{a}}{{}_{k}^{\dagger} \}\{}{\hat{a}} 
\hat{a}_{k}^{\dagger} \}\{ 
\bcontraction[1.5ex]{}{\hat{a}}{{}_p \hat{a}_q \hat{a}_r \} | N \rangle \langle N | \{\hat{a}_r^{\dagger}\hat{a}_q^{\dagger} }{\hat{a}}
\hat{a}_p 
\bcontraction[1ex]{}{\hat{a}}{{}_q \hat{a}_r \} | N \rangle  \langle N | \{ \hat{a}_r} {\hat{a}}
\hat{a}_q
\bcontraction[0.5ex]{}{\hat{a}}{{}_r \} | N \rangle  \langle N | \{}{\hat{a}}
\hat{a}_r \}
| N \rangle \langle N | \{ %
\contraction[1.5ex]{}{\hat{a}}{{}_r^{\dagger}\hat{a}_q^{\dagger} \hat{a}_p^{\dagger}\} \{\hat{a}_a \hat{a}_b}{\hat{a}}
\hat{a}_r^{\dagger}
\contraction[1ex]{}{\hat{a}}{{}_q^{\dagger} \hat{a}_p^{\dagger}\} \{\hat{a}_a}{\hat{a}}
\hat{a}_q^{\dagger}
\contraction[0.5ex]{}{\hat{a}}{{}_p^{\dagger}\} \{}{\hat{a}}
\hat{a}_p^{\dagger}\} \{\hat{a}_a \hat{a}_b \hat{a}_c \} W_{abc}| N \rangle }{\omega_p+\omega_q+\omega_r} \right] \times 36 \nonumber \\
& & +  \frac{1}{3!3!3!} \left[ \sum_{p,q,r}^{\text{denom.}\neq0} \sum_{\substack {i,j,k \\a,b,c}} \frac{ \langle N | W_{ijk}
\{\contraction[1.5ex]{}{\hat{a}}{{}_{i}\hat{a}_{j}\hat{a}_{k}\}\{\hat{a}_p^{\dagger}\hat{a}_q^{\dagger}}{\hat{a}} 
\hat{a}_{i} 
\contraction[1ex]{}{\hat{a}}{{}_q\hat{a}_r \}\{ \hat{a}_{i}^{\dagger}}{\hat{a}} 
\hat{a}_{j} 
\contraction[0.5ex]{}{\hat{a}}{{}_r \}\{}{\hat{a}}
\hat{a}_{k} \}\{ 
\bcontraction[1.5ex]{}{\hat{a}}{{}_p^{\dagger} \hat{a}_q^{\dagger} \hat{a}_r^{\dagger} \} | N \rangle \langle N | \{\hat{a}_r\hat{a}_q }{\hat{a}}
\hat{a}_p^{\dagger} 
\bcontraction[1ex]{}{\hat{a}}{{}_q^{\dagger} \hat{a}_r^{\dagger} \} | N \rangle  \langle N | \{ \hat{a}_r} {\hat{a}} 
\hat{a}_q^{\dagger}
\bcontraction[0.5ex]{}{\hat{a}}{{}_r^{\dagger} \} | N \rangle \langle N | \{}{\hat{a}}
\hat{a}_r^{\dagger} \} | N \rangle \langle N | \{ %
\contraction[1.5ex]{}{\hat{a}}{{}_r\hat{a}_q \hat{a}_p\} \{\hat{a}_a^{\dagger} \hat{a}_b^{\dagger}}{\hat{a}}
\hat{a}_r
\contraction[1ex]{}{\hat{a}}{{}_q \hat{a}_p\} \{\hat{a}_a^{\dagger}}{\hat{a}}
\hat{a}_q
\contraction[0.5ex]{}{\hat{a}}{{}_p\} \{}{\hat{a}}
\hat{a}_p\} \{\hat{a}_a^{\dagger} \hat{a}_b^{\dagger} \hat{a}_c^{\dagger}\} W_{abc}| N \rangle }{-\omega_p-\omega_q-\omega_r} \right ]\times 36 \nonumber \\
& & + \frac{1}{2!2!2! }\left[ \sum_{p,q,r}^{\text{denom.}\neq0} \sum_{\substack {i,j,k \\a,b,c}} \frac{ \langle N | W_{ijk} 
\{\contraction[1.5ex]{}{\hat{a}}{{}_{i}\hat{a}_{j}^{\dagger}\hat{a}_{k}^{\dagger}\}\{\hat{a}_p\hat{a}_q}{\hat{a}} 
\hat{a}_{i}
\contraction[1ex]{}{\hat{a}}{{}_{j}^{\dagger}\hat{a}_{k}^{\dagger} \}\{ \hat{a}_p}{\hat{a}}
\hat{a}_{j}^{\dagger} 
\contraction[0.5ex]{}{\hat{a}}{{}_r^{\dagger} \}\{}{\hat{a}}
\hat{a}_{k}^{\dagger} \}\{ 
\bcontraction[1.5ex]{}{\hat{a}}{{}_p \hat{a}_q \hat{a}_r^{\dagger} \} | N \rangle \langle N | \{\hat{a}_r\hat{a}_q^{\dagger} }{\hat{a}} 
\hat{a}_p 
\bcontraction[1ex]{}{\hat{a}}{{}_q \hat{a}_r^{\dagger} \} | N \rangle \langle N | \{ \hat{a}_r} {\hat{a}} 
\hat{a}_q
\bcontraction[0.5ex]{}{\hat{a}}{{}_r^{\dagger} \} | N \rangle  \langle N | \{}{\hat{a}}
\hat{a}_r^{\dagger} \} | N \rangle \langle N | \{ %
\contraction[1.5ex]{}{\hat{a}}{{}_r\hat{a}_q^{\dagger} \hat{a}_p^{\dagger}\} \{\hat{a}_a \hat{a}_b}{\hat{a}}
\hat{a}_r
\contraction[1ex]{}{\hat{a}}{{}_q^{\dagger} \hat{a}_p^{\dagger}\} \{\hat{a}_a}{\hat{a}}
\hat{a}_q^{\dagger}
\contraction[0.5ex]{}{\hat{a}}{{}_p^{\dagger}\} \{}{\hat{a}}
\hat{a}_p^{\dagger}\} \{\hat{a}_a \hat{a}_b \hat{a}_c^{\dagger}\} W_{abc}| N \rangle }{\omega_p+\omega_q-\omega_r} \right] \times 4 \nonumber \\
& &  + \frac{1}{2!2!2!} \left[\sum_{p,q,r}^{\text{denom.}\neq0} \sum_{\substack {i,j,k \\a,b,c}} \frac{ \langle N | W_{ijk}
\{\contraction[1.5ex]{}{\hat{a}}{{}_{i}^{\dagger}\hat{a}_{j}\hat{a}_{k}\}\{\hat{a}_p^{\dagger}\hat{a}_q^{\dagger}}{\hat{a}} 
\hat{a}_{i}^{\dagger}
\contraction[1ex]{}{\hat{a}}{{}_{j}\hat{a}_{k} \}\{ \hat{a}_p^{\dagger}}{\hat{a}}
\hat{a}_{j}
\contraction[0.5ex]{}{\hat{a}}{{}_{k} \}\{}{\hat{a}}
\hat{a}_{k} \}\{ 
\bcontraction[1.5ex]{}{\hat{a}}{{}_p^{\dagger} \hat{a}_q^{\dagger} \hat{a}_r \} | N \rangle \langle N | \{\hat{a}_r^{\dagger}\hat{a}_q }{\hat{a}} 
\hat{a}_p^{\dagger} 
\bcontraction[1ex]{}{\hat{a}}{{}_q^{\dagger} \hat{a}_r \} | N \rangle \langle N | \{ \hat{a}_r^{\dagger}} {\hat{a}} 
\hat{a}_q^{\dagger}
\bcontraction[0.5ex]{}{\hat{a}}{{}_r \} | N \rangle  \langle N | \{}{\hat{a}}
\hat{a}_r \} | N \rangle  \langle N | \{ %
\contraction[1.5ex]{}{\hat{a}}{{}_r^{\dagger}\hat{a}_q \hat{a}_p\} \{\hat{a}_a^{\dagger} \hat{a}_b^{\dagger}}{\hat{a}}
\hat{a}_r^{\dagger}
\contraction[1ex]{}{\hat{a}}{{}_q \hat{a}_p\} \{\hat{a}_a^{\dagger}}{\hat{a}}
\hat{a}_q
\contraction[0.5ex]{}{\hat{a}}{{}_p\} \{}{\hat{a}}
\hat{a}_p\} \{\hat{a}_a^{\dagger} \hat{a}_b^{\dagger} \hat{a}_c \} W_{abc} | N \rangle }{-\omega_p-\omega_q+\omega_r} \right] \times 4 \nonumber \\
&=&\frac{1}{6} \sum_{i,j,k}^{\mathrm{denom.\neq 0}}  \frac{W_{ijk} W_{ijk}}{\omega_i+\omega_j+\omega_k}f_if_jf_k + \frac{1}{6} \sum_{i,j,k}^{\mathrm{denom.\neq 0}}  \frac{W_{ijk} W_{ijk}}{-\omega_i-\omega_j-\omega_k}(f_i+1)(f_j+1)(f_k+1)   \nonumber\\
& & + \frac{1}{2} \sum_{i,j,k}^{\mathrm{denom.\neq 0}}  \frac{W_{ijk} W_{ijk}}{-\omega_i+\omega_j+\omega_k}(f_i+1)f_jf_k+ \frac{1}{2} \sum_{i,j,k}^{\mathrm{denom.\neq 0}}  \frac{W_{ijk} W_{ijk}}{\omega_i-\omega_j-\omega_k}f_i(f_j+1)(f_k+1) ,
%& =&  - \frac{1}{3!} \sum_{i,j,k}^{\mathrm{denom.\neq 0}} W_{ijk} W_{ijk} \frac{f_if_j+f_if_k+f_jf_k+f_i+f_j+f_k+1}{\omega_i+\omega_j+\omega_k} \nonumber\\
%& & + \frac{1}{2!} \sum_{i,j,k}^{\mathrm{denom.\neq 0}} W_{ijk} W_{ijk} \frac{f_jf_i - f_if_k - f_jf_k - f_k}{\omega_i+\omega_j-\omega_k}\nonumber\\
%&=& [ E_N^{(2)} ]_{\mathrm{T}} \{ \text{Eq.}(\ref{eq:EN2_AG_T}) \} \equiv  \{[ E_N^{(2)} ]_{\mathrm{T}}\}_{{L}},
\end{eqnarray}
which is equal to the sum of the sixteenth through nineteenth terms of Eq.\ (\ref{eq:Omg2_AG_final}).
The sixth term of Eq.\ (\ref{eq:Omg_SQ}) is the same as above, but without the denominators, which (i.e., fictitious denominators) are required to be zero.   

The seventh term of Eq.\ (\ref{eq:Omg_SQ}) is transformed similarly:
\begin{eqnarray}
\label{eq:EN2_SQ_Q}
%\resizebox{.9\hsize}{!}{
\Big[ \langle N | \hat{V}^{(1)} \hat{R}_4 \hat{V}^{(1)} | N \rangle \Big]
&=& 
\frac{1}{4!4!4!}  \left[ \sum_{p,q,r,s}^{\text{denom.}\neq 0} \sum_{\substack {i,j,k,l \\ a,b,c,d} } \frac{ \langle N | W_{ijkl}\{ %2
\contraction[2ex]{}{\hat{a}}{_i^{\dagger} \hat{a}_j^{\dagger}\hat{a}_k^{\dagger}\hat{a}_l^{\dagger} \}\{ \hat{a}_p \hat{a}_q \hat{a}_r}{\hat{a}}
\hat{a}_i^{\dagger}
\contraction[1.5ex]{}{\hat{a}}{_j^{\dagger}\hat{a}_k^{\dagger}\hat{a}_l^{\dagger} \}\{ \hat{a}_p \hat{a}_q }{\hat{a}}
\hat{a}_j^{\dagger}
\contraction[1ex]{}{\hat{a}}{_k^{\dagger}\hat{a}_l^{\dagger} \}\{ \hat{a}_p }{\hat{a}}
\hat{a}_k^{\dagger}
\contraction[0.5ex]{}{\hat{a}}{_l^{\dagger} \}\{ }{\hat{a}}
\hat{a}_l^{\dagger} \} \{
\bcontraction[2ex]{}{\hat{a}}{_p \hat{a}_q \hat{a}_r \hat{a}_s \} | N \rangle \langle N | \{ \hat{a}_s^{\dagger} \hat{a}_r^{\dagger} \hat{a}_q^{\dagger} }{\hat{a}}
\hat{a}_p
\bcontraction[1.5ex]{}{\hat{a}}{_q \hat{a}_r \hat{a}_s \} | N \rangle \langle N | \{ \hat{a}_s^{\dagger} \hat{a}_r^{\dagger} }{\hat{a}}
\hat{a}_q
\bcontraction[1ex]{}{\hat{a}}{_r \hat{a}_s \} | N \rangle \langle N | \{ \hat{a}_s^{\dagger}}{\hat{a}}
\hat{a}_r
\bcontraction[0.5ex]{}{\hat{a}}{_s \} | N \rangle \langle N | \{}{\hat{a}}
\hat{a}_s \} | N \rangle \langle N | \{
\contraction[2ex]{}{\hat{a}}{_s^{\dagger} \hat{a}_r^{\dagger} \hat{a}_q^{\dagger} \hat{a}_p^{\dagger} \}\{ \hat{a}_a \hat{a}_b \hat{a}_c  }{\hat{a}}
\hat{a}_s^{\dagger}
\contraction[1.5ex]{}{\hat{a}}{_r^{\dagger} \hat{a}_q^{\dagger} \hat{a}_p^{\dagger} \}\{ \hat{a}_a \hat{a}_b}{\hat{a}}
\hat{a}_r^{\dagger}
\contraction[1ex]{}{\hat{a}}{_q^{\dagger} \hat{a}_p^{\dagger} \}\{ \hat{a}_a }{\hat{a}}
\hat{a}_q^{\dagger}
\contraction[0.5ex]{}{\hat{a}}{_p^{\dagger} \}\{ }{\hat{a}}
\hat{a}_p^{\dagger} \}\{ \hat{a}_a \hat{a}_b \hat{a}_c \hat{a}_d \} W_{abcd} | N \rangle }{\omega_p + \omega_q +\omega_r+\omega_s} \right] \times 576 \nonumber\\
& &+ \frac{1}{4!4!4!} \left[ \sum_{p,q,r,s}^{\text{denom.}\neq 0} \sum_{\substack {i,j,k,l \\ a,b,c,d} } \frac{ \langle N |  W_{ijkl}\{ %1
\contraction[2ex]{}{\hat{a}}{_i \hat{a}_j\hat{a}_k\hat{a}_l \}\{ \hat{a}_p^{\dagger} \hat{a}_q^{\dagger} \hat{a}_r^{\dagger}}{\hat{a}}
\hat{a}_i
\contraction[1.5ex]{}{\hat{a}}{_j\hat{a}_k\hat{a}_l \}\{ \hat{a}_p^{\dagger} \hat{a}_q^{\dagger} }{\hat{a}}
\hat{a}_j
\contraction[1ex]{}{\hat{a}}{_k\hat{a}_l \}\{ \hat{a}_p^{\dagger} }{\hat{a}}
\hat{a}_k
\contraction[0.5ex]{}{\hat{a}}{_l \}\{ }{\hat{a}}
\hat{a}_l \} \{
\bcontraction[2ex]{}{\hat{a}}{_p^{\dagger} \hat{a}_q^{\dagger} \hat{a}_r^{\dagger} \hat{a}_s^{\dagger} \} | N \rangle \langle N | \{ \hat{a}_s \hat{a}_r \hat{a}_q }{\hat{a}}
\hat{a}_p^{\dagger}
\bcontraction[1.5ex]{}{\hat{a}}{_q^{\dagger} \hat{a}_r^{\dagger} \hat{a}_s^{\dagger} \} | N \rangle \langle N | \{ \hat{a}_s \hat{a}_r }{\hat{a}}
\hat{a}_q^{\dagger}
\bcontraction[1ex]{}{\hat{a}}{_r^{\dagger} \hat{a}_s^{\dagger} \} | N \rangle \langle N | \{ \hat{a}_s}{\hat{a}}
\hat{a}_r^{\dagger}
\bcontraction[0.5ex]{}{\hat{a}}{_s^{\dagger} \} | N \rangle \langle N | \{}{\hat{a}}
\hat{a}_s^{\dagger} \} | N \rangle \langle N | \{
\contraction[2ex]{}{\hat{a}}{_s \hat{a}_r \hat{a}_q \hat{a}_p \}\{ \hat{a}_a^{\dagger} \hat{a}_b^{\dagger} \hat{a}_c^{\dagger}  }{\hat{a}}
\hat{a}_s
\contraction[1.5ex]{}{\hat{a}}{_r \hat{a}_q \hat{a}_p \}\{ \hat{a}_a^{\dagger} \hat{a}_b^{\dagger}}{\hat{a}}
\hat{a}_r
\contraction[1ex]{}{\hat{a}}{_q \hat{a}_p \}\{ \hat{a}_a^{\dagger} }{\hat{a}}
\hat{a}_q
\contraction[0.5ex]{}{\hat{a}}{_p \}\{ }{\hat{a}}
\hat{a}_p\}\{ \hat{a}_a^{\dagger} \hat{a}_b^{\dagger} \hat{a}_c^{\dagger} \hat{a}_d^{\dagger}\} W_{abcd}| N \rangle }{-\omega_p - \omega_q - \omega_r - \omega_s} \right] \times 576 \nonumber\\
& & + \frac{1}{3!3!3!} \left[ \sum_{p,q,r,s}^{\text{denom.}\neq 0} \sum_{\substack {i,j,k,l \\ a,b,c,d} } \frac{  \langle N | \{ W_{ijkl} %3
\contraction[2ex]{}{\hat{a}}{_i \hat{a}_j^{\dagger}\hat{a}_k^{\dagger}\hat{a}_l^{\dagger} \}\{ \hat{a}_p \hat{a}_q \hat{a}_r}{\hat{a}}
\hat{a}_i
\contraction[1.5ex]{}{\hat{a}}{_i^{\dagger}\hat{a}_j^{\dagger}\hat{a}_k^{\dagger} \}\{ \hat{a}_p \hat{a}_q }{\hat{a}}
\hat{a}_j^{\dagger}
\contraction[1ex]{}{\hat{a}}{_k^{\dagger}\hat{a}_l^{\dagger} \}\{ \hat{a}_p }{\hat{a}}
\hat{a}_k^{\dagger}
\contraction[0.5ex]{}{\hat{a}}{_l^{\dagger} \}\{ }{\hat{a}}
\hat{a}_l^{\dagger} \} \{
\bcontraction[2ex]{}{\hat{a}}{_p \hat{a}_q \hat{a}_r \hat{a}_s \} | N \rangle \langle N | \{ \hat{a}_s^{\dagger} \hat{a}_r^{\dagger} \hat{a}_q^{\dagger} }{\hat{a}}
\hat{a}_p
\bcontraction[1.5ex]{}{\hat{a}}{_q \hat{a}_r \hat{a}_s \} | N \rangle \langle N | \{ \hat{a}_s^{\dagger} \hat{a}_r^{\dagger} }{\hat{a}}
\hat{a}_q
\bcontraction[1ex]{}{\hat{a}}{_r \hat{a}_s \} | N \rangle \langle N | \{ \hat{a}_s^{\dagger}}{\hat{a}}
\hat{a}_r
\bcontraction[0.5ex]{}{\hat{a}}{_s \} | N \rangle \langle N | \{}{\hat{a}}
\hat{a}_s^{\dagger} \} | N \rangle \langle N | \{
\contraction[2ex]{}{\hat{a}}{_s \hat{a}_r \hat{a}_q \hat{a}_p \}\{ \hat{a}_a \hat{a}_b \hat{a}_c  }{\hat{a}}
\hat{a}_s
\contraction[1.5ex]{}{\hat{a}}{_r^{\dagger} \hat{a}_q^{\dagger} \hat{a}_p^{\dagger} \}\{ \hat{a}_a \hat{a}_b}{\hat{a}}
\hat{a}_r^{\dagger}
\contraction[1ex]{}{\hat{a}}{_q^{\dagger} \hat{a}_p^{\dagger} \}\{ \hat{a}_a }{\hat{a}}
\hat{a}_q^{\dagger}
\contraction[0.5ex]{}{\hat{a}}{_p^{\dagger} \}\{ }{\hat{a}}
\hat{a}_p^{\dagger}\}\{ \hat{a}_a \hat{a}_b \hat{a}_c \hat{a}_d^{\dagger} \} W_{abcd}| N \rangle}{\omega_p + \omega_q +\omega_r-\omega_s} \right] \times 36 \nonumber\\
& &  + \frac{1}{3!3!3!} \left[ \sum_{p,q,r,s}^{\text{denom.}\neq 0} \sum_{\substack {i,j,k,l \\ a,b,c,d} } \frac{  \langle N | W_{ijkl}\{ %4
\contraction[2ex]{}{\hat{a}}{_i^{\dagger} \hat{a}_j \hat{a}_k \hat{a}_l \}\{ \hat{a}_p^{\dagger}  \hat{a}_q^{\dagger}  \hat{a}_r^{\dagger} }{\hat{a}}
\hat{a}_i^{\dagger} 
\contraction[1.5ex]{}{\hat{a}}{_j\hat{a}_k\hat{a}_l \}\{ \hat{a}_p^{\dagger}  \hat{a}_q^{\dagger}  }{\hat{a}}
\hat{a}_j
\contraction[1ex]{}{\hat{a}}{_k\hat{a}_l \}\{ \hat{a}_p^{\dagger}  }{\hat{a}}
\hat{a}_k
\contraction[0.5ex]{}{\hat{a}}{_l \}\{ }{\hat{a}}
\hat{a}_l \} \{
\bcontraction[2ex]{}{\hat{a}}{_p^{\dagger}  \hat{a}_q^{\dagger}  \hat{a}_r^{\dagger}  \hat{a}_s \} | N \rangle \langle N | \{ \hat{a}_s^{\dagger}  \hat{a}_r \hat{a}_q }{\hat{a}}
\hat{a}_p^{\dagger} 
\bcontraction[1.5ex]{}{\hat{a}}{_q^{\dagger}  \hat{a}_r^{\dagger}  \hat{a}_s \} | N \rangle \langle N | \{ \hat{a}_s^{\dagger}  \hat{a}_r }{\hat{a}}
\hat{a}_q^{\dagger} 
\bcontraction[1ex]{}{\hat{a}}{_r^{\dagger}  \hat{a}_s \} | N \rangle \langle N | \{ \hat{a}_s^{\dagger} }{\hat{a}}
\hat{a}_r^{\dagger} 
\bcontraction[0.5ex]{}{\hat{a}}{_s \} | N \rangle \langle N | \{}{\hat{a}}
\hat{a}_s \} | N \rangle \langle N | \{
\contraction[2ex]{}{\hat{a}}{_s \hat{a}_r \hat{a}_q \hat{a}_p \}\{ \hat{a}_a^{\dagger}  \hat{a}_b^{\dagger}  \hat{a}_c^{\dagger}   }{\hat{a}}
\hat{a}_s^{\dagger} 
\contraction[1.5ex]{}{\hat{a}}{_r \hat{a}_q \hat{a}_p \}\{ \hat{a}_a^{\dagger}  \hat{a}_b^{\dagger} }{\hat{a}}
\hat{a}_r
\contraction[1ex]{}{\hat{a}}{_q \hat{a}_p \}\{ \hat{a}_a^{\dagger}  }{\hat{a}}
\hat{a}_q
\contraction[0.5ex]{}{\hat{a}}{_p \}\{ }{\hat{a}}
\hat{a}_p\}\{ \hat{a}_a^{\dagger}  \hat{a}_b^{\dagger}  \hat{a}_c^{\dagger}  \hat{a}_d \} W_{abcd}| N \rangle}{-\omega_p - \omega_q -\omega_r+\omega_s} \right] \times 36 \nonumber\\
& &  +  \frac{1}{2!2!2!2!2!2!} \left[ \sum_{p,q,r,s}^{\text{denom.}\neq 0} \sum_{\substack {i,j,k,l \\ a,b,c,d} } \frac{  \langle N | W_{ijkl} \{ %5
\contraction[2ex]{}{\hat{a}}{_i \hat{a}_j \hat{a}_k^{\dagger} \hat{a}_l^{\dagger} \}\{ \hat{a}_p \hat{a}_q \hat{a}_r^{\dagger}}{\hat{a}}
\hat{a}_i
\contraction[1.5ex]{}{\hat{a}}{_j\hat{a}_k^{\dagger} \hat{a}_l^{\dagger} \}\{ \hat{a}_p \hat{a}_q }{\hat{a}}
\hat{a}_j
\contraction[1ex]{}{\hat{a}}{_k^{\dagger} \hat{a}_l^{\dagger} \}\{ \hat{a}_p }{\hat{a}}
\hat{a}_k^{\dagger}
\contraction[0.5ex]{}{\hat{a}}{_l^{\dagger} \}\{ }{\hat{a}}
\hat{a}_l^{\dagger} \} \{
\bcontraction[2ex]{}{\hat{a}}{_p \hat{a}_q \hat{a}_r^{\dagger} \hat{a}_s^{\dagger} \} | N \rangle \langle N | \{ \hat{a}_s \hat{a}_r \hat{a}_q^{\dagger} }{\hat{a}}
\hat{a}_p
\bcontraction[1.5ex]{}{\hat{a}}{_q \hat{a}_r^{\dagger} \hat{a}_s^{\dagger} \} | N \rangle \langle N | \{ \hat{a}_s \hat{a}_r }{\hat{a}}
\hat{a}_q
\bcontraction[1ex]{}{\hat{a}}{_r^{\dagger} \hat{a}_s^{\dagger} \} | N \rangle \langle N | \{ \hat{a}_s}{\hat{a}}
\hat{a}_r^{\dagger}
\bcontraction[0.5ex]{}{\hat{a}}{_s^{\dagger} \} | N \rangle\langle N | \{}{\hat{a}}
\hat{a}_s^{\dagger} \} | N \rangle \langle N | \{
\contraction[2ex]{}{\hat{a}}{_s \hat{a}_r \hat{a}_q^{\dagger} \hat{a}_p^{\dagger} \}\{ \hat{a}_a \hat{a}_b \hat{a}_c^{\dagger}  }{\hat{a}}
\hat{a}_s
\contraction[1.5ex]{}{\hat{a}}{_r \hat{a}_q^{\dagger} \hat{a}_p^{\dagger} \}\{ \hat{a}_a \hat{a}_b}{\hat{a}}
\hat{a}_r
\contraction[1ex]{}{\hat{a}}{_q^{\dagger} \hat{a}_p^{\dagger} \}\{ \hat{a}_a }{\hat{a}}
\hat{a}_q^{\dagger}
\contraction[0.5ex]{}{\hat{a}}{_p^{\dagger} \}\{ }{\hat{a}}
\hat{a}_p^{\dagger}\}\{ \hat{a}_a \hat{a}_b \hat{a}_c^{\dagger} \hat{a}_d^{\dagger} \}W_{abcd}| N \rangle}{\omega_p + \omega_q -\omega_r-\omega_s} \right] \times 16 \nonumber\\
& =&\frac{1}{24} \sum_{i,j,k,l}^{\mathrm{denom.}\neq 0} \frac{W_{ijkl} W_{ijkl}}{\omega_i+\omega_j+\omega_k+\omega_l} f_if_jf_kf_l+ \frac{1}{24} \sum_{i,j,k,l}^{\mathrm{denom.}\neq 0} \frac{ W_{ijkl} W_{ijkl}}{-\omega_i-\omega_j-\omega_k-\omega_l}(f_i+1)(f_j+1)(f_k+1)(f_l+1)\nonumber\\
& & + \frac{1}{6} \sum_{i,j,k,l}^{\mathrm{denom.}\neq 0}  \frac{W_{ijkl} W_{ijkl}}{-\omega_i+\omega_j+\omega_k+\omega_l} (f_i+1)f_jf_kf_l+ \frac{1}{6} \sum_{i,j,k,l}^{\mathrm{denom.}\neq 0}  \frac{ W_{ijkl} W_{ijkl}}{\omega_i-\omega_j-\omega_k-\omega_l}f_i(f_j+1)(f_k+1)(f_l+1)\nonumber\\
& & + \frac{1}{4} \sum_{i,j,k,l}^{\mathrm{denom.}\neq 0}  \frac{W_{ijkl} W_{ijkl}}{-\omega_i-\omega_j+\omega_k+\omega_l}(f_i+1)(f_j+1)f_kf_l,
%}
%&=& [ E_N^{(2)} ]_{\mathrm{Q}}\{ \text{Eq.}(\ref{eq:EN2_AG_Q})\} \equiv  \{[ E_N^{(2)} ]_{\mathrm{Q}}\}_{{L}},
\end{eqnarray}
\end{widetext}
which accounts for the rest of the terms with a denominator in Eq.\ (\ref{eq:Omg2_AG_final}). The eighth term of Eq.\ (\ref{eq:Omg_SQ}) can be evaluated analogously or inferred systematically from the above formula. 

In the thermal average of the last term of Eq.\ (\ref{eq:Omg_SQ}), only the constant part of the normal-ordered $\hat{V}^{(1)}$ [see Eq.\ (\ref{eq:VSQ})] survives. Therefore,
\begin{eqnarray}
&& -\frac{\beta}{2} \left( \Big[ \langle N | \hat{V}^{(1)} \hat{P}_0 \hat{V}^{(1)} | N \rangle \Big] -\Omega^{(1)}\Omega^{(1)} \right) \nonumber\\
&& = -\frac{\beta}{2} \left( \Big[ \langle N | \hat{V}^{(1)} |N \rangle \langle N | \hat{V}^{(1)} | N \rangle \Big] -\Omega^{(1)}\Omega^{(1)} \right) \nonumber\\
&& = -\frac{\beta}{2} \left( \left(E_{\text{XVSCF}}(T) - V_{\text{ref}} - \sum_i \omega_i (f_i + 1/2) \right)^2 -\Omega^{(1)}\Omega^{(1)} \right) \nonumber\\
&& = 0,
\end{eqnarray}
where Eq.\ (\ref{eq:Omg1_SQ}) was used.
Hence, the last line of Eq.\ (\ref{eq:Omg_SQ}) accounts for the  cancellation of all unlinked terms in $\Omega^{(2)}$.

Together, we arrive at the final reduced formula of $\Omega^{(2)}$, which reads
\begin{widetext}
\begin{eqnarray}
\label{eq:Omg_final_SQ}
 \Omega^{(2)} &=& 
 \sum_i^{\mathrm{denom.} \neq 0} \frac{W_i W_i}{\omega_i} f_i+ \sum_i^{\mathrm{denom.} \neq 0} \frac{W_i W_i }{-\omega_i} (f_i+1)
 + \frac{1}{2}\sum_{i, j}^{\mathrm{denom.} \neq 0} \frac{W_{ij} W_{ij}}{\omega_i+\omega_j} f_if_j  
 +  \frac{1}{2}\sum_{i, j}^{\mathrm{denom.} \neq 0} \frac{W_{ij} W_{ij}}{-\omega_i-\omega_j}(f_i+1)(f_j+1) 
 \nonumber\\& & 
 + \sum_{i, j}^{\mathrm{denom.} \neq 0} \frac{W_{ij} W_{ij}}{-\omega_i + \omega_j} (f_i+1)f_j
 +\frac{1}{6} \sum_{i,j,k}^{\mathrm{denom.\neq 0}}  \frac{W_{ijk} W_{ijk}}{\omega_i+\omega_j+\omega_k}f_if_jf_k
 + \frac{1}{6} \sum_{i,j,k}^{\mathrm{denom.\neq 0}}  \frac{W_{ijk} W_{ijk}}{-\omega_i-\omega_j-\omega_k} (f_i+1)(f_j+1)(f_k+1)  
 \nonumber\\& & 
+ \frac{1}{2} \sum_{i,j,k}^{\mathrm{denom.\neq 0}} \frac{W_{ijk} W_{ijk} }{-\omega_i+\omega_j+\omega_k}(f_i+1)f_jf_k
   + \frac{1}{2} \sum_{i,j,k}^{\mathrm{denom.\neq 0}}  \frac{W_{ijk} W_{ijk}}{\omega_i-\omega_j-\omega_k}f_i(f_j+1)(f_k+1)
    + \frac{1}{24} \sum_{i,j,k,l}^{\mathrm{denom.}\neq 0} \frac{W_{ijkl} W_{ijkl} }{\omega_i+\omega_j+\omega_k+\omega_l} f_if_jf_kf_l
  \nonumber\\& & 
   + \frac{1}{24} \sum_{i,j,k,l}^{\mathrm{denom.}\neq 0} \frac{W_{ijkl} W_{ijkl} }{-\omega_i-\omega_j-\omega_k-\omega_l}(f_i+1)(f_j+1)(f_k+1)(f_l+1)
   + \frac{1}{6} \sum_{i,j,k,l}^{\mathrm{denom.}\neq 0} \frac{W_{ijkl} W_{ijkl} }{-\omega_i+\omega_j+\omega_k+\omega_l} (f_i+1)f_jf_kf_l
   \nonumber\\& & 
+ \frac{1}{6} \sum_{i,j,k,l}^{\mathrm{denom.}\neq 0} \frac{W_{ijkl} W_{ijkl} }{\omega_i-\omega_j-\omega_k-\omega_l} f_i(f_j+1)(f_k+1)(f_l+1)
+ \frac{1}{4} \sum_{i,j,k,l}^{\mathrm{denom.}\neq 0} \frac{W_{ijkl} W_{ijkl} }{-\omega_i-\omega_j+\omega_k+\omega_l} (f_i+1)(f_j+1)f_kf_l
\nonumber\\& & 
- \frac{\beta}{2}\sum_i^{\mathrm{denom.}=  0} W_i W_i  {f_i} -  \frac{\beta}{2}\sum_i^{\mathrm{denom.}=  0} W_i W_i{(f_i+1}) 
-  \frac{\beta}{4}\sum_{i, j}^{\mathrm{denom.} = 0} W_{ij} W_{ij} f_i f_j 
- \frac{\beta}{4}\sum_{i, j}^{\mathrm{denom.} = 0} W_{ij} W_{ij} (f_i+1)(f_j+1) 
\nonumber\\& & 
- \frac{\beta}{2} \sum_{i, j}^{\mathrm{denom.} = 0} W_{ij} W_{ij} {(f_i+1)f_j}
- \frac{\beta}{12} \sum_{i,j,k}^{\mathrm{denom.= 0}} W_{ijk} W_{ijk} {f_if_jf_k} 
- \frac{\beta}{12} \sum_{i,j,k}^{\mathrm{denom.= 0}} W_{ijk} W_{ijk} {(f_i+1)(f_j+1)(f_k+1)} 
\nonumber\\& & 
- \frac{\beta}{4} \sum_{i,j,k}^{\mathrm{denom.= 0}} W_{ijk} W_{ijk} {(f_i+1)f_jf_k} 
- \frac{\beta}{4} \sum_{i,j,k}^{\mathrm{denom.= 0}} W_{ijk} W_{ijk} {f_i(f_j+1)(f_k+1)} 
- \frac{\beta}{48} \sum_{i,j,k,l}^{\mathrm{denom.}= 0} W_{ijkl} W_{ijkl}f_if_jf_kf_l  
\nonumber\\& & 
- \frac{\beta}{48} \sum_{i,j,k,l}^{\mathrm{denom.}= 0} W_{ijkl} W_{ijkl} (f_i+1)(f_j+1)(f_k+1)(f_l+1) 
- \frac{\beta}{12} \sum_{i,j,k,l}^{\mathrm{denom.}= 0} W_{ijkl} W_{ijkl} (f_i+1)f_jf_kf_l 
\nonumber\\& & 
- \frac{\beta}{12} \sum_{i,j,k,l}^{\mathrm{denom.}= 0} W_{ijkl} W_{ijkl} f_i(f_j+1)(f_k+1)(f_l+1)
- \frac{\beta}{8} \sum_{i,j,k,l}^{\mathrm{denom.}= 0} W_{ijkl} W_{ijkl} (f_i+1)(f_j+1)f_kf_l,
\end{eqnarray}
\end{widetext}
which is easily seen to be the same as Eq.\ (\ref{eq:Omg2_AG_final}) derived purely algebraically. 
Notice the vastly improved expediency of the normal-ordered reduction as compared to the algebraic reduction described in
Sec.\ \ref{section:AGderivation} and in Appendixes \ref{appendix:algebraicreduction} and \ref{appendix:algebraicreduction2}
using Tables \ref{table:thermalavg_EN1}--\ref{table:thermalavg_EN2}, which are furthermore limited to $\Omega^{(2)}$ (the reduction
of $U^{(2)}$ would require new tables). In addition, raising the rank of the force constants is much easier with the above expression because
the dressed force constants $W$ can include higher-order force constants systematically, i.e., without significantly 
altering the appearance of the above formula.

\subsection{Third-order correction to the thermal-averaged energy \label{section:thirdSQ}}

In the finite-temperature many-body perturbation theory for electrons,\cite{Hirata2021} 
 terms not seen in the zero-temperature theory emerge. 
One class of such terms is the anomalous terms,\cite{Kohn1960,Hirata2021} 
which lack a usual denominator of the corresponding ``normal'' terms, but instead are multiplied by a power of $\beta$.
Their diagrams are, therefore, missing one or more (up to all) resolvent lines (see Sec.\ \ref{section:Diagderivation}).

Another class,  appearing only at the third and higher orders, originates from the renormalization
terms of RSPT.\cite{Shavitt2009,Hirata2021} In the zero-temperature theory,\cite{Shavitt2009} these renormalization terms are 
wholly unlinked and cancel out the unlinked contributions of the parent term with the same magnitude but opposite sign, leaving 
only the linked contributions. It is well known\cite{szabo,Shavitt2009} that Brueckner explicitly confirmed exact cancellation of all unlinked contributions 
up to the sixth order, speculating on the same at all orders. This was then proved by Goldstone, which became the celebrated linked-diagram theorem.\cite{szabo,Shavitt2009} 

Unlike the zero-temperature RSPT,
the renormalization terms at finite temperature contain both linked and unlinked contributions. Only the unlinked contributions cancel
the same in the parent term, but the linked-renormalization contributions persist.\cite{Hirata2021} Their diagrams have one or more of their 
resolvent lines shifted relative to the corresponding normal diagrams, 
so that two or more resolvent lines pile up in between some pairs of adjacent vertexes, while there are none in between others (see Sec.\ \ref{section:Diagderivation}). 
Such shifting of resolvent lines is possible at the third and higher orders.\cite{Hirata2021} There are also anomalous, linked-renormalization terms at the fourth order and higher, in whose diagrams some resolvent
lines are shifted while others are erased at the same time. 

Given the parallel between the electronic and vibrational finite-temperature perturbation theories, the linked-renormalization terms must also appear in the vibrational theories.
In fact, in their $\beta$-dependent diagrammatic derivation of the fourth-order perturbation theory, Shukla and coworkers\cite{Shukla1971_2,Shukla1974,Shukla1985,Shukla1985_2}  
correctly recognized the emergence of such terms, but on the basis of a careful inspection of  index coincidence cases in each diagram. In this section, we succinctly summarize the mechanism by which the linked-renormalization terms emerge in a more transparent time-independent formalism.
This naturally leads to a simple and expedient set of rules for enumerating all types of diagrams in Sec.\ \ref{section:Diagderivation}. 

The linked-renormalization term appears for the first time in $[E_N^{(3)}]$. As per the recursion of degenerate RSPT [Eqs.\ (\ref{eq:E_In_degen}) and (\ref{eq:Phi_In_degen})], we write
\begin{eqnarray}
\Big[ E_N^{(3)} \Big] &=&  \Big[ \text{Tr}\left( \bm{E}^{(3)}\right) \Big] \nonumber\\
&=& \Big[ \langle N | \hat{V}^{(1)} \hat{R} \hat{V}^{(1)} \hat{R} \hat{V}^{(1)} | N \rangle \Big] \nonumber\\
&& - \Big[ \sum_{M}^{\text{degen.}} \langle N | \hat{V}^{(1)} \hat{R}\hat{R} \hat{V}^{(1)}| M \rangle \left( \bm{E}^{(1)}\right)_{NM} \Big] \nonumber\\
&=& \Big[ \langle N | \hat{V}^{(1)} \hat{R} \hat{V}^{(1)} \hat{R} \hat{V}^{(1)} | N \rangle \Big] \nonumber\\
&& - \Big[ \sum_{M}^{\text{degen.}}  \langle N | \hat{V}^{(1)} \hat{R}\hat{R} \hat{V}^{(1)}| M \rangle \langle M |\hat{V}^{(1)} | N \rangle \Big] \nonumber\\
&=& \Big[ \langle N | \hat{V}^{(1)} \hat{R} \hat{V}^{(1)} \hat{R} \hat{V}^{(1)} | N \rangle \Big] 
%\nonumber\\&& 
- \Big[ \langle N | \hat{V}^{(1)} \hat{R}\hat{R} \hat{V}^{(1)} \hat{P} \hat{V}^{(1)} | N \rangle \Big] , \nonumber\\ \label{eq:E3}
\end{eqnarray}
where $M$ runs over all Hartree-product states that are degenerate with the $N$th state, including $N$ itself. 
The second terms are the renormalization term of RSPT. In the zero-temperature nondegenerate RSPT, $|N\rangle = |M\rangle = |\Phi_0^{(0)}\rangle$, and, therefore,
it is a single product of two extensive scalars and is wholly unlinked.

The first term of the last line is decomposed into the linked and unlinked contributions.
\begin{eqnarray}
\Big[ \langle N | \hat{V}^{(1)} \hat{R} \hat{V}^{(1)} \hat{R} \hat{V}^{(1)} | N \rangle \Big] &=& 
\Big[ \langle N | \hat{V}^{(1)} \hat{R} \hat{V}^{(1)} \hat{R} \hat{V}^{(1)} | N \rangle \Big]_{{L}} \nonumber\\
 & & +\Big [ \langle N | 
 \contraction[0.5ex]{}{\hat{V}}{^{(1)}}{\hat{R}}
 \hat{V}^{(1)} 
 \bcontraction[1.2ex]{}{\hat{R}}{\Big[ E_N^{(1)} \Big]}{\hat{R}}
 \hat{R} \Big[ E_N^{(1)} \Big] 
 \contraction[0.5ex]{}{\hat{R}}{}{\hat{V}}
 \hat{R} \hat{V}^{(1)} | N \rangle \Big], \nonumber\\ \label{eq:E3_1}
\end{eqnarray}
where an overall full contraction pattern is outlined by the staple symbols in the second term, which is unlinked. In this term, the constant part of the uncontracted middle $\hat{V}^{(1)}$ 
is used in the full contraction, giving rise to the $[ E_N^{(1)}]$ factor. 

The renormalization term [the second term of Eq.\ (\ref{eq:E3})] is also broken down into the linked and unlinked parts:
\begin{eqnarray}
\Big[ \langle N | \hat{V}^{(1)} \hat{R} \hat{R} \hat{V}^{(1)} \hat{P} \hat{V}^{(1)} | N \rangle \Big] &=&\Big [ \langle N | \hat{V}^{(1)} \hat{R} \hat{R} \hat{V}^{(1)} \hat{P} \hat{V}^{(1)} | N \rangle \Big]_{{L}}\nonumber\\
& & + \Big[ \langle N | 
\contraction[0.5ex]{}{\hat{V}}{^{(1)}}{\hat{R}}
\hat{V}^{(1)}
\bcontraction[0.5ex]{}{\hat{R}}{}{\hat{R}}
\hat{R}
\contraction[0.5ex]{}{\hat{R}}{}{\hat{V}}
\hat{R} \hat{V}^{(1)} \Big[ E_N^{(1)} \Big] | N \rangle \Big], \nonumber\\ \label{eq:E3_2}
\end{eqnarray}
where the first, linked contribution arises from $\hat{P}_1 + \hat{P}_2 + \dots$, while the second, unlinked one from $\hat{P}_0$.
The former is identified as the linked-renormalization term, which is generally nonzero 
because there are states that are degenerate with the $N$th state. 

The unlinked contributions in Eqs.\ (\ref{eq:E3_1}) and (\ref{eq:E3_2}) clearly have an equal value, canceling each other in Eq.\ (\ref{eq:E3}).
The thermal average of $E_N^{(3)}$, therefore, consists of two linked contributions:
\begin{eqnarray}
\Big[ E_N^{(3)} \Big] &=& \Big[ \langle N | \hat{V}^{(1)} \hat{R} \hat{V}^{(1)} \hat{R} \hat{V}^{(1)} | N \rangle \Big]_{{L}}  
\nonumber\\
&&- \Big[ \langle N | \hat{V}^{(1)} \hat{R} \hat{R}  \hat{V}^{(1)} \hat{P} \hat{V}^{(1)} | N \rangle \Big]_{{L}} .
\end{eqnarray}
The first term is the normal term, which interleaves resolvent and perturbation operators. 
The corresponding diagram has a resolvent line in between every pair of adjacent vertexes. 
The second term is the linked-renormalization term, which has two resolvent operators in between the first and second perturbation operators, whereas
there is an inner projector in between the second and third perturbation operators.
Its diagram crams two resolvent lines in between a pair of vertexes and none in between the other pair. 

These terms occurring only at finite temperature and higher orders can be systematically and exhaustively included
with the aid of Feynman diagrams, which we now discuss.

% =====================================
% Diagrams
% =====================================

\section{Time-independent Feynman diagrams \label{section:Diagderivation}}

\subsection{Diagrammatic rules} 

The conventional derivation of finite-temperature perturbation theory 
begins with recognizing the isomorphism of the $\beta$-dependent Bloch equation and time-dependent Schr\"{o}dinger equation,\cite{march} 
and then reappropriates the time-dependent Feynman diagrammatic logic. In the zero-temperature case, the latter logic is fully established to be equivalent to 
the time-independent algebraic derivation of RSPT via the linked-diagram theorem.\cite{GellmannLow,brueckner,goldstone,hugenholtz,Frantz,Manne,Harris,Shavitt2009}

In a practical use of time-dependent diagrammatics, 
a user is asked to summon intuition to draw all possible, topologically distinct, closed, linked diagrams, each graphically representing a scattering 
process of mean-field particles.\cite{mattuck} The diagrams are then transformed into algebraic formulas. 
This transformation step has been justified mathematically rigorously,\cite{dyson_physicsworld} but is exceedingly complex and tedious, involving series of contour integrations over time or frequency, whose main outcome is to introduce
the denominators. This top-down approach to time-dependent diagrammatics was inevitable in quantum electrodynamics since it lacks the fundamental equation of motion in the style of the Schr\"{o}dinger or Dirac equation. 

For vibrations, starting with the equation of motion, we can instead adopt a bottom-up approach of time-independent diagrammatics. It is a set of expedient mnemonics of normal-ordered derivations, which are in turn firmly grounded on
the time-independent Schr\"{o}dinger equation and RSPT thereof. An algebraic formula for each diagram can be 
obtained instantaneously including its denominators by following a set of well-defined rules, which can furthermore be justified
by the normal-ordered derivations.
In this section, we document these rules and illustrate their utility for $\Omega^{(2)}$. 

\begin{table}
\begin{ruledtabular}
\caption{\label{table:DrawdigramRule}Rules to generate all $\Omega^{(n)}$ diagrams ($n \geq 2$).}
\begin{tabular}{rl}
(1) & Draw $n$ filled-circle or open-circle vertexes in an unambiguous \\
      & vertical order. A filled circle stands for a dressed force constant \\
      &  ($W$), while an open circle a scaled bare force constant ($\tilde{F}$ or $\tilde{\bar{F}}$).\\ 
(2) & Connect all vertexes with edges to form a closed, linked \\
     & diagram in all topologically distinct manners to generate all \\
     & skeleton diagrams. No short loop or ``ring'' (an edge connecting  \\
     & with one and the same vertex at both ends) shall be formed \\
     & from a  dressed-force-constant vertex; it must be formed  \\
     & from a bare-force-constant vertex. \\
(3) & Starting with each skeleton diagram, give each edge an  \\
     & upgoing or downgoing direction to generate all topologically   \\
     & distinct arrow diagrams.  \\
(4) & Insert zero through $n-1$ resolvent (wiggly) lines in all $2^{n-1}$ \\
      & ways with either zero or one resolvent line in between each pair \\
      & of adjacent vertexes. An arrow diagram with less than $n-1$ \\
      & resolvent lines is called an anomalous diagram.\\
(5) & For each arrow diagram, shift upward one or more resolvent \\
     & lines onto existing resolvents. An arrow diagram with at least \\
     & one resolvent line shifted is called a renormalization diagram. \\
     & An arrow diagram with missing and shifted resolvent lines is \\
     & called an anomalous-renormalization diagram. A diagram that \\
     & is neither anomalous nor renormalization one is called \\
     & a normal diagram. A normal diagram containing only upgoing \\
     & edges is called a zero-temperature diagram. 
\end{tabular}
\end{ruledtabular}
\end{table}

\begin{table}
\begin{ruledtabular}
\caption{\label{table:EvaluationRule}Rules to transform an $\Omega^{(n)}$ diagram to an algebraic formula ($n \geq 2$).}
\begin{tabular}{rl}
(1) & Label each edge with a mode index $i$, $j$, $k$, etc.\\
(2) & Associate an $n$-edged filled-circle vertex with an $n$th-order \\ 
     & dressed force constant ${W}$ with the $n$ connecting edge indexes \\
     & as subscripts in no particular order. \\
(3) & Associate an $n$-edged open-circle vertex with an $n$th-order \\ 
     & scaled bare force constant $\tilde{F}$ ($n\neq 2$) or $\tilde{\bar{F}}$ ($n=2$) with the $n$  \\
     & connecting  edge indexes as subscripts in no particular order. \\
(4) & Associate the $i$th downgoing edge with a Bose--Einstein  \\
     & distribution function $f_i$, and the $i$th upgoing edge with $f_i + 1$. \\
(5) & Associate the short loop made of the $i$th edge with $f_i + 1/2$.\\
(6) & Associate a resolvent line with $1/(\omega_i+\omega_j\cdots -\omega_k - \omega_l \ldots )$, \\
      & where $i,j,\dots$ ($k,l,\dots$) label downgoing (upgoing) edges \\
      & intersecting the resolvent line. No resolvent line shall  \\
      & intersect a short loop.\\
(7) & Sum over all edge indexes. Restrict the summation indexes to \\
     & only those cases in which denominator factors are nonzero  \\
     & and fictitious denominator factors are zero. \\
(8) & Multiply 1/$n!$ for each set of $n$ equivalent edges. Two edges  \\
      & are equivalent when they start from and end at the same two \\
      & vertexes and have the same arrow direction.\\
(9) & Multiply $(-\beta)^n/(n+1)!$ to an anomalous diagram with $n$ \\
     & missing resolvent lines. \\
(10) & Multiply $(-1)^n$ to a renormalization diagram with $n$ shifted \\
      & resolvent lines.
\end{tabular}
\end{ruledtabular}
\end{table}

The rules to generate all diagrams for $\Omega^{(n)}$ ($n \geq 2$) are given in Table \ref{table:DrawdigramRule}.
The rules to transform these diagrams into algebraic formulas are compiled in Table \ref{table:EvaluationRule}. 

Generally, all equal-valued full contractions in the normal-ordered derivation are many-to-one mapped onto a single arrow diagram.
Normal diagrams arise from the first term in the recursion of $\Omega^{(n)}$ [Eq.\ (\ref{eq:recursionOmg})]. The subsequent terms
multiplied by various powers of $\beta$ give rise to anomalous diagrams. Owing to the mechanism explained in Sec.\ \ref{section:thirdSQ}, linked-renormalization terms
can arise at the third and higher orders. They are represented by renormalization diagrams. 

In each arrow diagram, a filled-circle vertex graphically represents an amplitude (dressed force constant $W$) of the normal-ordered perturbation
operator $\hat{V}^{(1)}$, while an edge corresponds to a Wick binary contraction. 
The two directions (upgoing and downgoing) of an edge correspond to the two types of Wick contractions in Eqs.\ (\ref{eq:aiaidagger}) and (\ref{eq:aidaggerai}), respectively. 
That we only draw closed diagrams without short loops reflects the normal-ordered second-quantization rule that nonzero thermal averages arise only from full contractions
excluding internal contractions (and an internal contraction is a short loop). 
That we need to consider only linked diagrams is justified by the linked-diagram theorem 
at finite temperature proven in Sec.\ \ref{section:linkedtheorem}.

Once the diagrammatic rules based on the normal-ordered derivations are established, we can expand each dressed force constant $W$ (a filled-circle vertex) as a sum of  scaled bare force constants 
$\tilde{F}$ and $\tilde{\bar{F}}$ (open-circle vertexes) to arrive at an alternative set of rules for more widely used, but cluttered diagrams with short loops (rings). The rules in Tables \ref{table:DrawdigramRule} and \ref{table:EvaluationRule}
are applicable to both cases.

It is possible to devise diagrammatic rules for the perturbative internal energies and entropies.\cite{Hirata2021} This is not pursued in this study because
the derivative method of obtaining algebraic formulas for $U^{(n)}$ [see, e.g., Eq.\ (\ref{eq:U1derivative})] is judged far more expedient, once $\Omega^{(n)}$ is derived diagrammatically.
Then, $S^{(n)}$ is also readily available from Eq.\ (\ref{eq:S_exact}) without diagrams. It is also unnecessary to consider diagrams at the zeroth and first orders. 

\subsection{Second-order correction to the grand potential}

\begin{figure}
  \includegraphics[scale=0.8]{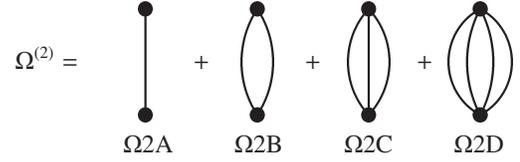}
\caption{Skeleton-diagram equation of $\Omega^{(2)}$ in a QFF.}
\label{fig:diagramskeleton}
\end{figure}
\begin{figure}
  \includegraphics[scale=0.8]{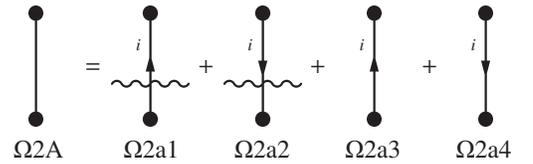}
\caption{The definition of the $\Omega2\text{A}$ skeleton diagram in terms of its arrow diagrams with edge indexes.}
\label{fig:diagram2A}
\end{figure}
\begin{figure}
  \includegraphics[scale=0.8]{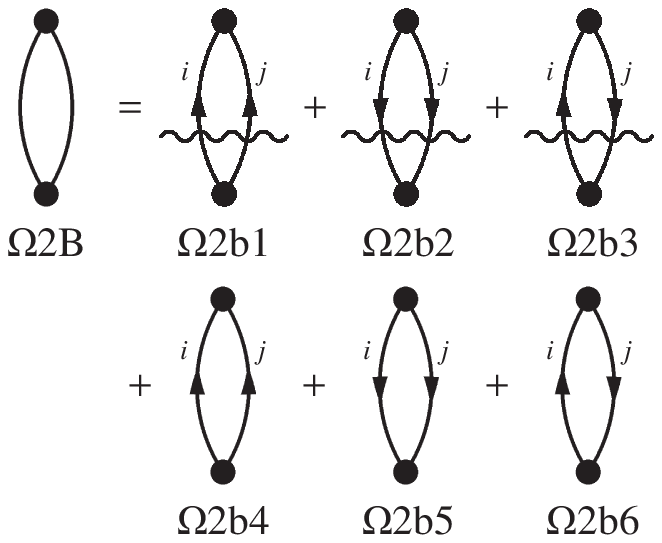}
\caption{Same as Fig.\ \ref{fig:diagram2A}, but for the $\Omega2\text{B}$ skeleton diagram.}
\label{fig:diagram2B}
\end{figure}
\begin{figure}
  \includegraphics[scale=0.8]{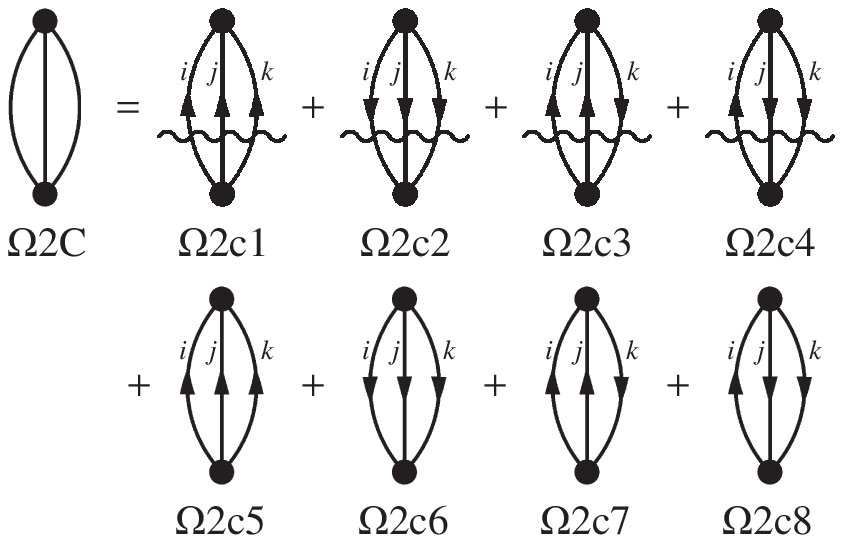}
\caption{Same as Fig.\ \ref{fig:diagram2A}, but for the $\Omega2\text{C}$ skeleton diagram.}
\label{fig:diagram2C}
\end{figure}
\begin{figure}
  \includegraphics[scale=0.8]{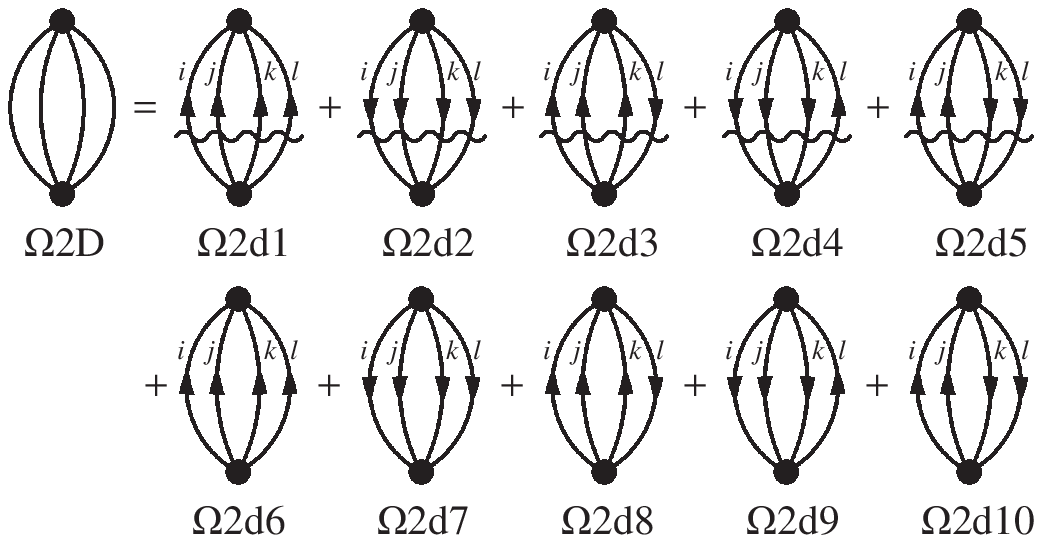}
\caption{Same as Fig.\ \ref{fig:diagram2A}, but for the $\Omega2\text{D}$ skeleton diagram.}
\label{fig:diagram2D}
\end{figure}

Here, we show the utility of the time-independent diagrammatic rules for $\Omega^{(2)}$ in a QFF. 
There are four skeleton diagrams for $\Omega^{(2)}$ shown in Fig.\ \ref{fig:diagramskeleton}. 
Each of them is broken down into arrow diagrams with index labels in Figs.\ \ref{fig:diagram2A}--\ref{fig:diagram2D},
which are easily transformed into algebraic formulas of Eq.\ (\ref{eq:Omg_final_SQ}) using Table \ref{table:EvaluationRule}.

Let us illustrate this step using diagram $\Omega$2B in Fig.\ \ref{fig:diagram2B}. 
As per Table \ref{table:EvaluationRule}, diagram $\Omega$2b1 is evaluated as
\begin{eqnarray}
\Omega 2\text{b}1 =  \frac{1}{2} \sum_{i,j}^{\mathrm{denom.} \neq 0} \frac{W_{ij} W_{ij} }{-\omega_i - \omega_j}(f_i+1)( f_j+1),
\end{eqnarray}
where the factors of $f_i+1$ and $f_j+1$ come from the $i$th and $j$th upgoing edges, the two $W_{ij}$ correspond to the two filled-circle vertexes connected with the $i$th and $j$th edges, 
the denominator $-\omega_i - \omega_j$ is rationalized by the intersections between the resolvent line and the $i$th and $j$th upgoing edges, and the summation needs to be
 restricted to the case that this denominator be nonzero. The factor of $1/2$ is due to the two edges being equivalent. 

Likewise, diagrams $\Omega$2b2  and $\Omega$2b3 are interpreted as
\begin{eqnarray}
\Omega 2\text{b}2 &=&\frac{1}{2} \sum_{i,j}^{\mathrm{denom.} \neq 0} \frac{W_{ij} W_{ij} }{\omega_i + \omega_j}f_i f_j, \\
\Omega2 \text{b}3 &=& \sum_{i,j}^{\mathrm{denom.} \neq 0} \frac{W_{ij} W_{ij} }{-\omega_i + \omega_j}(f_i+1) f_j.
\end{eqnarray}
In the former, the $i$th and $j$th edges are equivalent, hence the factor of $1/2$. 

In the second row of Fig.\ \ref{fig:diagram2B} are listed the corresponding anomalous diagrams with their resolvent lines erased. 
They are evaluated immediately as
\begin{eqnarray}
\Omega 2\text{b}4 &=& -\frac{\beta}{4} \sum_{i,j}^{\mathrm{denom.} = 0} {W_{ij} W_{ij} }(f_i+1)( f_j+1), \\
\Omega 2\text{b}5 &=& -\frac{\beta}{4} \sum_{i,j}^{\mathrm{denom.} = 0} {W_{ij} W_{ij} }f_i f_j, \\
\Omega2 \text{b}6 &=& -\frac{\beta}{2} \sum_{i,j}^{\mathrm{denom.} = 0} {W_{ij} W_{ij} }(f_i+1) f_j,
\end{eqnarray}
where the ``denom.=0'' restrictions demand the summations be limited to the cases $-\omega_i - \omega_j = 0$, $\omega_i + \omega_j = 0$, and $-\omega_i + \omega_j = 0$, respectively.

Together, we obtain
\begin{eqnarray}
\Omega\text{2B} &=& \Omega\text{2b1} + \Omega\text{2b2} + \Omega\text{2b3} + \Omega\text{2b4} + \Omega\text{2b5} + \Omega\text{2b6}  \nonumber\\
&=& 
 \frac{1}{2} \sum_{i,j}^{\mathrm{denom.} \neq 0} \frac{W_{ij} W_{ij} }{-\omega_i - \omega_j}(f_i+1)( f_j+1) 
\nonumber\\&&
+ \frac{1}{2} \sum_{i,j}^{\mathrm{denom.} \neq 0} \frac{W_{ij} W_{ij} }{\omega_i + \omega_j}f_i f_j
\nonumber\\&&
 + \sum_{i,j}^{\mathrm{denom.} \neq 0} \frac{W_{ij} W_{ij} }{-\omega_i + \omega_j}(f_i+1) f_j
 \nonumber\\&&
   -\frac{\beta}{4} \sum_{i,j}^{\mathrm{denom.} = 0} {W_{ij} W_{ij} }(f_i+1)( f_j+1) 
\nonumber\\&&
 -\frac{\beta}{4} \sum_{i,j}^{\mathrm{denom.} = 0} {W_{ij} W_{ij} }f_i f_j 
  \nonumber\\&&
 -\frac{\beta}{2} \sum_{i,j}^{\mathrm{denom.} = 0} {W_{ij} W_{ij} }(f_i+1) f_j,
\end{eqnarray}
which agrees with the sum of all $W_{ij}W_{ij}$ terms in Eq.\ (\ref{eq:Omg_final_SQ}). 
The same procedure can be repeated to Figs.\ \ref{fig:diagram2A}, \ref{fig:diagram2C}, and \ref{fig:diagram2D} to quickly arrive at the same $\Omega^{(2)}$ formula in Eq.\ (\ref{eq:Omg_final_SQ}).
This expedient time-independent diagrammatic mnemonic 
may be contrasted with a complex and lengthy time-dependent diagrammatic derivation,\cite{mattuck} which is furthermore, in practice, intuitive in its first step of diagram enumeration. 

In the zero-temperature limit, since $f_i \to 0$, only diagram $\Omega$2b1 survives among all arrow diagrams in Fig.\ \ref{fig:diagram2B} (and is hence called a zero-temperature diagram). It 
has the same algebraic expression and numerical value as the corresponding term of the zero-temperature perturbation theory.\cite{Hermes2013}
The first diagram in the right-hand side of each diagrammatic equation in Figs.\ \ref{fig:diagram2A}--{\ref{fig:diagram2D} is a zero-temperature diagram. 
 As will be discussed in Sec.\ \ref{section:zeroT}, all anomalous and renormalization diagrams vanish at zero temperature, insofar as $\omega_i > 0$.

\begin{figure}
  \includegraphics[scale=0.8]{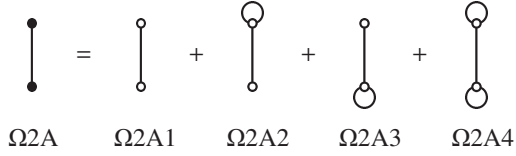}
\caption{The relationship between the dressed-force-constant ($W$) diagrams and the scaled-bare-force-constant ($\tilde{F}$) diagrams in a QFF.
A filled-circle vertex denotes $W$, while an open-circle vertex $\tilde{F}$.}
\label{fig:diagram2E}
\end{figure}

\begin{figure}
  \includegraphics[scale=0.8]{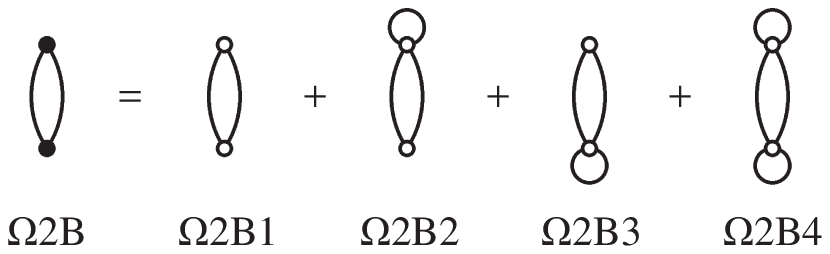}
\caption{The relationship between the dressed-force-constant ($W$) diagrams and the scaled-bare-force-constant ($\tilde{F}$) diagrams in a QFF.
A filled-circle vertex denotes $W$, while an open-circle vertex $\tilde{{F}}$ or $\tilde{\bar{F}}$.}
\label{fig:diagram2F}
\end{figure}

Every filled-circle vertex  represents a dressed force constant $W$, appearing in the finite-temperature XVSCF working equations (see Appendix \ref{appendix:normalorderedHam} and Sec.\ \ref{section:finiteT_XVSCF}).
When these $W$ vertexes (filled circles) are expanded in terms of
the $\tilde{F}$ vertexes (open circles), we arrive at diagrams made of scaled bare force constants. In a QFF,  skeleton diagrams $\Omega$2A and $\Omega$2B are reexpressed
as sums of scaled-bare-force-constant skeleton diagrams of Figs.\ \ref{fig:diagram2E} and \ref{fig:diagram2F}, respectively. 
It can be seen that each $W$ skeleton diagram  folds in itself ring diagrams made of higher-order scaled bare force constants. 

Diagrams $\Omega$2A1, $\Omega$2A2, $\Omega$2A3, and $\Omega$2A4 in Fig.\ \ref{fig:diagram2E} are isomorphic with diagrams 2G, 2F, 2E, and 2A, respectively, of XVMP2,\cite{Hermes2013} 
with the former reducing to the latter as $T \to 0$ both analytically and numerically (see also Sec.\ \ref{section:zeroT}). Likewise, the zero-temperature limits of diagrams $\Omega$2B1, $\Omega$2B2, $\Omega$2B3, and $\Omega$2B4
 in Fig.\ \ref{fig:diagram2F}
are diagrams 2J, 2I, 2H, and 2B of XVMP2.\cite{Hermes2013}

Diagram 1(c) of Shukla and Cowley\cite{Shukla1971_2} is identified as diagram $\Omega$2A4 of Fig. \ref{fig:diagram2E}. 
Diagram 2(b) of the same authors\cite{Shukla1971_2} is $\Omega$2B4 of Fig. \ref{fig:diagram2F}, but is, however, 
listed among the fourth-order perturbation correction ($\Omega$2B4 is second order in our theory) because the Van Hove ordering scheme was adopted by Shukla and Cowley. 
Other second-order diagrams in Figs.\ \ref{fig:diagram2E} and \ref{fig:diagram2F} are not seen in their figures, underscoring 
the generality and completeness of our perturbation treatment, although the basic physics is the same.

% =====================================
% Linked-diagram theorem
% =====================================

\section{Linked-diagram theorem\label{section:linkedtheorem}}

In Eq.\ (\ref{eq:linked}), we showed that $\Omega^{(1)}$ is linked and thus size-consistent, writing
\begin{eqnarray}
\Omega^{(1)} &=& \Big[ E_N^{(1)} \Big] = \Big[ E_N^{(1)} \Big]_{L},
\end{eqnarray}
where ``$L$'' denotes linkedness. 
Algebraically, the linkedness means that $\Omega^{(1)}$ is not a simple product of two or more extensive scalars.\cite{Hirata2011}
Diagrammatically, it means that the diagram does not consist of two or more disconnected parts, of which one or more are closed.\cite{Hirata2011}
In Eq.\ (\ref{eq:linked2}), likewise, we demonstrated the linkedness of $\Omega^{(2)}$, i.e.,
\begin{eqnarray}
     \Omega^{(2)} &=& \Big[ E_N^{(2)} \Big] - \frac{\beta}{2}  \left( \Big[ E_N^{(1)} E_N^{(1)} \Big] - \Big[ E_N^{(1)} \Big] \Big[ E_N^{(1)} \Big] \right) \nonumber\\
     &=& \Big[ E_N^{(2)} \Big]_{L} - \frac{\beta}{2}\Big[ E_N^{(1)} E_N^{(1)} \Big]_{L}.
\end{eqnarray}
The first term $[ E_N^{(2)}]_L$ is linked by itself. The second and third terms in the first line are
unlinked, but the unlinked part in the second term is canceled exactly by 
the wholly unlinked third term, leaving only the linked (anomalous-diagram) term, $-(\beta/2) [ E_N^{(1)} E_N^{(1)} ]_L$. 

We can generalize this observation to any order in the form of a linked-diagram theorem for the finite-temperature many-body perturbation theory
for vibrations. The proof is identical to the one for finite-temperature electrons.\cite{Hirata2021} For completeness, we shall succinctly summarize it. 

{\it Linked-diagram theorem.} $\Omega^{(n)}$, $U^{(n)}$, and $S^{(n)}$ are linked at any order $n$, and hence the finite-temperature many-body perturbation theory
for vibrations is size-consistent at all orders.

{\it Proof.} We begin with the recursion,
\begin{eqnarray}
\Omega^{(n)}&=&\Big[ E_{N}^{(n)} \Big]+\frac{(-\beta)}{2 !} \sum_{i=1}^{n-1}\left(\Big[ E_{N}^{(i)} E_{N}^{(n-i)}\Big]-\Omega^{(i)} \Omega^{(n-i)}\right) \nonumber\\
& &+\frac{(-\beta)^{2}}{3 !} \sum_{i=1}^{n-2} \sum_{j=1}^{n-i-1}\left( \Big[ E_{N}^{(i)} E_{N}^{(j)} E_{N}^{(n-i-j)} \Big]-\Omega^{(i)} \Omega^{(j)} \Omega^{(n-i-j)}\right) \nonumber\\
& &+\cdots+\frac{(-\beta)^{n-1}}{n !}\left\{ \Big[(E_{N}^{(1)})^{n}\Big]-(\Omega^{(1)})^{n}\right \}. \label{eq:Omega_recursion_again} 
\end{eqnarray}
The proof consists of two parts: (1) Prove that $[ E_N^{(n)}]$ is linked at any $n$; (2) Prove that the $[ E_N^{(i)} E_N^{(j)} \cdots E_N^{(k)}]$ terms
consist of linked and unlinked parts, and the unlinked part is canceled exactly by the unlinked $\Omega^{(i)}\Omega^{(j)}\cdots\Omega^{(k)}$ products. 

In part (1), we first rewrite the degenerate RSPT recursion [Eq.\ (\ref{eq:Phi_In_degen})] into
\begin{eqnarray}
\Phi_I^{(n)} &=& \hat{R} \left\{ \hat{V}^{(1)} \Phi_I^{(n-1)} - \sum_{i=1}^{n-1} \sum_{J \in \gamma} \left(\bm{E}_\gamma^{(i)}\right)_{IJ}\Phi_J^{(n-i)} \right\} \nonumber\\
&=& \hat{R} \left\{ \hat{V}^{(1)} \Phi_I^{(n-1)} - \sum_{i=1}^{n-1} \left(\bm{E}_\gamma^{(i)}\right)_{II}\Phi_I^{(n-i)} \right\} \nonumber\\
&& -  \hat{R} \sum_{i=1}^{n-1} \sum_{J \in \gamma}^{J \neq I } \left(\bm{E}_\gamma^{(i)}\right)_{IJ}\Phi_J^{(n-i)},
\end{eqnarray}
where $\gamma$ labels a set of states that are degenerate with the $I$th state. The second term is linked through index $J$, which gives rise 
to the linked-renormalization term (see Sec.\ \ref{section:thirdSQ}). The first term is isomorphic to the nondegenerate RSPT recursion, to which 
the linked-diagram theorem at zero temperature holds without modification.\cite{GellmannLow,brueckner,goldstone,hugenholtz,Frantz,Manne,Harris,Shavitt2009} 
Therefore, $[ E_N^{(n)}]$ is linked. 

In part (2), first, keep in mind that the cancellation does not complete within each parenthesis of the recursion [Eq.\ (\ref{eq:Omega_recursion_again})]. 
Let us define\cite{Hirata2021}  
\begin{eqnarray}
(-\beta) B^{(n)}&=&(-\beta)  \Big[ E_{N}^{(n)} \Big]_L +\frac{(-\beta)^2}{2 !} \sum_{i=1}^{n-1} \Big[ E_{N}^{(i)} E_{N}^{(n-i)}\Big]  \nonumber\\
& &+\frac{(-\beta)^{3}}{3 !} \sum_{i=1}^{n-2} \sum_{j=1}^{n-i-1}  \Big[ E_{N}^{(i)} E_{N}^{(j)} E_{N}^{(n-i-j)} \Big]  \nonumber\\
& &+\cdots+\frac{(-\beta)^{n}}{n !} \Big[(E_{N}^{(1)})^{n}\Big] , \label{eq:bigB} 
\end{eqnarray}
where we used the result of part (1) that the first term is linked.
We then divide $B^{(n)}$ into linked and unlinked parts:
\begin{eqnarray}
B^{(n)}&=&  B^{(n)}_L +  B^{(n)}_U. \label{eq:bigB_LU} 
\end{eqnarray}
The unlinked part will take an exponential form of the linked part,\cite{Shavitt2009} i.e.,
\begin{eqnarray}
 (-\beta) B^{(n)}_U %&=& \exp\left\{ (-\beta) B^{(n)}_L \right\} - 1 \nonumber\\
&=& \frac{(-\beta)^2}{2 !} \sum_{i=1}^{n-1}  B^{(i)}_L B^{(n-i)}_L \nonumber\\
&& +\frac{(-\beta)^3}{3 !} \sum_{i=1}^{n-2} \sum_{j=1}^{n-i-1}   B^{(i)}_L B^{(j)}_L B^{(n-i-j)}_L \nonumber\\
& &+\cdots+\frac{(-\beta)^n}{n !}  \left( B^{(1)}_L \right)^n . \label{eq:expB} 
\end{eqnarray}
Substituting the above two equations into Eq.\ (\ref{eq:bigB}), we obtain
\begin{eqnarray}
   B^{(n)} _L  &=& \Big[ E_{N}^{(n)} \Big]_L +\frac{(-\beta)}{2 !} \sum_{i=1}^{n-1}\left( \Big[ E_{N}^{(i)} E_{N}^{(n-i)}\Big]- B^{(i)} _L B^{(n-i)} _L \right) \nonumber\\
& &+\frac{(-\beta)^{2}}{3 !} \sum_{i=1}^{n-2} \sum_{j=1}^{n-i-1}\left( \Big[ E_{N}^{(i)} E_{N}^{(j)} E_{N}^{(n-i-j)} \Big]- B^{(i)} _LB^{(j)} _LB^{(n-i-j)} _L\right) \nonumber\\
& &+\cdots+\frac{(-\beta)^{n-1}}{n !}\left\{ \Big[(E_{N}^{(1)})^{n}\Big]-\left( B^{(1)} _L \right)^{n}\right \} .
\end{eqnarray}
Comparing this with Eq.\ (\ref{eq:Omega_recursion_again}), we infer $\Omega^{(n)} = B^{(n)}_L$, which is linked.
This proves the linkedness of $\Omega^{(n)}$, where mathematical induction is implicit. 

Since $U^{(n)}$ and $S^{(n)}$ are related to $\Omega^{(n)}$ by
\begin{eqnarray}
U^{(n)} &=& \Omega^{(n)} + \beta \frac{\partial \Omega^{(n)}}{\partial \beta}, \label{eq:U_from_deriv}\\
S^{(n)} &=& \frac{U^{(n)} - \Omega^{(n)}}{T},
\end{eqnarray}
and furthermore the $\beta$-derivative gives rise to only an intensive multiplier [see Eq.\ (\ref{eq:f_deriv})], they are also linked at any $n$.
This completes the proof of the linked-diagram theorem for finite-temperature vibrations.

% =====================================
% Zero-temperature limits
% =====================================

\section{Zero-temperature limits \label{section:zeroT}} 

In 1960, Kohn and Luttinger\cite{Kohn1960} pointed out that the finite-temperature many-body perturbation theory for electrons does not reduce to 
the zero-temperature counterpart as $T \to 0$ when the reference wave function is degenerate and also anisotropic. One of the present authors confirmed, both analytically and numerically, 
that the theory\cite{Hirata2021} indeed has zero radius 
of convergence at $T=0$ when the reference wave function is qualitatively different from the exact one.\cite{Hirata_KL2021} In particular, when the degeneracy of the reference
is lifted at the first order of degenerate RSPT,\cite{Hirschfelder1974} $\Omega^{(2)}$ and $U^{(2)}$ are divergent at $T = 0$ due to divergent anomalous diagrams (although $E^{(2)}$ is finite for the same reference). 
The root cause of this nonconvergence has been identified\cite{Hirata_KL2021} as the nonanalytic nature of the $\Omega$ and $U$ formulas, which contain
the Boltzmann factor, $\exp(-E/k_\text{B}T)$, a prime example of a nonanalytic function at $T=0$. 
A solution is to simply avoid such a problematic reference by adopting a symmetry-broken reference.\cite{Hirata_KL2022}

In the $T = 0$ limit of the present theory, in contrast, both $\Omega^{(n)}$ and $U^{(n)}$ are shown to converge at $E_0^{(n)}$ of RSPT (Refs.\ \onlinecite{Shavitt2009,Hirschfelder1974}) (subscript `0' for the zero-point state) 
under the assumption $\omega_i > 0$ for all $i$. This assumption is readily satisfied by choosing a reference geometry at or near a local minimum of the PES. Although a crystal in a local-minimum geometry does have zero-frequency acoustic modes at
the $\Gamma$ point, they are not vibrations but translations (or rotations in the case of a polymer) and must be excluded from the theoretical treatments. In other words, 
they have infinitesimally small volume elements and can be safely neglected in computational treatments also.\cite{Qin2020} Furthermore, when at least one $\omega_i$ is zero or imaginary, the Bose--Einstein theory is no longer valid, making the perturbation theory based on it moot.
That $\omega_i > 0$ for all $i$ is, therefore, a reasonable assumption, and is also implicit in the Born--Huang rules.

Below, we illustrate that $\Omega^{(0} + \Omega^{(1)}$ and $U^{(0)} + U^{(1)}$ reduce to $E_0^{(0)} + E_0^{(1)}$, i.e., the XVSCF zero-point  energy,\cite{Keceli2011,Hermes2012} as $T \to 0$, whereas
 $\Omega^{(2)}$ and $U^{(2)}$ converge at $E^{(2)}_0$, i.e., the XVMP2 correction,\cite{Hermes2013} under this assumption.

As per the Bose--Einstein theory [Eqs.\ (\ref{eq:Omg_0})--(\ref{eq:S_0})] and $f_i \asymp \exp(-\beta\omega_i) \to 0$ as $T \to 0$ ($\beta \to \infty$), we infer
\begin{eqnarray}
\lim_{T \to 0} \Omega^{(0)} &=& 
\lim_{T \to 0} U^{(0)} = V_\text{ref} + \sum_i \frac{\omega_i}{2} ,
\end{eqnarray}
which implies $S^{(0)} \to 0$. 
Using Eqs.\ (\ref{eq:Omg1_Ag}) and (\ref{eq:U1AG}), we obtain
\begin{eqnarray}
\lim_{T \to 0} \Omega^{(1)} =
\lim_{T \to 0} U^{(1)} &=& \frac{1}{2} \sum_i \tilde{\bar{F}}_{ii} + \frac{1}{8} \sum_i \tilde{{F}}_{iijj}  ,
\end{eqnarray}
for a QFF, because $\beta^n f_i \asymp \beta^n \exp(-\beta\omega_i) \to 0$ as $T \to 0$ for any $n$. This again implies $S^{(1)} \to 0$. 
Summing them together, we conclude
\begin{eqnarray}
\lim_{T \to 0} \left( \Omega^{(0)} + \Omega^{(1)} \right)&=& 
\lim_{T \to 0} \left( U^{(0)} + U^{(1)} \right) = E_{\text{XVSCF}}(0), \nonumber\\
\end{eqnarray}
where the zero-temperature XVSCF energy expression, $E_{\text{XVSCF}}(0)$, can be found in Refs.\ \onlinecite{Keceli2011,Hermes2012,Hermes2013_Dyson,Hirata2014} or obtained by substituting $f_i = 0$ in 
Eq.\ (\ref{eq:EXVSCFT}).  

From Eq.\ (\ref{eq:Omg2_AG_final}), we get
\begin{eqnarray}
\label{eq:Omg2_zeroT}
\lim_{T \to 0} \Omega^{(2)}& =& 
 \sum_{i} \frac{\tilde{F}_{i} \tilde{F}_{i}}{-\omega_{i}}
+ \sum_{i, j} \frac{\tilde{F}_{i} \tilde{F}_{i j j}}{-\omega_{i}}  
 +\frac{1}{4} \sum_{i, j, k}  \frac{\tilde{F}_{i j j} \tilde{F}_{i k k}}{-\omega_{i}} \nonumber\\
& &
+ \frac{1}{2} \sum_{i, j} \frac{\tilde{\bar{F}}_{i j} \tilde{\bar{F}}_{i j}}{-\omega_{i}-\omega_{j}} 
+  \frac{1}{2} \sum_{i, j, k} \frac{\tilde{\bar{F}}_{i j} \tilde{F}_{i j k k}}{-\omega_{i}-\omega_{j}}  \nonumber\\
&&
 + \frac{1}{8} \sum_{i ,j, k, l} \frac{\tilde{F}_{i j k k} \tilde{F}_{i j l l}}{-\omega_{i}-\omega_{j}} 
  + \frac{1}{6} \sum_{i,j,k}   \frac{\tilde{F}_{ijk} \tilde{F}_{ijk} } {-\omega_i-\omega_j-\omega_k} \nonumber\\
& &
+\frac{1}{24} \sum_{i,j,k,l}  \frac{\tilde{F}_{ijkl} \tilde{F}_{ijkl} }{{-\omega_i-\omega_j-\omega_k-\omega_l}} ,
\end{eqnarray}
where ``denom.$\neq$0'' restrictions are lifted as they are automatically fulfilled under the assumption $\omega_i > 0$. 
The above formula is equal to the XVMP2 correction.\cite{Hermes2013} 
All anomalous terms (as well as all renormalization terms at higher orders) vanish at $T = 0$ because with $f_i=0$ the diagrams that  contain upgoing arrows only (each giving rise to a $f_i + 1$ factor) can possibly survive, but
their denominator factors that take the form of $-\omega_i - \omega_j - \omega_k \dots$ cannot satisfy
the ``denom.=0'' restrictions under the assumption $\omega_i > 0$. 

From Eqs.\ (\ref{eq:U_from_deriv}) and (\ref{eq:f_deriv}), generally, we conclude
\begin{eqnarray}
\lim_{T \to 0} \Omega^{(n)} = \lim_{T \to 0} U^{(n)} &=& E_0^{(n)}, \label{eq:zeroT_1}\\
\lim_{T \to 0} S^{(n)} &=& 0. \label{eq:zeroT_2}
\end{eqnarray}

In the electronic case, the Kohn--Luttinger nonconvergence problem\cite{Kohn1960,Hirata_KL2021,Hirata_KL2022} occurs when the reference wave function is degenerate whereas the exact ground-state wave function is not or it  connects adiabatically to an exact excited state. 
In the vibrational case, a degenerate or excited reference is prohibited by the assumption $\omega_i > 0$ for all $i$. 
The Kohn--Luttinger nonconvergence problem,\cite{Hirata_KL2021,Hirata_KL2022} therefore,
does not exist in the present finite-temperature perturbation theory for vibrations.

% =====================================
% XVSCF(T) and references
% =====================================

\section{Finite-temperature XVSCF \label{section:finiteT_XVSCF}}

%============================================
% Table XVSCF
%============================================

In Appendix \ref{appendix:normalorderedHam}, the Hamiltonian is brought to a finite-temperature normal-ordered form.
Its constant part [Eq.\ (\ref{eq:EXVSCFT})], denoted $E_\text{XVSCF}(T)$, reduces to the XVSCF zero-point energy expression\cite{Keceli2011,Hermes2012,Hermes2013_Dyson,Hirata2014} 
as $T \to 0$. In the same limit, the dressed force constants $W_i$ [Eq.\ (\ref{eq:Wi_appendix})] and $W_{ij}$ [Eq.\ (\ref{eq:Wij_appendix})] 
become the XVSCF effective gradients and quadratic force constants, respectively,
defining the XVSCF working equation.\cite{Keceli2011,Hermes2012,Hermes2013_Dyson,Hirata2014} Hence, the finite-temperature normal-ordered Hamiltonian [Eq.\ (\ref{eq:normalorderedHam})] 
can be considered as postulating a finite-temperature extension of XVSCF. This mirrors the electronic case: The (finite-temperature) normal-ordered Hamiltonian 
naturally isolates the constant part that is the (finite-temperature) HF energy expression. Its one-electron operator becomes the (finite-temperature) Fock operator.\cite{Mermin,Shavitt2009,Hirata2021}

The finite-temperature XVSCF ``energy,'' $E_{\mathrm{XVSCF}}(T)$, is neither
a grand potential nor internal energy at some perturbation order. It instead bears the following relationship:
\begin{eqnarray}
    E_{\mathrm{XVSCF}}(T) &=& \Big[E_N^{(0)}\Big] + \Big[E_N^{(1)}\Big] \\
    &=& U^{(0)} + \Omega^{(1)} \\
&=& \Omega^{(0)} + \Omega^{(1)} + TS^{(0)} \\ 
&=& U^{(0)} + U^{(1)} - TS^{(1)}.
\end{eqnarray}
Modal frequencies of the finite-temperature XVSCF will differ from those at zero temperature, and their physical meaning is unknown at this point.
The situation is again analogous to the electronic case:\cite{Hirata2021} The constant part (``energy'') of the finite-temperature normal-ordered Hamiltonian is 
not equal to the first-order grand potential or internal energy. The physical meaning of its orbital energies 
is still unknown.\cite{Pain} Nevertheless, the fact that the quantities of XVSCF (but not of VSCF) and HF emerge naturally in the normal-ordered Hamiltonians underscores 
the central place XVSCF and HF theories occupy in the framework of {\it ab initio} vibrational and electronic structure theories, respectively.\cite{Shavitt2009,Hirata2021}

\begin{table}
\caption{\label{table:XVSCF}
Comparison of the harmonic approximation, XVSCF, and SCP methods at zero and nonzero temperatures, where 
$\nu_i$ is the $i$th modal anharmonic frequency of the respective methods, while $\omega_i$ is the $i$th modal harmonic frequency. 
See Eqs.\ (\ref{eq:sorted_V1})--(\ref{eq:sorted_V4}) for bare force constants ${F}$, and Eqs.\ (\ref{eq:Wi_appendix})--(\ref{eq:Wijkl_appendix}) for XVSCF dressed force constants $W$.
See Ref.\ \onlinecite{Hermes2013_Dyson} for the first-order Dyson geometry and coordinates.}
\begin{ruledtabular}
\begin{tabular}{lllll}
Method & Geometry & Gradient & Coordinate & Frequency \\ \hline
Harmonic      & Equilibrium & ${F}_i = 0$ & Normal & ${F}_{ij} = \delta_{ij} \omega_{i}^2$ \\ 
XVSCF($n$) & Equilibrium  & ${F}_i = 0$ & Normal & ${W}_{ii} + \omega_i / 2 = \nu_i$ \\
XVSCF[$n$] & Dyson & $W_i = 0$  & Normal & ${W}_{ii} + \omega_i / 2 = \nu_i$ \\
SCP($n$) & Equilibrium & ${F}_i = 0$ & Dyson & ${W}_{ij} + \delta_{ij} \omega_i / 2 = \delta_{ij} \nu_i$ \\
SCP[$n$] & Dyson & $W_i = 0$ & Dyson & ${W}_{ij} + \delta_{ij} \omega_i / 2 = \delta_{ij} \nu_i$ \\
\end{tabular}
\end{ruledtabular} 
\end{table}

The foregoing formulation of the finite-temperature many-body perturbation theory is valid, without modification, for any harmonic reference.
The latter includes the harmonic approximation and all incarnations of XVSCF,\cite{Hirata2010,Keceli2011,Hermes2012,Hermes2013_Dyson,Hirata2014} 
i.e., XVSCF($n$),\cite{Keceli2011} XVSCF[$n$],\cite{Hermes2012} SCP($n$),\cite{Hermes2013_Dyson,Hooton1958,Koehler1966,Choquard1967,Gillis1968} and SCP[$n$],\cite{Hermes2013_Dyson,Hooton1958,Koehler1966,Choquard1967,Gillis1968} (where $n$ is the truncation
rank of the force constants) as well as their finite-temperature extensions postulated above. 
Table \ref{table:XVSCF} summarizes the relationships among these mean-field methods. Reference \onlinecite{Hermes2013_Dyson} has a similar table.
It also gives the definitions of the first-order Dyson geometry and coordinates as well as the working equations of the respective methods. The new information provided in Table \ref{table:XVSCF} is
that each row now characterizes both finite- and zero-temperature methods. 

The XVSCF($n$) and XVSCF[$n$] methods rely on normal coordinates and leave
the dressed quadratic-force-constant matrix $(W_{ij})$ nondiagonal. Its diagonal elements report anharmonic modal frequencies. 
On the other hand, SCP($n$) and SCP[$n$] diagonalize $(W_{ij})$, determining the first-order Dyson coordinates,\cite{Hermes2013_Dyson} 
the vibrational analog of the Dyson orbitals.\cite{OrtizDyson} They account for the first-order anharmonic effects on the coordinates. The eigenvalues of $(W_{ij})$ 
lead to anharmonic modal frequencies.
The XVSCF[$n$] and SCP[$n$] methods, furthermore, shift the center of the PES from the equilibrium geometry to the first-order Dyson one,\cite{Hermes2013_Dyson} 
at which the dressed gradient vector $(W_i)$ is zero.
Therefore, with the finite-temperature XVSCF[$n$] reference, the first two terms and first two anomalous terms (multiplied by $\beta$) 
of $\Omega^{(2)}$ in Eq.\ (\ref{eq:Omg_final_SQ}) are zero because they contain a factor of $W_i$.
With the harmonic reference, all of the numerous terms of $\Omega^{(2)}$ in Eq.\ (\ref{eq:Omg2_AG_final}) that contain a factor of $\tilde{F}_i$ or $\tilde{\bar{F}}_{ij}$ vanish.
More importantly, the foregoing formulas retain all these potentially nonzero terms and are valid for any harmonic reference.

% =====================================
% General-order calculation
% =====================================

\section{General-order calculations}

\begin{table*}
\caption{\label{table:highorder10K} The $n$th-order perturbation corrections (in $E_\text{h}$) to the grand potential ($\Omega$) and internal energy ($U$) per molecule of an ideal gas of the identical, non-rotating water molecules at $T= 10$ K. }
\begin{ruledtabular}
\begin{tabular}{ldddddd}
  & \multicolumn{3}{c}{$\Omega^{(n)}$ / $E_{\mathrm{h}}$} & \multicolumn{3}{c}{$U^{(n)}$ / $E_{\mathrm{h}}$} \\\cline{2-4} \cline{5-7}
 \multicolumn{1}{l}{$n$} & \multicolumn{1}{r}{Recursion\footnotemark[1]} &  \multicolumn{1}{r}{$\lambda$-Variation\footnotemark[2]}  & \multicolumn{1}{r}{Analytical\footnotemark[3]} & \multicolumn{1}{r}{Recursion\footnotemark[1]}  &  \multicolumn{1}{r}{$\lambda$-Variation\footnotemark[2]} & \multicolumn{1}{r}{Analytical\footnotemark[3]} \\
\hline
0	&	0.021410	&	0.021410	&	0.021410	&	0.021410	&	0.021410	&	0.021410	\\
1	&	0.000234	&	0.000234	&	0.000234	&	0.000234	&	0.000234	&	0.000234	\\
2	&	-0.000540	&	-0.000540	&	-0.000540	&	-0.000540	&	-0.000540	&	-0.000540	\\
3	&	0.000121	&	0.000121	&	 	&	0.000121	&	0.000121	&	   	\\
4	&	-0.000098	&	-0.000098	&	  	&	-0.000098	&	-0.000098	&	 	\\
5	&	0.000058	&	0.000059	&	   	&	0.000058	&	0.000059	&	   	\\
6	&	-0.000049	&	-0.000049	&	  	&	-0.000049	&	-0.000049	&	 	\\
7	&	0.000041	&	  	&	  	&	0.000041	&	 	&	      	\\
8	&	-0.000040	&	 	&	   	&	-0.000040	&	  	&	   	\\
$\sum_0^8$	&	0.021137	&		&		&	0.021137	&		&		\\
FCI\footnotemark[4] 	&	0.021157	&		&		&	0.021157	&		&		\\
\end{tabular}
\end{ruledtabular}
\footnotetext[1]{Based on sum-over-states formulas from the recursions that are evaluated with a modified finite-temperature vibrational FCI code with 16 harmonic-oscillator basis functions. See Sec.\ \ref{section:alg:recursion}.}
\footnotetext[2]{Based on a central seven-point finite-difference approximation with $\Delta \lambda= 0.01$ to the finite-temperature vibrational FCI results obtained with $\hat{H} = \hat{H}^{(0)} + \lambda\hat{V}^{(1)}$ 
and 16 harmonic-oscillator basis functions. See Sec.\ \ref{section:alg:lambda}.}
\footnotetext[3]{Based on sum-over-modes formulas including the Bose--Einstein formulas, which are basis-set free. See Sec.\ \ref{section:alg:analytical}.}
\footnotetext[4]{Obtained with the finite-temperature vibrational FCI method\cite{Xiuyi2021} with 16 harmonic-oscillator basis functions.}
\end{table*}

\begin{table*}
\caption{\label{table:highorder103K} Same as Table \ref{table:highorder10K}, but at $T= 10^3$ K. }
\begin{ruledtabular}
\begin{tabular}{ldddddd}
  & \multicolumn{3}{c}{$\Omega^{(n)}$ / $E_{\mathrm{h}}$} & \multicolumn{3}{c}{$U^{(n)}$ / $E_{\mathrm{h}}$} \\\cline{2-4} \cline{5-7}
 \multicolumn{1}{l}{$n$} & \multicolumn{1}{r}{Recursion\footnotemark[1]} &  \multicolumn{1}{r}{$\lambda$-Variation\footnotemark[2]}  & \multicolumn{1}{r}{Analytical\footnotemark[3]} & \multicolumn{1}{r}{Recursion\footnotemark[1]}  &  \multicolumn{1}{r}{$\lambda$-Variation\footnotemark[2]} & \multicolumn{1}{r}{Analytical\footnotemark[3]} \\
\hline
0	&	0.021066	&	0.021066	&	0.021066	&	0.022331	&	0.022331	&	0.022331	\\
1	&	0.000195	&	0.000195	&	0.000195	&	0.000282	&	0.000282	&	0.000282	\\
2	&	-0.000547	&	-0.000547	&	-0.000547	&	-0.000491	&	-0.000491	&	-0.000491	\\
3	&	0.000123	&	0.000123	&	    	&	0.000098	&	0.000098	&		\\
4	&	-0.000106	&	-0.000106	&	    	&	-0.000062	&	-0.000062	&		\\
5	&	0.000066	&	0.000066	&	    	&	0.000025	&	0.000025	&		\\
6	&	-0.000058	&	-0.000058	&	    	&	-0.000007	&	-0.000007	&		\\
7	&	0.000052	&	   	&	   	&	-0.000008	&	   	&		\\
8	&	-0.000054	&	   	&	    	&	0.000024	&	   	&		\\
$\sum_0^8$  	&	0.020737	&	   	&	   	&	0.022194	&	   	&		\\
FCI\footnotemark[4]  	&	0.020757	&	   	&	    	&	0.022228	&	   	&		\\
\end{tabular}
\end{ruledtabular}
\footnotetext[1]{See the corresponding footnote of Table \ref{table:highorder10K}.}
\footnotetext[2]{See the corresponding footnote of Table \ref{table:highorder10K}.}
\footnotetext[3]{See the corresponding footnote of Table \ref{table:highorder10K}.}
\footnotetext[4]{See the corresponding footnote of Table \ref{table:highorder10K}.}
\end{table*}

\begin{table*}
\caption{\label{table:highorder105K} Same as Table \ref{table:highorder10K}, but at $T= 10^4$ K. }
\begin{ruledtabular}
\begin{tabular}{ldddddd}
  & \multicolumn{3}{c}{$\Omega^{(n)}$ / $E_{\mathrm{h}}$} & \multicolumn{3}{c}{$U^{(n)}$ / $E_{\mathrm{h}}$} \\\cline{2-4} \cline{5-7}
 \multicolumn{1}{l}{$n$} & \multicolumn{1}{r}{Recursion\footnotemark[1]} &  \multicolumn{1}{r}{$\lambda$-Variation\footnotemark[2]}  & \multicolumn{1}{r}{Analytical\footnotemark[3]} & \multicolumn{1}{r}{Recursion\footnotemark[1]}  &  \multicolumn{1}{r}{$\lambda$-Variation\footnotemark[2]} & \multicolumn{1}{r}{Analytical\footnotemark[3]} \\
\hline
0	&	-0.081153	&	-0.081153	&	0.081916	&	0.093852	&	0.093852	&	0.096790	\\
1	&	-0.000026	&	-0.000026	&	0.000672	&	-0.001401	&	-0.001401	&	0.000890	\\
2	&	-0.011827	&	-0.011827	&	-0.013770	&	0.010936	&	0.010936	&	0.016635	\\
3	&	0.013168	&	0.013167	&	                     	&	-0.023082	&	-0.023081	&	              	\\
4	&	-0.033952	&	-0.033951	&	                     	&	0.063492	&	0.063490	&	              	\\
5	&	0.086246	&	0.087326	&	                     	&	-0.179668	&	-0.181634	&	              	\\
6	&	-0.025221	&	-0.025553	&	                     	&	0.514059	&	0.518805	&	              	\\
7	&	0.763838	&	                      	&	                     	&	-1.397265	&	             	&	              	\\
8	&	-2.348398	&	                      	&	                     	&	3.386418	&	             	&	              	\\
$\sum_0^8$ 	&	-1.864313	&	                      	&	                     	&	2.467342	&	             	&	              	\\
FCI\footnotemark[4]        	&	-0.093747	&	                      	&	                     	&	0.093530	&	             	&	              	\\
\end{tabular}
\end{ruledtabular}
\footnotetext[1]{See the corresponding footnote of Table \ref{table:highorder10K}.}
\footnotetext[2]{See the corresponding footnote of Table \ref{table:highorder10K}.}
\footnotetext[3]{See the corresponding footnote of Table \ref{table:highorder10K}.}
\footnotetext[4]{See the corresponding footnote of Table \ref{table:highorder10K}.}
\end{table*}

We implemented three distinct and independent algorithms of the finite-temperature many-body perturbation theory
and compared their results. The algorithms developed are as follows:

\subsection{The $\lambda$-variation method\label{section:alg:lambda}}

Several lowest-order perturbation corrections of any quantity in any perturbation theory with any reference (or Hamiltonian partitioning) can be determined 
as the $\lambda$-derivatives of the same quantity calculated by 
an exact (FCI) method with a scaled Hamiltonian $\hat{H} = \hat{H}^{(0)} + \lambda\hat{V}^{(1)}$ [Eq.\ (\ref{eq:X_perturbed})]. This general numerical technique---the $\lambda$-variation method---was conceived
by Knowles {\it et al.}\cite{private2}\  in the course of their general-order many-body perturbation algorithm development.\cite{Knowles} It was adopted by one of the present authors with coauthors in
the time-independent formulation of one-particle many-body Green's-function theory\cite{Hirata2017} and 
of finite-temperature many-body perturbation theory for electrons.\cite{Jha2019,Hirata2019,Jha2020,Hirata_2020,Hirata2021} 
We also employ this method in this study, applying a finite-difference approximation to the $\lambda$-derivatives of the finite-temperature vibrational FCI results.\cite{Xiuyi2021,Qin_code2021}

A complication occurring uniquely for finite-temperature vibrations (not seen in electronic cases\cite{Hirata2017,Jha2019,Hirata2019,Jha2020,Hirata_2020,Hirata2021})
is the presence of finite-basis-set errors. They are substantial at higher temperatures.\cite{Xiuyi2021}
As emphasized in Introduction, the Bose--Einstein theory   is a basis-set-free theory because 
infinitely many energy levels can be analytically summed over. The present finite-temperature perturbation theory
inherits the basis-set-free nature of the Bose--Einstein reference. In contrast, the vibrational FCI method determines the energy levels as eigenvalues of a Hamiltonian matrix in Hartree-product states
spanned by a finite one-mode basis set. Therefore, at higher temperatures, as more high-lying vibrational states become populated, 
FCI begins to become less accurate than the perturbation theory or even the Bose--Einstein theory. 
For example, entropy in the high-temperature limit is determined solely by the number of available states, and this value is  infinite in the Bose--Einstein theory, in the perturbation theory thereof, and in reality.
It, however, plateaus at a small finite value according to the vibrational FCI method. 
Generally, the finite-temperature vibrational FCI results are 
reliable when  $k_{\text{B}}T <  \omega$, where $\omega$ is the lowest modal harmonic frequency.\cite{Xiuyi2021}

Our computer implementation of the $\lambda$-variation method capitalizes on the finite-temperature vibrational FCI program reported earlier.\cite{Xiuyi2021,Qin_code2021}
It was modified to accommodate a scaled Hamiltonian, $\hat{H} = \hat{H}^{(0)} + \lambda\hat{V}^{(1)}$, with an arbitrary value of $\lambda$. We executed the FCI calculations 
with seven values of $\lambda$ on an evenly spaced grid  ($\Delta\lambda=0.01$)  symmetrically about $\lambda=0$, so that we could approximate the $\lambda$-derivatives of Eq.\ (\ref{eq:X_perturbed})
by central seven-point finite-difference formulas.

\subsection{Sum-over-states analytical formulas from the recursions\label{section:alg:recursion}}

We literally translated the recursions for $\Omega^{(n)}$ [Eq.\ (\ref{eq:recursionOmg})] and $U^{(n)}$ [Eq.\ (\ref{eq:recursion_U})] into a general-order program by modifying 
the finite-temperature vibrational FCI program.\cite{Xiuyi2021,Qin_code2021}
The computational kernel of FCI is the action of $\hat{V}^{(1)}$ onto any wave function that is a linear combination of Hartree products, and it 
can be reused to evaluate the sum-over-states expressions of these perturbation
corrections generated by the recursions. 

The degenerate RSPT corrections to state energies form a block-diagonal matrix $\bm{E}^{(n)}$  [Eq.\ (\ref{eq:E_In_degen})], whose off-diagonal elements are nonzero within each degenerate subspace. These matrices may either be diagonalized or left undiagonalized, and correspondingly we can use the left- and right-hand sides of Eqs.\ (\ref{eq:traceinvariance})--(\ref{eq:traceinvariance3}) in the algorithms evaluating their thermal averages. 
Both are mathematically correct, but only the latter lends itself to algebraic or second-quantized reduction because the result of diagonalization cannot be expressed in a closed analytical form (see Secs.\ \ref{section:AGderivation} and \ref{section:2ndquan}).
In our implementation, we elected to leave the $\bm{E}^{(n)}$ matrices undiagonalized in order to numerically verify the validity of the nondiagonal route of evaluating the thermal averages.  

It is important to recognize that this general-order algorithm is also limited by 
the same finite-basis-set errors mentioned above, even though the recursions and sum-over-states formulas 
formally embody the basis-set-free perturbation theory. When implemented into a computer program, however,
the infinite sums over states implied in these formulas 
are approximated by short finite sums, thereby introducing the finite-basis-set errors.
This is in line with our observation\cite{Xiuyi2021} that the finite-temperature vibrational FCI calculation with $\lambda=0$ does not necessarily reproduce 
the Bose--Einstein theory at higher temperatures because the former is limited by a finite basis set, whereas the latter is not.

\subsection{Sum-over-modes analytical formulas\label{section:alg:analytical}}

We also implemented sum-over-modes analytical formulas of $\Omega^{(n)}$ and $U^{(n)}$ for $0 \leq n \leq 2$. They are given by Eqs.\ (\ref{eq:Omg_0}), (\ref{eq:U_0}), (\ref{eq:Omg1_Ag}), (\ref{eq:U1AG}), and 
(\ref{eq:Omg2_AG_final}). The $U^{(2)}$ working equation has not been given above because it is long, but can be easily generated via Eq.\ (\ref{eq:alternativeU2}) and then directly programmed.

Remarkably, these reduced formulas are not only inexpensive to evaluate but also basis-set-free. 
This is because the algebraic or second-quantized reduction process (Secs.\ \ref{section:AGderivation} and \ref{section:2ndquan}) achieves two feats: (1) It factorizes a thermal average of $m$-dimensional integrals over $m$-mode states into a product of thermal averages of one-dimensional integrals over one-mode states (where $m$ is the vibrational degrees of freedom); (2) It then carries out the summation over infinitely many one-mode states analytically and exactly. 

\subsection{Numerical results and discussion}

The values of $\Omega^{(n)}$ and $U^{(n)}$ per molecule were computed for an ideal gas of identical, non-rotating water molecules 
at $T = 10$, $10^3$, and $10^4$\,K. The results are summarized in Tables \ref{table:highorder10K}, \ref{table:highorder103K}, and \ref{table:highorder105K}.
The three algorithms described above, denoted $\lambda$-variation, `recursion,' and `analytical' in these tables, were used. 
An ideal gas means that there is no intermolecular interaction. The rotational and electronic temperatures were set to zero. The equilibrium geometry, normal coordinates,
and QFF of the water molecule were determined by the second-order M{\o}ller--Plesset perturbation theory with the aug-cc-pVTZ basis set. The harmonic approximation was adopted
as the reference. The same system was used in our finite-temperature vibrational FCI study previously.\cite{Xiuyi2021}

The temperature of 10\,K  is so low as compared with the vibrational frequencies of the water molecule that the data in Table \ref{table:highorder10K} may be viewed as the zero-temperature limits. 
The results from the three algorithms agree with one another for six decimal places for both $\Omega^{(n)}$ and $U^{(n)}$ at any order, mutually verifying their formalisms and computer codes. 
The finite-basis-set errors that are formally present in the $\lambda$-variation and recursion methods  are not noticeable in this case because the reference (zero-point) state dominates 
in every thermal average. The table also confirms the theoretically expected zero-temperature behaviors of Eqs.\ (\ref{eq:zeroT_1}) and (\ref{eq:zeroT_2}).   
The values of $\Omega^{(n)}$ or $U^{(n)}$ for $0 \leq n \leq 2$ agree with the results of the zero-temperature perturbation (XVH2) theory.\cite{Hermes2013} 
The accumulated perturbation corrections  at $n=8$ recover 99.9\% of the finite-temperature FCI value, suggesting that the perturbation series is convergent 
at exactness with no terms overlooked or neglected. While this is expected or even guaranteed when the theory is formulated bottom-up and algebraically,
it is not as self-evident when the theory is postulated diagrammatically and top-down.

Table \ref{table:highorder103K} shows that, at $T = 10^3$\,K, $\Omega^{(n)}$ and $U^{(n)}$ differ from each other by the entropy contribution. 
The three algorithms agree with one another for all shown digits, indicating that the formalisms are correct and that the finite-basis-set errors are still negligible.
The sum of  $\Omega^{(n)}$ or $U^{(n)}$ over $0 \leq n \leq 8$ accounts for 99.9\% of the respective finite-temperature FCI values. The perturbation theory
is, therefore, convergent at exactness. This would not be the case if the off-diagonal elements of $\bm{E}^{(n)}$ 
or any one of the anomalous and/or renormalization diagrams were neglected or overlooked. 

Table \ref{table:highorder105K}, on the other hand, illustrates the manifestation of the finite-basis-set effects at $T = 10^4$\,K, making the results of the $\lambda$-variation
and `recursion' methods differ visibly from those of the `analytical' method. Generally, the former two basis-set methods underestimate the magnitude of all thermodynamic functions because 
of the limited number of states spanned by a small basis set.\cite{Xiuyi2021} %The disagreement must not be taken as a formalism or programming error. Rather, the `analytical' formalisms
%that will be actually useful in molecular or solid-state applications have the remarkable ability to take into account infinitely many states at a low cost. The `recursion' method, which is used here only for benchmarking purposes, is based on a finite-basis-set vibrational FCI algorithm,\cite{Xiuyi2021} whose practical applicability is severely limited. 
At $T = 10^4$\,K, both $\Omega^{(n)}$ and $U^{(n)}$ series show signs of divergence, which are common in many perturbation theory applications.\cite{Olsen,HirataCC,Hirata2021}

\section{Conclusions}

Thermal effects are critically important on accurate vibrational structure calculations of molecules and solids.
Their inclusion requires accounting for anharmonic effects in a large or infinite number of vibrational states simultaneously, accurately, and size-consistently. These exacting requirements pose a severe constraint on what types of mathematical theories are viable. 
Since the Bose--Einstein theory solves the problem of statistical thermodynamics of harmonic oscillators exactly, a finite-temperature many-body perturbation theory 
that starts from it and systematically incorporates anharmonic effects as perturbation is the most promising approach.
% In fact, presently, it appears to be the only one established to work systematically accurately for thermodynamics of macroscopic systems.

This article is the most comprehensive and complete exposition of the finite-temperature many-body perturbation theory for anharmonic vibrations
based on the Bose--Einstein theory. It is a generalization of crystal phonon perturbation theory,\cite{Maradudin1961,Maradudin1961_2,Flinn1963,Cowley1963,Cowley1968,Shukla1971_2,Shukla1974,Shukla1985,Shukla1985_2} 
or a finite-temperature extension of the XVMP theory.\cite{Hermes2013} It may also be viewed as the vibrational analog of the finite-temperature many-body 
perturbation theory for electrons.\cite{Jha2019,Hirata2019,Jha2020,Hirata_2020,Hirata2021} It is diagrammatically linked and size-consistent at any order, 
as proved by the linked-diagram theorem in this study, and therefore applicable to molecules and solids on an equal footing. It inherits from the Bose--Einstein theory 
the remarkable ability of accumulating anharmonic effects from infinitely many states analytically. It is, therefore, a basis-set-free method, simultaneously 
achieving efficiency, accuracy, size-consistency, and systematic convergence. 

Unlike the conventional top-down diagrammatic derivation, % of a few lowest-order approximations in crystal phonon perturbation theory, 
we adopted a bottom-up approach to formulating the whole perturbation series starting with the Rayleigh--Schr\"{o}dinger-type recursions,  maintaining all algebraic terms so that the series converge at exactness. Many of these terms are graphically represented by counterintuitive diagrams such as
anomalous, renormalization, and anomalous-renormalization diagrams, whose resolvent lines are removed or shifted. We felt that appealing to human intuition to exhaustively enumerate all these 
diagrams is an unreliable starting point of a many-body theory formulation. We, therefore, resorted instead to a general and powerful strategy of deriving the recursions of any perturbation theory by way of 
taking the $\lambda$-derivatives.\cite{Hirata2017,Hirata2021}  

The recursions lead to sum-over-states analytical formulas for the perturbation corrections to thermodynamic functions. 
We developed three distinct methods of reducing them into sum-over-modes analytical formulas: 

The first method is a purely algebraic one, which factorizes each many-mode thermal average
into a product of one-mode thermal averages, and then evaluates the latter using the thermal Born--Huang rules. This leads to a combinatorial number of terms differing in the patterns of 
index coincidences, which are to be consolidated into much fewer terms. We introduced the canonical forms of the rules, which facilitate this otherwise intractable consolidation process. 
A major drawback of this method is that the thermal Born--Huang rules and their canonical forms need to be expanded as the perturbation order is raised. 
The value of this reduction method, nevertheless, lies in the fact that it shows that 
a complete set of sum-over-modes formulas at any order can, in principle, be obtained in a transparent and plain (albeit tedious) 
algebraic logic starting from the time-independent Schr\"{o}dinger equation and the partition function.

The second method uses the normal-ordered second quantization at finite temperature, which we fully developed in this study.
It is as transparent and straightforward as the algebraic reduction, but far more expedient. It furthermore lays a natural foundation for the subsequent graphical representations by Feynman diagrams.
In Appendix \ref{appendix:normalorderedSQ}, we stipulated the normal orders of harmonic-oscillator ladder operators at finite temperature, defined Wick contractions, and proved thermal Wick's theorem.
We then derived a finite-temperature normal-ordered form of the pure vibrational Hamiltonian, showing that its constant part defines a finite-temperature extension of the XVSCF energy while 
its force constants become the finite-temperature XVSCF force constants, which are dressed with higher-order force constants of the ring-diagram types. Hence, as a by-product of this study, we stipulated a finite-temperature extension
of XVSCF. 

The third and most expedient method is based on the time-independent Feynman diagrammatic rules. They are distinguished from those in crystal phonon perturbation theory by the presence of
the resolvent lines. Each diagram is merely a graphical representation of all equal-valued full contraction patterns in the finite-temperature normal-ordered reductions. The rules are, therefore, 
justified by the second method of reduction and include the need to draw anomalous, renormalization, and anomalous-renormalization diagrams mentioned above. The rule that only linked
diagrams need to be drawn is justified by the linked-diagram theorem for finite-temperature vibrations proved in this study. 

The zero-temperature limits of the perturbation corrections to the grand potential and internal energies are the corresponding corrections to the zero-point energy as determined by the zero-temperature
perturbation theory. The Kohn--Luttinger-type nonconvergence problem does not manifest itself for vibrations either in a molecule or crystal.

One general-order algorithm and two order-by-order algorithms were developed. The numerically exact agreement of the results among these three algorithms underscores the correctness of the formalisms
and computer implementations. The deviations that manifest at higher temperatures are ascribed to the finite-basis-set errors in the CI-based algorithms of the recursion and $\lambda$-variation methods, although the 
perturbation theory itself is fundamentally basis-set-free and can be implemented as such in the most efficient algorithm. 
The sums of the perturbation corrections converge at the finite-temperature vibrational FCI limits, numerically demonstrating that the perturbation theory is correct and complete with no term being 
overlooked or neglected. To achieve this convergence, it is crucial to treat the perturbation corrections to state energies as matrices (not scalars) 
and take into account their nonzero off-diagonal elements. 

\acknowledgments
This work was supported by the U.S. Department of Energy (DoE), Office of Science, Office of Basic Energy Sciences under Grant No.\ DE-SC0006028
and also by the Center for Scalable Predictive methods for Excitations and Correlated phenomena (SPEC), which is funded by the U.S. DoE, Office of Science, Office of Basic Energy Sciences, Division of Chemical Sciences, Geosciences and Biosciences as part of the Computational Chemical Sciences (CCS) program at Pacific Northwest National Laboratory (PNNL) under FWP 70942. PNNL is a multi-program national laboratory operated by Battelle Memorial Institute for the U.S. DoE.
S.H.\ is a Guggenheim Fellow of the John Simon Guggenheim Memorial Foundation.

\appendix

%============================================
% APPENDIX Thermal Born--Huang rules
%============================================

\section{Thermal Born--Huang rules \label{Appendix:BHrules}}

%============================================
% Table IX
%============================================
\begin{table*}
\caption{\label{table:matrix_ele}
The zero-temperature Born--Huang rules.\cite{born1988dynamical,Keceli2011}  $\langle \hat{F}(Q_i) \rangle_{\delta_i} = \langle n_i | \hat{F}(Q_i) | n_i-\delta_i \rangle$, where $\hat{F}(Q_i)$ 
is an operator made of the $i$th normal coordinate $Q_i$, and $|n_i\rangle$
is the harmonic oscillator wave function with quantum number $n_i$ and frequency $\omega_i$. All other cases are zero for a QFF.}
\begin{ruledtabular}
\begin{tabular}{cccccc}
 $ \delta_i   $ & $\langle  {\partial^2}/{\partial Q_i^2} \rangle_{\delta_i}$ &   $\langle  Q_i  \rangle_{\delta_i}$ &  $\langle Q_i^2 \rangle_{\delta_i}$  &  $\langle  Q_i^3 \rangle_{\delta_i}$ &  $\langle Q_i^4  \rangle_{\delta_i}$ \\
 \hline
0 & $-{\omega_i} (n_i+1/2)$ & 0 & ${(n_i+1/2)}/{\omega_i}$ & 0 & ${(6n_i^2+6n_i+3)}/{4\omega_i^2}$ \\
$1$ & 0 &$\sqrt{{n_i}/{2\omega_i}}$ & 0  &  $3({n_i}/{2\omega_i})^{3/2}$ & 0   \\
$2$ & ${\omega_i}\sqrt{n_i(n_i-1)}/2$ &  0 & ${\sqrt{n_i(n_i-1)}}/{2\omega_i}$ & 0 & ${(n_i-1/2)}\sqrt{n_i(n_i-1)}/{\omega_i^2}$ \\
$3$ & 0 &  0 & 0 & $\sqrt{{n_i(n_i-1)(n_i-2)}}/{(2\omega_i)^{3/2}}$ & 0  \\
$4$ & 0 &  0 & 0  & 0 & $\sqrt{n_i(n_i-1)(n_i-2)(n_i-3)}/{4\omega_i^2}$ \\
$-1$ & 0 &$\sqrt{(n_i+1)/{2\omega_i}}$ & 0  &  $3\{(n_i+1)/{2\omega_i}\}^{3/2}$ & 0   \\
$-2$ & ${\omega_i}\sqrt{(n_i+2)(n_i+1)}/2$ &  0 & ${\sqrt{(n_i+2)(n_i+1)}}/{2\omega_i}$ & 0 & ${(n_i+3/2)}\sqrt{(n_i+2)(n_i+1)}/{\omega_i^2}$ \\
$-3$ & 0 &  0 & 0 & $\sqrt{{(n_i+3)(n_i+2)(n_i+1)}}/{(2\omega_i)^{3/2}}$ & 0  \\
$-4$ & 0 &  0 & 0  & 0 & $\sqrt{(n_i+4)(n_i+3)(n_i+2)(n_i+1)}/{4\omega_i^2}$ \\
\end{tabular}
\end{ruledtabular} 
\end{table*}

The thermal Born--Huang rules summarized in Tables \ref{table:thermalavg_EN1} and \ref{table:thermalavg_EN2} were originally derived by Born and Huang\cite{born1988dynamical} on the basis of 
the zero-temperature rules also introduced by the same authors. The latter are reproduced in Table \ref{table:matrix_ele} for a QFF.\cite{born1988dynamical,Keceli2011} 
The thermal Born--Huang rules permit rapid evaluation of the thermal average of the expectation value for
operators made of the $i$th normal coordinate $Q_i$ or the products thereof [Eqs.\ (\ref{eq:fQnotation1}) and (\ref{eq:fQnotation2})]. These expectation values are, in turn, expressed by quantum number $n_i$ and frequency $\omega_i$ 
of the harmonic oscillator wave function.  In this Appendix, we show how these rules are systematically derived from the zero-temperature counterparts. 

The one-mode thermal averages considered in Table \ref{table:thermalavg_EN1} are expanded as
\begin{eqnarray}
\left[ \langle \hat{F}(Q_i) \rangle_{0} \right] &\equiv& \frac{\sum \limits_{n_i=0}^{\infty}\langle  \hat{F}(Q_i) \rangle_{0}\exp\left\{-\beta  (n_i +1/2) \omega_i\right\}}{ \sum \limits_{n_i=0}^{\infty} \exp\left\{-\beta  (n_i +1/2) \omega_i\right\}} 
\nonumber\\
&=&\frac{\sum \limits_{n_i=0}^{\infty} f(n_i; 0)\exp\left\{-\beta  (n_i +1/2) \omega_i\right\}}{ \sum \limits_{n_i=0}^{\infty} \exp\left\{-\beta  (n_i +1/2) \omega_i\right\}} 
\nonumber\\
&=& \Big[ f(n_i; 0) \Big]
\end{eqnarray}
where $f(n_i; 0) \equiv \langle \hat{F}(Q_i) \rangle_0$ is given in Table \ref{table:matrix_ele} and takes the form of a polynomial of $n_i$. 
For instance, for $\hat{F}(Q_i) = \partial^2/\partial Q_i^2$, $f(n_i; 0) = -\omega_i(n_i+1/2)$ as per the second column of Table \ref{table:matrix_ele}.

Because $f(n_i;0)$ is a polynomial of $n_i$, its thermal average is analytically evaluated with the thermal averages of various powers of $n_i$, e.g.,
\begin{eqnarray}
\label{eq:sum_I}
 \Big[{n_i}\Big] &=& \frac{\sum \limits_{n_i=0}^{\infty}n_i \exp\{ - \beta (n_i + 1/2) \omega_i \} }{\sum \limits_{n_i=0}^{\infty} \exp\{- \beta (n_i+ 1/2)\omega_i\} } 
% \nonumber\\ &=& 
= \frac{\sum \limits_{n_i=0}^{\infty}  n_i \exp( - \beta n_i  \omega_i) }{\sum \limits_{n_i=0}^{\infty} \exp( - \beta n_i \omega_i ) } 
 \nonumber\\
 &=& \frac{1}{z_i^{(0)}} \frac{\partial z_i^{(0)}}{\partial (-\beta \omega_i)} 
 %\nonumber\\&=& 
 = \frac{f_i (f_i+1)}{f_i + 1} = f_i, 
\end{eqnarray}
where $f_i$ is the Bose--Einstein distribution function [Eq.\ (\ref{eq:BEfunction})] and $z_i^{(0)}$ is defined by 
\begin{eqnarray}
\label{eq:z_i}
  z_i^{(0)} = \sum \limits_{n_i=0}^{\infty}\exp(-\beta n_i \omega_i  )  = \frac{1}{1 - \exp(-\beta\omega_i)} = f_i + 1,
\end{eqnarray} 
with
\begin{eqnarray}
\label{eq:deriv_f}
  \frac{\partial  z_i^{(0)}}{\partial (-\beta \omega_i)}= \frac{\partial (f_i+1)}{\partial (-\beta \omega_i)} = f_i (f_i + 1).
\end{eqnarray} 
The thermal averages of higher powers of $n_i$ are evaluated analytically in a similar way.
\begin{eqnarray}
 \label{eq:sum_II}
 \Big[n_i^2\Big] &=& 
%\frac{\sum \limits_{n_i=0}^{\infty}n_i^2 \exp\{ - \beta (n_i + 1/2) \omega_i \} }{\sum \limits_{n_i=0}^{\infty} \exp\{- \beta (n_i+ 1/2)\omega_i\} } 
% \nonumber\\ &=& 
% = \frac{\sum \limits_{n_i=0}^{\infty}  n_i^2 \exp( - \beta n_i  \omega_i) }{\sum \limits_{n_i=0}^{\infty} \exp( - \beta n_i \omega_i ) } \nonumber\\&=&  
\frac{1}{z_i^{(0)}} \frac{\partial^2 z_i^{(0)}}{\partial (-\beta \omega_i)^2}  = 2f_i^2+f_i, \\
   \label{eq:sum_III}
    \Big[{n_i^3}\Big] &=& %\frac{\sum \limits_{n_i=0}^{\infty}n_i^3 \exp(- n_i \beta \omega_i)}{\sum \limits_{n_i=0}^{\infty} \exp(- n_i\beta \omega_i)} = 
     \frac{1}{z_i^{(0)}} \frac{\partial^3 z_i^{(0)}}{\partial (-\beta \omega_i)^3} %= \frac{\exp(2\beta\omega_i) + 4\exp(\beta \omega_i) + 1}{\{\exp(\beta\omega_i) - 1\}^3} 
    %\nonumber\\
    = 6f_i^3+6f_i^2+f_i,\\
     \label{eq:sum_IV}
   \Big [{n_i^4}\Big] &=& %\frac{\sum \limits_{n_i=0}^{\infty}n_i^4 \exp(- n_i \beta \omega_i)}{\sum \limits_{n_i=0}^{\infty} \exp(- n_i\beta \omega_i)} = 
    \frac{1}{z_i^{(0)}} \frac{\partial^4 z_i^{(0)}}{\partial (-\beta \omega_i)^4} %\nonumber \\ &=& \frac{\exp(3\beta\omega_i)+11\exp(2\beta\omega_i)+ 11\exp(\beta\omega_i) + 1}{\{ \exp (\beta \omega_i) - 1 \}^4} 
 %   \nonumber\\
    = 24f_i^4 + 36f_i^3+14f_i^2+f_i.
\end{eqnarray}

For instance, the rule for $[\langle Q_i^4 \rangle_0]$ in Table \ref{table:thermalavg_EN1} can be derived from $\langle Q_i^4 \rangle_0$ in Table \ref{table:matrix_ele} as follows: 
\begin{eqnarray}
\left[\langle Q_i^4 \rangle_{0}\right] &=& \left[ \frac{6n_i^2+6n_i+3}{4\omega_i^2} \right] \nonumber\\
&=& \frac{1}{4\omega_i^2}\left( 6 \left[ n_i^2 \right] + 6 \Big[ n_i \Big] +3  \right) \nonumber\\
&=& \frac{3(f_i + 1/2)^2}{\omega_i^2},
\end{eqnarray}
where Eqs.\ (\ref{eq:sum_I}) and (\ref{eq:sum_II}) were used in the last equality. All the other rules in Table \ref{table:thermalavg_EN1} can be justified similarly.

The rules in Table \ref{table:thermalavg_EN2} are derived completely analogously, but care must be exercised to the different ranges of summation index depending on the sign of $\delta_i$.
For $\delta_i < 0$, 
\begin{widetext}
\begin{eqnarray}
\label{eq:minus}
\left[ \langle \hat{F}(Q_i) \rangle_{\delta_i}\langle \hat{G}(Q_i) \rangle_{-\delta_i} \right] &\equiv& \frac{\sum \limits_{n_i=0}^{\infty}\langle  n_i | \hat{F}(Q_i) | n_i -\delta_i \rangle \langle  n_i -\delta_i | \hat{G}(Q_i) | n_i  \rangle \exp\left\{-\beta  (n_i +1/2) \omega_i\right\}}{ \sum \limits_{n_i=0}^{\infty} \exp\left\{-\beta  (n_i +1/2) \omega_i\right\}} 
\nonumber\\
&=&\frac{\sum \limits_{n_i=0}^{\infty} f(n_i;\delta_i) g(n_i; \delta_i) \exp\left\{-\beta  (n_i +1/2) \omega_i\right\}}{ \sum \limits_{n_i=0}^{\infty} \exp\left\{-\beta  (n_i +1/2) \omega_i\right\}} = \Big[ f(n_i;\delta_i) g(n_i; \delta_i) \Big],
\end{eqnarray}
where $ f(n_i;\delta_i) \equiv \langle \hat{F}(Q_i) \rangle_{\delta_i}$ and $g(n_i; \delta_i) \equiv \langle \hat{G}(Q_i) \rangle_{\delta_i}$, which are polynomials of $n_i$ given in Table \ref{table:matrix_ele}. 
For $\delta_i \geq 0$, on the other hand, 
\begin{eqnarray}
\label{eq:plus}
\left[ \langle \hat{F}(Q_i) \rangle_{\delta_i}\langle \hat{G}(Q_i) \rangle_{-\delta_i} \right] &\equiv& \frac{\sum \limits_{n_i=\delta_i}^{\infty}\langle  n_i | \hat{F}(Q_i) | n_i -\delta_i \rangle \langle  n_i -\delta_i | \hat{G}(Q_i) | n_i  \rangle \exp\left\{-\beta  (n_i +1/2) \omega_i\right\}}{ \sum \limits_{n_i=0}^{\infty} \exp\left\{-\beta  (n_i +1/2) \omega_i\right\}} 
\nonumber\\
&=& \frac{\sum \limits_{n_i=0}^{\infty} \langle  n_i + \delta_i | \hat{F}(Q_i) | n_i  \rangle   \langle  n_i  | \hat{G}(Q_i) | n_i + \delta_i  \rangle \exp\left\{-\beta  (n_i +1/2) \omega_i\right\}\exp\left(-\beta \delta_i \omega_i\right)}{ \sum \limits_{n_i=0}^{\infty} \exp\left\{-\beta  (n_i +1/2) \omega_i\right\}} 
\nonumber\\
&=&\exp\left(-\beta \delta_i \omega_i\right) \Big[ f(n_i;-\delta_i) g(n_i; -\delta_i) \Big] = \left( \frac{f_i }{f_i+1} \right)^{\delta_i} \Big[ f(n_i;-\delta_i) g(n_i; -\delta_i) \Big] ,
\end{eqnarray}
\end{widetext}
where we used $\exp(-\beta\omega_i) = f_i/(f_i+1)$ [cf.\ Eq.\ (\ref{eq:BEfunction})]. Generally, for $\delta_i \neq 0$, 
\begin{eqnarray}
\left[ \langle \hat{F}(Q_i) \rangle_{\delta_i}\langle \hat{G}(Q_i) \rangle_{-\delta_i} \right]
&\neq& \left[ \langle \hat{F}(Q_i) \rangle_{-\delta_i}\langle \hat{G}(Q_i) \rangle_{\delta_i} \right], \\ 
\left[ \langle \hat{F}(Q_i) \rangle_{\delta_i}\langle \hat{G}(Q_i) \rangle_{-\delta_i} \right] &=& \left[ \langle \hat{G}(Q_i) \rangle_{\delta_i}\langle \hat{F}(Q_i) \rangle_{-\delta_i} \right].\label{eq:symmetry}
\end{eqnarray}

For instance, the rule for $[\langle Q_i^4 \rangle_{-2}\langle Q_i^2 \rangle_{2}]$ in Table \ref{table:thermalavg_EN2} can be derived as
\begin{eqnarray}
&& \left[\langle Q_i^4 \rangle_{-2}\langle Q_i^2 \rangle_{2}\right] \nonumber\\
&&= \left[ \frac{(n_i+3/2)\sqrt{(n_i+2)(n_i+1)}}{\omega_i^2}\frac{\sqrt{(n_i+2)(n_i+1)}}{2\omega_i} \right] \nonumber\\
&& =\left[ \frac{ 2 n_i^3 + 9n_i^2 + 13n_i + 6}{4\omega_i^3} \right] \nonumber\\
&& = \frac{ 2 \left[n_i^3\right] + 9\left[n_i^2\right] + 13\Big[n_i\Big] + 6}{4\omega_i^3}  \nonumber\\
%&& = \frac{ 2 \left(6 f_i^3 + 6f_i^2+f_i\right) + 9\left(2 f_i^2 + f_i \right) + 13 f_i + 6}{4\omega_i^3}  \nonumber\\
&& = \frac{ 6 f_i^3 + 15f_i^2+12 f_i + 3}{2\omega_i^3},
\end{eqnarray}
where Table \ref{table:matrix_ele} was used in the first equality and Eqs.\ (\ref{eq:sum_I})--(\ref{eq:sum_III}) in the last equality. 
$[\langle Q_i^4 \rangle_{2}\langle Q_i^2 \rangle_{-2}]$ can then be obtained readily according to Eq.\ (\ref{eq:plus}) as
\begin{eqnarray}
 \left[\langle Q_i^4 \rangle_{2}\langle Q_i^2 \rangle_{-2}\right] 
&=&\left( \frac{f_i }{f_i+1} \right)^{2}  \left[\langle Q_i^4 \rangle_{-2}\langle Q_i^2 \rangle_{2}\right] 
\nonumber\\
&=&\frac{ 6 f_i^3 + 3f_i^2}{2\omega_i^3}.
\end{eqnarray}
All the other rules in Table \ref{table:thermalavg_EN2} can be reproduced in the same manner.

%============================================
% APPENDIX Canonical forms 
%============================================

\section{Canonical forms \label{appendix:BHfactorization}}

The most serious weakness of the algebraic reduction based on the thermal Born--Huang rules is the need for enumerating all possible cases of summation index coincidences and 
having to use different rules depending on the cases. It causes proliferation of terms during the reduction process, even though they will be consolidated eventually with all summation index restrictions lifted.
This is analogous to extremely lengthy reductions of electron-correlated theories' formalisms using the Slater--Condon rules.\cite{Hirata_2020} 

The canonical forms of the thermal Born--Huang rules summarized in Table \ref{table:canonical} expedite this consolidation process. Generally, a canonical form is the way in which a thermal average is expressed as a sum of 
products of Wick contractions, which are 
\begin{eqnarray}
\left[\left\langle \frac{\partial^2}{\partial Q_i^2}  \right\rangle_0\right] &=& -\frac{
\contraction[1ex]{ \{ }{\hat{a}}{_i }{\hat{a}}
\contraction[1ex]{ \{ \hat{a}_i \hat{a}^\dagger_i \} + \{ }{\hat{a}}{^\dagger_i }{\hat{a}}
 \{ \hat{a}_i \hat{a}^\dagger_i \} + \{ \hat{a}^\dagger_i \hat{a}_i \}
 }{2\omega_i^{-1}}, \nonumber\\
\left[\langle Q_i^2  \rangle_0\right] &=& \frac{ 
\contraction[1ex]{ \{ }{\hat{a}}{_i }{\hat{a}}
\contraction[1ex]{ \{ \hat{a}_i \hat{a}^\dagger_i \} + \{ }{\hat{a}}{^\dagger_i }{\hat{a}}
 \{ \hat{a}_i \hat{a}^\dagger_i \} + \{ \hat{a}^\dagger_i \hat{a}_i \}
 }{2\omega_i}, \nonumber\\
\Big[\langle Q_i \rangle_{1} \langle Q_i \rangle_{-1}\Big] &=& \frac{ 
\contraction[1ex]{ \{ }{\hat{a}}{^\dagger_i \} \{ }{\hat{a}}
 \{ \hat{a}^\dagger_i \} \{ \hat{a}_i \}
 }{2\omega_i}, \nonumber\\
\Big[\langle Q_i \rangle_{-1} \langle Q_i \rangle_{1}\Big] &=& \frac{ 
\contraction[1ex]{ \{ }{\hat{a}}{_i \} \{ }{\hat{a}}
 \{ \hat{a}_i \} \{ \hat{a}^\dagger_i \}
 }{2\omega_i},
 \end{eqnarray}
 where curly brackets denote the finite-temperature normal ordering of second-quantized ladder operators discussed extensively in Appendix \ref{appendix:normalorderedSQ}. 
Therefore, the canonical forms partially impart  the normal-ordered second quantization logic (which does not have to enumerate the index coincidence cases) to the algebraic reduction.

Every canonical form in Table \ref{table:canonical} can be verified straightforwardly from the thermal Born--Huang rules in Tables \ref{table:thermalavg_EN1} and \ref{table:thermalavg_EN2},
which will not be repeated here. 

%============================================
% APPENDIX Algebraic reduction E(2)
%============================================

\section{Algebraic reduction of $[ E_N^{(2)} ]$\label{appendix:algebraicreduction}}

Here, we document the algebraic reduction of $[ E_N^{(2)} ]$ appearing in $\Omega^{(2)}$ [Eq.\ (\ref{eq:E2PT_recursion})] using the thermal Born--Huang rules and their canonical forms.

In a QFF, $[ E_N^{(2)} ]$ can be expanded as follows using the recursion of the degenerate RSPT [Eqs.\ (\ref{eq:E_In_degen}) and (\ref{eq:Phi_In_degen})] and the trace invariance [Eq.\ (\ref{eq:traceinvariance})]:
\begin{eqnarray}
\label{eq:E2nondegen}
    \left[E^{(2)}_N\right] &=&\left[ \mathrm{Tr}\left( \bm{E}^{(2)}\right) \right] \nonumber\\
    &=& \left[ \langle N | \hat{V}_1 \hat{R} \hat{V}_1 | N \rangle \right] + \left[ \langle N | \hat{V}_1 \hat{R} \hat{V}_3 | N \rangle \right]  \nonumber \\ 
    &&+ \left[ \langle N | \hat{V}_3 \hat{R} \hat{V}_1 | N \rangle \right] + \left[ \langle N | \hat{V}_2 \hat{R} \hat{V}_2 | N \rangle \right] \nonumber \\ 
    &&+ \left[ \langle N | \hat{V}_2 \hat{R} \hat{V}_4 | N \rangle \right] + \left[ \langle N | \hat{V}_4 \hat{R} \hat{V}_2 | N \rangle \right] \nonumber \\ 
    &&+ \left[ \langle N | \hat{V}_3 \hat{R} \hat{V}_3 | N \rangle \right] + \left[ \langle N | \hat{V}_4 \hat{R} \hat{V}_4 | N \rangle \right]
\end{eqnarray}
with the resolvent $\hat{R}$ being
\begin{eqnarray}
\hat{R} = \hat{R}_1 + \hat{R}_2 + \hat{R}_3 + \hat{R}_4,
\end{eqnarray}
where $\hat{V}_n$ is the $n$th-order force-constant term in the PES operator [Eq.\ (\ref{eq:sorted_V})]. 
Only nonvanishing combinations $(m,n)$ of $[\langle N |\hat{V}_m\hat{R}\hat{V}_n | N \rangle]$ in a QFF are shown above (see below for a justification).

%------
% 1 1
%------

\subsection{$[ \langle N | \hat{V}_1 \hat{R} \hat{V}_1 | N \rangle ]$ \label{appendix:V1V1}} 

The first term in the right-hand side of Eq.\ (\ref{eq:E2nondegen}) is evaluated as
\begin{eqnarray}
\label{eq:E11}
&& \left[ \langle N| \hat{V}_1 \hat{R} \hat{V}_1 |N\rangle \right] \nonumber\\
 &&= \sum_{i,j} F_i F_j \left[\langle N |  Q_i \hat{R} Q_j | N \rangle\right] \nonumber\\ 
 &&= \sum_{i,j} F_i F_j  \left[ \sum_M^{\text{denom.}\neq 0} \frac{\langle N | Q_i | M \rangle \langle M | Q_j | N \rangle}{E_N^{(0)} - E_{M}^{(0)}}\right] \nonumber\\ 
 &&= \sum^{\text{denom.}\neq 0}_{i} \frac{F_{i} F_{i}}{\omega_i} \Big[ \langle n_i | Q_i | n_i -1 \rangle \langle n_i-1 | Q_i | n_i \rangle\Big]   \nonumber\\
 &&+ \sum^{\text{denom.}\neq 0}_{i} \frac{F_{i} F_{i}}{-\omega_i} \Big[ \langle n_i | Q_i | n_i +1 \rangle \langle n_i + 1 | Q_i | n_i \rangle\Big]   \nonumber\\
 &&= \sum^{\text{denom.}\neq 0}_{i} \frac{F_{i} F_{i}}{\omega_i} \Big[ \langle Q_i \rangle_{1} \langle Q_i \rangle_{-1}\Big]   + \sum^{\text{denom.}\neq 0}_{i} \frac{F_{i} F_{i}}{-\omega_i} \Big[ \langle Q_i \rangle_{-1} \langle Q_i \rangle_{1}\Big]   \nonumber\\
 &&= \sum^{\text{denom.}\neq 0}_i \frac{F_{i} F_{i}}{\omega_i} \frac{f_i}{2\omega_i} + \sum^{\text{denom.}\neq 0}_i \frac{F_{i} F_{i}}{-\omega_i} \frac{f_i+1}{2\omega_i}\nonumber\\
 &&= \sum^{\text{denom.}\neq 0}_i \frac{\tilde{F}_{i} \tilde{F}_{i}}{\omega_i}{f_i} + \sum^{\text{denom.}\neq 0}_i \frac{\tilde{F}_{i} \tilde{F}_{i}}{-\omega_i}{(f_i+1)}. 
\end{eqnarray}
In the second equality, $M$ runs over all Hartree-product states that are not degenerate with the $N$th state. This is equivalent to $M$ running over all states with 
the restriction ``denom.$\neq$0'' requiring that 
$E_N^{(0)} - E_{M}^{(0)} \neq 0$. Furthermore, only when $i = j$ %and the $M$th state is either $| \dots (n_i \pm 1) \dots \rangle$ 
is the numerator nonzero. 
The penultimate equality used the thermal Born--Huang rules in Table \ref{table:thermalavg_EN2}. See Eq.\ (\ref{eq:F1tilde}) for the definition of $\tilde{F}$ appearing in the last equality. 
In the final expression, ``denom.$\neq$0'' means $\omega_i \neq 0$, and, therefore, ``a division by zero'' never occurs in this theory or its computer program. 
Both terms in the last line are linked through the common summation index $i$ and are, therefore, size-consistent.

%------
% 1 3
%------

\subsection{$[ \langle N | \hat{V}_1 \hat{R} \hat{V}_3 | N \rangle ]$ and $[ \langle N | \hat{V}_3 \hat{R} \hat{V}_1 | N \rangle ]$} 

Next, we consider the second term of Eq.\ (\ref{eq:E2nondegen}).
As per Tables \ref{table:thermalavg_EN1} and \ref{table:thermalavg_EN2}, only $\hat{R}_1$ will have a nonzero contribution to this term. 
Furthermore, since $\hat{R}_1 = \hat{R}_1^{(+1)} + \hat{R}_1^{(-1)}$ [cf.\ Eq.\ (\ref{eq:R2breakdown})], we write
\begin{eqnarray}
\left[ \langle N| \hat{V}_1 \hat{R} \hat{V}_3 |N\rangle \right] = \left[ \langle N| \hat{V}_1 \hat{R}_1^{(-1)} \hat{V}_3 |N\rangle \right]  + \left[ \langle N| \hat{V}_1 \hat{R}_1^{(+1)} \hat{V}_3 |N\rangle \right] . \nonumber\\
\end{eqnarray}

As inferred from Appendix \ref{appendix:V1V1}, the terms arising from $\hat{R}_1^{(+1)}$ are simply related to those from $\hat{R}_1^{(-1)}$, and the former are obtained by systematically flipping the signs of the $\omega$'s in the denominators
and of the subscripts of the thermal averages in the latter, and vice versa. Therefore, we initially focus on the latter, which is evaluated as
\begin{eqnarray}
\label{eq:E13}
&&  \left[ \langle N| \hat{V}_1 \hat{R}_1^{(-1)} \hat{V}_3 |N\rangle \right] \nonumber\\
 &&= \frac{1}{6} \sum_{i,j}^{} F_i F_{jjj} \left[\langle N |  Q_i \hat{R}_1^{(-1)} Q_j^3 | N \rangle\right] 
 \nonumber\\ &&
 + \frac{1}{2} \sum_{i,j,k}^{j\neq k} F_i F_{jkk} \left[\langle N |  Q_i \hat{R}_1^{(-1)} Q_j Q_k^2 | N \rangle\right] 
 \nonumber\\ &&
 + \frac{1}{6} \sum_{i,j,k,l}^{\substack{j\neq k, j \neq l \\  k \neq l}} F_i F_{jkl} \left[\langle N |  Q_i \hat{R}_1^{(-1)} Q_j Q_k Q_l | N \rangle\right] 
 \nonumber\\ 
 &&= \frac{1}{6} \sum_{i}^{\text{denom.}\neq 0} \frac{F_i F_{iii}}{\omega_i} \Big[ \langle Q_i \rangle_{1} \langle Q_i^3 \rangle_{-1}\Big]
 \nonumber\\ &&
 + \frac{1}{2} \sum_{i,j}^{\substack{\text{denom.}\neq 0\\i\neq j}} \frac{F_i F_{ijj}}{\omega_i} \Big[ \langle Q_i \rangle_{1} \langle Q_i \rangle_{-1}\Big]\left[\langle Q_j^2\rangle_0\right] 
 \nonumber\\ 
 &&= \frac{1}{2} \sum_{i}^{\text{denom.}\neq 0} \frac{F_i F_{iii}}{\omega_i} \Big[ \langle Q_i \rangle_{1} \langle Q_i \rangle_{-1}\Big] \Big[\langle Q_i^2 \rangle_{0}\Big]
 \nonumber\\ &&
 + \frac{1}{2} \sum_{i,j}^{\substack{\text{denom.}\neq 0\\i\neq j}} \frac{F_i F_{ijj}}{\omega_i} \Big[ \langle Q_i \rangle_{1} \langle Q_i \rangle_{-1}\Big]\left[\langle Q_j^2\rangle_0\right]  
 \nonumber\\ 
 &&= \frac{1}{2} \sum_{i,j}^{\text{denom.}\neq 0} \frac{F_i F_{ijj}}{\omega_i} \Big[ \langle Q_i \rangle_{1} \langle Q_i \rangle_{-1}\Big]\left[\langle Q_j^2\rangle_0\right]  
 \nonumber\\ 
&&=   \sum_{i,j}^{\text{denom.}\neq 0} \frac{\tilde{F}_i \tilde{F}_{ijj}}{\omega_i} f_i (f_j + 1/2).
\end{eqnarray}
In the right-hand side of the first equality, the third term vanishes according to Tables \ref{table:thermalavg_EN1} and \ref{table:thermalavg_EN2} because there is no thermal average of this type listed as having a nonzero value. 
[The same logic led to Eq.\ (\ref{eq:E2nondegen}) enumerating all nonvanishing thermal averages.] In the second equality,
the resolvent $\hat{R}$ introduces the ``denom.$\neq$0'' restrictions, which mean $\omega_i \neq 0$. In the third equality, canonical form (8) of Table \ref{table:canonical} was used to lift the $i \neq j$ summation index restrictions in the penultimate equality.
In the last equality, Tables \ref{table:thermalavg_EN1} and \ref{table:thermalavg_EN2} were again consulted with. 

The term arising from $\hat{R}_1^{(+1)}$ can then be inferred as
\begin{eqnarray}
\label{eq:E13_2}
  \left[ \langle N| \hat{V}_1 \hat{R}_1^{(+1)} \hat{V}_3 |N\rangle \right] %\nonumber\\
&=& \frac{1}{2} \sum_{i,j}^{\text{denom.}\neq 0} \frac{F_i F_{ijj}}{-\omega_i} \Big[ \langle Q_i \rangle_{-1} \langle Q_i \rangle_{1}\Big]\left[\langle Q_j^2\rangle_0\right]  
 \nonumber\\ 
&=&  \sum_{i,j}^{\text{denom.}\neq 0} \frac{\tilde{F}_i \tilde{F}_{ijj}}{-\omega_i} (f_i+1) (f_j + 1/2).
\end{eqnarray}
Together, we obtain
\begin{eqnarray}
\label{eq:E13_final}
\left[ \langle N| \hat{V}_1 \hat{R} \hat{V}_3 |N\rangle \right]
&=&  \sum_{i,j}^{\text{denom.}\neq 0} \frac{\tilde{F}_i \tilde{F}_{ijj} }{\omega_i} f_i (f_j + 1/2)\nonumber\\ 
&& + \sum_{i,j}^{\text{denom.}\neq 0} \frac{\tilde{F}_i \tilde{F}_{ijj}  }{-\omega_i} (f_i+1) (f_j + 1/2), \nonumber\\
\end{eqnarray}
which is linked through the common summation index $i$.

%------
% 3 1
%------

Following the same procedure,  the third term of Eq.\ (\ref{eq:E2nondegen}), $[ \langle N| \hat{V}_3 \hat{R} \hat{V}_1 |N\rangle ]$, is reduced to the same, linked formula as above.

%------
% 2 2
%------

\subsection{$[ \langle N | \hat{V}_2 \hat{R} \hat{V}_2 | N \rangle ]$} 

The fourth term of Eq.\ (\ref{eq:E2nondegen}) is decomposed as
\begin{eqnarray}
\label{eq:E22_0}
\left[ \langle N| \hat{V}_2 \hat{R} \hat{V}_2 |N\rangle \right] &= &\left[ \langle N| \hat{V}_2 \hat{R}_2^{(-2)} \hat{V}_2 |N\rangle \right] 
%\nonumber\\&&
+ \left[ \langle N| \hat{V}_2 \hat{R}_2^{(\pm0)} \hat{V}_2 |N\rangle \right]  \nonumber\\
&&+ \left[ \langle N| \hat{V}_2 \hat{R}_2^{(+2)} \hat{V}_2 |N\rangle \right] . 
\end{eqnarray}

We shall reduce the first two terms and infer the third  from the first. 
\begin{eqnarray}
\label{eq:E22}
&&  \left[ \langle N| \hat{V}_2 \hat{R}_2^{(-2)} \hat{V}_2 |N\rangle \right]  \nonumber\\
&&= \frac{1}{4} \sum_{i}^{\text{denom.}\neq 0} \frac{\bar{F}_{ii} \bar{F}_{ii}}{2 \omega_i } \Big[ \langle Q_i^2 \rangle_{2} \langle Q_i^2 \rangle_{-2}\Big] 
\nonumber\\ &&
+ \frac{1}{2} \sum_{i,j}^{\substack{\text{denom.}\neq 0\\ i\neq j}} \frac{\bar{F}_{ij} \bar{F}_{ij}}{\omega_i +\omega_j} \Big[ \langle Q_i \rangle_{1} \langle Q_i \rangle_{-1}\Big]\Big[ \langle Q_j \rangle_{1} \langle Q_j \rangle_{-1}\Big] 
\nonumber\\ 
&&= \frac{1}{2} \sum_{i}^{\text{denom.}\neq 0} \frac{\bar{F}_{ii} \bar{F}_{ii}}{2 \omega_i } \Big[ \langle Q_i \rangle_{1} \langle Q_i \rangle_{-1}\Big]\Big[ \langle Q_j \rangle_{1} \langle Q_j \rangle_{-1}\Big] 
\nonumber\\
&&+ \frac{1}{2} \sum_{i,j}^{\substack{\text{denom.}\neq 0\\ i\neq j}} \frac{\bar{F}_{ij} \bar{F}_{ij}}{\omega_i +\omega_j} \Big[ \langle Q_i \rangle_{1} \langle Q_i \rangle_{-1}\Big]\Big[ \langle Q_j \rangle_{1} \langle Q_j \rangle_{-1}\Big] 
\nonumber\\ 
&&= \frac{1}{2} \sum_{i,j}^{{\text{denom.}\neq 0}} \frac{\bar{F}_{ij} \bar{F}_{ij}}{\omega_i +\omega_j} \Big[ \langle Q_i \rangle_{1} \langle Q_i \rangle_{-1}\Big]\Big[ \langle Q_j \rangle_{1} \langle Q_j \rangle_{-1}\Big] 
\nonumber\\
&&= \frac{1}{2} \sum_{i,j}^{{\text{denom.}\neq 0}} \frac{\tilde{\bar{F}}_{ij} \tilde{\bar{F}}_{ij}}{\omega_i +\omega_j} f_i f_j. 
\end{eqnarray}
In the second equality, canonical form (3) was used, lifting the summation index restriction $i \neq j$ in the third equality. 

Likewise, we obtain
\begin{eqnarray}
\label{eq:E22_2}
&& \left[ \langle N| \hat{V}_2 \hat{R}_2^{(\pm0)} \hat{V}_2 |N\rangle \right]  \nonumber\\
&&=  \sum_{i,j}^{\substack{\text{denom.}\neq 0\\ i\neq j}} \frac{\bar{F}_{ij} \bar{F}_{ij}}{\omega_i-\omega_j} \Big[ \langle Q_i \rangle_{1} \langle Q_i \rangle_{-1}\Big]\Big[ \langle Q_j \rangle_{-1} \langle Q_j \rangle_{1}\Big] \nonumber\\ 
&&=  \sum_{i,j}^{\text{denom.}\neq 0} \frac{\bar{F}_{ij} \bar{F}_{ij}}{\omega_i-\omega_j} \Big[ \langle Q_i \rangle_{1} \langle Q_i \rangle_{-1}\Big]\Big[ \langle Q_j \rangle_{-1} \langle Q_j \rangle_{1}\Big] \nonumber\\ 
&&=  \sum_{i,j}^{\text{denom.}\neq 0} \frac{\tilde{\bar{F}}_{ij} \tilde{\bar{F}}_{ij}}{\omega_i-\omega_j} f_i (f_j + 1),
\end{eqnarray}
where the summation index restriction $i \neq j$ is absorbed by the stronger condition ``denom.$\neq$0'' (meaning $\omega_i - \omega_j \neq 0$).

The third term of Eq.\ (\ref{eq:E22_0}) can be inferred from Eq.\ (\ref{eq:E22}) as
\begin{eqnarray}
\label{eq:E22_3}
&& \left[ \langle N| \hat{V}_2 \hat{R}_2^{(+2)} \hat{V}_2 |N\rangle \right]  \nonumber\\
&&= \frac{1}{2} \sum_{i,j}^{{\text{denom.}\neq 0}} \frac{\bar{F}_{ij} \bar{F}_{ij}}{-\omega_i-\omega_j} \Big[ \langle Q_i \rangle_{-1} \langle Q_i \rangle_{1}\Big]\Big[ \langle Q_j \rangle_{-1} \langle Q_j \rangle_{1}\Big] \nonumber\\ 
&&=  \frac{1}{2} \sum_{i,j}^{{\text{denom.}\neq 0}} \frac{\tilde{\bar{F}}_{ij} \tilde{\bar{F}}_{ij}}{-\omega_i-\omega_j} (f_i + 1)(f_j + 1).
\end{eqnarray}

Therefore, we find
\begin{eqnarray}
\label{eq:E22_final}
 \left[ \langle N| \hat{V}_2 \hat{R} \hat{V}_2 |N\rangle \right]  
 &=& \frac{1}{2} \sum_{i,j}^{{\text{denom.}\neq 0}} \frac{\tilde{\bar{F}}_{ij} \tilde{\bar{F}}_{ij}}{\omega_i +\omega_j} f_i f_j 
\nonumber\\
&&+  \frac{1}{2} \sum_{i,j}^{{\text{denom.}\neq 0}} \frac{\tilde{\bar{F}}_{ij} \tilde{\bar{F}}_{ij}}{-\omega_i-\omega_j} (f_i + 1)(f_j + 1) 
\nonumber\\&&
+  \sum_{i,j}^{\text{denom.}\neq 0} \frac{\tilde{\bar{F}}_{ij} \tilde{\bar{F}}_{ij}}{\omega_i-\omega_j} f_i (f_j + 1).
\end{eqnarray}
Each term is linked through indexes $i$ and $j$.

%------
% 2 4
%------

\subsection{$[ \langle N | \hat{V}_2 \hat{R} \hat{V}_4 | N \rangle ]$ and  $[ \langle N | \hat{V}_4 \hat{R} \hat{V}_2 | N \rangle ]$} 

The fifth term of Eq.\ (\ref{eq:E2nondegen}) is evaluated as follows:
\begin{eqnarray}
\label{eq:E24_0}
 \left[ \langle N| \hat{V}_2 \hat{R} \hat{V}_4 |N\rangle \right] &=&  \left[ \langle N| \hat{V}_2 \hat{R}_2^{(-2)} \hat{V}_4 |N\rangle \right] 
  %\nonumber\\&&
   + \left[ \langle N| \hat{V}_2 \hat{R}_2^{(\pm0)} \hat{V}_4 |N\rangle \right]  \nonumber\\
 && + \left[ \langle N| \hat{V}_2 \hat{R}_2^{(+2)} \hat{V}_4 |N\rangle \right]
 \end{eqnarray}
 with
\begin{eqnarray}
\label{eq:E24}
&& \left[ \langle N| \hat{V}_2 \hat{R}_2^{(-2)} \hat{V}_4 |N\rangle \right] 
 \nonumber\\
 &&= \frac{1}{48} \sum_{i}^{\text{denom.}\neq 0} \frac{\bar{F}_{ii} {F}_{iiii}}{2\omega_i } \Big[ \langle Q_i^2 \rangle_{2} \langle Q_i^4 \rangle_{-2}\Big] 
  \nonumber\\ &&
 + \frac{1}{8} \sum_{i,k}^{\substack{\text{denom.}\neq 0\\ i\neq k}} \frac{\bar{F}_{ii} {F}_{iikk}}{2\omega_i} \Big[ \langle Q_i^2 \rangle_{2} \langle Q_i^2 \rangle_{-2}\Big]\Big[ \langle Q_k^2 \rangle_{0} \Big] 
  \nonumber\\ &&
 + \frac{1}{6} \sum_{i,j}^{\substack{\text{denom.}\neq 0\\ i\neq j}} \frac{\bar{F}_{ij} {F}_{ijjj}}{\omega_i +\omega_j} \Big[ \langle Q_i \rangle_{1} \langle Q_i \rangle_{-1}\Big]\Big[ \langle Q_j \rangle_{1} \langle Q_j^3 \rangle_{-1} \Big] 
  \nonumber\\ &&
 + \frac{1}{4} \sum_{i,j,k}^{\substack{\text{denom.}\neq 0\\ i\neq j, i\neq k \\ j\neq k}} \frac{\bar{F}_{ij} {F}_{ijkk}}{\omega_i +\omega_j} \Big[ \langle Q_i \rangle_{1} \langle Q_i \rangle_{-1}\Big]\Big[ \langle Q_j \rangle_{1} \langle Q_j \rangle_{-1}\Big]\Big[ \langle Q_k^2 \rangle_{0} \Big] 
  \nonumber\\ 
 &&= \frac{1}{4} \sum_{i}^{\text{denom.}\neq 0} \frac{\bar{F}_{ii} {F}_{iiii}}{2\omega_i} \Big[ \langle Q_i \rangle_{1} \langle Q_i \rangle_{-1}\Big]\Big[ \langle Q_i \rangle_{1} \langle Q_i \rangle_{-1}\Big]\Big[ \langle Q_i^2 \rangle_{0} \Big] 
  \nonumber\\ &&
 + \frac{1}{4} \sum_{i,k}^{\substack{\text{denom.}\neq 0\\ i\neq k}} \frac{\bar{F}_{ii} {F}_{iikk}}{2\omega_i} \Big[ \langle Q_i \rangle_{1} \langle Q_i \rangle_{-1}\Big]\Big[ \langle Q_i \rangle_{1} \langle Q_i \rangle_{-1}\Big]\Big[ \langle Q_k^2 \rangle_{0} \Big] 
  \nonumber\\ &&
 + \frac{1}{2} \sum_{i,j}^{\substack{\text{denom.}\neq 0\\ i\neq j}} \frac{\bar{F}_{ij} {F}_{ijjj}}{\omega_i +\omega_j} \Big[ \langle Q_i \rangle_{1} \langle Q_i \rangle_{-1}\Big]\Big[ \langle Q_j \rangle_{1} \langle Q_j \rangle_{-1}\Big]\Big[ \langle Q_j^2 \rangle_{0} \Big]  
  \nonumber\\ &&
 + \frac{1}{4} \sum_{i,j,k}^{\substack{\text{denom.}\neq 0\\ i\neq j, i\neq k \\ j\neq k}} \frac{\bar{F}_{ij} {F}_{ijkk}}{\omega_i +\omega_j} \Big[ \langle Q_i \rangle_{1} \langle Q_i \rangle_{-1}\Big]\Big[ \langle Q_j \rangle_{1} \langle Q_j \rangle_{-1}\Big]\Big[ \langle Q_k^2 \rangle_{0} \Big] 
  \nonumber\\ 
 &&= \frac{1}{4} \sum_{i,j,k}^{{\text{denom.}\neq 0}} \frac{\bar{F}_{ij} {F}_{ijkk}}{\omega_i +\omega_j} \Big[ \langle Q_i \rangle_{1} \langle Q_i \rangle_{-1}\Big]\Big[ \langle Q_j \rangle_{1} \langle Q_j \rangle_{-1}\Big]\Big[ \langle Q_k^2 \rangle_{0} \Big] 
  \nonumber\\ 
&&= \frac{1}{2} \sum_{i,j,k}^{{\text{denom.}\neq 0}} \frac{\tilde{\bar{F}}_{ij} \tilde{F}_{ijkk}}{\omega_i +\omega_j} f_i f_j (f_k + 1/2) ,
\end{eqnarray}
where canonical forms (3), (8), and (11) were used in the second equality as well as
\begin{eqnarray}
\label{eq:sums_mod1}
\sum_{i,j,k} X_{(ij)k} &=& \sum_{i,j,k}^{\substack{i\neq j, i\neq k \\ j \neq k}} X_{(ij)k} +  2 \sum_{i,j,k}^{\substack{i\neq j, i \neq  k \\ j = k}} X_{(ij)j} 
 +  \sum_{i,j,k}^{\substack{i =  j, i \neq k \\ j \neq k}} X_{(ii)k} 
 \nonumber\\&&
 +  \sum_{i,j,k}^{\substack{i =  j, i = k \\ j = k}} X_{(ii)i} , 
\end{eqnarray}
where $X_{(ij)k}$ has the index permutation symmetry of the form $X_{(ij)k} = X_{(ji)k}$.
All summation index restrictions are lifted except for ``denom.$\neq$0'' requiring 
$\omega_i + \omega_j \neq 0$. 

The second term of Eq.\ (\ref{eq:E24_0}) is reduced similarly:
\begin{eqnarray}
\label{eq:E24_2}
&& \left[ \langle N| \hat{V}_2 \hat{R}_2^{(\pm0)} \hat{V}_4 |N\rangle \right] 
 \nonumber\\&&
 =  \frac{1}{6} \sum_{i,j}^{\substack{\text{denom.}\neq 0\\ i\neq j}} \frac{\bar{F}_{ij} {F}_{ijjj}}{\omega_i - \omega_j} \Big[ \langle Q_i \rangle_{1} \langle Q_i \rangle_{-1}\Big]\Big[ \langle Q_j \rangle_{-1} \langle Q_j^3 \rangle_{1} \Big] 
 \nonumber\\&&
+ \frac{1}{6} \sum_{i,j}^{\substack{\text{denom.}\neq 0\\ i\neq j}} \frac{\bar{F}_{ij} {F}_{ijii}}{\omega_i - \omega_j} \Big[ \langle Q_i \rangle_{1} \langle Q_i^3 \rangle_{-1}\Big]\Big[ \langle Q_j \rangle_{-1} \langle Q_j \rangle_{1} \Big] 
 \nonumber\\ &&
 + \frac{1}{2} \sum_{i,j,k}^{\substack{\text{denom.}\neq 0\\ i\neq j, i\neq k \\ j\neq k}} \frac{\bar{F}_{ij} {F}_{ijkk}}{\omega_i - \omega_j} \Big[ \langle Q_i \rangle_{1} \langle Q_i \rangle_{-1}\Big]\Big[ \langle Q_j \rangle_{-1} \langle Q_j \rangle_{1}\Big]\Big[ \langle Q_k^2 \rangle_{0} \Big] \nonumber\\ 
 &&= \frac{1}{2} \sum_{i,j,k}^{\substack{\text{denom.}\neq 0 \\ i \neq j}} \frac{\bar{F}_{ij} {F}_{ijkk}}{\omega_i - \omega_j} \Big[ \langle Q_i \rangle_{1} \langle Q_i \rangle_{-1}\Big]\Big[ \langle Q_j \rangle_{-1} \langle Q_j \rangle_{1}\Big]\Big[ \langle Q_k^2 \rangle_{0} \Big] \nonumber\\ 
 &&= \frac{1}{2} \sum_{i,j,k}^{{\text{denom.}\neq 0}} \frac{\bar{F}_{ij} {F}_{ijkk}}{\omega_i - \omega_j} \Big[ \langle Q_i \rangle_{1} \langle Q_i \rangle_{-1}\Big]\Big[ \langle Q_j \rangle_{-1} \langle Q_j \rangle_{1}\Big]\Big[ \langle Q_k^2 \rangle_{0} \Big] \nonumber\\ 
&&=  \sum_{i,j,k}^{{\text{denom.}\neq 0}} \frac{\tilde{\bar{F}}_{ij} \tilde{F}_{ijkk}}{\omega_i - \omega_j} f_i (f_j+1)(f_k + 1/2) ,
\end{eqnarray}
where the $i\neq j$ restriction was absorbed by ``denom.$\neq$0'' requiring $\omega_i - \omega_j \neq 0$.
In the second equality, we used canonical forms (8) and (9) as well as
\begin{eqnarray}
\label{eq:sums_moda}
\sum_{i,j,k}^{i \neq j} X_{ijk} &=& \sum_{i,j,k}^{\substack{i\neq j, i\neq k \\ j \neq k}} X_{ijk} +   \sum_{i,j,k}^{\substack{i\neq j, i =  k \\ j \neq k}} X_{iji} +  \sum_{i,j,k}^{\substack{i\neq j, i \neq  k \\ j = k}} X_{ijj}  
\end{eqnarray}
in the third equality.

The third term of Eq.\ (\ref{eq:E24_0}) is inferred from the first:
\begin{eqnarray}
\label{eq:E24_3}
 && \left[ \langle N| \hat{V}_2 \hat{R}_2^{(+2)} \hat{V}_4 |N\rangle \right] 
 \nonumber\\&&
 = \frac{1}{4} \sum_{i,j,k}^{{\text{denom.}\neq 0}} \frac{\bar{F}_{ij} {F}_{ijkk}}{-\omega_i-\omega_j} \Big[ \langle Q_i \rangle_{-1} \langle Q_i \rangle_{1}\Big]\Big[ \langle Q_j \rangle_{-1} \langle Q_j \rangle_{1}\Big]\Big[ \langle Q_k^2 \rangle_{0} \Big] 
 \nonumber\\ &&
 = \frac{1}{2} \sum_{i,j,k}^{{\text{denom.}\neq 0}} \frac{\tilde{\bar{F}}_{ij} \tilde{{F}}_{ijkk}}{-\omega_i-\omega_j} (f_i + 1)(f_j+1)(f_k + 1/2).
\end{eqnarray}
We finally obtain
\begin{eqnarray}
\label{eq:E24_ZZZ}
&& \left[ \langle N| \hat{V}_2 \hat{R} \hat{V}_4 |N\rangle \right] \nonumber\\
 %\nonumber\\&&
 &&= \frac{1}{2} \sum_{i,j,k}^{{\text{denom.}\neq 0}} \frac{ \tilde{\bar{F}}_{ij} \tilde{F}_{ijkk}}{\omega_i +\omega_j} f_i f_j (f_k + 1/2) 
 \nonumber\\ &&
 + \frac{1}{2} \sum_{i,j,k}^{{\text{denom.}\neq 0}} \frac{\tilde{\bar{F}}_{ij} \tilde{{F}}_{ijkk}}{-\omega_i-\omega_j} (f_i + 1)(f_j+1)(f_k + 1/2)
 \nonumber\\ &&
  + \sum_{i,j,k}^{{\text{denom.}\neq 0}} \frac{\tilde{\bar{F}}_{ij}\tilde{F}_{ijkk} }{\omega_i - \omega_j} f_i (f_j+1)(f_k + 1/2) ,
\end{eqnarray}
which is linked through indexes $i$ and $j$. 

The sixth term of Eq.\ (\ref{eq:E2nondegen}), $[ \langle N| \hat{V}_4 \hat{R} \hat{V}_2 |N\rangle ]$, is reduced to the same, linked formula. 

%------
% 3 3
%------

\subsection{$[ \langle N | \hat{V}_3 \hat{R} \hat{V}_3 | N \rangle ]$} 

Next, consider the penultimate term of Eq.\ (\ref{eq:E2nondegen}). It is divided into groups as
\begin{eqnarray}
\label{eq:E33}
 \left[ \langle N| \hat{V}_3 \hat{R} \hat{V}_3 |N\rangle \right] 
 &=&  \left( \left[ \langle N| \hat{V}_3 \hat{R}_1^{(-1)} \hat{V}_3 |N\rangle \right] 
 %\nonumber\\&& 
 +\left[ \langle N| \hat{V}_3 \hat{R}_3^{(-1)} \hat{V}_3 |N\rangle \right] \right)
 \nonumber\\&& 
 + \left( \left[ \langle N| \hat{V}_3 \hat{R}_3^{(-3)} \hat{V}_3 |N\rangle \right] \right)
\nonumber\\&&
+ \left( \left[ \langle N| \hat{V}_3 \hat{R}_1^{(+1)} \hat{V}_3 |N\rangle \right] 
% \nonumber\\&& 
 +\left[ \langle N| \hat{V}_3 \hat{R}_3^{(+1)} \hat{V}_3 |N\rangle \right] \right)
 \nonumber\\&& 
 + \left( \left[ \langle N| \hat{V}_3 \hat{R}_3^{(+3)} \hat{V}_3 |N\rangle \right] \right). %\nonumber\\
 \end{eqnarray}
The first two terms (the first parenthesized group) are expanded together as 
\begin{widetext}
\begin{eqnarray}
\label{eq:E33_1}
&& \left[ \langle N| \hat{V}_3 \hat{R}_1^{(-1)} \hat{V}_3 |N\rangle \right] + \left[ \langle N| \hat{V}_3 \hat{R}_3^{(-1)} \hat{V}_3 |N\rangle \right] 
 \nonumber\\&&
 = \frac{1}{36} \sum_{i}^{\text{denom.}\neq 0} \frac{{F}_{iii} {F}_{iii}}{\omega_i } \Big[ \langle Q_i^3 \rangle_{1} \langle Q_i^3 \rangle_{-1}\Big] 
 %\nonumber\\ &&
 + \frac{1}{4} \sum_{i,j}^{\substack{\text{denom.}\neq 0\\ i\neq j}} \frac{{F}_{ijj} {F}_{ijj}}{\omega_i} \Big[ \langle Q_i \rangle_{1} \langle Q_i \rangle_{-1}\Big]\Big[ \langle Q_j^2 \rangle_{0}\langle Q_j^2 \rangle_{0} \Big] 
 %\nonumber\\ &&
 + \frac{1}{6} \sum_{i,j}^{\substack{\text{denom.}\neq 0\\ i\neq k}} \frac{{F}_{iii} {F}_{ikk}}{\omega_i} \Big[ \langle Q_i^3 \rangle_{1} \langle Q_i \rangle_{-1}\Big]\Big[ \langle Q_k^2 \rangle_{0} \Big] 
 \nonumber\\ &&
 + \frac{1}{4} \sum_{i,j,k}^{\substack{\text{denom.}\neq 0\\ i\neq j, i\neq k \\ j\neq k}} \frac{{F}_{ijj} {F}_{ikk}}{\omega_i} \Big[ \langle Q_i \rangle_{1} \langle Q_i \rangle_{-1}\Big]\Big[ \langle Q_j^2 \rangle_{0} \Big]\Big[ \langle Q_k^2 \rangle_{0} \Big] 
% \nonumber\\ &&
 + \frac{1}{4} \sum_{i,j}^{\substack{\text{denom.}\neq 0\\ i\neq j}} \frac{{F}_{ijj} {F}_{ijj}}{-\omega_i+2\omega_j} \Big[ \langle Q_i \rangle_{-1} \langle Q_i \rangle_{1}\Big]\Big[ \langle Q_j^2 \rangle_{2}\langle Q_j^2 \rangle_{-2} \Big] 
 \nonumber\\ &&
  + \frac{1}{2} \sum_{i,j,k}^{\substack{\text{denom.}\neq 0\\ i\neq j, i\neq k \\ j\neq k}} \frac{{F}_{ijk} {F}_{ijk}}{\omega_i+\omega_j-\omega_k} \Big[ \langle Q_i \rangle_{1} \langle Q_i \rangle_{-1}\Big]\Big[ \langle Q_j \rangle_{1} \langle Q_j \rangle_{-1}\Big]\Big[ \langle Q_k \rangle_{-1} \langle Q_k \rangle_{1}\Big]
 \nonumber\\ 
&&= \frac{1}{4} \sum_{i}^{\text{denom.}\neq 0} \frac{{F}_{iii} {F}_{iii}}{\omega_i }\Big[ \langle Q_i \rangle_{1} \langle Q_i \rangle_{-1}\Big]\Big[ \langle Q_i^2 \rangle_{0} \Big]\Big[ \langle Q_i^2 \rangle_{0} \Big]
 %\nonumber\\ &&
 + \frac{1}{2} \sum_{i}^{\text{denom.}\neq 0} \frac{{F}_{iii} {F}_{iii}}{\omega_i }\Big[ \langle Q_i \rangle_{1} \langle Q_i \rangle_{-1}\Big]\Big[ \langle Q_i \rangle_{1} \langle Q_i \rangle_{-1}\Big]\Big[ \langle Q_i \rangle_{-1} \langle Q_i \rangle_{1}\Big]
 \nonumber\\ &&
+ \frac{1}{4} \sum_{i,j}^{\substack{\text{denom.}\neq 0\\ i\neq j}} \frac{{F}_{ijj} {F}_{ijj}}{\omega_i} \Big[ \langle Q_i \rangle_{1} \langle Q_i \rangle_{-1}\Big]\Big[ \langle Q_j^2 \rangle_{0}\Big]\Big[\langle Q_j^2 \rangle_{0} \Big] 
 %\nonumber\\ &&
+  \sum_{i,j}^{\substack{\text{denom.}\neq 0\\ i\neq j}} \frac{{F}_{ijj} {F}_{ijj}}{\omega_i} \Big[ \langle Q_i \rangle_{1} \langle Q_i \rangle_{-1}\Big]\Big[ \langle Q_j \rangle_{1} \langle Q_j \rangle_{-1}\Big]\Big[ \langle Q_j \rangle_{-1} \langle Q_j \rangle_{1}\Big]
 \nonumber\\ &&
 + \frac{1}{2} \sum_{i,j}^{\substack{\text{denom.}\neq 0\\ i\neq k}} \frac{{F}_{iii} {F}_{ikk}}{\omega_i} \Big[ \langle Q_i \rangle_{1} \langle Q_i \rangle_{-1}\Big]\Big[ \langle Q_i^2 \rangle_{0} \Big]\Big[ \langle Q_k^2 \rangle_{0} \Big] 
% \nonumber\\ &&
 + \frac{1}{4} \sum_{i,j,k}^{\substack{\text{denom.}\neq 0\\ i\neq j, i\neq k \\ j\neq k}} \frac{{F}_{ijj} {F}_{ikk}}{\omega_i} \Big[ \langle Q_i \rangle_{1} \langle Q_i \rangle_{-1}\Big]\Big[ \langle Q_j^2 \rangle_{0} \Big]\Big[ \langle Q_k^2 \rangle_{0} \Big] 
\nonumber\\ &&
+ \frac{1}{2} \sum_{i,j}^{\substack{\text{denom.}\neq 0\\ i\neq j}} \frac{{F}_{ijj} {F}_{ijj}}{-\omega_i+2\omega_j} \Big[ \langle Q_i \rangle_{-1} \langle Q_i \rangle_{1}\Big]
\Big[ \langle Q_j \rangle_{1} \langle Q_j \rangle_{-1}\Big]\Big[ \langle Q_j \rangle_{1} \langle Q_j \rangle_{-1}\Big]
 \nonumber\\ &&
  + \frac{1}{2} \sum_{i,j,k}^{\substack{\text{denom.}\neq 0\\ i\neq j, i\neq k \\ j\neq k}} \frac{{F}_{ijk} {F}_{ijk}}{\omega_i+\omega_j-\omega_k} \Big[ \langle Q_i \rangle_{1} \langle Q_i \rangle_{-1}\Big]\Big[ \langle Q_j \rangle_{1} \langle Q_j \rangle_{-1}\Big]\Big[ \langle Q_k \rangle_{-1} \langle Q_k \rangle_{1}\Big]
\nonumber\\
&&= \frac{1}{4} \sum_{i,j,k}^{{\text{denom.}\neq 0}} \frac{{F}_{ijj} {F}_{ikk}}{\omega_i} \Big[ \langle Q_i \rangle_{1} \langle Q_i \rangle_{-1}\Big]\Big[ \langle Q_j^2 \rangle_{0} \Big]\Big[ \langle Q_k^2 \rangle_{0} \Big]  %\nonumber\\ &&
+ \frac{1}{2} \sum_{i,j,k}^{\text{denom.}\neq 0} \frac{{F}_{ijk} {F}_{ijk}}{\omega_i+\omega_j-\omega_k} \Big[ \langle Q_i \rangle_{1} \langle Q_i \rangle_{-1}\Big]\Big[ \langle Q_j \rangle_{1} \langle Q_j \rangle_{-1}\Big]\Big[ \langle Q_k \rangle_{-1} \langle Q_k \rangle_{1}\Big]
\nonumber\\
&&=  \sum_{i,j,k}^{{\text{denom.}\neq 0}} \frac{{F}_{ijj} {F}_{ikk}}{\omega_i} f_i (f_j + 1/2)(f_k + 1/2)
 %\nonumber\\ &&
+ \frac{1}{2} \sum_{i,j,k}^{\text{denom.}\neq 0} \frac{{F}_{ijk} {F}_{ijk}}{\omega_i+\omega_j-\omega_k} f_i f_j (f_k+1) ,
\end{eqnarray}
where canonical forms (2), (3), (8), and (16) were used in the second equality as well as Eq.\ (\ref{eq:sums_mod1}) in the last equality. Likewise, the third term of Eq.\ (\ref{eq:E33}) is reduced as
\begin{eqnarray}
\label{eq:E33_2}
&& \left[ \langle N| \hat{V}_3 \hat{R}_3^{(-3)} \hat{V}_3 |N\rangle \right]  
\nonumber\\  
&&= \frac{1}{36} \sum_{i}^{\text{denom.}\neq 0} \frac{{F}_{iii} {F}_{iii}}{3\omega_i } \Big[ \langle Q_i^3 \rangle_{3} \langle Q_i^3 \rangle_{-3}\Big] 
% \nonumber\\&&
 + \frac{1}{4} \sum_{i,j}^{\substack{\text{denom.}\neq 0\\ i\neq j}} \frac{{F}_{ijj} {F}_{ijj}}{\omega_i+2\omega_j} \Big[ \langle Q_i \rangle_{1} \langle Q_i \rangle_{-1}\Big]\Big[ \langle Q_j^2 \rangle_{2}\langle Q_j^2 \rangle_{-2} \Big] 
 \nonumber\\ &&
 + \frac{1}{6} \sum_{i,j,k}^{\substack{\text{denom.}\neq 0\\ i\neq j, i\neq k \\ j\neq k}} \frac{{F}_{ijk} {F}_{ijk}}{\omega_i+\omega_j+\omega_k} \Big[ \langle Q_i \rangle_{1} \langle Q_i \rangle_{-1}\Big]\Big[ \langle Q_j \rangle_{1} \langle Q_j \rangle_{-1}\Big]\Big[ \langle Q_k \rangle_{1} \langle Q_k \rangle_{-1}\Big]
 \nonumber\\ 
 &&= \frac{1}{6} \sum_{i}^{\text{denom.}\neq 0} \frac{{F}_{iii} {F}_{iii}}{3\omega_i }\Big[ \langle Q_i \rangle_{1} \langle Q_i \rangle_{-1}\Big]\Big[ \langle Q_i \rangle_{1} \langle Q_i \rangle_{-1}\Big]\Big[ \langle Q_i \rangle_{1} \langle Q_i \rangle_{-1}\Big]
 %\nonumber\\&&
 + \frac{1}{2} \sum_{i,j}^{\substack{\text{denom.}\neq 0\\ i\neq j}} \frac{{F}_{ijj} {F}_{ijj}}{\omega_i+2\omega_j} \Big[ \langle Q_i \rangle_{1} \langle Q_i \rangle_{-1}\Big]\Big[ \langle Q_j \rangle_{1} \langle Q_j \rangle_{-1}\Big]\Big[ \langle Q_j \rangle_{1} \langle Q_j \rangle_{-1}\Big]
\nonumber\\ &&
 + \frac{1}{6} \sum_{i,j,k}^{\substack{\text{denom.}\neq 0\\ i\neq j, i\neq k \\ j\neq k}} \frac{{F}_{ijk} {F}_{ijk}}{\omega_i+\omega_j+\omega_k} \Big[ \langle Q_i \rangle_{1} \langle Q_i \rangle_{-1}\Big]\Big[ \langle Q_j \rangle_{1} \langle Q_j \rangle_{-1}\Big]\Big[ \langle Q_k \rangle_{1} \langle Q_k \rangle_{-1}\Big]
 \nonumber\\
  &&= \frac{1}{6} \sum_{i,j,k}^{{\text{denom.}\neq 0}} \frac{{F}_{ijk} {F}_{ijk}}{\omega_i+\omega_j+\omega_k} \Big[ \langle Q_i \rangle_{1} \langle Q_i \rangle_{-1}\Big]\Big[ \langle Q_j \rangle_{1} \langle Q_j \rangle_{-1}\Big]\Big[ \langle Q_k \rangle_{1} \langle Q_k \rangle_{-1}\Big]
 \nonumber\\  &&
 = \frac{1}{6} \sum_{i,j,k}^{\text{denom.}\neq 0} \frac{{F}_{ijk} {F}_{ijk}}{\omega_i+\omega_j+\omega_k} f_i f_j f_k ,
\end{eqnarray}
where we used canonical forms (3) and (17) as well as 
\begin{eqnarray}
\label{eq:sums_mod2}
\sum_{i,j,k} X_{(ijk)} &=& \sum_{i,j,k}^{\substack{i\neq j, i\neq k \\ j \neq k}} X_{(ijk)} +  3 \sum_{i,j,k}^{\substack{i\neq j, i \neq  k \\ j = k}} X_{(ijj)} 
 +  \sum_{i,j,k}^{\substack{i =  j, i = k \\ j = k}} X_{(iii)} , 
\end{eqnarray}
where $X_{(ijk)}$ has the full index permutation symmetry, i.e., $X_{(ijk)} = X_{(ikj)}= X_{(jik)}=X_{(jki)}=X_{(kij)}=X_{(kji)}$.

From these, we can readily infer the remaining terms of Eq.\ (\ref{eq:E33}):
\begin{eqnarray}
\label{eq:E33_3}
&&  \left[ \langle N| \hat{V}_3 \hat{R}_1^{(+1)} \hat{V}_3 |N\rangle \right] + \left[ \langle N| \hat{V}_3 \hat{R}_3^{(+1)} \hat{V}_3 |N\rangle \right] 
\nonumber\\
 &&= \frac{1}{4} \sum_{i,j,k}^{{\text{denom.}\neq 0}} \frac{{F}_{ijj} {F}_{ikk}}{-\omega_i} \Big[ \langle Q_i \rangle_{-1} \langle Q_i \rangle_{1}\Big]\Big[ \langle Q_j^2 \rangle_{0} \Big]\Big[ \langle Q_k^2 \rangle_{0} \Big]  
 %\nonumber\\ &&
+ \frac{1}{2} \sum_{i,j,k}^{\text{denom.}\neq 0} \frac{{F}_{ijk} {F}_{ijk}}{-\omega_i-\omega_j+\omega_k} \Big[ \langle Q_i \rangle_{-1} \langle Q_i \rangle_{1}\Big]\Big[ \langle Q_j \rangle_{-1} \langle Q_j \rangle_{1}\Big]\Big[ \langle Q_k \rangle_{1} \langle Q_k \rangle_{-1}\Big]
\nonumber\\
&&=  \sum_{i,j,k}^{{\text{denom.}\neq 0}} \frac{{F}_{ijj} {F}_{ikk}}{-\omega_i} (f_i+1) (f_j + 1/2)(f_k + 1/2)
 %\nonumber\\ &&
+ \frac{1}{2} \sum_{i,j,k}^{\text{denom.}\neq 0} \frac{{F}_{ijk} {F}_{ijk}}{-\omega_i-\omega_j+\omega_k} (f_i+1) (f_j+1) f_k ,
\end{eqnarray}
and
\begin{eqnarray}
\label{eq:E33_4}
 \left[ \langle N| \hat{V}_3 \hat{R}_3^{(+3)} \hat{V}_3 |N\rangle \right]  
&=& \frac{1}{6} \sum_{i,j,k}^{{\text{denom.}\neq 0}} \frac{{F}_{ijk} {F}_{ijk}}{-\omega_i-\omega_j-\omega_k} \Big[ \langle Q_i \rangle_{-1} \langle Q_i \rangle_{1}\Big]\Big[ \langle Q_j \rangle_{-1} \langle Q_j \rangle_{1}\Big]\Big[ \langle Q_k \rangle_{-1} \langle Q_k \rangle_{1}\Big]
  \nonumber\\ 
 &=& \frac{1}{6} \sum_{i,j,k}^{\text{denom.}\neq 0} \frac{{F}_{ijk} {F}_{ijk}}{-\omega_i-\omega_j-\omega_k} (f_i+1) (f_j+1)(f_k+1).
\end{eqnarray}
%\end{widetext}

Together, we obtain
\begin{eqnarray}
%&& 
\left[ \langle N| \hat{V}_3 \hat{R} \hat{V}_3 |N\rangle \right] 
%\nonumber\\&&
&=& \sum_{i,j,k}^{{\text{denom.}\neq 0}} \frac{{F}_{ijj} {F}_{ikk}}{\omega_i} f_i (f_j + 1/2)(f_k + 1/2)
 %\nonumber\\ &&
+ \sum_{i,j,k}^{{\text{denom.}\neq 0}} \frac{{F}_{ijj} {F}_{ikk}}{-\omega_i} (f_i + 1) (f_j + 1/2)(f_k + 1/2)
 \nonumber\\&&
+ \frac{1}{6} \sum_{i,j,k}^{\text{denom.}\neq 0} \frac{{F}_{ijk} {F}_{ijk}}{\omega_i+\omega_j+\omega_k} f_i f_j f_k 
%  \nonumber\\&&
+ \frac{1}{6} \sum_{i,j,k}^{\text{denom.}\neq 0} \frac{{F}_{ijk} {F}_{ijk}}{-\omega_i-\omega_j-\omega_k} (f_i+1)(f_j + 1)(f_k+1) 
 \nonumber\\&&
 + \frac{1}{2} \sum_{i,j,k}^{\text{denom.}\neq 0} \frac{{F}_{ijk} {F}_{ijk}}{\omega_i+\omega_j-\omega_k} f_i f_j (f_k+1)
% \nonumber\\&&
 + \frac{1}{2} \sum_{i,j,k}^{\text{denom.}\neq 0} \frac{{F}_{ijk} {F}_{ijk}}{-\omega_i-\omega_j+\omega_k} (f_i+1)(f_j+1) f_k.
\end{eqnarray}
Every term is linked.

%------
% 4 4
%------

\subsection{$[ \langle N | \hat{V}_4 \hat{R} \hat{V}_4 | N \rangle ]$} 

The last term of Eq.\ (\ref{eq:E2nondegen}) will be split and then grouped as
\begin{eqnarray}
\label{eq:E44}
%&& 
\left[ \langle N| \hat{V}_4 \hat{R} \hat{V}_4 |N\rangle \right] 
% \nonumber\\&&
&=&  \left( \left[ \langle N| \hat{V}_4 \hat{R}_4^{(-4)} \hat{V}_4 |N\rangle \right] \right) 
 %\nonumber\\ && 
 + \left( \left[ \langle N| \hat{V}_4 \hat{R}_4^{(-2)} \hat{V}_4 |N\rangle \right] + \left[ \langle N| \hat{V}_4 \hat{R}_2^{(-2)} \hat{V}_4 |N\rangle \right] \right) 
 %\nonumber\\&& 
 + \left( \left[ \langle N| \hat{V}_4 \hat{R}_4^{(\pm0)} \hat{V}_4 |N\rangle \right] + \left[ \langle N| \hat{V}_4 \hat{R}_2^{(\pm0)} \hat{V}_4 |N\rangle \right] \right) 
\nonumber\\&& 
+ \left( \left[ \langle N| \hat{V}_4 \hat{R}_4^{(+2)} \hat{V}_4 |N\rangle \right] + \left[ \langle N| \hat{V}_4 \hat{R}_2^{(+2)} \hat{V}_4 |N\rangle \right] \right) 
%\nonumber\\ && 
+ \left( \left[ \langle N| \hat{V}_4 \hat{R}_4^{(+4)} \hat{V}_4 |N\rangle \right]  \right). 
 \end{eqnarray}
 In the following, we shall reduce the first three parenthesized groups explicitly and  infer the rest. 
 We also use  
 %\begin{widetext}
 \begin{eqnarray}
 \label{eq:sums2}
&& \sum_{i,j,k,l} X_{ijkl} = \left\{ \sum_{i,j,k,l}^{\substack{i\neq j, i\neq k \\ i\neq l, j \neq k \\ j \neq l, k \neq l}}  X_{ijkl}\right\}
+\left\{  \sum_{i,j,k,l}^{\substack{i\neq j, i\neq k \\ i\neq l, j \neq k \\ j \neq l, k = l}} X_{ijkk}
+\sum_{i,j,k,l}^{\substack{i\neq j, i\neq k \\ i\neq l, j \neq k \\ j = l, k \neq l}} X_{ijkj}
+\sum_{i,j,k,l}^{\substack{i\neq j, i\neq k \\ i\neq l, j = k \\ j \neq l, k \neq l}} X_{ijjl}
+\sum_{i,j,k,l}^{\substack{i\neq j, i\neq k \\ i= l, j \neq k \\ j \neq l, k \neq l}} X_{ijki}
+\sum_{i,j,k,l}^{\substack{i\neq j, i= k \\ i\neq l, j \neq k \\ j \neq l, k \neq l}} X_{ijil}
+\sum_{i,j,k,l}^{\substack{i= j, i\neq k \\ i\neq l, j \neq k \\ j \neq l, k \neq l}} X_{iikl} \right\} 
\nonumber\\&& 
+ \left\{ \sum_{i,j,k,l}^{\substack{i\neq j, i\neq k \\ i\neq l, j = k \\ j = l, k = l}} X_{ijjj}
+ \sum_{i,j,k,l}^{\substack{i \neq j, i= k \\ i= l, j \neq k \\ j \neq l, k = l}} X_{ijii}
+ \sum_{i,j,k,l}^{\substack{i= j, i\neq k \\ i= l, j \neq k \\ j = l, k \neq l}} X_{iiki}
+ \sum_{i,j,k,l}^{\substack{i= j, i=k \\ i\neq l, j = k \\ j \neq l, k \neq l}} X_{iiil} \right\} 
+ \left\{ \sum_{i,j,k,l}^{\substack{i= j, i\neq k \\ i\neq l, j \neq k \\ j \neq l, k = l}} X_{iikk}
+\sum_{i,j,k,l}^{\substack{i\neq j, i= k \\ i\neq l, j \neq k \\ j = l, k \neq l}} X_{ijij}
+\sum_{i,j,k,l}^{\substack{i\neq j, i\neq k \\ i= l, j = k \\ j \neq l, k \neq l}} X_{ijji} \right\} 
+  \left\{ \sum_{i,j,k,l}^{\substack{i = j, i= k \\ i= l, j = k \\ j = l, k = l}}  X_{iiii} \right\},
\end{eqnarray}
%\end{widetext}
where no index permutation symmetry is assumed for $X_{ijkl}$. 
The first term of Eq.\ (\ref{eq:E44}) is transformed as
%\begin{widetext}
\begin{eqnarray}
\label{eq:E44_minus4}
&&\left[ \langle N| \hat{V}_4 \hat{R}_4^{(-4)} \hat{V}_4 |N\rangle \right]
\nonumber\\
&& = \frac{1}{576}\sum^{\text{denom.}\neq 0}_{i} \frac{F_{iiii} F_{iiii} }{4\omega_i}   \left[\langle Q_i^4 \rangle_{4} \langle Q_i^4 \rangle_{-4} \right]
+ \frac{1}{32}\sum_{i,k}^{\substack{\text{denom.}\neq 0 \\ i\neq k}} \frac{F_{iikk} F_{iikk}}{2\omega_i+2\omega_k}   \Big[\langle Q_i^2 \rangle_{2} \langle Q_i^2 \rangle_{-2}  \Big]\Big[ \langle Q_k^2 \rangle_{2} \langle Q_k^2 \rangle_{-2} \Big] 
\nonumber\\&&
+\frac{1}{36}\sum_{i,j}^{\substack{\text{denom.}\neq 0 \\ i\neq j}} \frac{F_{ijjj} F_{ijjj}}{\omega_i+3\omega_j}   \Big[ \langle Q_i \rangle_{1} \langle Q_i \rangle_{-1}  \Big]\Big[ \langle Q_j^3 \rangle_{3} \langle Q_j^3 \rangle_{-3}  \Big]
+\frac{1}{8}\sum_{i,j,k}^{\substack{\text{denom.}\neq 0 \\ i\neq j, i\neq k \\ j\neq k}} \frac{F_{ijkk} F_{ijkk} }{\omega_i+\omega_j+2\omega_k}   \Big[ \langle Q_i \rangle_{1} \langle Q_i \rangle_{-1}  \Big]\Big[ \langle Q_j \rangle_{1} \langle Q_j \rangle_{-1} \Big] \Big[ \langle Q_k^2 \rangle_{2} \langle Q_k^2 \rangle_{-2}\Big]
\nonumber\\&&
+\frac{1}{24}\sum_{i,j,k,l}^{\substack{\text{denom.}\neq 0 \\ i\neq j, i\neq k \\ i\neq l, j\neq k \\ j\neq l, k\neq l}} \frac{F_{ijkl} F_{ijkl}}{\omega_i+\omega_j+\omega_k+\omega_l}   \Big[ \langle Q_i \rangle_{1} \langle Q_i \rangle_{-1}  \Big]\Big[ \langle Q_j \rangle_{1} \langle Q_j \rangle_{-1} \Big] \Big[ \langle Q_k \rangle_{1} \langle Q_k \rangle_{-1}\Big] \Big[ \langle Q_l \rangle_{1} \langle Q_l \rangle_{-1}  \Big]
\nonumber\\
&& = \frac{1}{24}\sum^{\text{denom.}\neq 0}_{i} \frac{F_{iiii} F_{iiii} }{4\omega_i}  \Big[ \langle Q_i \rangle_{1} \langle Q_i \rangle_{-1}  \Big]\Big[ \langle Q_i \rangle_{1} \langle Q_i \rangle_{-1} \Big] \Big[ \langle Q_i \rangle_{1} \langle Q_i \rangle_{-1}\Big] \Big[ \langle Q_i \rangle_{1} \langle Q_i \rangle_{-1}  \Big]
\nonumber\\
&&+ \frac{1}{8}\sum_{i,k}^{\substack{\text{denom.}\neq 0 \\ i\neq k}} \frac{F_{iikk} F_{iikk}}{2\omega_i+2\omega_k}   \Big[ \langle Q_i \rangle_{1} \langle Q_i \rangle_{-1}  \Big]\Big[ \langle Q_i \rangle_{1} \langle Q_i \rangle_{-1} \Big] \Big[ \langle Q_k \rangle_{1} \langle Q_k \rangle_{-1}\Big] \Big[ \langle Q_k \rangle_{1} \langle Q_k \rangle_{-1}  \Big]
\nonumber\\&&
+\frac{1}{6}\sum_{i,j}^{\substack{\text{denom.}\neq 0 \\ i\neq j}} \frac{F_{ijjj} F_{ijjj}}{\omega_i+3\omega_j}  \Big[ \langle Q_i \rangle_{1} \langle Q_i \rangle_{-1}  \Big]\Big[ \langle Q_j \rangle_{1} \langle Q_j \rangle_{-1} \Big] \Big[ \langle Q_j \rangle_{1} \langle Q_j \rangle_{-1}\Big] \Big[ \langle Q_j \rangle_{1} \langle Q_j \rangle_{-1}  \Big]
\nonumber\\&&
+\frac{1}{4}\sum_{i,j,k}^{\substack{\text{denom.}\neq 0 \\ i\neq j, i\neq k \\ j\neq k}} \frac{F_{ijkk} F_{ijkk} }{\omega_i+\omega_j+2\omega_k}   \Big[ \langle Q_i \rangle_{1} \langle Q_i \rangle_{-1}  \Big]\Big[ \langle Q_j \rangle_{1} \langle Q_j \rangle_{-1} \Big] \Big[ \langle Q_k \rangle_{1} \langle Q_k \rangle_{-1}\Big] \Big[ \langle Q_k \rangle_{1} \langle Q_k \rangle_{-1}  \Big]
\nonumber\\&&
+\frac{1}{24}\sum_{i,j,k,l}^{\substack{\text{denom.}\neq 0 \\ i\neq j, i\neq k \\ i\neq l, j\neq k \\ j\neq l, k\neq l}} \frac{F_{ijkl} F_{ijkl}}{\omega_i+\omega_j+\omega_k+\omega_l}   \Big[ \langle Q_i \rangle_{1} \langle Q_i \rangle_{-1}  \Big]\Big[ \langle Q_j \rangle_{1} \langle Q_j \rangle_{-1} \Big] \Big[ \langle Q_k \rangle_{1} \langle Q_k \rangle_{-1}\Big] \Big[ \langle Q_l \rangle_{1} \langle Q_l \rangle_{-1}  \Big]
\nonumber\\&&
=\frac{1}{24}\sum_{i,j,k,l}^{{\text{denom.}\neq 0 }} \frac{F_{ijkl} F_{ijkl}}{\omega_i+\omega_j+\omega_k+\omega_l}   \Big[ \langle Q_i \rangle_{1} \langle Q_i \rangle_{-1}  \Big]\Big[ \langle Q_j \rangle_{1} \langle Q_j \rangle_{-1} \Big] \Big[ \langle Q_k \rangle_{1} \langle Q_k \rangle_{-1}\Big] \Big[ \langle Q_l \rangle_{1} \langle Q_l \rangle_{-1}  \Big]
\nonumber\\&&
=\frac{1}{24}\sum_{i,j,k,l}^{{\text{denom.}\neq 0 }} \frac{\tilde{F}_{ijkl} \tilde{F}_{ijkl}}{\omega_i+\omega_j+\omega_k+\omega_l} f_i f_j f_k f_l,
\end{eqnarray}
%\end{widetext}
where we used canonical forms (3), (17), and (22) as well as Eq.\ (\ref{eq:sums2}) to consolidate terms. 

We can then also obtain
\begin{eqnarray}
\label{eq:E44_plus4}
\left[ \langle N| \hat{V}_4 \hat{R}_4^{(+4)} \hat{V}_4 |N\rangle \right]
&=&\frac{1}{24}\sum_{i,j,k,l}^{{\text{denom.}\neq 0 }} \frac{F_{ijkl} F_{ijkl}}{-\omega_i-\omega_j-\omega_k-\omega_l}   \Big[ \langle Q_i \rangle_{-1} \langle Q_i \rangle_{1}  \Big]\Big[ \langle Q_j \rangle_{-1} \langle Q_j \rangle_{1} \Big] \Big[ \langle Q_k \rangle_{-1} \langle Q_k \rangle_{1}\Big] \Big[ \langle Q_l \rangle_{-1} \langle Q_l \rangle_{1}  \Big]
\nonumber\\
&=&\frac{1}{24}\sum_{i,j,k,l}^{{\text{denom.}\neq 0 }} \frac{\tilde{F}_{ijkl} \tilde{F}_{ijkl}}{-\omega_i-\omega_j-\omega_k-\omega_l} (f_i+1)( f_j +1)(f_k +1)(f_l+1),
\end{eqnarray}

% ----------
% E44(-2)
% ----------

%\begin{widetext}
The second parenthesized group of terms in Eq.\ (\ref{eq:E44}) is processed as
\begin{eqnarray}
\label{eq:E44_minus2}
&& \left[ \langle N| \hat{V}_4 \hat{R}_4^{(-2)} \hat{V}_4 |N\rangle \right] + \left[ \langle N| \hat{V}_4 \hat{R}_2^{(-2)} \hat{V}_4 |N\rangle \right]
\nonumber\\
&& = \frac{1}{576}\sum^{\text{denom.}\neq 0}_{i} \frac{F_{iiii} F_{iiii} }{2\omega_i}   \left[\langle Q_i^4 \rangle_{2} \langle Q_i^4 \rangle_{-2} \right]
+ \frac{1}{16}\sum_{i,k}^{\substack{\text{denom.}\neq 0 \\ i\neq k}} \frac{F_{iikk} F_{iikk}}{2\omega_i}   \Big[\langle Q_i^2 \rangle_{2} \langle Q_i^2 \rangle_{-2}  \Big]\Big[ \langle Q_k^2 \rangle_{0} \langle Q_k^2 \rangle_{0} \Big] 
\nonumber\\&&
+\frac{1}{36}\sum_{i,j}^{\substack{\text{denom.}\neq 0 \\ i\neq j}} \frac{F_{ijjj} F_{ijjj}}{\omega_i+\omega_j}   \Big[ \langle Q_i \rangle_{1} \langle Q_i \rangle_{-1}  \Big]\Big[ \langle Q_j^3 \rangle_{1} \langle Q_j^3 \rangle_{-1}  \Big]
+\frac{1}{8}\sum_{i,j,k}^{\substack{\text{denom.}\neq 0 \\ i\neq j, i\neq k \\ j\neq k}} \frac{F_{ijkk} F_{ijkk} }{\omega_i+\omega_j}   \Big[ \langle Q_i \rangle_{1} \langle Q_i \rangle_{-1}  \Big]\Big[ \langle Q_j \rangle_{1} \langle Q_j \rangle_{-1} \Big] \Big[ \langle Q_k^2 \rangle_{0} \langle Q_k^2 \rangle_{0}\Big]
\nonumber\\&&
+\frac{1}{8}\sum_{i,j,k,l}^{\substack{\text{denom.}\neq 0 \\ i\neq j, i\neq k \\ i \neq l, j\neq k \\ j \neq l, k\neq l}} \frac{F_{ijkk} F_{ijll} }{\omega_i+\omega_j}   \Big[ \langle Q_i \rangle_{1} \langle Q_i \rangle_{-1}  \Big]\Big[ \langle Q_j \rangle_{1} \langle Q_j \rangle_{-1} \Big] \Big[ \langle Q_k^2 \rangle_{0} \Big]\Big[\langle Q_l^2 \rangle_{0}\Big]
+ \frac{1}{16}\sum_{i,k,l}^{\substack{\text{denom.}\neq 0 \\ i\neq k, i\neq l \\ k \neq l}} \frac{F_{iikk} F_{iill}}{2\omega_i}   \Big[\langle Q_i^2 \rangle_{2} \langle Q_i^2 \rangle_{-2}  \Big]\Big[ \langle Q_k^2 \rangle_{0}\Big]\Big[ \langle Q_l^2 \rangle_{0} \Big] 
\nonumber\\&&
+ \frac{1}{48}\sum_{i,l}^{\substack{\text{denom.}\neq 0 \\ i\neq l }} \frac{F_{iiii} F_{iill}}{2\omega_i}   \Big[\langle Q_i^4 \rangle_{2} \langle Q_i^2 \rangle_{-2}  \Big] \Big[ \langle Q_l^2 \rangle_{0} \Big] 
+ \frac{1}{6}\sum_{i,j,l}^{\substack{\text{denom.}\neq 0 \\ i\neq j, i \neq l \\ j \neq l }} \frac{F_{ijjj} F_{ijll}}{\omega_i + \omega_j}  \Big[\langle Q_i \rangle_{1} \langle Q_i \rangle_{-1}  \Big]   \Big[\langle Q_j^3 \rangle_{1} \langle Q_j \rangle_{-1}  \Big] \Big[ \langle Q_l^2 \rangle_{0} \Big] 
\nonumber\\&&
+ \frac{1}{36}\sum_{i,j}^{\substack{\text{denom.}\neq 0 \\ i\neq j}} \frac{F_{jiii} F_{ijjj}}{\omega_i + \omega_j}  \Big[\langle Q_i \rangle_{1} \langle Q_i^3 \rangle_{-1}  \Big]   \Big[\langle Q_j \rangle_{1} \langle Q_j^3 \rangle_{-1}  \Big] 
+ \frac{1}{36}\sum_{i,l}^{\substack{\text{denom.}\neq 0 \\ i\neq l }} \frac{F_{iiil} F_{iiil}}{3\omega_i  - \omega_l} 
 \Big[\langle Q_i^3 \rangle_{3} \langle Q_i^3 \rangle_{-3}  \Big]  \Big[\langle Q_l \rangle_{-1} \langle Q_l \rangle_{1}  \Big]
\nonumber\\&&
+ \frac{1}{4}\sum_{i,k,l}^{\substack{\text{denom.}\neq 0 \\ i\neq k, i\neq l \\ k \neq l }} \frac{F_{iikl} F_{iikl}}{2\omega_i + \omega_k - \omega_l} 
 \Big[\langle Q_i^2 \rangle_{2} \langle Q_i^2 \rangle_{-2}  \Big]   \Big[\langle Q_k \rangle_{1} \langle Q_k \rangle_{-1}  \Big]  \Big[\langle Q_l \rangle_{-1} \langle Q_l \rangle_{1}  \Big]
\nonumber\\&&
+ \frac{1}{6}\sum_{i,j,k,l}^{\substack{\text{denom.}\neq 0 \\ i\neq j, i\neq k \\ i \neq l, j \neq k \\ j \neq l, k \neq l}} \frac{F_{ijkl} F_{ijkl}}{\omega_i + \omega_j + \omega_k - \omega_l} 
 \Big[\langle Q_i \rangle_{1} \langle Q_i \rangle_{-1}  \Big]    \Big[\langle Q_j \rangle_{1} \langle Q_j \rangle_{-1}  \Big]  \Big[\langle Q_k \rangle_{1} \langle Q_k \rangle_{-1}  \Big]  \Big[\langle Q_l \rangle_{-1} \langle Q_l \rangle_{1}  \Big] 
 \nonumber\\
 && =\frac{1}{8}\sum_{i,j,k,l}^{{\text{denom.}\neq 0 }} \frac{F_{ijkk} F_{ijll} }{\omega_i+\omega_j}   \Big[ \langle Q_i \rangle_{1} \langle Q_i \rangle_{-1}  \Big]\Big[ \langle Q_j \rangle_{1} \langle Q_j \rangle_{-1} \Big] \Big[ \langle Q_k^2 \rangle_{0} \Big]\Big[\langle Q_l^2 \rangle_{0}\Big]
\nonumber\\&&
+ \frac{1}{6}\sum_{i,j,k,l}^{{\text{denom.}\neq 0 }} \frac{F_{ijkl} F_{ijkl}}{\omega_i + \omega_j + \omega_k - \omega_l} 
 \Big[\langle Q_i \rangle_{1} \langle Q_i \rangle_{-1}  \Big]    \Big[\langle Q_j \rangle_{1} \langle Q_j \rangle_{-1}  \Big]  \Big[\langle Q_k \rangle_{1} \langle Q_k \rangle_{-1}  \Big]  \Big[\langle Q_l \rangle_{-1} \langle Q_l \rangle_{1}  \Big] 
 \nonumber\\
 && = \frac{1}{2}\sum^{\text{denom.}\neq 0}_{i,j,k,l}  \frac{\tilde{F}_{ijkk} \tilde{F}_{ijll}}{\omega_i+\omega_j} {f_i} {f_j} {(f_k+1/2)}{} {(f_l+1/2)}{}
 + \frac{1}{6}\sum^{\text{denom.}\neq 0}_{i,j,k,l}  \frac{\tilde{F}_{ijkl} \tilde{F}_{ijkl} }{\omega_i+\omega_j+\omega_k-\omega_l} {f_i}  {f_j}  {f_k}(f_l+1),
\end{eqnarray}
where canonical forms (2), (3), (8), (11), (16), (17), and (21) as well as Eq.\ (\ref{eq:sums2}) were invoked to fully remove summation index restrictions in the second equality
except ``denom.$\neq$0'' meaning either $\omega_i + \omega_j \neq 0$ or $\omega_i + \omega_j + \omega_k - \omega_l \neq 0$. In the last equality, the thermal Born--Huang rules were used. 

We can then infer
\begin{eqnarray}
\label{eq:E44_2}
&&\left[ \langle N| \hat{V}_4 \hat{R}_4^{(+2)} \hat{V}_4 |N\rangle \right] + \left[ \langle N| \hat{V}_4 \hat{R}_2^{(+2)} \hat{V}_4 |N\rangle \right] 
\nonumber\\
 && =\frac{1}{8}\sum_{i,j,k,l}^{{\text{denom.}\neq 0 }} \frac{F_{ijkk} F_{ijll} }{-\omega_i-\omega_j}   \Big[ \langle Q_i \rangle_{-1} \langle Q_i \rangle_{1}  \Big]\Big[ \langle Q_j \rangle_{-1} \langle Q_j \rangle_{1} \Big] \Big[ \langle Q_k^2 \rangle_{0} \Big]\Big[\langle Q_l^2 \rangle_{0}\Big]
\nonumber\\&&
+ \frac{1}{6}\sum_{i,j,k,l}^{{\text{denom.}\neq 0 }} \frac{F_{ijkl} F_{ijkl}}{-\omega_i - \omega_j - \omega_k + \omega_l} 
 \Big[\langle Q_i \rangle_{-1} \langle Q_i \rangle_{1}  \Big]    \Big[\langle Q_j \rangle_{-1} \langle Q_j \rangle_{1}  \Big]  \Big[\langle Q_k \rangle_{-1} \langle Q_k \rangle_{1}  \Big]  \Big[\langle Q_l \rangle_{1} \langle Q_l \rangle_{-1}  \Big] 
 \nonumber\\
 && = \frac{1}{2}\sum^{\text{denom.}\neq 0}_{i,j,k,l}  \frac{\tilde{F}_{ijkk} \tilde{F}_{ijll}}{-\omega_i-\omega_j} {(f_i+1)} {(f_j+1)} {(f_k+1/2)}{} {(f_l+1/2)}{}
 + \frac{1}{6}\sum^{\text{denom.}\neq 0}_{i,j,k,l}  \frac{\tilde{F}_{ijkl} \tilde{F}_{ijkl} }{-\omega_i-\omega_j-\omega_k+\omega_l} {(f_i+1)}  {(f_j+1)}  {(f_k+1)}f_l.
\end{eqnarray}

% ----------
% E44(0)
% ----------

The third parenthesized group of terms in  Eq.\ (\ref{eq:E44}) is evaluated analogously.
\begin{eqnarray}
\label{eq:E44_0} 
&&\left[ \langle N| \hat{V}_4 \hat{R}_4^{(\pm0)} \hat{V}_4 |N\rangle \right] + \left[ \langle N| \hat{V}_4 \hat{R}_2^{(\pm0)} \hat{V}_4 |N\rangle \right]
\nonumber\\
&& = \frac{1}{18}\sum_{i,j}^{\substack{\text{denom.}\neq 0 \\ i\neq j}} \frac{F_{ijjj} F_{ijjj}}{\omega_i-\omega_j}   \Big[ \langle Q_i \rangle_{1} \langle Q_i \rangle_{-1}  \Big]\Big[ \langle Q_j^3 \rangle_{-1} \langle Q_j^3 \rangle_{1}  \Big]
+\frac{1}{4}\sum_{i,j,k}^{\substack{\text{denom.}\neq 0 \\ i\neq j, i\neq k \\ j\neq k}} \frac{F_{ijkk} F_{ijkk} }{\omega_i-\omega_j}   \Big[ \langle Q_i \rangle_{1} \langle Q_i \rangle_{-1}  \Big]\Big[ \langle Q_j \rangle_{-1} \langle Q_j \rangle_{1} \Big] \Big[ \langle Q_k^2 \rangle_{0} \langle Q_k^2 \rangle_{0}\Big]
\nonumber\\&&
+\frac{1}{4}\sum_{i,j,k,l}^{\substack{\text{denom.}\neq 0 \\ i\neq j, i\neq k \\ i \neq l, j\neq k \\ j \neq l, k\neq l}} \frac{F_{ijkk} F_{ijll} }{\omega_i-\omega_j}   \Big[ \langle Q_i \rangle_{1} \langle Q_i \rangle_{-1}  \Big]\Big[ \langle Q_j \rangle_{-1} \langle Q_j \rangle_{1} \Big] \Big[ \langle Q_k^2 \rangle_{0} \Big]\Big[\langle Q_l^2 \rangle_{0}\Big]
+ \frac{1}{3}\sum_{i,j,l}^{\substack{\text{denom.}\neq 0 \\ i\neq j, i \neq l \\ j \neq l }} \frac{F_{ijjj} F_{ijll}}{\omega_i - \omega_j}  \Big[\langle Q_i \rangle_{1} \langle Q_i \rangle_{-1}  \Big]   \Big[\langle Q_j^3 \rangle_{-1} \langle Q_j \rangle_{1}  \Big] \Big[ \langle Q_l^2 \rangle_{0} \Big] 
\nonumber\\&&
+ \frac{1}{18}\sum_{i,j}^{\substack{\text{denom.}\neq 0 \\ i\neq j}} \frac{F_{jiii} F_{ijjj}}{\omega_i - \omega_j}  \Big[\langle Q_i^3 \rangle_{1} \langle Q_i \rangle_{-1}  \Big]   \Big[\langle Q_j \rangle_{-1} \langle Q_j^3 \rangle_{1}  \Big] 
+ \frac{1}{8}\sum_{i,k,l}^{\substack{\text{denom.}\neq 0 \\ i\neq k, i\neq l \\ k \neq l }} \frac{F_{iikl} F_{iikl}}{2\omega_i - \omega_k - \omega_l} 
 \Big[\langle Q_i^2 \rangle_{2} \langle Q_i^2 \rangle_{-2}  \Big]   \Big[\langle Q_k \rangle_{-1} \langle Q_k \rangle_{1}  \Big]  \Big[\langle Q_l \rangle_{-1} \langle Q_l \rangle_{1}  \Big] 
\nonumber\\&&
+ \frac{1}{8}\sum_{i,j,k}^{\substack{\text{denom.}\neq 0 \\ i\neq j, i\neq k \\ j \neq k }} \frac{F_{ijkk} F_{ijkk}}{\omega_i + \omega_j -2 \omega_k } 
 \Big[\langle Q_i \rangle_{1} \langle Q_i \rangle_{-1}  \Big]    \Big[\langle Q_j \rangle_{1} \langle Q_j \rangle_{-1}  \Big]  \Big[\langle Q_k^2 \rangle_{-2} \langle Q_k^2 \rangle_{2}  \Big] 
\nonumber\\&&
+ \frac{1}{4}\sum_{i,j,k,l}^{\substack{\text{denom.}\neq 0 \\ i\neq j, i\neq k \\ i \neq l, j \neq k \\ j \neq l, k \neq l}} \frac{F_{ijkl} F_{ijkl}}{\omega_i + \omega_j - \omega_k - \omega_l} 
 \Big[\langle Q_i \rangle_{1} \langle Q_i \rangle_{-1}  \Big]    \Big[\langle Q_j \rangle_{1} \langle Q_j \rangle_{-1}  \Big]  \Big[\langle Q_k \rangle_{-1} \langle Q_k \rangle_{1}  \Big]  \Big[\langle Q_l \rangle_{-1} \langle Q_l \rangle_{1}  \Big] 
 \nonumber\\
 && =\frac{1}{4}\sum_{i,j,k,l}^{{\text{denom.}\neq 0 }} \frac{F_{ijkk} F_{ijll} }{\omega_i-\omega_j}   \Big[ \langle Q_i \rangle_{1} \langle Q_i \rangle_{-1}  \Big]\Big[ \langle Q_j \rangle_{-1} \langle Q_j \rangle_{1} \Big] \Big[ \langle Q_k^2 \rangle_{0} \Big]\Big[\langle Q_l^2 \rangle_{0}\Big]
\nonumber\\&&
+ \frac{1}{4}\sum_{i,j,k,l}^{{\text{denom.}\neq 0 }} \frac{F_{ijkl} F_{ijkl}}{\omega_i + \omega_j - \omega_k - \omega_l} 
 \Big[\langle Q_i \rangle_{1} \langle Q_i \rangle_{-1}  \Big]    \Big[\langle Q_j \rangle_{1} \langle Q_j \rangle_{-1}  \Big]  \Big[\langle Q_k \rangle_{-1} \langle Q_k \rangle_{1}  \Big]  \Big[\langle Q_l \rangle_{-1} \langle Q_l \rangle_{1}  \Big] 
 \nonumber\\
 && =\sum^{\text{denom.}\neq 0}_{i,j,k,l}  \frac{\tilde{F}_{ijkk} \tilde{F}_{ijll}}{\omega_i-\omega_j} {f_i} {(f_j+1)} {(f_k+1/2)}{} {(f_l+1/2)}{}
 + \frac{1}{4}\sum^{\text{denom.}\neq 0}_{i,j,k,l}  \frac{\tilde{F}_{ijkl} \tilde{F}_{ijkl} }{\omega_i+\omega_j-\omega_k-\omega_l} {f_i}  {f_j}  {(f_k+1)}(f_l+1),
\end{eqnarray}
where canonical forms (2), (3), (4), (8), (9), and (18) as well as Eq.\ (\ref{eq:sums2}) were used.
%\begin{eqnarray}
%\label{eq:sums2_mod4}
%\sum_{i,j,k,l} X_{[ij](kl)} &=& \sum_{i,j,k,l}^{\substack{i\neq j, i\neq k \\ i\neq l, j \neq k \\ j \neq l, k \neq l}}  X_{[ij](kl)} 
%+ \sum_{i,j,k,l}^{\substack{i\neq j, i\neq k \\ i\neq l, j \neq k \\ j \neq l, k = l}} X_{[ij](kk)}
%+4 \sum_{i,j,k,l}^{\substack{i\neq j, i\neq k \\ i\neq l, j \neq k \\ j = l, k \neq l}} X_{[ij](kj)}
%\nonumber\\&& 
%+ 2 \sum_{i,j,k,l}^{\substack{i\neq j, i\neq k \\ i\neq l, j = k \\ j = l, k = l}} X_{[ij](jj)}
%+2 \sum_{i,j,k,l}^{\substack{i\neq j, i= k \\ i\neq l, j \neq k \\ j = l, k \neq l}} X_{[ij](ij)},
%\end{eqnarray}
%for summand $X_{[ij](kl)}$ that is antisymmetric with $i\leftrightarrow j$ and is symmetric with $k\leftrightarrow l$, and 
%bears the following relationship:\ $X_{[ij](kl)} = -X_{[ji](kl)} = X_{[ij](lk)} = -X_{[ji](lk)}$. For summand $X_{ijkl}$ that is symmetric with $i\leftrightarrow j$ and $k\leftrightarrow l$, but 
%vanishes when $(i,j) = (k,l)$, we used
%\begin{eqnarray}
%\label{eq:sums2_mod5}
%\sum_{i,j,k,l} X_{(ij)(kl)} &=&  \sum_{i,j,k,l}^{\substack{i\neq j, i\neq k \\ i\neq l, j \neq k \\ j \neq l, k \neq l}}  X_{(ij)(kl)}
%+ 4  \sum_{i,j,k,l}^{\substack{i\neq j, i\neq k \\ i\neq l, j \neq k \\ j \neq l, k = l}} X_{(ij)(kk)}
%+ \sum_{i,j,k,l}^{\substack{i\neq j, i\neq k \\ i\neq l, j \neq k \\ j = l, k \neq l}} X_{(ij)(kj)}
%\nonumber\\&& 
%+\sum_{i,j,k,l}^{\substack{i\neq j, i= k \\ i\neq l, j \neq k \\ j \neq l, k \neq l}} X_{(ij)(il)}
%+ 4 \sum_{i,j,k,l}^{\substack{i\neq j, i\neq k \\ i\neq l, j = k \\ j = l, k = l}} X_{(ij)(jj)}.
%\end{eqnarray}

Putting these together, we obtain
\begin{eqnarray}
&&\left[ \langle N| \hat{V}_4 \hat{R} \hat{V}_4 |N\rangle \right]
\nonumber\\&& 
= \frac{1}{24}\sum_{i,j,k,l}^{{\text{denom.}\neq 0 }} \frac{\tilde{F}_{ijkl} \tilde{F}_{ijkl}}{\omega_i+\omega_j+\omega_k+\omega_l} f_i f_j f_k f_l
%\nonumber\\&& 
+ \frac{1}{24}\sum_{i,j,k,l}^{{\text{denom.}\neq 0 }} \frac{\tilde{F}_{ijkl} \tilde{F}_{ijkl}}{-\omega_i-\omega_j-\omega_k-\omega_l} (f_i+1)( f_j +1)(f_k +1)(f_l+1)
\nonumber\\ && 
+ \frac{1}{2}\sum^{\text{denom.}\neq 0}_{i,j,k,l}  \frac{\tilde{F}_{ijkk} \tilde{F}_{ijll}}{\omega_i+\omega_j} {f_i} {f_j} {(f_k+1/2)}{} {(f_l+1/2)}{}
% \nonumber\\ &&
 + \frac{1}{6}\sum^{\text{denom.}\neq 0}_{i,j,k,l}  \frac{\tilde{F}_{ijkl} \tilde{F}_{ijkl} }{\omega_i+\omega_j+\omega_k-\omega_l} {f_i}  {f_j}  {f_k}(f_l+1)
\nonumber\\&& 
+ \frac{1}{2}\sum^{\text{denom.}\neq 0}_{i,j,k,l}  \frac{\tilde{F}_{ijkk} \tilde{F}_{ijll}}{-\omega_i-\omega_j} {(f_i+1)} {(f_j+1)} {(f_k+1/2)}{} {(f_l+1/2)}{}
% \nonumber\\ && 
 + \frac{1}{6}\sum^{\text{denom.}\neq 0}_{i,j,k,l}  \frac{\tilde{F}_{ijkl} \tilde{F}_{ijkl} }{-\omega_i-\omega_j-\omega_k+\omega_l} {(f_i+1)}  {(f_j+1)}  {(f_k+1)}f_l
\nonumber\\&& 
+\sum^{\text{denom.}\neq 0}_{i,j,k,l}  \frac{\tilde{F}_{ijkk} \tilde{F}_{ijll}}{\omega_i-\omega_j} {f_i} {(f_j+1)} {(f_k+1/2)}{} {(f_l+1/2)}{}
% \nonumber\\ && 
 + \frac{1}{4}\sum^{\text{denom.}\neq 0}_{i,j,k,l}  \frac{\tilde{F}_{ijkl} \tilde{F}_{ijkl} }{\omega_i+\omega_j-\omega_k-\omega_l} {f_i}  {f_j}  {(f_k+1)}(f_l+1),
\end{eqnarray}
which is linked.
\end{widetext}

Since all terms of $[E_N^{(2)}]$ are linked, we write
\begin{eqnarray}
\Big[ E_N^{(2)} \Big] = \Big[ E_N^{(2)} \Big]_L. \label{eq:E2linked}
\end{eqnarray}

%============================================
% APPENDIX Algebraic reduction E(1)E(1)
%============================================

\section{Algebraic reduction of $[ E_N^{(1)} E_N^{(1)} ]$\label{appendix:algebraicreduction2}}

Here, we illustrate an algebraic reduction of $[ E_N^{(1)} E_N^{(1)} ]$ entering $\Omega^{(2)}$ [Eq.\ (\ref{eq:E2PT_recursion})].
In parallel with Eq.\ (\ref{eq:E2nondegen}), we first divide it into its components as follows:
\begin{eqnarray}
\label{eq:E1E1}
    \left[E_N^{(1)} E_N^{(1)}\right] &=&\left[ \mathrm{Tr}\left( \bm{E}^{(1)}\bm{E}^{(1)}\right) \right] \nonumber\\
    &=& \left[ \langle N | \hat{V}_1 \hat{P} \hat{V}_1 | N \rangle \right] + \left[ \langle N | \hat{V}_1 \hat{P} \hat{V}_3 | N \rangle \right]  \nonumber \\ 
    &&+ \left[ \langle N | \hat{V}_3 \hat{P} \hat{V}_1 | N \rangle \right] + \left[ \langle N | \hat{V}_2 \hat{P} \hat{V}_2 | N \rangle \right] \nonumber \\ 
    &&+ \left[ \langle N | \hat{V}_2 \hat{P} \hat{V}_4 | N \rangle \right] + \left[ \langle N | \hat{V}_4 \hat{P} \hat{V}_2 | N \rangle \right] \nonumber \\ 
    &&+ \left[ \langle N | \hat{V}_3 \hat{P} \hat{V}_3 | N \rangle \right] + \left[ \langle N | \hat{V}_4 \hat{P} \hat{V}_4 | N \rangle \right],
\end{eqnarray}
where the inner projector $\hat{P}$ consists of
\begin{eqnarray}
\label{eq:P_breakdown}
\hat{P} = \hat{P}_0 + \hat{P}_1 + \hat{P}_2 + \hat{P}_3 + \hat{P}_4,
\end{eqnarray}
in a QFF. 
Recall that $E_N^{(1)}$ is an eigenvalue of the matrix $\bm{E}^{(1)}$ and  cannot generally be expressed in a closed form. 
However, as per the trace invariance [Eq.\ (\ref{eq:traceinvariance2})], the thermal average of the squares of these eigenvalues is equal to the trace
of the matrix product $\bm{E}^{(1)}\bm{E}^{(1)}$, which can be written in a closed form according to the degenerate RSPT [Eqs.\ (\ref{eq:E_In_degen}) and (\ref{eq:Phi_In_degen})].
Through this matrix product $\bm{E}^{(1)}\bm{E}^{(1)}$, off-diagonal as well as diagonal elements of $\bm{E}^{(1)}$ enter the thermal average.

With Eq.\ (\ref{eq:P_breakdown}), we can further divide each term as
\begin{eqnarray}
 \left[ \langle N | \hat{V}_m \hat{P} \hat{V}_n | N \rangle \right]  &=&  \left[ \langle N | \hat{V}_m \hat{P}_0 \hat{V}_n | N \rangle \right]  
 + \left[ \langle N | \hat{V}_m \hat{P}_1 \hat{V}_n | N \rangle \right] 
  \nonumber\\   &&+ \left[ \langle N | \hat{V}_m \hat{P}_2 \hat{V}_n | N \rangle \right] 
   + \left[ \langle N | \hat{V}_m \hat{P}_3 \hat{V}_n | N \rangle \right] 
      \nonumber \\   &&
+ \left[ \langle N | \hat{V}_m \hat{P}_4 \hat{V}_n | N \rangle \right]
 \end{eqnarray}
 with
 \begin{eqnarray}
  \left[ \langle N | \hat{V}_m \hat{P}_1 \hat{V}_n | N \rangle \right] &=& \left[ \langle N | \hat{V}_m \hat{P}_1^{(+1)} \hat{V}_n | N \rangle \right] 
 %\nonumber\\&& 
 +  \left[ \langle N | \hat{V}_m \hat{P}_1^{(-1)} \hat{V}_n | N \rangle \right] , \nonumber\\ 
 \\
  \left[ \langle N | \hat{V}_m \hat{P}_2 \hat{V}_n | N \rangle \right] &=& \left[ \langle N | \hat{V}_m \hat{P}_2^{(+2)} \hat{V}_n | N \rangle \right] 
  %\nonumber\\&& 
  +  \left[ \langle N | \hat{V}_m \hat{P}_2^{(\pm0)} \hat{V}_n | N \rangle \right] 
 \nonumber\\&& +  \left[ \langle N | \hat{V}_m \hat{P}_2^{(-2)} \hat{V}_n | N \rangle \right] , \\
  \left[ \langle N | \hat{V}_m \hat{P}_3 \hat{V}_n | N \rangle \right] &=& \left[ \langle N | \hat{V}_m \hat{P}_3^{(+3)} \hat{V}_n | N \rangle \right] 
  %\nonumber\\&& 
  +  \left[ \langle N | \hat{V}_m \hat{P}_3^{(+1)} \hat{V}_n | N \rangle \right] 
  \nonumber\\&& +  \left[ \langle N | \hat{V}_m \hat{P}_3^{(-1)} \hat{V}_n | N \rangle \right] 
 \nonumber\\&& 
 +  \left[ \langle N | \hat{V}_m \hat{P}_3^{(-3)} \hat{V}_n | N \rangle \right] ,  \\
  \left[ \langle N | \hat{V}_m \hat{P}_4 \hat{V}_n | N \rangle \right] &=& \left[ \langle N | \hat{V}_m \hat{P}_4^{(+4)} \hat{V}_n | N \rangle \right] 
  %\nonumber\\&& 
  +  \left[ \langle N | \hat{V}_m \hat{P}_4^{(+2)} \hat{V}_n | N \rangle \right] 
  \nonumber\\&& +  \left[ \langle N | \hat{V}_m \hat{P}_4^{(\pm0)} \hat{V}_n | N \rangle \right] 
  %\nonumber\\&& 
  +  \left[ \langle N | \hat{V}_m \hat{P}_4^{(-2)} \hat{V}_n | N \rangle \right] 
 \nonumber\\&& +  \left[ \langle N | \hat{V}_m \hat{P}_4^{(-4)} \hat{V}_n | N \rangle \right] . 
  \end{eqnarray}
 An algebraic reduction of $[ \langle N | \hat{V}_m \hat{P}_x^{(y)} \hat{V}_n | N \rangle ]$ ($y\neq 0$) is essentially the same 
 as that of $[ \langle N | \hat{V}_m \hat{R}_x^{(y)} \hat{V}_n | N \rangle ]$ with the only difference being that $\hat{R}$ introduces 
 a denominator with the summation index restriction requiring the denominator to be nonzero, while $\hat{P}$ instead injects the opposite restriction
 that the corresponding fictitious denominator be zero. 
 
 For example, $[ \langle N| \hat{V}_1 \hat{P}_1 \hat{V}_1 |N\rangle ]$ is reduced, using the identical logic as Eq.\ (\ref{eq:E11}), as 
 \begin{eqnarray}
\label{eq:E11_P}
&& \left[ \langle N| \hat{V}_1 \hat{P}_1 \hat{V}_1 |N\rangle \right] \nonumber\\
 &&= \sum^{\text{denom.}= 0}_{i} {F_{i} F_{i}} \Big[ \langle Q_i \rangle_{1} \langle Q_i \rangle_{-1}\Big]   + \sum^{\text{denom.}= 0}_{i} {F_{i} F_{i}} \Big[ \langle Q_i \rangle_{-1} \langle Q_i \rangle_{1}\Big]   \nonumber\\
 &&= \sum^{\text{denom.}= 0}_i {\tilde{F}_{i} \tilde{F}_{i}}{}{f_i} + \sum^{\text{denom.}= 0}_i {\tilde{F}_{i} \tilde{F}_{i}}{}{(f_i+1)},
\end{eqnarray}
where ``denom.=0'' means $\omega_i = 0$. These fictitious denominators can be inferred from the corresponding thermal average having 
$\hat{R}$ in place of $\hat{P}$, i.e., Eq.\ (\ref{eq:E11}). We shall not repeat essentially the same reduction of $[ \langle N | \hat{V}_m \hat{P}_x^{(y)} \hat{V}_n | N \rangle ]$ ($y \neq 0$) and refer the reader to Appendix \ref{appendix:algebraicreduction}.

In this Appendix, we instead focus on the terms that require special attention, i.e., $[ \langle N | \hat{V}_m \hat{P}_x^{(y)} \hat{V}_n | N \rangle ]$ with $x=0$ or $y=0$.
The term in Eq.\ (\ref{eq:E1E1}) involving $\hat{P}_0$ consists of
\begin{eqnarray}
\label{eq:E1E1_0}
 \left[ \mathrm{Tr}\left( \bm{E}^{(1)}\hat{P}_0 \bm{E}^{(1)}\right) \right] &=& \left(  \left[ \langle N | \hat{V}_2 \hat{P}_0 \hat{V}_2 | N \rangle \right]   \right)
   \nonumber\\   &&+\left( \left[ \langle N | \hat{V}_2 \hat{P}_0 \hat{V}_4 | N \rangle \right] 
+ \left[ \langle N | \hat{V}_4 \hat{P}_0 \hat{V}_2 | N \rangle \right] \right)
  \nonumber\\   &&   +\left(  \left[ \langle N | \hat{V}_4 \hat{P}_0 \hat{V}_4 | N \rangle \right] \right).
 \end{eqnarray}
Only these combinations of $\hat{V}_m$ and $\hat{V}_n$ lead to nonzero results, as inferred from Tables \ref{table:thermalavg_EN1} and \ref{table:thermalavg_EN2}.
 Each term in the right-hand side together with the associated contributions from $[ \langle N | \hat{V}_m \hat{P}_x^{(\pm0)} \hat{V}_n | N \rangle ]$ ($x = 2$ or $4$) will be evaluated in the following.

\subsection{$[\langle N | \hat{V}_2 \hat{P}_0 \hat{V}_2 | N \rangle]+[\langle N | \hat{V}_2 \hat{P}_2^{(\pm 0)} \hat{V}_2 | N \rangle]$} 

The first term of Eq.\ (\ref{eq:E1E1_0}) needs to be reduced in conjunction with $[\langle N | \hat{V}_2 \hat{P}_2^{(\pm 0)} \hat{V}_2 | N \rangle]$.
\begin{eqnarray}
\label{eq:P_E22_2}
&&  \left[ \langle N| \hat{V}_2 \hat{P}_0 \hat{V}_2 |N\rangle \right]+ \left[ \langle N| \hat{V}_2 \hat{P}_2^{(\pm0)} \hat{V}_2 |N\rangle \right]  \nonumber\\
&&=  \frac{1}{4} \sum_{i,j}^{{ i\neq j}} {\bar{F}_{ii} \bar{F}_{jj}} \Big[ \langle Q_i^2 \rangle_{0} \Big]\Big[ \langle Q_j^2\rangle_{0} \Big] 
%\nonumber\\ && 
+  \frac{1}{4} \sum_{i} {\bar{F}_{ii} \bar{F}_{ii}} \Big[ \langle Q_i^2 \rangle_{0} \langle Q_i^2\rangle_{0} \Big] \nonumber\\ 
&& +  \sum_{i,j}^{\substack{\text{denom.}= 0\\ i\neq j}} {\bar{F}_{ij} \bar{F}_{ij}} \Big[ \langle Q_i \rangle_{1} \langle Q_i \rangle_{-1}\Big]\Big[ \langle Q_j \rangle_{-1} \langle Q_j \rangle_{1}\Big] \nonumber\\ 
&&=  \frac{1}{4} \sum_{i,j}^{{ i\neq j}} {\bar{F}_{ii} \bar{F}_{jj}} \Big[ \langle Q_i^2 \rangle_{0} \Big]\Big[ \langle Q_j^2\rangle_{0} \Big] 
%\nonumber\\ && 
+  \frac{1}{4} \sum_{i} {\bar{F}_{ii} \bar{F}_{ii}} \Big[ \langle Q_i^2 \rangle_{0}\Big]\Big[ \langle Q_i^2\rangle_{0} \Big] \nonumber\\ 
&& +  \sum_{i} {\bar{F}_{ii} \bar{F}_{ii}} \Big[ \langle Q_i \rangle_{1} \langle Q_i \rangle_{-1}\Big]\Big[ \langle Q_i \rangle_{-1} \langle Q_i \rangle_{1}\Big] \nonumber\\ 
&& +  \sum_{i,j}^{\substack{\text{denom.}= 0\\ i\neq j}} {\bar{F}_{ij} \bar{F}_{ij}} \Big[ \langle Q_i \rangle_{1} \langle Q_i \rangle_{-1}\Big]\Big[ \langle Q_j \rangle_{-1} \langle Q_j \rangle_{1}\Big] \nonumber\\ 
&&=  \frac{1}{4} \sum_{i,j} {\bar{F}_{ii} \bar{F}_{jj}} \Big[ \langle Q_i^2 \rangle_{0} \Big]\Big[ \langle Q_j^2\rangle_{0} \Big] \nonumber\\
&& +  \sum_{i,j}^{{\text{denom.}= 0}} {\bar{F}_{ij} \bar{F}_{ij}} \Big[ \langle Q_i \rangle_{1} \langle Q_i \rangle_{-1}\Big]\Big[ \langle Q_j \rangle_{-1} \langle Q_j \rangle_{1}\Big] \nonumber\\ 
&&=   \sum_{i,j} {\tilde{\bar{F}}_{ii} \tilde{\bar{F}}_{jj}} (f_i +1/2) (f_j + 1/2)
+ \sum_{i,j}^{\text{denom.}= 0} {\tilde{\bar{F}}_{ij} \tilde{\bar{F}}_{ij}} f_i (f_j + 1), \nonumber\\
\end{eqnarray}
where $\hat{P}_0$ (unlike $\hat{P}_x$ with $x \geq 1$) does not introduce the ``denom.=0'' restriction. Canonical form (2) of Table \ref{table:canonical} was used in the second equality to consolidate terms and thereby eliminate the $i\neq j$ restrictions. This is contrasted with Eq.\ (\ref{eq:E22_2}), where the $i\neq j$ restriction was removed because it is implied by the stronger condition ``denom.$\neq$0''. 
The ``denom.=0'' restriction in the second term of the final expression 
originates from $\hat{P}_2^{(\pm 0)}$ and it demands that the fictitious denominator ${\omega_i-\omega_j}$ be zero. 

The first term of the final reduced formula is unlinked because it is a product of two extensive scalars with no common summation index. 
It will be canceled out by another unlinked term of the same magnitude  in $\Omega^{(2)}$ (see the main text). The second term is linked through indexes $i$ and $j$.

\subsection{$[\langle N | \hat{V}_2 \hat{P}_0 \hat{V}_4 | N \rangle] + [\langle N | \hat{V}_2 \hat{P}_2^{(\pm0)} \hat{V}_4 | N \rangle]$ and $[\langle N | \hat{V}_4 \hat{P}_0 \hat{V}_2 | N \rangle]+ [\langle N | \hat{V}_4 \hat{P}_2^{(\pm0)} \hat{V}_2 | N \rangle]$} 

The second term of Eq.\ (\ref{eq:E1E1_0}) and $[\langle N | \hat{V}_2 \hat{P}_2^{(\pm0)} \hat{V}_4 | N \rangle]$ are reduced together.
\begin{eqnarray}
\label{eq:P_E24_2}
&& \left[ \langle N| \hat{V}_2 \hat{P}_0 \hat{V}_4 |N\rangle \right] +\left[ \langle N| \hat{V}_2 \hat{P}_2^{(\pm0)} \hat{V}_4 |N\rangle \right] 
 \nonumber\\&&
 =  \frac{1}{16} \sum_{i,j,k}^{\substack{ i\neq j, i\neq k \\ j\neq k}} {\bar{F}_{ii} {F}_{jjkk}} \Big[ \langle Q_i^2\rangle_{0} \Big]\Big[ \langle Q_j^2 \rangle_{0} \Big] \Big[ \langle Q_k^2 \rangle_{0} \Big] 
 \nonumber\\&&
 + \frac{1}{48} \sum_{i,j}^{{ i\neq j}} {\bar{F}_{ii} {F}_{jjjj}} \Big[ \langle Q_i^2\rangle_{0} \Big]\Big[ \langle Q_j^4 \rangle_{0} \Big] 
% \nonumber\\&&
 + \frac{1}{48} \sum_{i} {\bar{F}_{ii} {F}_{iiii}} \Big[ \langle Q_i^2\rangle_{0}  \langle Q_i^4 \rangle_{0} \Big] 
 \nonumber\\&&
+  \frac{1}{8} \sum_{i,k}^{{ i\neq k }} {\bar{F}_{ii} {F}_{iikk}} \Big[ \langle Q_i^2\rangle_{0}  \langle Q_i^2 \rangle_{0} \Big] \Big[ \langle Q_k^2 \rangle_{0} \Big] 
\nonumber\\&&
 +  \frac{1}{6} \sum_{i,j}^{\substack{\text{denom.}= 0\\ i\neq j}} {\bar{F}_{ij} {F}_{ijjj}} \Big[ \langle Q_i \rangle_{1} \langle Q_i \rangle_{-1}\Big]\Big[ \langle Q_j \rangle_{-1} \langle Q_j^3 \rangle_{1} \Big] 
\nonumber\\&&
 +  \frac{1}{6} \sum_{i,j}^{\substack{\text{denom.}= 0\\ i\neq j}} {\bar{F}_{ij} {F}_{ijii}} \Big[ \langle Q_i \rangle_{1} \langle Q_i^3 \rangle_{-1}\Big]\Big[ \langle Q_j \rangle_{-1} \langle Q_j \rangle_{1} \Big] 
 \nonumber\\ &&
 + \frac{1}{2} \sum_{i,j,k}^{\substack{\text{denom.}= 0\\ i\neq j, i\neq k \\ j\neq k}} {\bar{F}_{ij} {F}_{ijkk}} \Big[ \langle Q_i \rangle_{1} \langle Q_i \rangle_{-1}\Big]\Big[ \langle Q_j \rangle_{-1} \langle Q_j \rangle_{1}\Big]\Big[ \langle Q_k^2 \rangle_{0} \Big] \nonumber\\ 
 &&= \frac{1}{16} \sum_{i,j,k} {\bar{F}_{ii} {F}_{jjkk}} \Big[ \langle Q_i^2\rangle_{0} \Big]\Big[ \langle Q_j^2 \rangle_{0} \Big] \Big[ \langle Q_k^2 \rangle_{0} \Big]  \nonumber\\ 
 &&+ \frac{1}{2} \sum_{i,j,k}^{{\text{denom.}= 0}} {\bar{F}_{ij} {F}_{ijkk}} \Big[ \langle Q_i \rangle_{1} \langle Q_i \rangle_{-1}\Big]\Big[ \langle Q_j \rangle_{-1} \langle Q_j \rangle_{1}\Big]\Big[ \langle Q_k^2 \rangle_{0} \Big] \nonumber\\ 
&&=  \frac{1}{2}\sum_{i,j,k} {\tilde{\bar{F}}_{ii} \tilde{F}_{jjkk}} (f_i+1/2) (f_j+1/2)(f_k + 1/2) \nonumber\\
&& + \sum_{i,j,k}^{{\text{denom.}= 0}} {\tilde{\bar{F}}_{ij} \tilde{F}_{ijkk}} f_i (f_j+1)(f_k + 1/2) , 
\end{eqnarray}
which may be contrasted with Eq.\ (\ref{eq:E24_2}). The terms with the ``denom.=0'' restriction (meaning $\omega_i - \omega_j = 0$) originate from $\hat{P}_2^{(\pm0)}$, whereas
those without the restriction come from $\hat{P}_0$. 
In the second equality, we used canonical forms (1), (2), (8), (9), and (10) as well as Eq.\ (\ref{eq:sums}) and 
\begin{eqnarray}
\label{eq:sums_modb}
\sum_{i,j,k} X_{i(jk)} &=&  \sum_{i,j,k}^{\substack{i\neq j, i\neq k \\ j \neq k}} X_{i(jk)} 
+ 2  \sum_{i,j,k}^{\substack{i\neq j, i =  k \\ j \neq k}} X_{i(ji)} 
+  \sum_{i,j,k}^{\substack{i\neq j, i \neq  k \\ j = k}} X_{i(jj)} \nonumber\\
&& +  \sum_{i,j,k}^{\substack{i =  j, i = k \\ j = k}} X_{i(ii)},
\end{eqnarray}
where $X_{i(jk)}$ has the index permutation symmetry of the form $X_{i(jk)} = X_{i(kj)}$.
The first term in the final reduced expression is unlinked, while the second term is linked.

$[\langle N | \hat{V}_4 \hat{P}_0 \hat{V}_2 | N \rangle]+ [\langle N | \hat{V}_4 \hat{P}_2^{(\pm0)} \hat{V}_2 | N \rangle]$ is the same as above.

\subsection{$[\langle N | \hat{V}_4 \hat{P}_0 \hat{V}_4 | N \rangle]+[\langle N | \hat{V}_4 \hat{P}_2^{(\pm0)} \hat{V}_4 | N \rangle] + [\langle N | \hat{V}_4 \hat{P}_4^{(\pm0)} \hat{V}_4 | N \rangle]$}

Capitalizing on Eq.\ (\ref{eq:E44_0}), we write
\begin{widetext}
\begin{eqnarray}
\label{eq:P_E44_0} 
&&\left[\langle N | \hat{V}_4 \hat{P}_0 \hat{V}_4 | N \rangle\right] + \left[ \langle N| \hat{V}_4 \hat{P}_2^{(\pm0)} \hat{V}_4 |N\rangle \right]  + \left[ \langle N| \hat{V}_4 \hat{P}_4^{(\pm0)} \hat{V}_4 |N\rangle \right]
\nonumber\\&&
= \frac{1}{64}\sum_{i,j,k,l}^{\substack{ i\neq j, i\neq k \\ i \neq l, j\neq k \\ j \neq l, k\neq l}} {F_{iijj} F_{kkll} }   \Big[ \langle Q_i^2 \rangle_{0}  \Big] \Big[ \langle Q_j^2 \rangle_{0}  \Big] \Big[ \langle Q_k^2 \rangle_{0} \Big]\Big[\langle Q_l^2 \rangle_{0}\Big]
+ \frac{1}{16}\sum_{i,k,l}^{\substack{ i\neq k, i\neq l \\ k\neq l}} {F_{iikk} F_{iill} }   \Big[ \langle Q_i^2 \rangle_{0} \langle Q_i^2 \rangle_{0}  \Big] \Big[ \langle Q_k^2 \rangle_{0} \Big]\Big[\langle Q_l^2 \rangle_{0}\Big]
\nonumber\\&&
 + \frac{1}{32}\sum_{i,j}^{{ i\neq j}} {F_{iijj} F_{iijj} }   \Big[ \langle Q_i^2 \rangle_{0}  \langle Q_i^2 \rangle_{0}  \Big] \Big[ \langle Q_j^2 \rangle_{0} \langle Q_j^2 \rangle_{0}  \Big] 
+ \frac{1}{48}\sum_{i,l}^{{ i\neq l}} {F_{iiii} F_{iill} }   \Big[ \langle Q_i^2 \rangle_{0} \langle Q_i^4 \rangle_{0}  \Big] \Big[ \langle Q_l^2 \rangle_{0} \Big]
+ \frac{1}{576}\sum_{i} {F_{iiii} F_{iiii} }   \Big[ \langle Q_i^4 \rangle_{0} \langle Q_i^4 \rangle_{0}  \Big] 
\nonumber\\&&
+ \frac{1}{96}\sum_{i,j,k}^{\substack{ i\neq j, i \neq k \\ j \neq k}} {F_{iijj} F_{kkkk} }   \Big[ \langle Q_i^2 \rangle_{0} \Big]\Big[\langle Q_j^2 \rangle_{0}  \Big] \Big[ \langle Q_k^4 \rangle_{0} \Big]
+ \frac{1}{576}\sum_{i,k}^{i\neq k} {F_{iiii} F_{kkkk} }   \Big[ \langle Q_i^4 \rangle_{0} \Big] \Big[ \langle Q_k^4 \rangle_{0}  \Big] 
\nonumber\\&&
 + \frac{1}{18}\sum_{i,j}^{\substack{\text{denom.}= 0 \\ i\neq j}} {F_{ijjj} F_{ijjj}}   \Big[ \langle Q_i \rangle_{1} \langle Q_i \rangle_{-1}  \Big]\Big[ \langle Q_j^3 \rangle_{-1} \langle Q_j^3 \rangle_{1}  \Big]
+\frac{1}{4}\sum_{i,j,k}^{\substack{\text{denom.}= 0 \\ i\neq j, i\neq k \\ j\neq k}} {F_{ijkk} F_{ijkk} }  \Big[ \langle Q_i \rangle_{1} \langle Q_i \rangle_{-1}  \Big]\Big[ \langle Q_j \rangle_{-1} \langle Q_j \rangle_{1} \Big] \Big[ \langle Q_k^2 \rangle_{0} \langle Q_k^2 \rangle_{0}\Big]
\nonumber\\&&
+\frac{1}{4}\sum_{i,j,k,l}^{\substack{\text{denom.}= 0 \\ i\neq j, i\neq k \\ i \neq l, j\neq k \\ j \neq l, k\neq l}} {F_{ijkk} F_{ijll} }   \Big[ \langle Q_i \rangle_{1} \langle Q_i \rangle_{-1}  \Big]\Big[ \langle Q_j \rangle_{-1} \langle Q_j \rangle_{1} \Big] \Big[ \langle Q_k^2 \rangle_{0} \Big]\Big[\langle Q_l^2 \rangle_{0}\Big]
+ \frac{1}{3}\sum_{i,j,l}^{\substack{\text{denom.}= 0 \\ i\neq j, i \neq l \\ j \neq l }} {F_{ijjj} F_{ijll}}  \Big[\langle Q_i \rangle_{1} \langle Q_i \rangle_{-1}  \Big]   \Big[\langle Q_j^3 \rangle_{-1} \langle Q_j \rangle_{1}  \Big] \Big[ \langle Q_l^2 \rangle_{0} \Big] 
\nonumber\\&&
+ \frac{1}{18}\sum_{i,j}^{\substack{\text{denom.}= 0 \\ i\neq j}} {F_{jiii} F_{ijjj}}  \Big[\langle Q_i^3 \rangle_{1} \langle Q_i \rangle_{-1}  \Big]   \Big[\langle Q_j \rangle_{-1} \langle Q_j^3 \rangle_{1}  \Big] 
+ \frac{1}{8}\sum_{i,k,l}^{\substack{\text{denom.}= 0 \\ i\neq k, i\neq l \\ k \neq l }} {F_{iikl} F_{iikl}}
 \Big[\langle Q_i^2 \rangle_{2} \langle Q_i^2 \rangle_{-2}  \Big]   \Big[\langle Q_k \rangle_{-1} \langle Q_k \rangle_{1}  \Big]  \Big[\langle Q_l \rangle_{-1} \langle Q_l \rangle_{1}  \Big] 
\nonumber\\&&
+ \frac{1}{8}\sum_{i,j,k}^{\substack{\text{denom.}= 0 \\ i\neq j, i\neq k \\ j \neq k }} {F_{ijkk} F_{ijkk}}
 \Big[\langle Q_i \rangle_{1} \langle Q_i \rangle_{-1}  \Big]    \Big[\langle Q_j \rangle_{1} \langle Q_j \rangle_{-1}  \Big]  \Big[\langle Q_k^2 \rangle_{-2} \langle Q_k^2 \rangle_{2}  \Big] 
 + \frac{1}{16}\sum_{i,j}^{\substack{\text{denom.}= 0 \\ i\neq j}} {F_{iijj} F_{iijj} }   \Big[ \langle Q_i^2 \rangle_{2}  \langle Q_i^2 \rangle_{-2}  \Big] \Big[ \langle Q_j^2 \rangle_{-2} \langle Q_j^2 \rangle_{2}  \Big] 
\nonumber\\&&
+ \frac{1}{4}\sum_{i,j,k,l}^{\substack{\text{denom.}= 0 \\ i\neq j, i\neq k \\ i \neq l, j \neq k \\ j \neq l, k \neq l}} {F_{ijkl} F_{ijkl}}
 \Big[\langle Q_i \rangle_{1} \langle Q_i \rangle_{-1}  \Big]    \Big[\langle Q_j \rangle_{1} \langle Q_j \rangle_{-1}  \Big]  \Big[\langle Q_k \rangle_{-1} \langle Q_k \rangle_{1}  \Big]  \Big[\langle Q_l \rangle_{-1} \langle Q_l \rangle_{1}  \Big] 
 \nonumber\\
 && =\frac{1}{64}\sum_{i,j,k,l} {F_{iijj} F_{kkll} }   \Big[ \langle Q_i^2 \rangle_{0}  \Big] \Big[ \langle Q_j^2 \rangle_{0}  \Big] \Big[ \langle Q_k^2 \rangle_{0} \Big]\Big[\langle Q_l^2 \rangle_{0}\Big]
% \nonumber\\&&
+ \frac{1}{4}\sum_{i,j,k,l}^{{\text{denom.}= 0 }} {F_{ijkk} F_{ijll} }   \Big[ \langle Q_i \rangle_{1} \langle Q_i \rangle_{-1}  \Big]\Big[ \langle Q_j \rangle_{-1} \langle Q_j \rangle_{1} \Big] \Big[ \langle Q_k^2 \rangle_{0} \Big]\Big[\langle Q_l^2 \rangle_{0}\Big]
\nonumber\\&&
+ \frac{1}{4}\sum_{i,j,k,l}^{{\text{denom.}= 0 }} {F_{ijkl} F_{ijkl}}
 \Big[\langle Q_i \rangle_{1} \langle Q_i \rangle_{-1}  \Big]    \Big[\langle Q_j \rangle_{1} \langle Q_j \rangle_{-1}  \Big]  \Big[\langle Q_k \rangle_{-1} \langle Q_k \rangle_{1}  \Big]  \Big[\langle Q_l \rangle_{-1} \langle Q_l \rangle_{1}  \Big] 
 \nonumber\\
 && =\frac{1}{4} \sum_{i,j,k,l}  {\tilde{F}_{iijj} \tilde{F}_{kkll}}  {(f_i + 1/2)} {(f_j+1/2)} {(f_k+1/2)} {(f_l+1/2)}
 + \sum^{\text{denom.}= 0}_{i,j,k,l}  {\tilde{F}_{ijkk} \tilde{F}_{ijll}}  {f_i} {(f_j+1)} {(f_k+1/2)}{} {(f_l+1/2)}{}
 \nonumber\\&&
 + \frac{1}{4}\sum^{\text{denom.}= 0}_{i,j,k,l}  {\tilde{F}_{ijkl} \tilde{F}_{ijkl} } {f_i}  {f_j}  {(f_k+1)}(f_l+1),
\end{eqnarray}
\end{widetext}
where canonical forms (1), (2), (3), (4), (8), (9), (10), (18), and (20) as well as Eq.\ (\ref{eq:sums2}) were used.
In the last expression, ``denom.=0'' means $\omega_i - \omega_j = 0$ in the second term and $\omega_i + \omega_j - \omega_k - \omega_l = 0$ in the third term.
The first term is unlinked, while the second and third terms are linked. 

Combining all these, we obtain $[ E_N^{(1)} E_N^{(1)} ]$ given in the main text. It consists of both linked and unlinked contributions and is thus written as
\begin{eqnarray}
&& \Big [ E_N^{(1)} E_N^{(1)}  \Big] 
%\nonumber\\&&
= \Big [ E_N^{(1)} E_N^{(1)}  \Big]_L 
+ \sum_{i,j} {\tilde{\bar{F}}_{ii} \tilde{\bar{F}}_{jj}} (f_i +1/2) (f_j + 1/2)
\nonumber\\&&
+\sum_{i,j,k} {\tilde{\bar{F}}_{ii} \tilde{F}_{jjkk}} (f_i+1/2) (f_j+1/2)(f_k + 1/2)
\nonumber\\&&
+\frac{1}{4} \sum_{i,j,k,l}  {\tilde{F}_{iijj} \tilde{F}_{kkll}}  {(f_i + 1/2)} {(f_j+1/2)} {(f_k+1/2)} {(f_l+1/2)}
\nonumber\\&&
= \Big [ E_N^{(1)} E_N^{(1)}  \Big]_L  + \Big[ E_N^{(1)} \Big]_L \Big[ E_N^{(1)} \Big]_L, \label{eq:E1E1linked}
\end{eqnarray}
where $[E_N^{(1)}]_L$ is given by Eq.\ (\ref{eq:Omg1_Ag}).

%============================================
% APPENDIX Normal ordering
%============================================

\section{Normal-ordered second quantization at finite temperature \label{appendix:normalorderedSQ}}

This Appendix gives a derivation of the rules of normal-ordered second quantization for vibrations at finite temperature, including
a proof of thermal Wick's theorem.
The reader is referred to Shavitt and Bartlett\cite{Shavitt2009} for normal ordering for zero-temperature electrons and Wick's theorem, to Hirata and Hermes\cite{Hirata2014} for normal ordering for zero-temperature vibrations, and 
to Hirata\cite{Hirata2021} for normal ordering for finite-temperature electrons. 
Normal ordering for finite-temperature vibrations was also considered by Nooijen and Bao.\cite{Nooijen2021}
 
\subsection{Second quantization for vibrations}
 The second-quantized creation ($\hat{a}_i^{\dagger}$) and annihilation ($\hat{a}_i$) operators (“ladder operators”) for the $i$th normal mode are defined\cite{Hirata2014} by
\begin{eqnarray}
\label{eq:ladder_rule1}
\hat{a}_{i}| n_{i}\rangle&=&n_{i}^{1 / 2}| n_{i}-1 \rangle, \\
\label{eq:ladder_rule2}
\hat{a}_{i}^{\dagger}| n_{i} \rangle&=&\left(n_{i}+1\right)^{1 / 2} | n_{i}+1 \rangle,
\end{eqnarray}
where $| n_i \rangle$ is the harmonic oscillator wave function with quantum number $n_i$. 
They are related to the normal-coordinate operators by
\begin{eqnarray}
Q_{i} &=&\left(2 \omega_{i}\right)^{-1 / 2}\left(\hat{a}_{i}+\hat{a}_{i}^{\dagger}\right), \\
-i \frac{\partial}{\partial Q_{i}} &=&i\left(\frac{\omega_{i}}{2}\right)^{1 / 2}\left(\hat{a}_{i}^{\dagger}-\hat{a}_{i}\right).
\end{eqnarray}
They obey the boson commutation rules, i.e.,
\begin{eqnarray}
\hat{a}_i \hat{a}_j - \hat{a}_j \hat{a}_i &=& 0, \\
\hat{a}_i^\dagger  \hat{a}_j^\dagger - \hat{a}_j^\dagger \hat{a}_i^\dagger  &=& 0, \\
\hat{a}_i \hat{a}_j^\dagger - \hat{a}_j^\dagger \hat{a}_i &=& \delta_{ij}. \label{eq:boson}
\end{eqnarray}

The pure vibrational Hamiltonian can then be expressed\cite{Hirata2014} as 
\begin{eqnarray}
\label{eq:2ndquatHam}
\hat{H}&=& \frac{1}{4} \sum_{i} \omega_{i}\left(-\hat{a}_{i} \hat{a}_{i}+\hat{a}_{i} \hat{a}_{i}^{\dagger}+\hat{a}_{i}^{\dagger} \hat{a}_{i}-\hat{a}_{i}^{\dagger} \hat{a}_{i}^{\dagger}\right) 
\nonumber\\&&
+V_{\mathrm{ref}}
+\sum_{i} \tilde{F}_{i}\left(\hat{a}_{i}+\hat{a}_{i}^{\dagger}\right) 
\nonumber\\&&
+\frac{1}{2!} \sum_{i, j} \tilde{F}_{i j}\left(\hat{a}_{i} \hat{a}_{j}+\hat{a}_{i} \hat{a}_{j}^{\dagger}+\hat{a}_{i}^{\dagger} \hat{a}_{j}+\hat{a}_{i}^{\dagger} \hat{a}_{j}^{\dagger}\right)\nonumber \\
& & +\frac{1}{3 !} \sum_{i, j, k} \tilde{F}_{i j k}\left(\hat{a}_{i} \hat{a}_{j} \hat{a}_{k}+\hat{a}_{i} \hat{a}_{j} \hat{a}_{k}^{\dagger}+\hat{a}_{i} \hat{a}_{j}^{\dagger} \hat{a}_{k}+\hat{a}_{i} \hat{a}_{j}^{\dagger} \hat{a}_{k}^{\dagger}\right.\nonumber\\
& &\left.+\hat{a}_{i}^{\dagger} \hat{a}_{j} \hat{a}_{k}+\hat{a}_{i}^{\dagger} \hat{a}_{j} \hat{a}_{k}^{\dagger}+\hat{a}_{i}^{\dagger} \hat{a}_{j}^{\dagger} \hat{a}_{k}+\hat{a}_{i}^{\dagger} \hat{a}_{j}^{\dagger} \hat{a}_{k}^{\dagger}\right)\nonumber \\
& &+\frac{1}{4 !} \sum_{i, j, k, l} \tilde{F}_{i j k l}\left(\hat{a}_{i} \hat{a}_{j} \hat{a}_{k} \hat{a}_{l}+\hat{a}_{i} \hat{a}_{j} \hat{a}_{k} \hat{a}_{l}^{\dagger}+\hat{a}_{i} \hat{a}_{j} \hat{a}_{k}^{\dagger} \hat{a}_{l}\right.\nonumber \\
& &\left.+\hat{a}_{i} \hat{a}_{j} \hat{a}_{k}^{\dagger} \hat{a}_{l}^{\dagger}+\cdots+\hat{a}_{i}^{\dagger} \hat{a}_{j}^{\dagger} \hat{a}_{k}^{\dagger} \hat{a}_{l}^{\dagger}\right) + \dots,
\end{eqnarray}
where terms that exist in a QFF are shown explicitly.

\subsection{Normal ordering at finite temperature \label{appendix:normalordering}}

Normal-ordered products of creation and annihilation operators at finite temperature are defined by
\begin{eqnarray}
\label{eq:normal}
\{\hat{a}_i\hat{a}_i^{\dagger}\} = \{\hat{a}_i^{\dagger}\hat{a}_i\} &\equiv& (f_i+1)\,\hat{a}_i^{\dagger}\hat{a}_i - f_i\,\hat{a}_i\hat{a}_i^{\dagger} \nonumber\\
&=& \hat{a}_i^{\dagger}\hat{a}_i - f_i,
\end{eqnarray}
where $f_i$ is the Bose--Einstein distribution function [Eq.\ (\ref{eq:BEfunction})]. 
When $i \neq j$,
\begin{eqnarray}
\{\hat{a}_i^{\dagger} \hat{a}_j\} &\equiv& \hat{a}_i^{\dagger} \hat{a}_j,\\
 \{\hat{a}_i \hat{a}_j^{\dagger}\} &\equiv& \hat{a}_i \hat{a}_j^{\dagger}.
\end{eqnarray}
Two annihilation or two creation operators are also unchanged by normal ordering, hence, 
\begin{eqnarray}
\{\hat{a}_i \hat{a}_j\} &\equiv& \hat{a}_i \hat{a}_j,\\
 \{\hat{a}_i^\dagger \hat{a}_j^{\dagger}\} &\equiv& \hat{a}_i^\dagger \hat{a}_j^{\dagger},
\end{eqnarray}
for both $i = j$ and $i \neq j$. As $T \to 0$ ($f_i \to 0$), the above reduces to the zero-temperature normal ordering.\cite{Hirata2014}

These  orders are defined such that a thermal average of a normal-ordered product of operators is always zero. For instance,
\begin{eqnarray}
\label{eq:normal_confirm}
 \left[\langle N | \{\hat{a}_i^{\dagger} \hat{a}_i\} | N \rangle \right] 
%&=& \left[\langle N | \{\hat{a}_i^{\dagger} \hat{a}_i\} | N \rangle \right] \nonumber\\
&=&  \left[ \langle N| \hat{a}_i^{\dagger}\hat{a}_i| N \rangle\right] - \left[ \langle N| f_i | N \rangle\right] \nonumber\\
&=&  \Big[ n_i \Big] - f_i  \nonumber\\
%&=&  f_i -  f_i  \nonumber\\
&=& 0,
\end{eqnarray}
where Eqs.\ (\ref{eq:ladder_rule1}) and (\ref{eq:ladder_rule2}) were used in the penultimate equality, and Eq.\ (\ref{eq:sum_I}) in the last equality. 

Normal ordering can be extended to a product of three operators. For instance,
\begin{eqnarray}
\label{eq:normal3}
\{\hat{a}_i^{\dagger} \hat{a}_i \hat{a}_j \} &=& \{\hat{a}_i^{\dagger} \hat{a}_j \hat{a}_i \} =  \{\hat{a}_i \hat{a}_i^{\dagger}  \hat{a}_j \} 
\nonumber\\ &=&  \{\hat{a}_i  \hat{a}_j \hat{a}_i^{\dagger} \} =  \{\hat{a}_j  \hat{a}_i^{\dagger} \hat{a}_i \} =  \{\hat{a}_j  \hat{a}_i \hat{a}_i^{\dagger} \} \nonumber\\
&\equiv& \{\hat{a}_i^{\dagger} \hat{a}_i \} \{ \hat{a}_j \} \nonumber\\
&=& \left( \hat{a}_i^{\dagger}\hat{a}_i  - f_i \right) \hat{a}_j ,
\end{eqnarray}
for $i\neq j$. The first two lines indicate that the original order is immaterial, and the third line uses the fact that operators of different modes commute and
are separately normal ordered.  
For $i=j$, we instead have
\begin{eqnarray}
\label{eq:normal3_2}
\{\hat{a}_i^{\dagger} \hat{a}_i \hat{a}_i \} &\equiv& 
\hat{a}_i^{\dagger}\hat{a}_i \hat{a}_i - 2 f_i\, \hat{a}_i.
\end{eqnarray}
The logic behind these definitions is given in Appendix \ref{appendix:Wicktheorem}.

For a product of four operators, we stipulate, e.g., 
\begin{eqnarray}
\label{eq:normal4}
    %\{\hat{a}_{i}^{\dagger} \hat{a}_{j}^{\dagger}\hat{a}_{i} \hat{a}_{j} \} &=&  \{\hat{a}_{i}^{\dagger} \hat{a}_{j}^{\dagger}\hat{a}_{j} \hat{a}_{i} \} = 
    \{\hat{a}_{i}^{\dagger} \hat{a}_{i}\hat{a}_{j}^{\dagger} \hat{a}_{j} \} &=& \{\hat{a}_{i}^{\dagger} \hat{a}_{i} \hat{a}_{j}\hat{a}_{j}^{\dagger} \} = \{\hat{a}_{i}^{\dagger} \hat{a}_{j}^{\dagger}  \hat{a}_{i} \hat{a}_{j}\} = \dots \nonumber\\
    & \equiv & \{\hat{a}_{i}^{\dagger} \hat{a}_{i}\} \{ \hat{a}_{j}^{\dagger} \hat{a}_{j} \} \nonumber\\
    &=&  \left( \hat{a}_i^{\dagger} \hat{a}_i - f_i\right) \left( \hat{a}_j^{\dagger} \hat{a}_j - f_j\right) \nonumber\\
    &=&     \hat{a}_i^{\dagger} \hat{a}_j^{\dagger}  \hat{a}_i  \hat{a}_j - f_j\, \hat{a}_i^{\dagger}  \hat{a}_i - f_i\, \hat{a}_j^{\dagger}  \hat{a}_j + f_i f_j, 
\end{eqnarray}
for $i\neq j$. For $i=j$, we instead postulate
\begin{eqnarray}
\label{eq:normal4_2}
    %\{\hat{a}_{i}^{\dagger} \hat{a}_{j}^{\dagger}\hat{a}_{i} \hat{a}_{j} \} &=&  \{\hat{a}_{i}^{\dagger} \hat{a}_{j}^{\dagger}\hat{a}_{j} \hat{a}_{i} \} = 
    \{\hat{a}_{i}^{\dagger} \hat{a}_{i}\hat{a}_{i}^{\dagger} \hat{a}_{i} \}  &\equiv& \hat{a}_i^{\dagger} \hat{a}_i^{\dagger}  \hat{a}_i  \hat{a}_i- 4 f_i\, \hat{a}_i^{\dagger}  \hat{a}_i  + 2 f_i^2.
\end{eqnarray}
See Appendix \ref{appendix:Wicktheorem} for a justification.

These longer normal-ordered products are also defined in such a way that their thermal averages are guaranteed to vanish. 
Since there are an odd number of operators in Eqs.\ (\ref{eq:normal3}) and (\ref{eq:normal3_2}), their thermal averages are trivially zero. 
For Eq.\ (\ref{eq:normal4}) with $i \neq j$, we find
\begin{eqnarray}
\label{eq:normal4_confirm}
  \left[ \langle N |  \{\hat{a}_{i}^{\dagger} \hat{a}_{i}\hat{a}_{j}^{\dagger} \hat{a}_{j} \} | N \rangle \right]  
&=& \Big[ n_i \Big] \Big[ n_j \Big]
%\nonumber\\&&
    -  f_j \Big[ n_i \Big]  
%    \nonumber\\&&
    -  f_i  \Big[ n_j \Big] 
 %  \nonumber\\&&
    +f_if_j
    \nonumber\\&=&  0,
\end{eqnarray}
where Eq.\ (\ref{eq:sum_I}) was used. We can also confirm that a thermal average of Eq.\ (\ref{eq:normal4_2}) is zero:
\begin{eqnarray}
\label{eq:normal4_2_confirm}
  \left[ \langle N | \{\hat{a}_{i}^{\dagger} \hat{a}_{i}\hat{a}_{i}^{\dagger} \hat{a}_{i} \} | N \rangle \right]  
&=&  \Big[ n_i (n_i -1) \Big]
%\nonumber\\&&
    -4 f_i \Big[ n_i \Big] 
%\nonumber\\&&
+ 2 f_i^2
%\nonumber\\&&
\nonumber\\ &=& 0,
\end{eqnarray}
where Eqs.\ (\ref{eq:sum_I}) and (\ref{eq:sum_II}) were used.

\subsection{Thermal Wick's theorem\label{appendix:Wicktheorem}}

A Wick (binary) contraction is defined as the difference between the normal-ordered and original products of operators. It is designated by a staple symbol. 
\begin{eqnarray}
\label{eq:aidaggerai}
\contraction{}{\hat{a}}{{}_i^{\dagger}}{\hat{a}}\hat{a}_i^{\dagger} \hat{a}_i &\equiv& \hat{a}_i^{\dagger} \hat{a}_i - \{\hat{a}_i^{\dagger} \hat{a}_i\} 
%= f_i \left( \hat{a}_i \hat{a}_i^{\dagger} -\hat{a}_i^{\dagger}\hat{a}_i \right) 
= f_i, \\
\label{eq:aiaidagger}
\contraction{}{\hat{a}}{{}_i}{\hat{a}}\hat{a}_i \hat{a}_i^{\dagger} &\equiv& \hat{a}_i \hat{a}_i^{\dagger} - \{\hat{a}_i \hat{a}_i^{\dagger}\} %=(f_i+1)\left(\hat{a}_i \hat{a}_i^{\dagger} -\hat{a}_i^{\dagger}\hat{a}_i\right) 
= f_i + 1,
\end{eqnarray}
where we used Eq.\ (\ref{eq:normal}) and the boson commutation rule [Eq.\ (\ref{eq:boson})]. All other binary contractions are zero.

{\it Thermal Wick's theorem.}\cite{Shavitt2009,Hirata2014,Hirata2021} A product of creation and annihilation operators is the sum of its normal-ordered product plus all its partial and full contractions.

See Shavitt and Bartlett\cite{Shavitt2009} for the definition of partial and full contractions for zero-temperature electrons and Ref.\ \onlinecite{Hirata2021} for finite-temperature electrons. 
The only major difference from these electronic cases is that there is no parity involved in vibrations. See also Shavitt and Bartlett\cite{Shavitt2009}
for a proof of the theorem for zero-temperature electrons, which is rather different from the proof we provide below for finite-temperature vibrations.

For a product of two operators, the theorem is just a restatement of the definitions of Wick contractions:
\begin{eqnarray}
\hat{a}_i^{\dagger} \hat{a}_j &=& \{\hat{a}_i^{\dagger} \hat{a}_j\} + \contraction{}{\hat{a}}{{}_i^{\dagger}}{\hat{a}} \hat{a}_i^{\dagger} \hat{a}_j, \\
\hat{a}_i \hat{a}_j^{\dagger} &=& \{\hat{a}_i \hat{a}_j^{\dagger}\} + \contraction{}{\hat{a}}{{}_i}{\hat{a}} \hat{a}_i \hat{a}_j^{\dagger} .
\end{eqnarray}
For a product of three operators, the theorem asserts
\begin{eqnarray}
\hat{a}_{i}^{\dagger} \hat{a}_{j} \hat{a}_{k} = \{\hat{a}_{i}^{\dagger} \hat{a}_{j} \hat{a}_{k}\} + \{\contraction{}{\hat{a}}{{}_{i}^{\dagger}}{\hat{a}} \hat{a}_{i}^{\dagger} \hat{a}_{j} \hat{a}_{k}\} +
\{\contraction{}{\hat{a}}{{}_{i}^{\dagger}\hat{a}_{j}}{\hat{a}} \hat{a}_{i}^{\dagger} \hat{a}_{j} \hat{a}_{k}\},
\end{eqnarray}
where only potentially nonzero partial contractions are shown.
Likewise, for a product of four operators,
\begin{eqnarray}
\hat{a}_{i} \hat{a}_{j} \hat{a}_{k}^{\dagger} \hat{a}_{l}^{\dagger} &=& \{\hat{a}_{i} \hat{a}_{j} \hat{a}_{k}^{\dagger} \hat{a}_{l}^{\dagger}\} + \{\contraction{}{\hat{a}}{_{i} \hat{a}_{j} \hat{a}_{k}^{\dagger}}{\hat{a}}\hat{a}_{i} \hat{a}_{j} \hat{a}_{k}^{\dagger} \hat{a}_{l}^{\dagger}\} + 
\{\hat{a}_{i} \contraction{}{\hat{a}}{_{j} \hat{a}_{k}^{\dagger}}{\hat{a}}\hat{a}_{j} \hat{a}_{k}^{\dagger} \hat{a}_{l}^{\dagger}\} + 
\{\hat{a}_{i} \contraction{}{\hat{a}}{_{j}}{\hat{a}}\hat{a}_{j} \hat{a}_{k}^{\dagger} \hat{a}_{l}^{\dagger}\} \nonumber\\
& &+ 
\{\contraction{}{\hat{a}}{_{i} \hat{a}_{j}}{\hat{a}}\hat{a}_{i} \hat{a}_{j} \hat{a}_{k}^{\dagger} \hat{a}_{l}^{\dagger}\} + \{\contraction{}{\hat{a}}{_{i} \hat{a}_{j} \hat{a}_{k}^{\dagger}}{\hat{a}} \hat{a}_{i} \contraction[0.5ex]{}{\hat{a}}{_{j}}{\hat{a}}\hat{a}_{j} \hat{a}_{k}^{\dagger} \hat{a}_{l}^{\dagger}\} + \{\contraction{}{\hat{a}}{_{i} \hat{a}_{j}}{\hat{a}}\hat{a}_{i} \contraction[0.5ex]{}{\hat{a}}{_{j} \hat{a}_{k}^{\dagger}}{\hat{a}}\hat{a}_{j} \hat{a}_{k}^{\dagger} \hat{a}_{l}^{\dagger}\}.
\end{eqnarray}
The first term and its partial contractions (the second through fifth terms) are normal-ordered operators (multiplied by the value of the binary contraction). All of these 
will vanish when a thermal average is taken. The last two terms (the full contractions) are real 
numbers (a product of the values of the two binary contractions) and will be the only ones that survive upon thermal averaging.

{\it Proof.} The definitions of normal orders for more than two operators given in Appendix \ref{appendix:normalordering} were deduced from thermal Wick's theorem:\ The normal-ordered product
was defined as the difference between the original product and all of its partial and full contractions. Here, we prove that the normal ordering thus defined gives a zero thermal average, i.e., the converse
of the theorem. Since normal ordering is not unique and its usefulness comes from its ability to make its thermal average zero, this
constitutes a rigorous proof of thermal Wick's theorem. 

Recognize that a thermal average is zero unless there are an equal number of creation and annihilation operators for each mode in a normal-ordered product. 
Therefore, only nontrivial normal-ordered product that needs to be considered consists of an even number of operators per each mode such as
\begin{eqnarray}
&& \{ (\hat{a}_{i}^\dagger\hat{a}_{i}^\dagger\hat{a}_{i}^\dagger\hat{a}_{i}\hat{a}_{i}\hat{a}_{i} )( \hat{a}_{j}^\dagger\hat{a}_{j}^\dagger\hat{a}_{j}\hat{a}_{j} )( \hat{a}_{k}^\dagger\hat{a}_{k} )\} \nonumber\\
&&=\{ \hat{a}_{i}^\dagger\hat{a}_{i}^\dagger\hat{a}_{i}^\dagger\hat{a}_{i}\hat{a}_{i}\hat{a}_{i} \}\{ \hat{a}_{j}^\dagger\hat{a}_{j}^\dagger\hat{a}_{j}\hat{a}_{j} \}\{ \hat{a}_{k}^\dagger\hat{a}_{k} \} ,
\end{eqnarray}
where we used the fact that the normal ordering of the whole product is just the
product of normal-ordered operators of separate modes. 
See Eqs.\ (\ref{eq:normal3}) and (\ref{eq:normal4}) for examples of such constructions. Therefore, we only need to prove that a normal-ordered product of an even number of operators for a single mode
gives a zero thermal average. Such normal-ordered products are 
\begin{eqnarray}
&& \{ \hat{a}_{i}^\dagger\hat{a}_{i} \}, \\
&& \{ \hat{a}_{i}^\dagger\hat{a}_{i}^\dagger\hat{a}_{i}\hat{a}_{i} \}, \\
&& \{ \hat{a}_{i}^\dagger\hat{a}_{i}^\dagger\hat{a}_{i}^\dagger\hat{a}_{i}\hat{a}_{i}\hat{a}_{i} \},
\end{eqnarray}
etc. 

As per thermal Wick's theorem, the normal ordering of $2n$ one-mode operators is defined by
\begin{eqnarray}
\label{eq:induction}
 \{ (\hat{a}_{i}^\dagger)^n(\hat{a}_{i})^n \} &=& (\hat{a}_{i}^\dagger)^n(\hat{a}_{i})^n - n^2\, \contraction{}{\hat{a}}{_i^\dagger }{\hat{a}} \hat{a}_i^\dagger \hat{a}_i \{ (\hat{a}_{i}^\dagger)^{n-1}(\hat{a}_{i})^{n-1} \} \nonumber\\
%&& - \contraction{}{\hat{a}}{_i^\dagger }{\hat{a}} \hat{a}_i^\dagger \hat{a}_i \contraction{}{\hat{a}}{_i^\dagger }{\hat{a}} \hat{a}_i^\dagger \hat{a}_i  \{ (\hat{a}_{i}^\dagger)^{n-2}(\hat{a}_{i})^{n-2} \} \nonumber\\
%&& - \contraction{}{\hat{a}}{_i^\dagger }{\hat{a}} \hat{a}_i^\dagger \hat{a}_i \contraction{}{\hat{a}}{_i^\dagger }{\hat{a}} \hat{a}_i^\dagger \hat{a}_i \contraction{}{\hat{a}}{_i^\dagger }{\hat{a}} \hat{a}_i^\dagger \hat{a}_i  \{ (\hat{a}_{i}^\dagger)^{n-3}(\hat{a}_{i})^{n-3} \} \nonumber\\
&& - \dots - n! \,\Big( \contraction{}{\hat{a}}{_i^\dagger }{\hat{a}} \hat{a}_i^\dagger \hat{a}_i \Big)^n,
\end{eqnarray}
where the first term is the original order, the second  term is a partial contraction (there are $n^2$ distinct ways of contracting a pair of $\hat{a}_i^\dagger$ and $\hat{a}_i$), and the last term is the full contraction (there are $n!$ distinct ways of contracting $n$ pairs of $\hat{a}_i^\dagger$ and $\hat{a}_i$). 
Our goal is to show that the thermal average of this normal-ordered product is zero. 
It should be recalled that the original order of operators in the left-hand side is immaterial. The order of operators in the right-hand side is also arbitrary, and different choices lead to different normal-ordering 
definitions, but they are mathematically equal to one another. The above choice is the most convenient for the following proof.

We use mathematical induction: The thermal average of the normal-ordered product as defined by Eq.\ (\ref{eq:induction}) is zero for $n=1$ as per the definition of Wick contractions [Eq.\ (\ref{eq:normal_confirm})].
We will show that if the thermal average is zero for $n = k-1$ then the same is true for $n = k$. 

Taking a thermal average of Eq.\ (\ref{eq:induction}), we see that only the first and last terms in the right-hand side persist, i.e.,
\begin{eqnarray}
\label{eq:induction2}
\Big[  \langle N | \{ (\hat{a}_{i}^\dagger)^{k}(\hat{a}_{i})^{k} \} |N \rangle \Big] &=& \Big[ \langle N | (\hat{a}_{i}^\dagger)^{k}(\hat{a}_{i})^{k} |N\rangle \Big] - k! \Big( \contraction{}{\hat{a}}{_i^\dagger }{\hat{a}} \hat{a}_i^\dagger \hat{a}_i \Big)^{k} \nonumber\\
&=& \Big[ n_i (n_i -1) (n_i - 2) \dots (n_i - k+1) \Big] \nonumber\\
&& - k!   f_i ^{k}, 
\end{eqnarray}
because the thermal averages of partial contractions, which are by themselves normal-ordered operators of length $n = k-1$ or shorter, vanish as per the assumption of this induction proof. 
Equations (\ref{eq:ladder_rule1}) and (\ref{eq:ladder_rule2}) as well as Eq.\ (\ref{eq:aidaggerai}) were used in the last equality.

We can furthermore evaluate $[ n_i (n_i -1) (n_i - 2) \dots (n_i - k+1) ]$ by a recursive method analogous to the one in Appendix \ref{Appendix:BHrules}. We write
\begin{eqnarray}
z_i^{(0)} &=& \sum_{n_i=0}^\infty \exp(-\beta n_i \omega_i) = f_i + 1, \\
z_i^{(1)} &=& \sum_{n_i=0}^\infty n_i \exp(-\beta n_i \omega_i) \nonumber\\
&=& \exp(- 0\, \beta \omega_i) \frac{\partial }{\partial (-\beta\omega_i)}   z_i^{(0)} \exp(0\, \beta \omega_i) \nonumber\\
&=&  f_i (f_i + 1), \\
z_i^{(2)} &=& \sum_{n_i=0}^\infty n_i(n_i -1)  \exp(-\beta n_i \omega_i) \nonumber\\
&=& \exp(- 1\, \beta \omega_i) \frac{\partial }{\partial (-\beta\omega_i)}   z_i^{(1)} \exp(1\, \beta \omega_i) \nonumber\\
&=& 2!\,f_i ^2 (f_i + 1), \\
z_i^{(3)} &=& \sum_{n_i=0}^\infty n_i(n_i -1)(n_i-2)  \exp(-\beta n_i \omega_i) \nonumber\\
&=& \exp(- 2\, \beta \omega_i) \frac{\partial }{\partial (-\beta\omega_i)}   z_i^{(2)} \exp(2\, \beta \omega_i) \nonumber\\
&=& 3!\, f_i ^3 (f_i + 1), 
\end{eqnarray}
and so on, where $f_i \exp(\beta\omega_i) = f_i + 1$ and Eq.\ (\ref{eq:deriv_f}) were used. Generally, 
\begin{eqnarray}
z_i^{(k)} &=& k! f_i^k (f_i+1).
\end{eqnarray}
Finally, substituting 
\begin{eqnarray}
\Big[ n_i (n_i -1) (n_i - 2) \dots (n_i - k+1) \Big] = \frac{z_i^{(k)} }{z_i^{(0)}} = k!  f_i  ^k,
\end{eqnarray}
into Eq.\ (\ref{eq:induction2}), we prove
\begin{eqnarray}
\label{eq:induction3}
\Big[ \langle N |  \{ (\hat{a}_{i}^\dagger)^{k}(\hat{a}_{i})^{k} \} |N \rangle \Big] &=& 0.
\end{eqnarray}
This completes the induction proof, establishing that the thermal average of a normal-ordered product as defined by thermal Wick's theorem is zero.

\subsection{Hamiltonian \label{appendix:normalorderedHam}}

Here, using thermal Wick's theorem, we will bring the pure vibrational Hamiltonian [Eq.\ (\ref{eq:2ndquatHam})] in a normal-ordered form. 

The first term of Eq.\ (\ref{eq:2ndquatHam}) is the kinetic-energy operator, which is normal ordered as
\begin{eqnarray}
\label{eq:kineticSQ}
&& \frac{1}{4} \sum_{i} \omega_{i}\left(-\hat{a}_{i} \hat{a}_{i}+\hat{a}_{i} \hat{a}_{i}^{\dagger}+\hat{a}_{i}^{\dagger} \hat{a}_{i}-\hat{a}_{i}^{\dagger} \hat{a}_{i}^{\dagger}\right) \nonumber\\
&&=  - \frac{1}{4} \sum_{i} \omega_{i}\{\hat{a}_{i}\hat{a}_{i}\}+\frac{1}{4} \sum_{i} \omega_{i} \left(\{ \hat{a}_{i}\hat{a}_{i}^{\dagger}\} + \{\contraction{}{\hat{a}}{_{i}}{\hat{a}}\hat{a}_{i}\hat{a}_{i}^{\dagger}\} \right)\nonumber\\
& & +\frac{1}{4} \sum_{i} \omega_{i} \left(\{\hat{a}_{i}^{\dagger} \hat{a}_{i}\} + \{\contraction{}{\hat{a}}{_{i}^{\dagger}}{\hat{a}}\hat{a}_{i}^{\dagger} \hat{a}_{i}\}\right)-\frac{1}{4} \sum_{i} \omega_{i} \{\hat{a}_{i}^{\dagger} \hat{a}_{i}^{\dagger}\} \nonumber\\
&&= -\frac{1}{4} \sum_i \omega_i \{\hat{a}_i \hat{a}_i\} + \frac{1}{2} \sum_i \omega_i  \{\hat{a}_i \hat{a}_i^{\dagger}\} - \frac{1}{4} \sum_i \omega_i\{\hat{a}_i^{\dagger} \hat{a}_i^{\dagger} \}  \nonumber\\
&  & + \frac{1}{2} \sum_i \omega_i (f_i + 1/2).
\end{eqnarray}
The linear- and quadratic-force-constant terms of Eq.\ (\ref{eq:2ndquatHam}) are rewritten as
\begin{eqnarray}
\label{eq:FiSQ}
\sum_i \tilde{F}_{i}\left(\hat{a}_{i}+\hat{a}_{i}^{\dagger}\right) &=& \sum_i \tilde{F}_{i}\{\hat{a}_{i}\}+ \sum_i \tilde{F}_i \{\hat{a}_{i}^{\dagger}\},
\end{eqnarray}
and
\begin{eqnarray}
\label{eq:FijSQ}
&& \frac{1}{2} \sum_{i, j} \tilde{F}_{i j}\left(\hat{a}_{i} \hat{a}_{j}+\hat{a}_{i} \hat{a}_{j}^{\dagger}+\hat{a}_{i}^{\dagger} \hat{a}_{j}+\hat{a}_{i}^{\dagger} \hat{a}_{j}^{\dagger}\right) \nonumber\\
&&= \frac{1}{2} \sum_{i, j} \tilde{F}_{i j}\{ \hat{a}_{i} \hat{a}_{j} \}  + \frac{1}{2} \sum_{i, j} \tilde{F}_{i j} \left( \{ \hat{a}_{i} \hat{a}_{j}^{\dagger} \}  + \{\contraction{}{\hat{a}}{_{i}}{\hat{a}}\hat{a}_{i} \hat{a}_{j}^{\dagger}\} \right)  \nonumber\\
& & + \frac{1}{2} \sum_{i, j} \tilde{F}_{i j} \left( \{ \hat{a}_{i}^{\dagger} \hat{a}_{j} \}  + \{\contraction{}{\hat{a}}{_{i}^{\dagger}}{\hat{a}}\hat{a}_{i}^{\dagger} \hat{a}_{j}\} \right) 
+ \frac{1}{2} \sum_{i, j} \tilde{F}_{i j}  \{\hat{a}_{i}^{\dagger} \hat{a}_{j}^{\dagger}\} \nonumber\\
&&=  \frac{1}{2} \sum_{i,j} \tilde{F}_{ij} \{\hat{a}_{i} \hat{a}_{j}\} + \sum_{i,j} \tilde{F}_{ij}\{\hat{a}_{i} \hat{a}_{j}^{\dagger}\}    +\frac{1}{2} \sum_{i,j} \tilde{F}_{ij} \{ \hat{a}_{i}^{\dagger} \hat{a}_{j}^{\dagger} \} \nonumber\\ 
& & +  \sum_{i} \tilde{F}_{ii} (f_i+1/2).
\end{eqnarray}
The same, but  lengthier process leads to the normal-ordered form of the cubic- and quartic-force-constant terms, which read
\begin{eqnarray}
\label{eq:fc3_contraction}
&& \frac{1}{3 !} \sum_{i, j, k} \tilde{F}_{i j k}\left(\hat{a}_{i} \hat{a}_{j} \hat{a}_{k}+\hat{a}_{i} \hat{a}_{j} \hat{a}_{k}^{\dagger}+\hat{a}_{i} \hat{a}_{j}^{\dagger} \hat{a}_{k}+\hat{a}_{i} \hat{a}_{j}^{\dagger} \hat{a}_{k}^{\dagger}\right.\nonumber\\
& &\left.+\hat{a}_{i}^{\dagger} \hat{a}_{j} \hat{a}_{k}+\hat{a}_{i}^{\dagger} \hat{a}_{j} \hat{a}_{k}^{\dagger}+\hat{a}_{i}^{\dagger} \hat{a}_{j}^{\dagger} \hat{a}_{k}+\hat{a}_{i}^{\dagger} \hat{a}_{j}^{\dagger} \hat{a}_{k}^{\dagger}\right)\nonumber \\
&&=   \sum_{i,j} \tilde{F}_{ijj} (f_j+1/2) \{\hat{a}_i\}   +  \sum_{i,j} \tilde{F}_{ijj} (f_j+1/2)  \{\hat{a}_i^{\dagger}\} \nonumber\\
&& + \frac{1}{3!} \sum_{i, j, k} \tilde{F}_{i j k} \{\hat{a}_{i}\hat{a}_j \hat{a}_k\}   + \frac{1}{2} \sum_{i, j, k} \tilde{F}_{i j k}\{\hat{a}_{i} \hat{a}_{j} \hat{a}_{k}^{\dagger}\} \nonumber\\
& & + \frac{1}{2} \sum_{i, j, k} \tilde{F}_{i j k} \{\hat{a}_{i} \hat{a}_{j}^{\dagger} \hat{a}_{k}^{\dagger}\} + \frac{1}{3!} \sum_{i, j, k} \tilde{F}_{i j k} \{ \hat{a}_{i}^{\dagger}\hat{a}_j^{\dagger} \hat{a}_k^{\dagger}\}, 
\end{eqnarray}
and
\begin{eqnarray}
\label{eq:fc4_contraction}
& &\frac{1}{4 !} \sum_{i, j, k, l} \tilde{F}_{i j k l}\left(\hat{a}_{i} \hat{a}_{j} \hat{a}_{k} \hat{a}_{l}+\hat{a}_{i} \hat{a}_{j} \hat{a}_{k} \hat{a}_{l}^{\dagger}+\hat{a}_{i} \hat{a}_{j} \hat{a}_{k}^{\dagger} \hat{a}_{l}\right.\nonumber \\
& &\left.+\hat{a}_{i} \hat{a}_{j} \hat{a}_{k}^{\dagger} \hat{a}_{l}^{\dagger}+\cdots+\hat{a}_{i}^{\dagger} \hat{a}_{j}^{\dagger} \hat{a}_{k}^{\dagger} \hat{a}_{l}^{\dagger}\right) \nonumber\\
&&= \frac{1}{4!} \sum_{i,j,k,l} \tilde{F}_{ijkl}\{\hat{a}_{i} \hat{a}_{j} \hat{a}_{k} \hat{a}_{l}\}   + \frac{1}{3!} \sum_{i,j,k,l}\tilde{F}_{ijkl} \{\hat{a}_{i} \hat{a}_{j} \hat{a}_{k} \hat{a}_{l}^{\dagger}\} \nonumber\\
&& + \frac{1}{2!2!} \sum_{i,j,k,l} \tilde{F}_{ijkl} \{\hat{a}_{i} \hat{a}_{j} \hat{a}_{k}^{\dagger} \hat{a}_{l}^{\dagger} \}  +  \frac{1}{3!} \sum_{i,j,k,l}\tilde{F}_{ijkl} \{\hat{a}_{i}^{\dagger} \hat{a}_{j}^{\dagger} \hat{a}_{k}^{\dagger} \hat{a}_{l}\} \nonumber\\
& & + \frac{1}{4!} \sum_{i,j,k,l} \tilde{F}_{ijkl} \{\hat{a}_{i}^{\dagger} \hat{a}_{j}^{\dagger} \hat{a}_{k}^{\dagger} \hat{a}_{l}^{\dagger}\}  
 + \frac{1}{2}\sum_{i,j,k} \tilde{F}_{ijkk} (f_k + 1/2)  \{\hat{a}_{i} \hat{a}_{j}\} \nonumber\\
& &
 + \sum_{i,j,k} \tilde{F}_{ijkk} (f_k + 1/2)  \{\hat{a}_i \hat{a}_j^{\dagger} \} + \frac{1}{2}\sum_{i,j,k} \tilde{F}_{ijkk} (f_k + 1/2)  \{\hat{a}_i^{\dagger} \hat{a}_j^{\dagger} \} \nonumber\\
&& 
+ \frac{1}{2} \sum_{i,j}\tilde{F}_{iijj} (f_i+1/2)(f_j+1/2).
\end{eqnarray}
Higher-order potential operators can  be systematically normal ordered analogously. 

Together, we obtain the Hamiltonian in a finite-temperature normal-ordered form as
\begin{eqnarray}
\label{eq:normalorderedHam}
\hat{H} & = & E_{\mathrm{XVSCF}}(T) + \sum_i W_i \{\hat{a}_i\}  
+ \sum_i W_i \{\hat{a}_i^{\dagger}\} + \frac{1}{2!}\sum_{i,j} W_{ij}  \{\hat{a}_i \hat{a}_j\} \nonumber\\
& &+\frac{1}{2!}\sum_{i,j} W_{ij} \{\hat{a}_i^{\dagger}\hat{a}_j^{\dagger}\}   
+  \sum_{i,j} (W_{ij}+\delta_{ij} \omega_i) \{\hat{a}_i\hat{a}_j^{\dagger}\}  \nonumber\\
& & + \frac{1}{3!} \sum_{i, j, k} W_{i j k} \{\hat{a}_{i}\hat{a}_j \hat{a}_k\} 
+  \frac{1}{3!} \sum_{i, j, k} W_{i j k} \{ \hat{a}_{i}^{\dagger}\hat{a}_j^{\dagger} \hat{a}_k^{\dagger}\}  \nonumber\\
& &+ \frac{1}{2!} \sum_{i, j, k} W_{i j k} \{\hat{a}_{i} \hat{a}_{j} \hat{a}_{k}^{\dagger}\} 
+\frac{1}{2!} \sum_{i, j, k} W_{i j k}  \{\hat{a}_{i}^{\dagger} \hat{a}_{j}^{\dagger} \hat{a}_{k}\} \nonumber\\
& & +\frac{1}{4!} \sum_{i,j,k,l} W_{ijkl}  \{\hat{a}_{i} \hat{a}_{j} \hat{a}_{k} \hat{a}_{l}\} 
+ \frac{1}{4!} \sum_{i,j,k,l} W_{ijkl} \{\hat{a}_{i}^{\dagger} \hat{a}_{j}^{\dagger} \hat{a}_{k}^{\dagger} \hat{a}_{l}^{\dagger}\} \nonumber\\
& &
+ \frac{1}{3!} \sum_{i,j,k,l}W_{ijkl} \{\hat{a}_{i} \hat{a}_{j} \hat{a}_{k} \hat{a}_{l}^{\dagger}\} 
 + \frac{1}{3!} \sum_{i,j,k,l}W_{ijkl} \{\hat{a}_{i}^{\dagger} \hat{a}_{j}^{\dagger} \hat{a}_{k}^{\dagger} \hat{a}_{l}\}  \nonumber\\
& &
+ \frac{1}{2!2!} \sum_{i,j,k,l}W_{ijkl} \{\hat{a}_{i} \hat{a}_{j} \hat{a}_{k}^{\dagger} \hat{a}_{l}^{\dagger} \}  + \dots,
\end{eqnarray}
where
\begin{eqnarray}
\label{eq:EXVSCFT}
 E_{\mathrm{XVSCF}}(T) & = & V_{\mathrm{ref}} + \frac{1}{2} \sum_i \omega_i (f_i+1/2) + \sum_i \tilde{F}_{ii}(f_i+1/2) \nonumber\\ 
& & + \frac{1}{2!}\sum_{i,j} \tilde{F}_{iijj}(f_i+1/2)(f_j+1/2) \nonumber\\
&& + \frac{1}{3!}\sum_{i,j,k} \tilde{F}_{iijjkk}(f_i+1/2)(f_j+1/2)(f_k+1/2) \nonumber\\ 
&&+ \dots, 
\end{eqnarray}
and
\begin{eqnarray}
\label{eq:Wi_appendix}
W_i & = & \tilde{F}_i +  \sum_j \tilde{F}_{ijj} (f_j+1/2) \nonumber\\
&& + \frac{1}{2!} \sum_{j,k} \tilde{F}_{ijjkk}(f_j+1/2)(f_k+1/2) \nonumber\\
&& + \frac{1}{3!} \sum_{j,k,l} \tilde{F}_{ijjkkll}(f_j+1/2)(f_k+1/2)(f_l+1/2) \nonumber\\
&& +\dots,  
\\
\label{eq:Wij_appendix}
W_{ij} & = &  \tilde{\bar{F}}_{ij} +  \sum_k \tilde{F}_{ijkk}(f_k+1/2) \nonumber\\
&& + \frac{1}{2!} \sum_{k,l} \tilde{F}_{ijkkll}(f_k+1/2)(f_l+1/2) \nonumber\\ 
&& + \frac{1}{3!} \sum_{k,l,m} \tilde{F}_{ijkkllmm}(f_k+1/2)(f_l+1/2)(f_m+1/2) \nonumber\\
&&+\dots, 
\\
\label{eq:Wijk_appendix}
W_{ijk} & = & \tilde{F}_{ijk} +  \sum_l \tilde{F}_{ijkll}(f_l+1/2) \nonumber\\
&& + \frac{1}{2!} \sum_{l,m}  \tilde{F}_{ijkllmm}(f_l+1/2)(f_m+1/2)+ \dots ,
\\
\label{eq:Wijkl_appendix}
W_{ijkl} & = & \tilde{F}_{ijkl} + \sum_{m} \tilde{F}_{ijklmm}(f_m+1/2) \nonumber\\
&& +\frac{1}{2!} \sum_{m,n} \tilde{F}_{ijklmmnn}(f_m+1/2)(f_n+1/2)+ \dots, 
\end{eqnarray}
etc. The scaled bare force constants, $\tilde{{F}}$ and $\tilde{\bar{F}}$, are given by Eqs.\ (\ref{eq:F1tilde})--(\ref{eq:Ftildebar}). 
The dressed force constants, $W$, have the same permutation symmetry as the bare force constants.

These are the quantities defining a finite-temperature extension of XVSCF as discussed in Sec.\ \ref{section:finiteT_XVSCF}.
Note that they emerge naturally by normal ordering $\hat{H}$.

\subsection{Perturbation operator}

The zeroth-order Hamiltonian must take a harmonic form [cf.\ Eq.\ (\ref{eq:H0})]. In second quantization, it can be written as
\begin{eqnarray}
\label{eq:H0definitionSQ}
\hat{H}^{(0)} &=& V_{\text{ref}} +  \frac{1}{2} \sum_i \omega_i \left( \hat{a}_i \hat{a}_i^{\dagger} +   \hat{a}_i^{\dagger} \hat{a}_i \right) \nonumber\\
 &=&V_{\text{ref}} +  \sum_i {{\omega}_i}\left(  f_i+ {1}/{2} \right) +  \sum_i {{\omega}_i} \{\hat{a}_i \hat{a}_i^{\dagger} \} , 
\end{eqnarray}
so that the zeroth-order thermodynamics is exactly solvable by the Bose--Einstein theory. 
The definitions of Wick contractions [Eqs.\ (\ref{eq:aidaggerai}) and (\ref{eq:aiaidagger})] were used in the last equality.

Here, the ``harmonic'' form is not limited to the harmonic approximation to the PES. It can correspond to a zero- or nonzero-temperature XVSCF reference wave function, for instance,
in which case $\omega_i$ and $Q_i$ are the $i$th modal frequency and coordinate of XVSCF. Generally, the values of $V_\text{ref}$ and all force constants change with
the reference, but the foregoing perturbation theoretical formalisms are unchanged. 
See Sec.\ \ref{section:finiteT_XVSCF} for a salient discussion.

The perturbation operator, $\hat{V}^{(1)} \equiv \hat{H} - \hat{H}^{(0)}$, can then be written in a finite-temperature normal-ordered form as
\begin{eqnarray}
\label{eq:VSQ}
\hat{V}^{(1)} 
&=& E_{\mathrm{XVSCF}}(T)  - V_{\text{ref}}  - \sum_i {\omega_i} \left(f_i+{1}/{2}\right) \nonumber\\
&&+ \sum_i W_i\{\hat{a}_i\}   + \sum_i W_i\{\hat{a}_i^{\dagger}\}  \nonumber\\%+\sum_i \left(W_{ii}^{+} - {\omega}_i\right) \{ \hat{a}_i \hat{a}_i^{\dagger}\} 
& &   + \frac{1}{2!}\sum_{i, j} W_{ij}  \{\hat{a}_i \hat{a}_j\} +\frac{1}{2!}\sum_{i , j} W_{ij} \{\hat{a}_i^{\dagger}\hat{a}_j^{\dagger}\}  +  \sum_{i, j} W_{ij} \{\hat{a}_i\hat{a}_j^{\dagger}\}  \nonumber\\
& & + \frac{1}{3!} \sum_{i, j, k} W_{i j k} \{\hat{a}_{i}\hat{a}_j \hat{a}_k\} +  \frac{1}{3!} \sum_{i, j, k} W_{i j k} \{ \hat{a}_{i}^{\dagger}\hat{a}_j^{\dagger} \hat{a}_k^{\dagger}\}  \nonumber\\
&& + \frac{1}{2!} \sum_{i, j, k} W_{i j k} \{\hat{a}_{i} \hat{a}_{j} \hat{a}_{k}^{\dagger}\} +\frac{1}{2!} \sum_{i, j, k} W_{i j k}  \{\hat{a}_{i}^{\dagger} \hat{a}_{j}^{\dagger} \hat{a}_{k}\} \nonumber\\
& & +\frac{1}{4!} \sum_{i,j,k,l} W_{ijkl}  \{\hat{a}_{i} \hat{a}_{j} \hat{a}_{k} \hat{a}_{l}\} + \frac{1}{4!} \sum_{i,j,k,l} W_{ijkl} \{\hat{a}_{i}^{\dagger} \hat{a}_{j}^{\dagger} \hat{a}_{k}^{\dagger} \hat{a}_{l}^{\dagger}\} \nonumber\\
&& + \frac{1}{3!} \sum_{i,j,k,l}W_{ijkl} \{\hat{a}_{i} \hat{a}_{j} \hat{a}_{k} \hat{a}_{l}^{\dagger}\}  
+ \frac{1}{3!} \sum_{i,j,k,l}W_{ijkl} \{\hat{a}_{i}^{\dagger} \hat{a}_{j}^{\dagger} \hat{a}_{k}^{\dagger} \hat{a}_{l}\}  \nonumber\\
& &
+ \frac{1}{2!2!} \sum_{i,j,k,l}W_{ijkl} \{\hat{a}_{i} \hat{a}_{j} \hat{a}_{k}^{\dagger} \hat{a}_{l}^{\dagger} \} + \dots,
\end{eqnarray}
where terms that are pertinent to a QFF are shown explicitly.

\subsection{Inner projection and resolvent operators \label{appendix:PR}}

The utility of normal ordering lies in the fact that a thermal average of a normal-ordered product is zero.  
A product of two or more normal-ordered products can also be thermal-averaged expediently into the sum of all full contractions
excluding internal contractions.\cite{Shavitt2009,Hirata2014,Hirata2021}
For general matrices $\bm{x}$ and $\bm{y}$, therefore, we can quickly evaluate
\begin{eqnarray}
&& \Big [\sum_{i,j,k,l}  \langle N| x_{ij} \{\hat{a}_i\hat{a}_j^{\dagger}\}\{\hat{a}_k\hat{a}_l^{\dagger}\} y_{kl} | N\rangle \Big] \nonumber\\
&&= \Big [ \sum_{i,j,k,l}   \langle N|x_{ij}\{
\contraction[1ex]{}{\hat{a}}{_i\hat{a}_j^{\dagger}\}\{\hat{a}_k}{\hat{a}}
\hat{a}_i
\contraction[0.5ex]{}{\hat{a}}{_j^{\dagger}\}\{}{\hat{a}}
\hat{a}_j^{\dagger}\}\{ 
\hat{a}_k \hat{a}_l^{\dagger} \}
  y_{kl} | N\rangle \Big] \nonumber\\
&&= \sum\limits_{i,j}x_{ij}y_{ji}(f_i+1)f_j,
\end{eqnarray}
from which the internal contraction,
\begin{eqnarray}
    \Big[ \langle N | \sum_{i,j,k,l}  x_{ij} \{ \contraction[1ex]{}{\hat{a}}{_i}{\hat{a}}\hat{a}_i \hat{a}_j^{\dagger} \} \{ \contraction[1ex]{}{\hat{a}}{_k^{\dagger}}{\hat{a}} \hat{a}_k^{\dagger} \hat{a}_l \} y_{kl}  | N \rangle \Big],
\end{eqnarray}
was excluded.

Some care needs to be exercised when there is a projection operator involved:
\begin{eqnarray}
\label{eq:degen_example}
&& \Big [\sum_{i,j,k,l} \langle N|  x_{ij} \{\hat{a}_i\hat{a}_j^{\dagger}\}\hat{P} \{\hat{a}_k\hat{a}_l^{\dagger}\}  y_{kl}| N\rangle \Big] \nonumber\\
&&= \left[  \sum_M^{\text{denom.}=0}\sum_{i,j,k,l} \langle N|  x_{ij}\{\hat{a}_i\hat{a}_j^{\dagger}\}| M \rangle \langle M |\{\hat{a}_k\hat{a}_l^{\dagger}\}y_{kl} | N\rangle \right] \nonumber\\
&&= \left[ \sum_{p,q}^{\text{denom.}=0} \sum_{i,j,k,l}\langle N| x_{ij}\{\contraction[1ex]{}{\hat{a}}{_i\hat{a}_j^{\dagger}\}\{\hat{a}_p}{\hat{a}}
\hat{a}_i
\contraction[0.5ex]{}{\hat{a}}{_j^{\dagger}\}\{}{\hat{a}}\hat{a}_j^{\dagger}\}\{
\bcontraction[1ex]{}{\hat{a}}{_q\hat{a}_p^{\dagger}\}| N\rangle\langle N| \{\hat{a}_p}{\hat{a}}
\hat{a}_p
\bcontraction[0.5ex]{}{\hat{a}}{_q^{\dagger}\}| N\rangle\langle N| \{ }{\hat{a}}
\hat{a}_q^{\dagger}\}| N\rangle\langle N| \{
\contraction[1ex]{}{\hat{a}}{_q\hat{a}_p^{\dagger}\} \{\hat{a}_k}{\hat{a}}
\hat{a}_q
\contraction[0.5ex]{}{\hat{a}}{_p^{\dagger}\} \{}{\hat{a}}
\hat{a}_p^{\dagger}\} \{\hat{a}_k\hat{a}_l^{\dagger}\}  y_{kl} | N\rangle \right]\nonumber\\
&& = \sum_{i,j}^{\text{denom.}=0} x_{ij}y_{ji}(f_i+1)f_j,
\end{eqnarray}
where each chain of contractions, $\hat{a}_i$-$\hat{a}_q^{\dagger}$-$\hat{a}_{q}$-$\hat{a}_l^{\dagger}$ or 
$\hat{a}_j^{\dagger}$-$\hat{a}_p$-$\hat{a}_{p}^{\dagger}$-$\hat{a}_k$, should be regarded\cite{Hirata2021} as a single contraction and 
evaluated by Eq.\ (\ref{eq:aiaidagger}) or (\ref{eq:aidaggerai}), respectively. In the second line, ``denom.=0'' indicates that 
$M$ runs over all Hartree-product states that are degenerate with $N$th state. 
The same restrictions in the third and fourth lines 
demand that the fictitious denominators, $\omega_p - \omega_q$ and $\omega_j - \omega_i$, be zero.

A resolvent operator is handled in an analogous manner. 
%\begin{widetext}
\begin{eqnarray}
\label{eq:nondegen_example}
&& \Big [\sum_{i,j,k,l}\langle N|   x_{ij} \{\hat{a}_i^{\dagger}\hat{a}_j^{\dagger}\}\hat{R} \{\hat{a}_k\hat{a}_l\}  y_{kl}| N\rangle \Big] \nonumber\\
&&= \left[  \sum_M^{\text{denom.}\neq0}\sum_{i,j,k,l} \frac{ \langle N|  x_{ij}\{\hat{a}_i^{\dagger}\hat{a}_j^{\dagger}\}| M \rangle  \langle M |\{\hat{a}_k\hat{a}_l\}y_{kl} | N\rangle }{E_N^{(0)} - E_M^{(0)}} \right] \nonumber\\
&&= \left[  \sum_{p \leq q}^{\text{denom.}\neq0}\sum_{i,j,k,l} \frac{ \langle N|  x_{ij}\{\hat{a}_i^{\dagger}\hat{a}_j^{\dagger}\} \{\hat{a}_p\hat{a}_q\} | N \rangle  \langle N |\{\hat{a}_q^{\dagger}\hat{a}_p^{\dagger}\} \{\hat{a}_k\hat{a}_l\}y_{kl} | N\rangle }{\omega_p + \omega_q} \right] \nonumber\\
&&= \frac{1}{2} \left[  \sum_{p \neq q}^{\text{denom.}\neq0} \sum_{i,j,k,l} \frac{ \langle N| x_{ij}\{\contraction[1ex]{}{\hat{a}}{_i^{\dagger}\hat{a}_j^{\dagger}\}\{\hat{a}_p}{\hat{a}}
\hat{a}_i^{\dagger}
\contraction[0.5ex]{}{\hat{a}}{_j^{\dagger}\}\{}{\hat{a}}\hat{a}_j^{\dagger}\}\{
\bcontraction[1ex]{}{\hat{a}}{_q\hat{a}_p\}| N\rangle  \langle N| \{\hat{a}_p}{\hat{a}}
\hat{a}_p
\bcontraction[0.5ex]{}{\hat{a}}{_q^{\dagger}\}| N\rangle \langle N| \{ }{\hat{a}}
\hat{a}_q\}| N\rangle \langle N| \{
\contraction[1ex]{}{\hat{a}}{_q^{\dagger}\hat{a}_p^{\dagger}\} \{\hat{a}_k}{\hat{a}}
\hat{a}_q^{\dagger}
\contraction[0.5ex]{}{\hat{a}}{_p^{\dagger}\} \{}{\hat{a}}
\hat{a}_p^{\dagger}\} \{\hat{a}_k\hat{a}_l\}  y_{kl} | N\rangle }{\omega_p + \omega_q}  \right]\nonumber\\
&&\,\,\,\,\, \times 4 \nonumber\\
&&+   \left[ \sum_{p}^{\text{denom.}\neq0} \sum_{i,j,k,l} \frac{ \langle N| x_{ij}\{\contraction[1ex]{}{\hat{a}}{_i^{\dagger}\hat{a}_j^{\dagger}\}\{\hat{a}_p}{\hat{a}}
\hat{a}_i^{\dagger}
\contraction[0.5ex]{}{\hat{a}}{_j^{\dagger}\}\{}{\hat{a}}\hat{a}_j^{\dagger}\}\{
\bcontraction[1ex]{}{\hat{a}}{_p\hat{a}_p\}| N\rangle  \langle N| \{\hat{a}_p}{\hat{a}}
\hat{a}_p
\bcontraction[0.5ex]{}{\hat{a}}{_p^{\dagger}\}| N\rangle \langle N| \{ }{\hat{a}}
\hat{a}_p\}| N\rangle \langle N| \{
\contraction[1ex]{}{\hat{a}}{_p^{\dagger}\hat{a}_p^{\dagger}\} \{\hat{a}_k}{\hat{a}}
\hat{a}_p^{\dagger}
\contraction[0.5ex]{}{\hat{a}}{_p^{\dagger}\} \{}{\hat{a}}
\hat{a}_p^{\dagger}\} \{\hat{a}_k\hat{a}_l\}  y_{kl} | N\rangle }{\omega_p + \omega_p}  \right]\nonumber\\
&&\,\,\,\,\, \times 2 \nonumber\\
&&=\frac{1}{2}  \left[  \sum_{p,q}^{\text{denom.}\neq0} \sum_{i,j,k,l} \frac{ \langle N| x_{ij}\{\contraction[1ex]{}{\hat{a}}{_i^{\dagger}\hat{a}_j^{\dagger}\}\{\hat{a}_p}{\hat{a}}
\hat{a}_i^{\dagger}
\contraction[0.5ex]{}{\hat{a}}{_j^{\dagger}\}\{}{\hat{a}}\hat{a}_j^{\dagger}\}\{
\bcontraction[1ex]{}{\hat{a}}{_q\hat{a}_p\}| N\rangle  \langle N| \{\hat{a}_p}{\hat{a}}
\hat{a}_p
\bcontraction[0.5ex]{}{\hat{a}}{_q^{\dagger}\}| N\rangle \langle N| \{ }{\hat{a}}
\hat{a}_q\}| N\rangle \langle N| \{
\contraction[1ex]{}{\hat{a}}{_q^{\dagger}\hat{a}_p^{\dagger}\} \{\hat{a}_k}{\hat{a}}
\hat{a}_q^{\dagger}
\contraction[0.5ex]{}{\hat{a}}{_p^{\dagger}\} \{}{\hat{a}}
\hat{a}_p^{\dagger}\} \{\hat{a}_k\hat{a}_l\}  y_{kl} | N\rangle }{\omega_p + \omega_q}  \right]\nonumber\\
&&\,\,\,\,\, \times 4 \nonumber\\
&& = 2 \sum_{i,j}^{\text{denom.}\neq 0} \frac{x_{ij}y_{ji}  }{\omega_j + \omega_i}f_i f_j ,
\end{eqnarray}
where $\bm{x}$ and $\bm{y}$ are assumed to have index permutation symmetry, and we used $\sum_{p \leq q} = (1/2) \sum_{p \neq q} + \sum_{p = q}$.
We also meant by ``$\times 4$'' and ``$\times 2$'' that there are four and two equal-valued full contractions, respectively.
In the penultimate equality, while initially we need to consider the $p \neq q$ and $p = q$ cases separately, we arrive at the same, correct result
by treating $p$ and $q$ as distinct indexes. 
%\end{widetext}

\section*{Data availability}
The data that support the findings of this study are available within the article.

%\bibliography{Accepted.bib}

\begin{thebibliography}{115}%
\makeatletter
\providecommand \@ifxundefined [1]{%
 \@ifx{#1\undefined}
}%
\providecommand \@ifnum [1]{%
 \ifnum #1\expandafter \@firstoftwo
 \else \expandafter \@secondoftwo
 \fi
}%
\providecommand \@ifx [1]{%
 \ifx #1\expandafter \@firstoftwo
 \else \expandafter \@secondoftwo
 \fi
}%
\providecommand \natexlab [1]{#1}%
\providecommand \enquote  [1]{``#1''}%
\providecommand \bibnamefont  [1]{#1}%
\providecommand \bibfnamefont [1]{#1}%
\providecommand \citenamefont [1]{#1}%
\providecommand \href@noop [0]{\@secondoftwo}%
\providecommand \href [0]{\begingroup \@sanitize@url \@href}%
\providecommand \@href[1]{\@@startlink{#1}\@@href}%
\providecommand \@@href[1]{\endgroup#1\@@endlink}%
\providecommand \@sanitize@url [0]{\catcode `\\12\catcode `\$12\catcode
  `\&12\catcode `\#12\catcode `\^12\catcode `\_12\catcode `\%12\relax}%
\providecommand \@@startlink[1]{}%
\providecommand \@@endlink[0]{}%
\providecommand \url  [0]{\begingroup\@sanitize@url \@url }%
\providecommand \@url [1]{\endgroup\@href {#1}{\urlprefix }}%
\providecommand \urlprefix  [0]{URL }%
\providecommand \Eprint [0]{\href }%
\providecommand \doibase [0]{https://doi.org/}%
\providecommand \selectlanguage [0]{\@gobble}%
\providecommand \bibinfo  [0]{\@secondoftwo}%
\providecommand \bibfield  [0]{\@secondoftwo}%
\providecommand \translation [1]{[#1]}%
\providecommand \BibitemOpen [0]{}%
\providecommand \bibitemStop [0]{}%
\providecommand \bibitemNoStop [0]{.\EOS\space}%
\providecommand \EOS [0]{\spacefactor3000\relax}%
\providecommand \BibitemShut  [1]{\csname bibitem#1\endcsname}%
\let\auto@bib@innerbib\@empty
%</preamble>
\bibitem [{\citenamefont {He}\ \emph {et~al.}(2012)\citenamefont {He},
  \citenamefont {Sode}, \citenamefont {Xantheas},\ and\ \citenamefont
  {Hirata}}]{He2012}%
  \BibitemOpen
  \bibfield  {author} {\bibinfo {author} {\bibfnamefont {X.}~\bibnamefont
  {He}}, \bibinfo {author} {\bibfnamefont {O.}~\bibnamefont {Sode}}, \bibinfo
  {author} {\bibfnamefont {S.~S.}\ \bibnamefont {Xantheas}},\ and\ \bibinfo
  {author} {\bibfnamefont {S.}~\bibnamefont {Hirata}},\ }\bibfield  {title}
  {\enquote {\bibinfo {title} {Second-order many-body perturbation study of ice
  {Ih}},}\ }\href@noop {} {\bibfield  {journal} {\bibinfo  {journal} {J. Chem.
  Phys.}\ }\textbf {\bibinfo {volume} {137}},\ \bibinfo {pages} {204505}
  (\bibinfo {year} {2012})}\BibitemShut {NoStop}%
\bibitem [{\citenamefont {White}(1965)}]{White1965}%
  \BibitemOpen
  \bibfield  {author} {\bibinfo {author} {\bibfnamefont {G.~K.}\ \bibnamefont
  {White}},\ }\bibfield  {title} {\enquote {\bibinfo {title} {Thermal expansion
  of alkali halides at low temperatures},}\ }\href@noop {} {\bibfield
  {journal} {\bibinfo  {journal} {Proc. R. Soc. A (London)}\ }\textbf {\bibinfo
  {volume} {286}},\ \bibinfo {pages} {204--217} (\bibinfo {year}
  {1965})}\BibitemShut {NoStop}%
\bibitem [{\citenamefont {Li}, \citenamefont {Sode},\ and\ \citenamefont
  {Hirata}(2015)}]{Li2015}%
  \BibitemOpen
  \bibfield  {author} {\bibinfo {author} {\bibfnamefont {J.~J.}\ \bibnamefont
  {Li}}, \bibinfo {author} {\bibfnamefont {O.}~\bibnamefont {Sode}},\ and\
  \bibinfo {author} {\bibfnamefont {S.}~\bibnamefont {Hirata}},\ }\bibfield
  {title} {\enquote {\bibinfo {title} {Second-order many-body perturbation
  study on thermal expansion of solid carbon dioxide},}\ }\href@noop {}
  {\bibfield  {journal} {\bibinfo  {journal} {J. Chem. Theory Comput.}\
  }\textbf {\bibinfo {volume} {11}},\ \bibinfo {pages} {224--229} (\bibinfo
  {year} {2015})}\BibitemShut {NoStop}%
\bibitem [{\citenamefont {Salim}, \citenamefont {Willow},\ and\ \citenamefont
  {Hirata}(2016)}]{Salim2016}%
  \BibitemOpen
  \bibfield  {author} {\bibinfo {author} {\bibfnamefont {M.~A.}\ \bibnamefont
  {Salim}}, \bibinfo {author} {\bibfnamefont {S.~Y.}\ \bibnamefont {Willow}},\
  and\ \bibinfo {author} {\bibfnamefont {S.}~\bibnamefont {Hirata}},\
  }\bibfield  {title} {\enquote {\bibinfo {title} {Ice {Ih} anomalies: Thermal
  contraction, anomalous volume isotope effect, and pressure-induced
  amorphization},}\ }\href {https://doi.org/10.1063/1.4951687} {\bibfield
  {journal} {\bibinfo  {journal} {J. Chem. Phys.}\ }\textbf {\bibinfo {volume}
  {144}},\ \bibinfo {pages} {204503} (\bibinfo {year} {2016})}\BibitemShut
  {NoStop}%
\bibitem [{\citenamefont {Herring}(1954)}]{Herring1954}%
  \BibitemOpen
  \bibfield  {author} {\bibinfo {author} {\bibfnamefont {C.}~\bibnamefont
  {Herring}},\ }\bibfield  {title} {\enquote {\bibinfo {title} {Role of
  low-energy phonons in thermal conduction},}\ }\href@noop {} {\bibfield
  {journal} {\bibinfo  {journal} {Phys. Rev.}\ }\textbf {\bibinfo {volume}
  {95}},\ \bibinfo {pages} {954--965} (\bibinfo {year} {1954})}\BibitemShut
  {NoStop}%
\bibitem [{\citenamefont {Wilson}\ and\ \citenamefont
  {Kim}(1973)}]{Wilson1973}%
  \BibitemOpen
  \bibfield  {author} {\bibinfo {author} {\bibfnamefont {R.~S.}\ \bibnamefont
  {Wilson}}\ and\ \bibinfo {author} {\bibfnamefont {S.~K.}\ \bibnamefont
  {Kim}},\ }\bibfield  {title} {\enquote {\bibinfo {title} {Theory of thermal
  conductivity of anharmonic crystals},}\ }\href@noop {} {\bibfield  {journal}
  {\bibinfo  {journal} {Phys. Rev. B}\ }\textbf {\bibinfo {volume} {7}},\
  \bibinfo {pages} {4674--4677} (\bibinfo {year} {1973})}\BibitemShut {NoStop}%
\bibitem [{\citenamefont {Knauss}\ and\ \citenamefont
  {Wilson}(1974)}]{Knauss1974}%
  \BibitemOpen
  \bibfield  {author} {\bibinfo {author} {\bibfnamefont {D.~C.}\ \bibnamefont
  {Knauss}}\ and\ \bibinfo {author} {\bibfnamefont {R.~S.}\ \bibnamefont
  {Wilson}},\ }\bibfield  {title} {\enquote {\bibinfo {title} {Theory of
  thermal conductivity of anharmonic crystals: Nondiagonal {Peierls}
  contribution},}\ }\href@noop {} {\bibfield  {journal} {\bibinfo  {journal}
  {Phys. Rev. B}\ }\textbf {\bibinfo {volume} {10}},\ \bibinfo {pages}
  {4383--4387} (\bibinfo {year} {1974})}\BibitemShut {NoStop}%
\bibitem [{\citenamefont {Ackerman}\ and\ \citenamefont
  {Guyer}(1968)}]{Ackerman1968}%
  \BibitemOpen
  \bibfield  {author} {\bibinfo {author} {\bibfnamefont {C.~C.}\ \bibnamefont
  {Ackerman}}\ and\ \bibinfo {author} {\bibfnamefont {R.~A.}\ \bibnamefont
  {Guyer}},\ }\bibfield  {title} {\enquote {\bibinfo {title} {Temperature
  pulses in dielectric solids},}\ }\href@noop {} {\bibfield  {journal}
  {\bibinfo  {journal} {Ann. Phys.}\ }\textbf {\bibinfo {volume} {50}},\
  \bibinfo {pages} {128--185} (\bibinfo {year} {1968})}\BibitemShut {NoStop}%
\bibitem [{\citenamefont {Shukla}\ and\ \citenamefont
  {Cowley}(1971)}]{Shukla1971_2}%
  \BibitemOpen
  \bibfield  {author} {\bibinfo {author} {\bibfnamefont {R.~C.}\ \bibnamefont
  {Shukla}}\ and\ \bibinfo {author} {\bibfnamefont {E.~R.}\ \bibnamefont
  {Cowley}},\ }\bibfield  {title} {\enquote {\bibinfo {title} {{Helmholtz} free
  energy of an anharmonic crystal to ${O}(\lambda^4)$},}\ }\href
  {https://doi.org/10.1103/physrevb.3.4055} {\bibfield  {journal} {\bibinfo
  {journal} {Phys. Rev. B}\ }\textbf {\bibinfo {volume} {3}},\ \bibinfo {pages}
  {4055--4065} (\bibinfo {year} {1971})}\BibitemShut {NoStop}%
\bibitem [{\citenamefont {Sode}\ \emph {et~al.}(2013)\citenamefont {Sode},
  \citenamefont {Ke{\c{c}}eli}, \citenamefont {Yagi},\ and\ \citenamefont
  {Hirata}}]{Sode2013}%
  \BibitemOpen
  \bibfield  {author} {\bibinfo {author} {\bibfnamefont {O.}~\bibnamefont
  {Sode}}, \bibinfo {author} {\bibfnamefont {M.}~\bibnamefont {Ke{\c{c}}eli}},
  \bibinfo {author} {\bibfnamefont {K.}~\bibnamefont {Yagi}},\ and\ \bibinfo
  {author} {\bibfnamefont {S.}~\bibnamefont {Hirata}},\ }\bibfield  {title}
  {\enquote {\bibinfo {title} {Fermi resonance in solid {CO}$_2$ under
  pressure},}\ }\href {https://doi.org/10.1063/1.4790537} {\bibfield  {journal}
  {\bibinfo  {journal} {J. Chem. Phys.}\ }\textbf {\bibinfo {volume} {138}},\
  \bibinfo {pages} {074501} (\bibinfo {year} {2013})}\BibitemShut {NoStop}%
\bibitem [{\citenamefont {Hirata}\ \emph {et~al.}(2014)\citenamefont {Hirata},
  \citenamefont {Sode}, \citenamefont {Ke\c{c}eli}, \citenamefont {Yagi},\ and\
  \citenamefont {Li}}]{HirataSode2014}%
  \BibitemOpen
  \bibfield  {author} {\bibinfo {author} {\bibfnamefont {S.}~\bibnamefont
  {Hirata}}, \bibinfo {author} {\bibfnamefont {O.}~\bibnamefont {Sode}},
  \bibinfo {author} {\bibfnamefont {M.}~\bibnamefont {Ke\c{c}eli}}, \bibinfo
  {author} {\bibfnamefont {K.}~\bibnamefont {Yagi}},\ and\ \bibinfo {author}
  {\bibfnamefont {J.~J.}\ \bibnamefont {Li}},\ }\bibfield  {title} {\enquote
  {\bibinfo {title} {Response to ``{Comment} on `{Fermi} resonance in solid
  {CO}$_2$ under pressure''' [{J. Chem. Phys.} 140, 177101 (2014)]},}\
  }\href@noop {} {\bibfield  {journal} {\bibinfo  {journal} {J. Chem. Phys.}\
  }\textbf {\bibinfo {volume} {140}},\ \bibinfo {pages} {177102} (\bibinfo
  {year} {2014})}\BibitemShut {NoStop}%
\bibitem [{\citenamefont {Qin}\ and\ \citenamefont {Hirata}(2020)}]{Qin2020}%
  \BibitemOpen
  \bibfield  {author} {\bibinfo {author} {\bibfnamefont {X.}~\bibnamefont
  {Qin}}\ and\ \bibinfo {author} {\bibfnamefont {S.}~\bibnamefont {Hirata}},\
  }\bibfield  {title} {\enquote {\bibinfo {title} {Anharmonic phonon dispersion
  in polyethylene},}\ }\href {https://doi.org/10.1021/acs.jpcb.0c08493}
  {\bibfield  {journal} {\bibinfo  {journal} {J. Phys. Chem. B}\ }\textbf
  {\bibinfo {volume} {124}},\ \bibinfo {pages} {10477--10485} (\bibinfo {year}
  {2020})}\BibitemShut {NoStop}%
\bibitem [{\citenamefont {Born}\ and\ \citenamefont
  {Huang}(1954)}]{born1988dynamical}%
  \BibitemOpen
  \bibfield  {author} {\bibinfo {author} {\bibfnamefont {M.}~\bibnamefont
  {Born}}\ and\ \bibinfo {author} {\bibfnamefont {K.}~\bibnamefont {Huang}},\
  }\href@noop {} {\emph {\bibinfo {title} {Dynamical Theory of Crystal
  Lattices}}}\ (\bibinfo  {publisher} {Oxford University Press},\ \bibinfo
  {address} {Oxford},\ \bibinfo {year} {1954})\BibitemShut {NoStop}%
\bibitem [{\citenamefont {Hirata}(2011)}]{Hirata2011}%
  \BibitemOpen
  \bibfield  {author} {\bibinfo {author} {\bibfnamefont {S.}~\bibnamefont
  {Hirata}},\ }\bibfield  {title} {\enquote {\bibinfo {title} {Thermodynamic
  limit and size-consistent design},}\ }\href@noop {} {\bibfield  {journal}
  {\bibinfo  {journal} {Theor. Chem. Acc.}\ }\textbf {\bibinfo {volume}
  {129}},\ \bibinfo {pages} {727--746} (\bibinfo {year} {2011})}\BibitemShut
  {NoStop}%
\bibitem [{\citenamefont {Hirata}\ \emph {et~al.}(2012)\citenamefont {Hirata},
  \citenamefont {Ke\c{c}eli}, \citenamefont {Ohnishi}, \citenamefont {Sode},\
  and\ \citenamefont {Yagi}}]{HirataARPC2012}%
  \BibitemOpen
  \bibfield  {author} {\bibinfo {author} {\bibfnamefont {S.}~\bibnamefont
  {Hirata}}, \bibinfo {author} {\bibfnamefont {M.}~\bibnamefont {Ke\c{c}eli}},
  \bibinfo {author} {\bibfnamefont {Y.}~\bibnamefont {Ohnishi}}, \bibinfo
  {author} {\bibfnamefont {O.}~\bibnamefont {Sode}},\ and\ \bibinfo {author}
  {\bibfnamefont {K.}~\bibnamefont {Yagi}},\ }\bibfield  {title} {\enquote
  {\bibinfo {title} {Extensivity of energy and electronic and vibrational
  structure methods for crystals},}\ }\href@noop {} {\bibfield  {journal}
  {\bibinfo  {journal} {Annu. Rev. Phys. Chem.}\ }\textbf {\bibinfo {volume}
  {63}},\ \bibinfo {pages} {131--153} (\bibinfo {year} {2012})}\BibitemShut
  {NoStop}%
\bibitem [{\citenamefont {Hirata}\ and\ \citenamefont
  {Grabowski}(2014)}]{Hirata_conjecture2014}%
  \BibitemOpen
  \bibfield  {author} {\bibinfo {author} {\bibfnamefont {S.}~\bibnamefont
  {Hirata}}\ and\ \bibinfo {author} {\bibfnamefont {I.}~\bibnamefont
  {Grabowski}},\ }\bibfield  {title} {\enquote {\bibinfo {title} {On the mutual
  exclusion of variationality and size consistency},}\ }\href@noop {}
  {\bibfield  {journal} {\bibinfo  {journal} {Theor. Chem. Acc.}\ }\textbf
  {\bibinfo {volume} {133}},\ \bibinfo {pages} {1440} (\bibinfo {year}
  {2014})}\BibitemShut {NoStop}%
\bibitem [{\citenamefont {Hooton}(1958)}]{Hooton1958}%
  \BibitemOpen
  \bibfield  {author} {\bibinfo {author} {\bibfnamefont {D.~J.}\ \bibnamefont
  {Hooton}},\ }\bibfield  {title} {\enquote {\bibinfo {title} {The use of a
  model in anharmonic lattice dynamics},}\ }\href
  {https://doi.org/10.1080/14786435808243224} {\bibfield  {journal} {\bibinfo
  {journal} {Philos. Mag.}\ }\textbf {\bibinfo {volume} {3}},\ \bibinfo {pages}
  {49--54} (\bibinfo {year} {1958})}\BibitemShut {NoStop}%
\bibitem [{\citenamefont {Koehler}(1966)}]{Koehler1966}%
  \BibitemOpen
  \bibfield  {author} {\bibinfo {author} {\bibfnamefont {T.~R.}\ \bibnamefont
  {Koehler}},\ }\bibfield  {title} {\enquote {\bibinfo {title} {Theory of the
  self-consistent harmonic approximation with application to solid neon},}\
  }\href {https://doi.org/10.1103/PhysRevLett.17.89} {\bibfield  {journal}
  {\bibinfo  {journal} {Phys. Rev. Lett.}\ }\textbf {\bibinfo {volume} {17}},\
  \bibinfo {pages} {89--91} (\bibinfo {year} {1966})}\BibitemShut {NoStop}%
\bibitem [{\citenamefont {Choquard}(1967)}]{Choquard1967}%
  \BibitemOpen
  \bibfield  {author} {\bibinfo {author} {\bibfnamefont {P.~F.}\ \bibnamefont
  {Choquard}},\ }\href@noop {} {\emph {\bibinfo {title} {The Anharmonic
  Crystal}}}\ (\bibinfo  {publisher} {Benjamin, New York},\ \bibinfo {year}
  {1967})\BibitemShut {NoStop}%
\bibitem [{\citenamefont {Gillis}, \citenamefont {Werthamer},\ and\
  \citenamefont {Koehler}(1968)}]{Gillis1968}%
  \BibitemOpen
  \bibfield  {author} {\bibinfo {author} {\bibfnamefont {N.~S.}\ \bibnamefont
  {Gillis}}, \bibinfo {author} {\bibfnamefont {N.~R.}\ \bibnamefont
  {Werthamer}},\ and\ \bibinfo {author} {\bibfnamefont {T.~R.}\ \bibnamefont
  {Koehler}},\ }\bibfield  {title} {\enquote {\bibinfo {title} {Properties of
  crystalline argon and neon in the self-consistent phonon approximation},}\
  }\href {https://doi.org/10.1103/PhysRev.165.951} {\bibfield  {journal}
  {\bibinfo  {journal} {Phys. Rev.}\ }\textbf {\bibinfo {volume} {165}},\
  \bibinfo {pages} {951--959} (\bibinfo {year} {1968})}\BibitemShut {NoStop}%
\bibitem [{\citenamefont {Hermes}\ and\ \citenamefont
  {Hirata}(2013{\natexlab{a}})}]{Hermes2013_Dyson}%
  \BibitemOpen
  \bibfield  {author} {\bibinfo {author} {\bibfnamefont {M.~R.}\ \bibnamefont
  {Hermes}}\ and\ \bibinfo {author} {\bibfnamefont {S.}~\bibnamefont
  {Hirata}},\ }\bibfield  {title} {\enquote {\bibinfo {title} {First-order
  {Dyson} coordinates and geometry},}\ }\href
  {https://doi.org/10.1021/jp4008834} {\bibfield  {journal} {\bibinfo
  {journal} {J. Phys. Chem. A}\ }\textbf {\bibinfo {volume} {117}},\ \bibinfo
  {pages} {7179--7189} (\bibinfo {year} {2013}{\natexlab{a}})}\BibitemShut
  {NoStop}%
\bibitem [{\citenamefont {Shukla}\ and\ \citenamefont
  {Wilk}(1974)}]{Shukla1974}%
  \BibitemOpen
  \bibfield  {author} {\bibinfo {author} {\bibfnamefont {R.~C.}\ \bibnamefont
  {Shukla}}\ and\ \bibinfo {author} {\bibfnamefont {L.}~\bibnamefont {Wilk}},\
  }\bibfield  {title} {\enquote {\bibinfo {title} {Helmholtz free energy of an
  anharmonic crystal to ${O}(\lambda^4)$. {II}},}\ }\href
  {https://doi.org/10.1103/physrevb.10.3660} {\bibfield  {journal} {\bibinfo
  {journal} {Phys. Rev. B}\ }\textbf {\bibinfo {volume} {10}},\ \bibinfo
  {pages} {3660--3666} (\bibinfo {year} {1974})}\BibitemShut {NoStop}%
\bibitem [{\citenamefont {March}, \citenamefont {Young},\ and\ \citenamefont
  {Sampanthar}(1967)}]{march}%
  \BibitemOpen
  \bibfield  {author} {\bibinfo {author} {\bibfnamefont {N.~H.}\ \bibnamefont
  {March}}, \bibinfo {author} {\bibfnamefont {W.~H.}\ \bibnamefont {Young}},\
  and\ \bibinfo {author} {\bibfnamefont {S.}~\bibnamefont {Sampanthar}},\
  }\href@noop {} {\emph {\bibinfo {title} {The Many-Body Problem in Quantum
  Mechanics}}}\ (\bibinfo  {publisher} {Cambridge University Press},\ \bibinfo
  {address} {Cambridge},\ \bibinfo {year} {1967})\BibitemShut {NoStop}%
\bibitem [{\citenamefont {Fetter}\ and\ \citenamefont
  {Walecka}(1971)}]{Fetter1971}%
  \BibitemOpen
  \bibfield  {author} {\bibinfo {author} {\bibfnamefont {A.~L.}\ \bibnamefont
  {Fetter}}\ and\ \bibinfo {author} {\bibfnamefont {J.~D.}\ \bibnamefont
  {Walecka}},\ }\href@noop {} {\emph {\bibinfo {title} {Quantum Theory of
  Many-Particle Systems}}}\ (\bibinfo  {publisher} {McGraw-Hill},\ \bibinfo
  {address} {Boston},\ \bibinfo {year} {1971})\BibitemShut {NoStop}%
\bibitem [{\citenamefont {Mattuck}(1992)}]{mattuck}%
  \BibitemOpen
  \bibfield  {author} {\bibinfo {author} {\bibfnamefont {R.~D.}\ \bibnamefont
  {Mattuck}},\ }\href@noop {} {\emph {\bibinfo {title} {A Guide to Feynman
  Diagrams in the Many-Body Problem}}}\ (\bibinfo  {publisher} {Dover},\
  \bibinfo {address} {New York},\ \bibinfo {year} {1992})\BibitemShut {NoStop}%
\bibitem [{\citenamefont {Maradudin}, \citenamefont {Flinn},\ and\
  \citenamefont {Coldwell-Horsfall}(1961{\natexlab{a}})}]{Maradudin1961}%
  \BibitemOpen
  \bibfield  {author} {\bibinfo {author} {\bibfnamefont {A.~A.}\ \bibnamefont
  {Maradudin}}, \bibinfo {author} {\bibfnamefont {P.~A.}\ \bibnamefont
  {Flinn}},\ and\ \bibinfo {author} {\bibfnamefont {R.~A.}\ \bibnamefont
  {Coldwell-Horsfall}},\ }\bibfield  {title} {\enquote {\bibinfo {title}
  {Anharmonic contributions to vibrational thermodynamic properties of solids.
  {Part} {I}. {General} formulation and application to the linear chain},}\
  }\href {https://doi.org/10.1016/0003-4916(61)90188-9} {\bibfield  {journal}
  {\bibinfo  {journal} {Ann. Phys.}\ }\textbf {\bibinfo {volume} {15}},\
  \bibinfo {pages} {337--359} (\bibinfo {year}
  {1961}{\natexlab{a}})}\BibitemShut {NoStop}%
\bibitem [{\citenamefont {Maradudin}, \citenamefont {Flinn},\ and\
  \citenamefont {Coldwell-Horsfall}(1961{\natexlab{b}})}]{Maradudin1961_2}%
  \BibitemOpen
  \bibfield  {author} {\bibinfo {author} {\bibfnamefont {A.~A.}\ \bibnamefont
  {Maradudin}}, \bibinfo {author} {\bibfnamefont {P.~A.}\ \bibnamefont
  {Flinn}},\ and\ \bibinfo {author} {\bibfnamefont {R.~A.}\ \bibnamefont
  {Coldwell-Horsfall}},\ }\bibfield  {title} {\enquote {\bibinfo {title}
  {Anharmonic contributions to vibrational thermodynamic properties of solids.
  {Part} {II}. {The} high temperature limit},}\ }\href
  {https://doi.org/10.1016/0003-4916(61)90189-0} {\bibfield  {journal}
  {\bibinfo  {journal} {Ann. Phys.}\ }\textbf {\bibinfo {volume} {15}},\
  \bibinfo {pages} {360--386} (\bibinfo {year}
  {1961}{\natexlab{b}})}\BibitemShut {NoStop}%
\bibitem [{\citenamefont {Flinn}\ and\ \citenamefont
  {Maradudin}(1963)}]{Flinn1963}%
  \BibitemOpen
  \bibfield  {author} {\bibinfo {author} {\bibfnamefont {P.~A.}\ \bibnamefont
  {Flinn}}\ and\ \bibinfo {author} {\bibfnamefont {A.~A.}\ \bibnamefont
  {Maradudin}},\ }\bibfield  {title} {\enquote {\bibinfo {title} {Anharmonic
  contributions to vibrational thermodynamic properties of solids},}\ }\href
  {https://doi.org/10.1016/0003-4916(63)90054-x} {\bibfield  {journal}
  {\bibinfo  {journal} {Ann. Phys.}\ }\textbf {\bibinfo {volume} {22}},\
  \bibinfo {pages} {223--238} (\bibinfo {year} {1963})}\BibitemShut {NoStop}%
\bibitem [{\citenamefont {Cowley}(1963)}]{Cowley1963}%
  \BibitemOpen
  \bibfield  {author} {\bibinfo {author} {\bibfnamefont {R.~A.}\ \bibnamefont
  {Cowley}},\ }\bibfield  {title} {\enquote {\bibinfo {title} {The lattice
  dynamics of an anharmonic crystal},}\ }\href
  {https://doi.org/10.1080/00018736300101333} {\bibfield  {journal} {\bibinfo
  {journal} {Adv. Phys.}\ }\textbf {\bibinfo {volume} {12}},\ \bibinfo {pages}
  {421--480} (\bibinfo {year} {1963})}\BibitemShut {NoStop}%
\bibitem [{\citenamefont {Cowley}(1968)}]{Cowley1968}%
  \BibitemOpen
  \bibfield  {author} {\bibinfo {author} {\bibfnamefont {R.~A.}\ \bibnamefont
  {Cowley}},\ }\bibfield  {title} {\enquote {\bibinfo {title} {Anharmonic
  crystals},}\ }\href {https://doi.org/10.1088/0034-4885/31/1/303} {\bibfield
  {journal} {\bibinfo  {journal} {Rep. Prog. Phys.}\ }\textbf {\bibinfo
  {volume} {31}},\ \bibinfo {pages} {123--166} (\bibinfo {year}
  {1968})}\BibitemShut {NoStop}%
\bibitem [{\citenamefont {Shukla}\ and\ \citenamefont
  {Cowley}(1985)}]{Shukla1985}%
  \BibitemOpen
  \bibfield  {author} {\bibinfo {author} {\bibfnamefont {R.~C.}\ \bibnamefont
  {Shukla}}\ and\ \bibinfo {author} {\bibfnamefont {E.~R.}\ \bibnamefont
  {Cowley}},\ }\bibfield  {title} {\enquote {\bibinfo {title} {{Helmholtz} free
  energy of an anharmonic crystal to ${O}(\lambda^4)$. {III}. {Equation} of
  state for the {Lennard}--{Jones} solid},}\ }\href
  {https://doi.org/10.1103/physrevb.31.372} {\bibfield  {journal} {\bibinfo
  {journal} {Phys. Rev. B}\ }\textbf {\bibinfo {volume} {31}},\ \bibinfo
  {pages} {372--378} (\bibinfo {year} {1985})}\BibitemShut {NoStop}%
\bibitem [{\citenamefont {Shukla}\ and\ \citenamefont
  {Shanes}(1985)}]{Shukla1985_2}%
  \BibitemOpen
  \bibfield  {author} {\bibinfo {author} {\bibfnamefont {R.~C.}\ \bibnamefont
  {Shukla}}\ and\ \bibinfo {author} {\bibfnamefont {F.}~\bibnamefont
  {Shanes}},\ }\bibfield  {title} {\enquote {\bibinfo {title} {Helmholtz free
  energy of an anharmonic crystal to ${O}(\lambda^4)$. {IV}. {Thermodynamic}
  properties of {Kr} and {Xe} for the {Lennard}--{Jones}, {Morse}, and
  {Rydberg} potentials},}\ }\href {https://doi.org/10.1103/physrevb.32.2513}
  {\bibfield  {journal} {\bibinfo  {journal} {Phys. Rev. B}\ }\textbf {\bibinfo
  {volume} {32}},\ \bibinfo {pages} {2513--2521} (\bibinfo {year}
  {1985})}\BibitemShut {NoStop}%
\bibitem [{\citenamefont {Kobashi}(1978)}]{Kobashi1978_1}%
  \BibitemOpen
  \bibfield  {author} {\bibinfo {author} {\bibfnamefont {K.}~\bibnamefont
  {Kobashi}},\ }\bibfield  {title} {\enquote {\bibinfo {title} {Anharmonic
  lattice vibrations in solid {N}$_2$. {I}. $\alpha$-phase},}\ }\href@noop {}
  {\bibfield  {journal} {\bibinfo  {journal} {Mol. Phys.}\ }\textbf {\bibinfo
  {volume} {36}},\ \bibinfo {pages} {225--240} (\bibinfo {year}
  {1978})}\BibitemShut {NoStop}%
\bibitem [{\citenamefont {Kobashi}\ and\ \citenamefont
  {Chandrasekharan}(1978)}]{Kobashi1978_2}%
  \BibitemOpen
  \bibfield  {author} {\bibinfo {author} {\bibfnamefont {K.}~\bibnamefont
  {Kobashi}}\ and\ \bibinfo {author} {\bibfnamefont {V.}~\bibnamefont
  {Chandrasekharan}},\ }\bibfield  {title} {\enquote {\bibinfo {title}
  {Anharmonic lattice vibrations in solid {N}$_2$ .{II}. the $\gamma$-phase},}\
  }\href@noop {} {\bibfield  {journal} {\bibinfo  {journal} {Mol. Phys.}\
  }\textbf {\bibinfo {volume} {36}},\ \bibinfo {pages} {1645--1659} (\bibinfo
  {year} {1978})}\BibitemShut {NoStop}%
\bibitem [{\citenamefont {Bloch}\ and\ \citenamefont
  {De~Dominicis}(1958)}]{bloch}%
  \BibitemOpen
  \bibfield  {author} {\bibinfo {author} {\bibfnamefont {C.}~\bibnamefont
  {Bloch}}\ and\ \bibinfo {author} {\bibfnamefont {C.}~\bibnamefont
  {De~Dominicis}},\ }\bibfield  {title} {\enquote {\bibinfo {title} {Un
  d\'{e}veloppement du potentiel de {Gibbs} d'un syst\`{e}me quantique
  compos\'{e} d'un grand nombre de particules},}\ }\href@noop {} {\bibfield
  {journal} {\bibinfo  {journal} {Nucl. Phys.}\ }\textbf {\bibinfo {volume}
  {7}},\ \bibinfo {pages} {459--479} (\bibinfo {year} {1958})}\BibitemShut
  {NoStop}%
\bibitem [{\citenamefont {Kohn}\ and\ \citenamefont
  {Luttinger}(1960)}]{Kohn1960}%
  \BibitemOpen
  \bibfield  {author} {\bibinfo {author} {\bibfnamefont {W.}~\bibnamefont
  {Kohn}}\ and\ \bibinfo {author} {\bibfnamefont {J.~M.}\ \bibnamefont
  {Luttinger}},\ }\bibfield  {title} {\enquote {\bibinfo {title} {Ground-state
  energy of a many-fermion system},}\ }\href
  {https://doi.org/10.1103/PhysRev.118.41} {\bibfield  {journal} {\bibinfo
  {journal} {Phys. Rev.}\ }\textbf {\bibinfo {volume} {118}},\ \bibinfo {pages}
  {41--45} (\bibinfo {year} {1960})}\BibitemShut {NoStop}%
\bibitem [{\citenamefont {Luttinger}\ and\ \citenamefont
  {Ward}(1960)}]{Luttinger1960}%
  \BibitemOpen
  \bibfield  {author} {\bibinfo {author} {\bibfnamefont {J.~M.}\ \bibnamefont
  {Luttinger}}\ and\ \bibinfo {author} {\bibfnamefont {J.~C.}\ \bibnamefont
  {Ward}},\ }\bibfield  {title} {\enquote {\bibinfo {title} {Ground-state
  energy of a many-fermion system. {II}},}\ }\href
  {https://doi.org/10.1103/PhysRev.118.1417} {\bibfield  {journal} {\bibinfo
  {journal} {Phys. Rev.}\ }\textbf {\bibinfo {volume} {118}},\ \bibinfo {pages}
  {1417--1427} (\bibinfo {year} {1960})}\BibitemShut {NoStop}%
\bibitem [{\citenamefont {Balian}, \citenamefont {Bloch},\ and\ \citenamefont
  {De~Dominicis}(1961)}]{balian}%
  \BibitemOpen
  \bibfield  {author} {\bibinfo {author} {\bibfnamefont {R.}~\bibnamefont
  {Balian}}, \bibinfo {author} {\bibfnamefont {C.}~\bibnamefont {Bloch}},\ and\
  \bibinfo {author} {\bibfnamefont {C.}~\bibnamefont {De~Dominicis}},\
  }\bibfield  {title} {\enquote {\bibinfo {title} {Formulation de la
  m\'{e}canique statistique en termes de nombres d'occupation {(I)}},}\
  }\href@noop {} {\bibfield  {journal} {\bibinfo  {journal} {Nucl. Phys.}\
  }\textbf {\bibinfo {volume} {25}},\ \bibinfo {pages} {529--567} (\bibinfo
  {year} {1961})}\BibitemShut {NoStop}%
\bibitem [{\citenamefont {Bloch}(1965)}]{blochbook}%
  \BibitemOpen
  \bibfield  {author} {\bibinfo {author} {\bibfnamefont {C.}~\bibnamefont
  {Bloch}},\ }in\ \href@noop {} {\emph {\bibinfo {booktitle} {Studies in
  Statistical Mechanics}}},\ \bibinfo {editor} {edited by\ \bibinfo {editor}
  {\bibfnamefont {J.}~\bibnamefont {De~Boer}}\ and\ \bibinfo {editor}
  {\bibfnamefont {G.~E.}\ \bibnamefont {Uhlenbeck}}}\ (\bibinfo  {publisher}
  {North Holland},\ \bibinfo {address} {Amsterdam},\ \bibinfo {year} {1965})\
  pp.\ \bibinfo {pages} {3--211}\BibitemShut {NoStop}%
\bibitem [{\citenamefont {Matsubara}(1955)}]{Matsubara1955}%
  \BibitemOpen
  \bibfield  {author} {\bibinfo {author} {\bibfnamefont {T.}~\bibnamefont
  {Matsubara}},\ }\bibfield  {title} {\enquote {\bibinfo {title} {A new
  approach to quantum-statistical mechanics},}\ }\href
  {https://doi.org/10.1143/PTP.14.351} {\bibfield  {journal} {\bibinfo
  {journal} {Prog. Theor. Phys.}\ }\textbf {\bibinfo {volume} {14}},\ \bibinfo
  {pages} {351--378} (\bibinfo {year} {1955})}\BibitemShut {NoStop}%
\bibitem [{\citenamefont {Hirata}(2021{\natexlab{a}})}]{Hirata2021}%
  \BibitemOpen
  \bibfield  {author} {\bibinfo {author} {\bibfnamefont {S.}~\bibnamefont
  {Hirata}},\ }\bibfield  {title} {\enquote {\bibinfo {title}
  {Finite-temperature many-body perturbation theory for electrons: Algebraic
  recursive definitions, second-quantized derivation, linked-diagram theorem,
  general-order algorithms, and grand canonical and canonical ensembles},}\
  }\href {https://doi.org/10.1063/5.0061384} {\bibfield  {journal} {\bibinfo
  {journal} {J. Chem. Phys.}\ }\textbf {\bibinfo {volume} {155}},\ \bibinfo
  {pages} {094106} (\bibinfo {year} {2021}{\natexlab{a}})}\BibitemShut
  {NoStop}%
\bibitem [{\citenamefont {Van~Hove}(1961)}]{vanhove}%
  \BibitemOpen
  \bibfield  {author} {\bibinfo {author} {\bibfnamefont {L.}~\bibnamefont
  {Van~Hove}},\ }\href@noop {} {\emph {\bibinfo {title} {Quantum Theory of Many
  Particle Systems}}}\ (\bibinfo  {publisher} {Benjamin},\ \bibinfo {address}
  {New York},\ \bibinfo {year} {1961})\BibitemShut {NoStop}%
\bibitem [{\citenamefont {Bloino}, \citenamefont {Biczysko},\ and\
  \citenamefont {Barone}(2012)}]{Bloino2012}%
  \BibitemOpen
  \bibfield  {author} {\bibinfo {author} {\bibfnamefont {J.}~\bibnamefont
  {Bloino}}, \bibinfo {author} {\bibfnamefont {M.}~\bibnamefont {Biczysko}},\
  and\ \bibinfo {author} {\bibfnamefont {V.}~\bibnamefont {Barone}},\
  }\bibfield  {title} {\enquote {\bibinfo {title} {General perturbative
  approach for spectroscopy, thermodynamics, and kinetics: Methodological
  background and benchmark studies},}\ }\href@noop {} {\bibfield  {journal}
  {\bibinfo  {journal} {J. Chem. Theory Comput.}\ }\textbf {\bibinfo {volume}
  {8}},\ \bibinfo {pages} {1015--1036} (\bibinfo {year} {2012})}\BibitemShut
  {NoStop}%
\bibitem [{\citenamefont {Tajti}\ \emph {et~al.}(2004)\citenamefont {Tajti},
  \citenamefont {Szalay}, \citenamefont {Cs\'{a}sz\'{a}r}, \citenamefont
  {K\'{a}llay}, \citenamefont {Gauss}, \citenamefont {Valeev}, \citenamefont
  {Flowers}, \citenamefont {V\'{a}zquez},\ and\ \citenamefont
  {Stanton}}]{HEAT2004}%
  \BibitemOpen
  \bibfield  {author} {\bibinfo {author} {\bibfnamefont {A.}~\bibnamefont
  {Tajti}}, \bibinfo {author} {\bibfnamefont {P.~G.}\ \bibnamefont {Szalay}},
  \bibinfo {author} {\bibfnamefont {A.~G.}\ \bibnamefont {Cs\'{a}sz\'{a}r}},
  \bibinfo {author} {\bibfnamefont {M.}~\bibnamefont {K\'{a}llay}}, \bibinfo
  {author} {\bibfnamefont {J.}~\bibnamefont {Gauss}}, \bibinfo {author}
  {\bibfnamefont {E.~F.}\ \bibnamefont {Valeev}}, \bibinfo {author}
  {\bibfnamefont {B.~A.}\ \bibnamefont {Flowers}}, \bibinfo {author}
  {\bibfnamefont {J.}~\bibnamefont {V\'{a}zquez}},\ and\ \bibinfo {author}
  {\bibfnamefont {J.~F.}\ \bibnamefont {Stanton}},\ }\bibfield  {title}
  {\enquote {\bibinfo {title} {{HEAT}: High accuracy extrapolated {\it ab
  initio} thermochemistry},}\ }\href@noop {} {\bibfield  {journal} {\bibinfo
  {journal} {J. Chem. Phys.}\ }\textbf {\bibinfo {volume} {121}},\ \bibinfo
  {pages} {11599--11613} (\bibinfo {year} {2004})}\BibitemShut {NoStop}%
\bibitem [{\citenamefont {Hirata}\ and\ \citenamefont
  {Yagi}(2008)}]{HirataYagiCPL2008}%
  \BibitemOpen
  \bibfield  {author} {\bibinfo {author} {\bibfnamefont {S.}~\bibnamefont
  {Hirata}}\ and\ \bibinfo {author} {\bibfnamefont {K.}~\bibnamefont {Yagi}},\
  }\bibfield  {title} {\enquote {\bibinfo {title} {Predictive electronic and
  vibrational many-body methods for molecules and macromolecules},}\
  }\href@noop {} {\bibfield  {journal} {\bibinfo  {journal} {Chem. Phys.
  Lett.}\ }\textbf {\bibinfo {volume} {464}},\ \bibinfo {pages} {123--134}
  (\bibinfo {year} {2008})}\BibitemShut {NoStop}%
\bibitem [{\citenamefont {Helgaker}, \citenamefont {Klopper},\ and\
  \citenamefont {Tew}(2008)}]{Helgaker2008}%
  \BibitemOpen
  \bibfield  {author} {\bibinfo {author} {\bibfnamefont {T.}~\bibnamefont
  {Helgaker}}, \bibinfo {author} {\bibfnamefont {W.}~\bibnamefont {Klopper}},\
  and\ \bibinfo {author} {\bibfnamefont {D.~P.}\ \bibnamefont {Tew}},\
  }\bibfield  {title} {\enquote {\bibinfo {title} {Quantitative quantum
  chemistry},}\ }\href@noop {} {\bibfield  {journal} {\bibinfo  {journal} {Mol.
  Phys.}\ }\textbf {\bibinfo {volume} {106}},\ \bibinfo {pages} {2107--2143}
  (\bibinfo {year} {2008})}\BibitemShut {NoStop}%
\bibitem [{\citenamefont {Rodriguez-Garcia}\ \emph {et~al.}(2007)\citenamefont
  {Rodriguez-Garcia}, \citenamefont {Hirata}, \citenamefont {Yagi},
  \citenamefont {Hirao}, \citenamefont {Taketsugu}, \citenamefont
  {Schweigert},\ and\ \citenamefont {Tasumi}}]{RodriguezGarcia2007}%
  \BibitemOpen
  \bibfield  {author} {\bibinfo {author} {\bibfnamefont {V.}~\bibnamefont
  {Rodriguez-Garcia}}, \bibinfo {author} {\bibfnamefont {S.}~\bibnamefont
  {Hirata}}, \bibinfo {author} {\bibfnamefont {K.}~\bibnamefont {Yagi}},
  \bibinfo {author} {\bibfnamefont {K.}~\bibnamefont {Hirao}}, \bibinfo
  {author} {\bibfnamefont {T.}~\bibnamefont {Taketsugu}}, \bibinfo {author}
  {\bibfnamefont {I.}~\bibnamefont {Schweigert}},\ and\ \bibinfo {author}
  {\bibfnamefont {M.}~\bibnamefont {Tasumi}},\ }\bibfield  {title} {\enquote
  {\bibinfo {title} {{Fermi} resonance in {CO}$_2$: A combined electronic
  coupled-cluster and vibrational configuration-interaction prediction},}\
  }\href {https://doi.org/10.1063/1.2710256} {\bibfield  {journal} {\bibinfo
  {journal} {J. Chem. Phys.}\ }\textbf {\bibinfo {volume} {126}},\ \bibinfo
  {pages} {124303} (\bibinfo {year} {2007})}\BibitemShut {NoStop}%
\bibitem [{\citenamefont {Amos}\ \emph {et~al.}(1991)\citenamefont {Amos},
  \citenamefont {Handy}, \citenamefont {Green}, \citenamefont {Jayatilaka},
  \citenamefont {Willetts},\ and\ \citenamefont {Palmieri}}]{Amos1991}%
  \BibitemOpen
  \bibfield  {author} {\bibinfo {author} {\bibfnamefont {R.~D.}\ \bibnamefont
  {Amos}}, \bibinfo {author} {\bibfnamefont {N.~C.}\ \bibnamefont {Handy}},
  \bibinfo {author} {\bibfnamefont {W.~H.}\ \bibnamefont {Green}}, \bibinfo
  {author} {\bibfnamefont {D.}~\bibnamefont {Jayatilaka}}, \bibinfo {author}
  {\bibfnamefont {A.}~\bibnamefont {Willetts}},\ and\ \bibinfo {author}
  {\bibfnamefont {P.}~\bibnamefont {Palmieri}},\ }\bibfield  {title} {\enquote
  {\bibinfo {title} {Anharmonic vibrational properties of {CH}$_2${F}$_2$: A
  comparison of theory and experiment},}\ }\href@noop {} {\bibfield  {journal}
  {\bibinfo  {journal} {J. Chem. Phys.}\ }\textbf {\bibinfo {volume} {95}},\
  \bibinfo {pages} {8323--8336} (\bibinfo {year} {1991})}\BibitemShut {NoStop}%
\bibitem [{\citenamefont {Martin}\ \emph {et~al.}(1995)\citenamefont {Martin},
  \citenamefont {Lee}, \citenamefont {Taylor},\ and\ \citenamefont
  {Francois}}]{Martin1995}%
  \BibitemOpen
  \bibfield  {author} {\bibinfo {author} {\bibfnamefont {J.~M.~L.}\
  \bibnamefont {Martin}}, \bibinfo {author} {\bibfnamefont {T.~J.}\
  \bibnamefont {Lee}}, \bibinfo {author} {\bibfnamefont {P.~R.}\ \bibnamefont
  {Taylor}},\ and\ \bibinfo {author} {\bibfnamefont {J.~P.}\ \bibnamefont
  {Francois}},\ }\bibfield  {title} {\enquote {\bibinfo {title} {The anharmonic
  force field of ethylene, {C}$_2${H}$_4$, by means of accurate ab initio
  calculations},}\ }\href@noop {} {\bibfield  {journal} {\bibinfo  {journal}
  {J. Chem. Phys.}\ }\textbf {\bibinfo {volume} {103}},\ \bibinfo {pages}
  {2589--2602} (\bibinfo {year} {1995})}\BibitemShut {NoStop}%
\bibitem [{\citenamefont {Kuhler}, \citenamefont {Truhlar},\ and\ \citenamefont
  {Isaacson}(1996)}]{Kuhler1996}%
  \BibitemOpen
  \bibfield  {author} {\bibinfo {author} {\bibfnamefont {K.~M.}\ \bibnamefont
  {Kuhler}}, \bibinfo {author} {\bibfnamefont {D.~G.}\ \bibnamefont
  {Truhlar}},\ and\ \bibinfo {author} {\bibfnamefont {A.~D.}\ \bibnamefont
  {Isaacson}},\ }\bibfield  {title} {\enquote {\bibinfo {title} {General method
  for removing resonance singularities in quantum mechanical perturbation
  theory},}\ }\href@noop {} {\bibfield  {journal} {\bibinfo  {journal} {J.
  Chem. Phys.}\ }\textbf {\bibinfo {volume} {104}},\ \bibinfo {pages}
  {4664--4671} (\bibinfo {year} {1996})}\BibitemShut {NoStop}%
\bibitem [{\citenamefont {Stanton}\ and\ \citenamefont
  {Gauss}(1998)}]{Stanton1998}%
  \BibitemOpen
  \bibfield  {author} {\bibinfo {author} {\bibfnamefont {J.~F.}\ \bibnamefont
  {Stanton}}\ and\ \bibinfo {author} {\bibfnamefont {J.}~\bibnamefont
  {Gauss}},\ }\bibfield  {title} {\enquote {\bibinfo {title} {Anharmonicity in
  the ring stretching modes of diborane},}\ }\href@noop {} {\bibfield
  {journal} {\bibinfo  {journal} {J. Chem. Phys.}\ }\textbf {\bibinfo {volume}
  {108}},\ \bibinfo {pages} {9218--9220} (\bibinfo {year} {1998})}\BibitemShut
  {NoStop}%
\bibitem [{\citenamefont {Ruud}, \citenamefont {{\AA}strand},\ and\
  \citenamefont {Taylor}(2000)}]{Ruud2000}%
  \BibitemOpen
  \bibfield  {author} {\bibinfo {author} {\bibfnamefont {K.}~\bibnamefont
  {Ruud}}, \bibinfo {author} {\bibfnamefont {P.~O.}\ \bibnamefont
  {{\AA}strand}},\ and\ \bibinfo {author} {\bibfnamefont {P.~R.}\ \bibnamefont
  {Taylor}},\ }\bibfield  {title} {\enquote {\bibinfo {title} {An efficient
  approach for calculating vibrational wave functions and zero-point
  vibrational corrections to molecular properties of polyatomic molecules},}\
  }\href@noop {} {\bibfield  {journal} {\bibinfo  {journal} {J. Chem. Phys.}\
  }\textbf {\bibinfo {volume} {112}},\ \bibinfo {pages} {2668--2683} (\bibinfo
  {year} {2000})}\BibitemShut {NoStop}%
\bibitem [{\citenamefont {Barone}(2004)}]{Barone2004}%
  \BibitemOpen
  \bibfield  {author} {\bibinfo {author} {\bibfnamefont {V.}~\bibnamefont
  {Barone}},\ }\bibfield  {title} {\enquote {\bibinfo {title} {Vibrational
  zero-point energies and thermodynamic functions beyond the harmonic
  approximation},}\ }\href@noop {} {\bibfield  {journal} {\bibinfo  {journal}
  {J. Chem. Phys.}\ }\textbf {\bibinfo {volume} {120}},\ \bibinfo {pages}
  {3059--3065} (\bibinfo {year} {2004})}\BibitemShut {NoStop}%
\bibitem [{\citenamefont {Barone}(2005)}]{Barone2005}%
  \BibitemOpen
  \bibfield  {author} {\bibinfo {author} {\bibfnamefont {V.}~\bibnamefont
  {Barone}},\ }\bibfield  {title} {\enquote {\bibinfo {title} {Anharmonic
  vibrational properties by a fully automated second-order perturbative
  approach},}\ }\href@noop {} {\bibfield  {journal} {\bibinfo  {journal} {J.
  Chem. Phys.}\ }\textbf {\bibinfo {volume} {122}},\ \bibinfo {pages} {014108}
  (\bibinfo {year} {2005})}\BibitemShut {NoStop}%
\bibitem [{\citenamefont {V\'{a}zquez}\ and\ \citenamefont
  {Stanton}(2006)}]{Vazquez2005}%
  \BibitemOpen
  \bibfield  {author} {\bibinfo {author} {\bibfnamefont {J.}~\bibnamefont
  {V\'{a}zquez}}\ and\ \bibinfo {author} {\bibfnamefont {J.~F.}\ \bibnamefont
  {Stanton}},\ }\bibfield  {title} {\enquote {\bibinfo {title} {Simple(r)
  algebraic equation for transition moments of fundamental transitions in
  vibrational second-order perturbation theory},}\ }\href@noop {} {\bibfield
  {journal} {\bibinfo  {journal} {Mol. Phys.}\ }\textbf {\bibinfo {volume}
  {104}},\ \bibinfo {pages} {377--388} (\bibinfo {year} {2006})}\BibitemShut
  {NoStop}%
\bibitem [{\citenamefont {V\'{a}zquez}\ and\ \citenamefont
  {Stanton}(2007)}]{Vazquez2006}%
  \BibitemOpen
  \bibfield  {author} {\bibinfo {author} {\bibfnamefont {J.}~\bibnamefont
  {V\'{a}zquez}}\ and\ \bibinfo {author} {\bibfnamefont {J.~F.}\ \bibnamefont
  {Stanton}},\ }\bibfield  {title} {\enquote {\bibinfo {title} {Treatment of
  {Fermi} resonance effects on transition moments in vibrational perturbation
  theory},}\ }\href@noop {} {\bibfield  {journal} {\bibinfo  {journal} {Mol.
  Phys.}\ }\textbf {\bibinfo {volume} {105}},\ \bibinfo {pages} {101--109}
  (\bibinfo {year} {2007})}\BibitemShut {NoStop}%
\bibitem [{\citenamefont {Barone}\ \emph {et~al.}(2010)\citenamefont {Barone},
  \citenamefont {Bloino}, \citenamefont {Guido},\ and\ \citenamefont
  {Lipparini}}]{Barone2010}%
  \BibitemOpen
  \bibfield  {author} {\bibinfo {author} {\bibfnamefont {V.}~\bibnamefont
  {Barone}}, \bibinfo {author} {\bibfnamefont {J.}~\bibnamefont {Bloino}},
  \bibinfo {author} {\bibfnamefont {C.~A.}\ \bibnamefont {Guido}},\ and\
  \bibinfo {author} {\bibfnamefont {F.}~\bibnamefont {Lipparini}},\ }\bibfield
  {title} {\enquote {\bibinfo {title} {A fully automated implementation of
  {VPT2} infrared intensities},}\ }\href@noop {} {\bibfield  {journal}
  {\bibinfo  {journal} {Chem. Phys. Lett.}\ }\textbf {\bibinfo {volume}
  {496}},\ \bibinfo {pages} {157--161} (\bibinfo {year} {2010})}\BibitemShut
  {NoStop}%
\bibitem [{\citenamefont {Truhlar}\ and\ \citenamefont
  {Isaacson}(1991)}]{Truhlar1991}%
  \BibitemOpen
  \bibfield  {author} {\bibinfo {author} {\bibfnamefont {D.~G.}\ \bibnamefont
  {Truhlar}}\ and\ \bibinfo {author} {\bibfnamefont {A.~D.}\ \bibnamefont
  {Isaacson}},\ }\bibfield  {title} {\enquote {\bibinfo {title} {Simple
  perturbation theory estimates of equilibrium constants from force fields},}\
  }\href@noop {} {\bibfield  {journal} {\bibinfo  {journal} {J. Chem. Phys.}\
  }\textbf {\bibinfo {volume} {94}},\ \bibinfo {pages} {357--359} (\bibinfo
  {year} {1991})}\BibitemShut {NoStop}%
\bibitem [{\citenamefont {Makri}\ and\ \citenamefont
  {Miller}(2002)}]{Makri2002}%
  \BibitemOpen
  \bibfield  {author} {\bibinfo {author} {\bibfnamefont {N.}~\bibnamefont
  {Makri}}\ and\ \bibinfo {author} {\bibfnamefont {W.~H.}\ \bibnamefont
  {Miller}},\ }\bibfield  {title} {\enquote {\bibinfo {title} {Coherent state
  semiclassical initial value representation for the {Boltzmann} operator in
  thermal correlation functions},}\ }\href@noop {} {\bibfield  {journal}
  {\bibinfo  {journal} {J. Chem. Phys.}\ }\textbf {\bibinfo {volume} {116}},\
  \bibinfo {pages} {9207--9212} (\bibinfo {year} {2002})}\BibitemShut {NoStop}%
\bibitem [{\citenamefont {Bowman}(1986)}]{Bowman1986}%
  \BibitemOpen
  \bibfield  {author} {\bibinfo {author} {\bibfnamefont {J.~M.}\ \bibnamefont
  {Bowman}},\ }\bibfield  {title} {\enquote {\bibinfo {title} {The
  self-consistent-field approach to polyatomic vibrations},}\ }\href
  {https://doi.org/10.1021/ar00127a002} {\bibfield  {journal} {\bibinfo
  {journal} {Acc. Chem. Res.}\ }\textbf {\bibinfo {volume} {19}},\ \bibinfo
  {pages} {202--208} (\bibinfo {year} {1986})}\BibitemShut {NoStop}%
\bibitem [{\citenamefont {Bowman}(1978)}]{Bowman1978}%
  \BibitemOpen
  \bibfield  {author} {\bibinfo {author} {\bibfnamefont {J.~M.}\ \bibnamefont
  {Bowman}},\ }\bibfield  {title} {\enquote {\bibinfo {title} {Self-consistent
  field energies and wavefunctions for coupled oscillators},}\ }\href
  {https://doi.org/10.1063/1.435782} {\bibfield  {journal} {\bibinfo  {journal}
  {J. Chem. Phys.}\ }\textbf {\bibinfo {volume} {68}},\ \bibinfo {pages}
  {608--610} (\bibinfo {year} {1978})}\BibitemShut {NoStop}%
\bibitem [{\citenamefont {Ratner}\ and\ \citenamefont
  {Gerber}(1986)}]{Ratner1986}%
  \BibitemOpen
  \bibfield  {author} {\bibinfo {author} {\bibfnamefont {M.~A.}\ \bibnamefont
  {Ratner}}\ and\ \bibinfo {author} {\bibfnamefont {R.~B.}\ \bibnamefont
  {Gerber}},\ }\bibfield  {title} {\enquote {\bibinfo {title} {Excited
  vibrational states of polyatomic molecules: The semiclassical self-consistent
  field approach},}\ }\href {https://doi.org/10.1021/j100273a008} {\bibfield
  {journal} {\bibinfo  {journal} {J. Phys. Chem.}\ }\textbf {\bibinfo {volume}
  {90}},\ \bibinfo {pages} {20--30} (\bibinfo {year} {1986})}\BibitemShut
  {NoStop}%
\bibitem [{\citenamefont {Christoffel}\ and\ \citenamefont
  {Bowman}(1982)}]{Christoffel1982}%
  \BibitemOpen
  \bibfield  {author} {\bibinfo {author} {\bibfnamefont {K.~M.}\ \bibnamefont
  {Christoffel}}\ and\ \bibinfo {author} {\bibfnamefont {J.~M.}\ \bibnamefont
  {Bowman}},\ }\bibfield  {title} {\enquote {\bibinfo {title} {Investigations
  of self-consistent field, {SCF CI} and virtual state configuration
  interaction vibrational energies for a model three-mode system},}\ }\href
  {https://doi.org/10.1016/0009-2614(82)80335-7} {\bibfield  {journal}
  {\bibinfo  {journal} {Chem. Phys. Lett.}\ }\textbf {\bibinfo {volume} {85}},\
  \bibinfo {pages} {220--224} (\bibinfo {year} {1982})}\BibitemShut {NoStop}%
\bibitem [{\citenamefont {Norris}\ \emph {et~al.}(1996)\citenamefont {Norris},
  \citenamefont {Ratner}, \citenamefont {Roitberg},\ and\ \citenamefont
  {Gerber}}]{Norris1996}%
  \BibitemOpen
  \bibfield  {author} {\bibinfo {author} {\bibfnamefont {L.~S.}\ \bibnamefont
  {Norris}}, \bibinfo {author} {\bibfnamefont {M.~A.}\ \bibnamefont {Ratner}},
  \bibinfo {author} {\bibfnamefont {A.~E.}\ \bibnamefont {Roitberg}},\ and\
  \bibinfo {author} {\bibfnamefont {R.~B.}\ \bibnamefont {Gerber}},\ }\bibfield
   {title} {\enquote {\bibinfo {title} {M{\o}ller--plesset perturbation theory
  applied to vibrational problems},}\ }\href {https://doi.org/10.1063/1.472922}
  {\bibfield  {journal} {\bibinfo  {journal} {J. Chem. Phys.}\ }\textbf
  {\bibinfo {volume} {105}},\ \bibinfo {pages} {11261--11267} (\bibinfo {year}
  {1996})}\BibitemShut {NoStop}%
\bibitem [{\citenamefont {Christiansen}(2004)}]{Christiansen2004}%
  \BibitemOpen
  \bibfield  {author} {\bibinfo {author} {\bibfnamefont {O.}~\bibnamefont
  {Christiansen}},\ }\bibfield  {title} {\enquote {\bibinfo {title}
  {Vibrational coupled cluster theory},}\ }\href
  {https://doi.org/10.1063/1.1637579} {\bibfield  {journal} {\bibinfo
  {journal} {J. Chem. Phys.}\ }\textbf {\bibinfo {volume} {120}},\ \bibinfo
  {pages} {2149--2159} (\bibinfo {year} {2004})}\BibitemShut {NoStop}%
\bibitem [{\citenamefont {Szabo}\ and\ \citenamefont {Ostlund}(1982)}]{szabo}%
  \BibitemOpen
  \bibfield  {author} {\bibinfo {author} {\bibfnamefont {A.}~\bibnamefont
  {Szabo}}\ and\ \bibinfo {author} {\bibfnamefont {N.~S.}\ \bibnamefont
  {Ostlund}},\ }\href@noop {} {\emph {\bibinfo {title} {Modern Quantum
  Chemistry}}}\ (\bibinfo  {publisher} {MacMillan},\ \bibinfo {address} {New
  York, NY},\ \bibinfo {year} {1982})\BibitemShut {NoStop}%
\bibitem [{\citenamefont {Shavitt}\ and\ \citenamefont
  {Bartlett}(2009)}]{Shavitt2009}%
  \BibitemOpen
  \bibfield  {author} {\bibinfo {author} {\bibfnamefont {I.}~\bibnamefont
  {Shavitt}}\ and\ \bibinfo {author} {\bibfnamefont {R.}~\bibnamefont
  {Bartlett}},\ }\href@noop {} {\emph {\bibinfo {title} {Many-Body Methods in
  Chemistry and Physics}}}\ (\bibinfo  {publisher} {Cambridge University
  Press},\ \bibinfo {address} {Cambridge},\ \bibinfo {year} {2009})\BibitemShut
  {NoStop}%
\bibitem [{\citenamefont {Njegic}\ and\ \citenamefont
  {Gordon}(2006)}]{Njegic2006}%
  \BibitemOpen
  \bibfield  {author} {\bibinfo {author} {\bibfnamefont {B.}~\bibnamefont
  {Njegic}}\ and\ \bibinfo {author} {\bibfnamefont {M.~S.}\ \bibnamefont
  {Gordon}},\ }\bibfield  {title} {\enquote {\bibinfo {title} {Exploring the
  effect of anharmonicity of molecular vibrations on thermodynamic
  properties},}\ }\href@noop {} {\bibfield  {journal} {\bibinfo  {journal} {J.
  Chem. Phys.}\ }\textbf {\bibinfo {volume} {125}},\ \bibinfo {pages} {224102}
  (\bibinfo {year} {2006})}\BibitemShut {NoStop}%
\bibitem [{\citenamefont {Hansen}\ \emph {et~al.}(2008)\citenamefont {Hansen},
  \citenamefont {Christiansen}, \citenamefont {Toffoli},\ and\ \citenamefont
  {Kongsted}}]{Hansen2008}%
  \BibitemOpen
  \bibfield  {author} {\bibinfo {author} {\bibfnamefont {M.~B.}\ \bibnamefont
  {Hansen}}, \bibinfo {author} {\bibfnamefont {O.}~\bibnamefont
  {Christiansen}}, \bibinfo {author} {\bibfnamefont {D.}~\bibnamefont
  {Toffoli}},\ and\ \bibinfo {author} {\bibfnamefont {J.}~\bibnamefont
  {Kongsted}},\ }\bibfield  {title} {\enquote {\bibinfo {title} {A virtual
  vibrational self-consistent-field method for efficient calculation of
  molecular vibrational partition functions and thermal effects on molecular
  properties},}\ }\href@noop {} {\bibfield  {journal} {\bibinfo  {journal} {J.
  Chem. Phys.}\ }\textbf {\bibinfo {volume} {128}},\ \bibinfo {pages} {174106}
  (\bibinfo {year} {2008})}\BibitemShut {NoStop}%
\bibitem [{\citenamefont {Roy}\ and\ \citenamefont {Prasad}(2009)}]{Roy2009}%
  \BibitemOpen
  \bibfield  {author} {\bibinfo {author} {\bibfnamefont {T.~K.}\ \bibnamefont
  {Roy}}\ and\ \bibinfo {author} {\bibfnamefont {M.~D.}\ \bibnamefont
  {Prasad}},\ }\bibfield  {title} {\enquote {\bibinfo {title} {A thermal
  self-consistent field theory for the calculation of molecular vibrational
  partition functions},}\ }\href@noop {} {\bibfield  {journal} {\bibinfo
  {journal} {J. Chem. Phys.}\ }\textbf {\bibinfo {volume} {131}},\ \bibinfo
  {pages} {114102} (\bibinfo {year} {2009})}\BibitemShut {NoStop}%
\bibitem [{\citenamefont {Qin}\ and\ \citenamefont {Hirata}(2021)}]{Xiuyi2021}%
  \BibitemOpen
  \bibfield  {author} {\bibinfo {author} {\bibfnamefont {X.}~\bibnamefont
  {Qin}}\ and\ \bibinfo {author} {\bibfnamefont {S.}~\bibnamefont {Hirata}},\
  }\bibfield  {title} {\enquote {\bibinfo {title} {Finite-temperature
  vibrational full configuration interaction},}\ }\href
  {https://doi.org/10.1080/00268976.2021.1949503} {\bibfield  {journal}
  {\bibinfo  {journal} {Mol. Phys.}\ }\textbf {\bibinfo {volume} {119}},\
  \bibinfo {pages} {e1949503} (\bibinfo {year} {2021}); The ZPE of XVH2 in Table 1
  should read `$-0.00054$' instead of `$-0.00052$'}\BibitemShut {NoStop}%
\bibitem [{\citenamefont {Hirata}, \citenamefont {Ke{\c{c}}eli},\ and\
  \citenamefont {Yagi}(2010)}]{Hirata2010}%
  \BibitemOpen
  \bibfield  {author} {\bibinfo {author} {\bibfnamefont {S.}~\bibnamefont
  {Hirata}}, \bibinfo {author} {\bibfnamefont {M.}~\bibnamefont
  {Ke{\c{c}}eli}},\ and\ \bibinfo {author} {\bibfnamefont {K.}~\bibnamefont
  {Yagi}},\ }\bibfield  {title} {\enquote {\bibinfo {title} {First-principles
  theories for anharmonic lattice vibrations},}\ }\href
  {https://doi.org/10.1063/1.3462237} {\bibfield  {journal} {\bibinfo
  {journal} {J. Chem. Phys.}\ }\textbf {\bibinfo {volume} {133}},\ \bibinfo
  {pages} {034109} (\bibinfo {year} {2010})}\BibitemShut {NoStop}%
\bibitem [{\citenamefont {Ke\c{c}eli}\ and\ \citenamefont
  {Hirata}(2011)}]{Keceli2011}%
  \BibitemOpen
  \bibfield  {author} {\bibinfo {author} {\bibfnamefont {M.}~\bibnamefont
  {Ke\c{c}eli}}\ and\ \bibinfo {author} {\bibfnamefont {S.}~\bibnamefont
  {Hirata}},\ }\bibfield  {title} {\enquote {\bibinfo {title} {Size-extensive
  vibrational self-consistent field method},}\ }\href
  {https://doi.org/10.1063/1.3644895} {\bibfield  {journal} {\bibinfo
  {journal} {J. Chem. Phys.}\ }\textbf {\bibinfo {volume} {135}},\ \bibinfo
  {pages} {134108} (\bibinfo {year} {2011})}\BibitemShut {NoStop}%
\bibitem [{\citenamefont {Ke\c{c}eli}\ \emph {et~al.}(2009)\citenamefont
  {Ke\c{c}eli}, \citenamefont {Shiozaki}, \citenamefont {Yagi},\ and\
  \citenamefont {Hirata}}]{KeceliShio2009}%
  \BibitemOpen
  \bibfield  {author} {\bibinfo {author} {\bibfnamefont {M.}~\bibnamefont
  {Ke\c{c}eli}}, \bibinfo {author} {\bibfnamefont {T.}~\bibnamefont
  {Shiozaki}}, \bibinfo {author} {\bibfnamefont {K.}~\bibnamefont {Yagi}},\
  and\ \bibinfo {author} {\bibfnamefont {S.}~\bibnamefont {Hirata}},\
  }\bibfield  {title} {\enquote {\bibinfo {title} {Anharmonic vibrational
  frequencies and vibrationally-averaged structures of key species in
  hydrocarbon combustion: {HCO}$^+$, {HCO}, {HNO}, {HOO}, {HOO}$^-$,
  {CH}$_3^+$, and {CH}$_3$},}\ }\href@noop {} {\bibfield  {journal} {\bibinfo
  {journal} {Mol. Phys.}\ }\textbf {\bibinfo {volume} {107}},\ \bibinfo {pages}
  {1283--1301} (\bibinfo {year} {2009})}\BibitemShut {NoStop}%
\bibitem [{\citenamefont {Makri}(1999)}]{Makri1999}%
  \BibitemOpen
  \bibfield  {author} {\bibinfo {author} {\bibfnamefont {N.}~\bibnamefont
  {Makri}},\ }\bibfield  {title} {\enquote {\bibinfo {title} {The linear
  response approximation and its lowest order corrections: An influence
  functional approach},}\ }\href@noop {} {\bibfield  {journal} {\bibinfo
  {journal} {J. Phys. Chem. B}\ }\textbf {\bibinfo {volume} {103}},\ \bibinfo
  {pages} {2823--2829} (\bibinfo {year} {1999})}\BibitemShut {NoStop}%
\bibitem [{\citenamefont {Hermes}, \citenamefont {Ke\c{c}eli},\ and\
  \citenamefont {Hirata}(2012)}]{Hermes2012}%
  \BibitemOpen
  \bibfield  {author} {\bibinfo {author} {\bibfnamefont {M.~R.}\ \bibnamefont
  {Hermes}}, \bibinfo {author} {\bibfnamefont {M.}~\bibnamefont {Ke\c{c}eli}},\
  and\ \bibinfo {author} {\bibfnamefont {S.}~\bibnamefont {Hirata}},\
  }\bibfield  {title} {\enquote {\bibinfo {title} {Size-extensive vibrational
  self-consistent field methods with anharmonic geometry corrections},}\ }\href
  {https://doi.org/10.1063/1.4729602} {\bibfield  {journal} {\bibinfo
  {journal} {J. Chem. Phys.}\ }\textbf {\bibinfo {volume} {136}},\ \bibinfo
  {pages} {234109} (\bibinfo {year} {2012})}\BibitemShut {NoStop}%
\bibitem [{\citenamefont {Trickey}(2011)}]{private}%
  \BibitemOpen
  \bibfield  {author} {\bibinfo {author} {\bibfnamefont {S.~B.}\ \bibnamefont
  {Trickey}},\ }\href@noop {} {\enquote {\bibinfo {title} {private
  communications},}\ } (\bibinfo {year} {2011})\BibitemShut {NoStop}%
\bibitem [{\citenamefont {Hermes}\ and\ \citenamefont
  {Hirata}(2013{\natexlab{b}})}]{Hermes2013}%
  \BibitemOpen
  \bibfield  {author} {\bibinfo {author} {\bibfnamefont {M.~R.}\ \bibnamefont
  {Hermes}}\ and\ \bibinfo {author} {\bibfnamefont {S.}~\bibnamefont
  {Hirata}},\ }\bibfield  {title} {\enquote {\bibinfo {title} {Second-order
  many-body perturbation expansions of vibrational {Dyson} self-energies},}\
  }\href {https://doi.org/10.1063/1.4813123} {\bibfield  {journal} {\bibinfo
  {journal} {J. Chem. Phys.}\ }\textbf {\bibinfo {volume} {139}},\ \bibinfo
  {pages} {034111} (\bibinfo {year} {2013}{\natexlab{b}}); Equation (53) is missing a prefactor of $1/2$}\BibitemShut
  {NoStop}%
\bibitem [{\citenamefont {Faucheaux}\ and\ \citenamefont
  {Hirata}(2015)}]{Faucheaux2015}%
  \BibitemOpen
  \bibfield  {author} {\bibinfo {author} {\bibfnamefont {J.~A.}\ \bibnamefont
  {Faucheaux}}\ and\ \bibinfo {author} {\bibfnamefont {S.}~\bibnamefont
  {Hirata}},\ }\bibfield  {title} {\enquote {\bibinfo {title} {Higher-order
  diagrammatic vibrational coupled-cluster theory},}\ }\href
  {https://doi.org/10.1063/1.4931472} {\bibfield  {journal} {\bibinfo
  {journal} {J. Chem. Phys.}\ }\textbf {\bibinfo {volume} {143}},\ \bibinfo
  {pages} {134105} (\bibinfo {year} {2015})}\BibitemShut {NoStop}%
\bibitem [{\citenamefont {Faucheaux}, \citenamefont {Nooijen},\ and\
  \citenamefont {Hirata}(2018)}]{Faucheaux2018}%
  \BibitemOpen
  \bibfield  {author} {\bibinfo {author} {\bibfnamefont {J.~A.}\ \bibnamefont
  {Faucheaux}}, \bibinfo {author} {\bibfnamefont {M.}~\bibnamefont {Nooijen}},\
  and\ \bibinfo {author} {\bibfnamefont {S.}~\bibnamefont {Hirata}},\
  }\bibfield  {title} {\enquote {\bibinfo {title} {Similarity-transformed
  equation-of-motion vibrational coupled-cluster theory},}\ }\href
  {https://doi.org/10.1063/1.5004151} {\bibfield  {journal} {\bibinfo
  {journal} {J. Chem. Phys.}\ }\textbf {\bibinfo {volume} {148}},\ \bibinfo
  {pages} {054104} (\bibinfo {year} {2018})}\BibitemShut {NoStop}%
\bibitem [{\citenamefont {Cao}\ and\ \citenamefont {Voth}(1995)}]{Cao1995}%
  \BibitemOpen
  \bibfield  {author} {\bibinfo {author} {\bibfnamefont {J.~S.}\ \bibnamefont
  {Cao}}\ and\ \bibinfo {author} {\bibfnamefont {G.~A.}\ \bibnamefont {Voth}},\
  }\bibfield  {title} {\enquote {\bibinfo {title} {Modeling physical systems by
  effective harmonic oscillators:\ {The} optimized quadratic approximation},}\
  }\href@noop {} {\bibfield  {journal} {\bibinfo  {journal} {J. Chem. Phys.}\
  }\textbf {\bibinfo {volume} {102}},\ \bibinfo {pages} {3337--3348} (\bibinfo
  {year} {1995})}\BibitemShut {NoStop}%
\bibitem [{\citenamefont {Prasad}(1988)}]{Prasad1988}%
  \BibitemOpen
  \bibfield  {author} {\bibinfo {author} {\bibfnamefont {M.~D.}\ \bibnamefont
  {Prasad}},\ }\bibfield  {title} {\enquote {\bibinfo {title} {Time-dependent
  coupled cluster method: A new approach to the calculation of molecular
  absorption spectra},}\ }\href@noop {} {\bibfield  {journal} {\bibinfo
  {journal} {J. Chem. Phys.}\ }\textbf {\bibinfo {volume} {88}},\ \bibinfo
  {pages} {7005--7010} (\bibinfo {year} {1988})}\BibitemShut {NoStop}%
\bibitem [{\citenamefont {Nagalakshmi}\ \emph {et~al.}(1994)\citenamefont
  {Nagalakshmi}, \citenamefont {Lakshminarayana}, \citenamefont {Sumithra},\
  and\ \citenamefont {Prasad}}]{Prasad1994}%
  \BibitemOpen
  \bibfield  {author} {\bibinfo {author} {\bibfnamefont {V.}~\bibnamefont
  {Nagalakshmi}}, \bibinfo {author} {\bibfnamefont {V.}~\bibnamefont
  {Lakshminarayana}}, \bibinfo {author} {\bibfnamefont {G.}~\bibnamefont
  {Sumithra}},\ and\ \bibinfo {author} {\bibfnamefont {M.~D.}\ \bibnamefont
  {Prasad}},\ }\bibfield  {title} {\enquote {\bibinfo {title} {Coupled-cluster
  description of anharmonic molecular vibrations: Application to {O}$_3$ and
  {SO}$_2$},}\ }\href@noop {} {\bibfield  {journal} {\bibinfo  {journal} {Chem.
  Phys. Lett.}\ }\textbf {\bibinfo {volume} {217}},\ \bibinfo {pages}
  {279--282} (\bibinfo {year} {1994})}\BibitemShut {NoStop}%
\bibitem [{\citenamefont {Banik}, \citenamefont {Pal},\ and\ \citenamefont
  {Prasad}(2008)}]{Banik2008}%
  \BibitemOpen
  \bibfield  {author} {\bibinfo {author} {\bibfnamefont {S.}~\bibnamefont
  {Banik}}, \bibinfo {author} {\bibfnamefont {S.}~\bibnamefont {Pal}},\ and\
  \bibinfo {author} {\bibfnamefont {M.~D.}\ \bibnamefont {Prasad}},\ }\bibfield
   {title} {\enquote {\bibinfo {title} {Calculation of vibrational energy of
  molecule using coupled cluster linear response theory in bosonic
  representation: Convergence studies},}\ }\href@noop {} {\bibfield  {journal}
  {\bibinfo  {journal} {J. Chem. Phys.}\ }\textbf {\bibinfo {volume} {129}},\
  \bibinfo {pages} {134111} (\bibinfo {year} {2008})}\BibitemShut {NoStop}%
\bibitem [{\citenamefont {Hermes}\ and\ \citenamefont
  {Hirata}(2015)}]{Hermes_review2015}%
  \BibitemOpen
  \bibfield  {author} {\bibinfo {author} {\bibfnamefont {M.~R.}\ \bibnamefont
  {Hermes}}\ and\ \bibinfo {author} {\bibfnamefont {S.}~\bibnamefont
  {Hirata}},\ }\bibfield  {title} {\enquote {\bibinfo {title} {Diagrammatic
  theories of anharmonic molecular vibrations},}\ }\href@noop {} {\bibfield
  {journal} {\bibinfo  {journal} {Int. Rev. Phys. Chem.}\ }\textbf {\bibinfo
  {volume} {34}},\ \bibinfo {pages} {71--97} (\bibinfo {year}
  {2015})}\BibitemShut {NoStop}%
\bibitem [{\citenamefont {Hirata}\ and\ \citenamefont
  {Hermes}(2014)}]{Hirata2014}%
  \BibitemOpen
  \bibfield  {author} {\bibinfo {author} {\bibfnamefont {S.}~\bibnamefont
  {Hirata}}\ and\ \bibinfo {author} {\bibfnamefont {M.~R.}\ \bibnamefont
  {Hermes}},\ }\bibfield  {title} {\enquote {\bibinfo {title} {Normal-ordered
  second-quantized {Hamiltonian} for molecular vibrations},}\ }\href
  {https://doi.org/10.1063/1.4901061} {\bibfield  {journal} {\bibinfo
  {journal} {J. Chem. Phys.}\ }\textbf {\bibinfo {volume} {141}},\ \bibinfo
  {pages} {184111} (\bibinfo {year} {2014})}\BibitemShut {NoStop}%
\bibitem [{\citenamefont {Jha}\ and\ \citenamefont {Hirata}(2019)}]{Jha2019}%
  \BibitemOpen
  \bibfield  {author} {\bibinfo {author} {\bibfnamefont {P.~K.}\ \bibnamefont
  {Jha}}\ and\ \bibinfo {author} {\bibfnamefont {S.}~\bibnamefont {Hirata}},\
  }\bibfield  {title} {\enquote {\bibinfo {title} {Numerical evidence
  invalidating finite-temperature many-body perturbation theory},}\ }\href
  {https://doi.org/10.1016/bs.arcc.2019.08.002} {\bibfield  {journal} {\bibinfo
   {journal} {Annu. Rep. Comput. Chem.}\ }\textbf {\bibinfo {volume} {15}},\
  \bibinfo {pages} {3--15} (\bibinfo {year} {2019})}\BibitemShut {NoStop}%
\bibitem [{\citenamefont {Hirata}\ and\ \citenamefont
  {Jha}(2019)}]{Hirata2019}%
  \BibitemOpen
  \bibfield  {author} {\bibinfo {author} {\bibfnamefont {S.}~\bibnamefont
  {Hirata}}\ and\ \bibinfo {author} {\bibfnamefont {P.~K.}\ \bibnamefont
  {Jha}},\ }\bibfield  {title} {\enquote {\bibinfo {title} {Converging
  finite-temperature many-body perturbation theory in the grand canonical
  ensemble that conserves the average number of electrons},}\ }\href
  {https://doi.org/10.1016/bs.arcc.2019.08.003} {\bibfield  {journal} {\bibinfo
   {journal} {Annu. Rep. Comput. Chem.}\ }\textbf {\bibinfo {volume} {15}},\
  \bibinfo {pages} {17--37} (\bibinfo {year} {2019})}\BibitemShut {NoStop}%
\bibitem [{\citenamefont {Jha}\ and\ \citenamefont {Hirata}(2020)}]{Jha2020}%
  \BibitemOpen
  \bibfield  {author} {\bibinfo {author} {\bibfnamefont {P.~K.}\ \bibnamefont
  {Jha}}\ and\ \bibinfo {author} {\bibfnamefont {S.}~\bibnamefont {Hirata}},\
  }\bibfield  {title} {\enquote {\bibinfo {title} {Finite-temperature many-body
  perturbation theory in the canonical ensemble},}\ }\href
  {https://doi.org/10.1103/PhysRevE.101.022106} {\bibfield  {journal} {\bibinfo
   {journal} {Phys. Rev. E}\ }\textbf {\bibinfo {volume} {101}},\ \bibinfo
  {pages} {022106} (\bibinfo {year} {2020})}\BibitemShut {NoStop}%
\bibitem [{\citenamefont {Hirata}\ and\ \citenamefont
  {Jha}(2020)}]{Hirata_2020}%
  \BibitemOpen
  \bibfield  {author} {\bibinfo {author} {\bibfnamefont {S.}~\bibnamefont
  {Hirata}}\ and\ \bibinfo {author} {\bibfnamefont {P.~K.}\ \bibnamefont
  {Jha}},\ }\bibfield  {title} {\enquote {\bibinfo {title} {Finite-temperature
  many-body perturbation theory in the grand canonical ensemble},}\ }\href
  {https://doi.org/10.1063/5.0009679} {\bibfield  {journal} {\bibinfo
  {journal} {J. Chem. Phys.}\ }\textbf {\bibinfo {volume} {153}},\ \bibinfo
  {pages} {014103} (\bibinfo {year} {2020})}\BibitemShut {NoStop}%
\bibitem [{\citenamefont {Hirschfelder}\ and\ \citenamefont
  {Certain}(1974)}]{Hirschfelder1974}%
  \BibitemOpen
  \bibfield  {author} {\bibinfo {author} {\bibfnamefont {J.~O.}\ \bibnamefont
  {Hirschfelder}}\ and\ \bibinfo {author} {\bibfnamefont {P.~R.}\ \bibnamefont
  {Certain}},\ }\bibfield  {title} {\enquote {\bibinfo {title} {Degenerate {RS}
  perturbation theory},}\ }\href {https://doi.org/10.1063/1.1681123} {\bibfield
   {journal} {\bibinfo  {journal} {J. Chem. Phys.}\ }\textbf {\bibinfo {volume}
  {60}},\ \bibinfo {pages} {1118--1137} (\bibinfo {year} {1974})}\BibitemShut
  {NoStop}%
\bibitem [{\citenamefont {Nooijen}\ and\ \citenamefont
  {Bao}(2021)}]{Nooijen2021}%
  \BibitemOpen
  \bibfield  {author} {\bibinfo {author} {\bibfnamefont {M.}~\bibnamefont
  {Nooijen}}\ and\ \bibinfo {author} {\bibfnamefont {S.}~\bibnamefont {Bao}},\
  }\bibfield  {title} {\enquote {\bibinfo {title} {Normal ordered exponential
  approach to thermal properties and time-correlation functions: general theory
  and simple examples},}\ }\href
  {https://doi.org/10.1080/00268976.2021.1980832} {\bibfield  {journal}
  {\bibinfo  {journal} {Mol. Phys.}\ }\textbf {\bibinfo {volume} {119}},\
  \bibinfo {pages} {e1980832} (\bibinfo {year} {2021})}\BibitemShut {NoStop}%
\bibitem [{\citenamefont {Hirata}\ \emph {et~al.}(2017)\citenamefont {Hirata},
  \citenamefont {Doran}, \citenamefont {Knowles},\ and\ \citenamefont
  {Ortiz}}]{Hirata2017}%
  \BibitemOpen
  \bibfield  {author} {\bibinfo {author} {\bibfnamefont {S.}~\bibnamefont
  {Hirata}}, \bibinfo {author} {\bibfnamefont {A.~E.}\ \bibnamefont {Doran}},
  \bibinfo {author} {\bibfnamefont {P.~J.}\ \bibnamefont {Knowles}},\ and\
  \bibinfo {author} {\bibfnamefont {J.~V.}\ \bibnamefont {Ortiz}},\ }\bibfield
  {title} {\enquote {\bibinfo {title} {One-particle many-body {Green’s}
  function theory: Algebraic recursive definitions, linked-diagram theorem,
  irreducible-diagram theorem, and general-order algorithms},}\ }\href
  {https://doi.org/10.1063/1.4994837} {\bibfield  {journal} {\bibinfo
  {journal} {J. Chem. Phys.}\ }\textbf {\bibinfo {volume} {147}},\ \bibinfo
  {pages} {044108} (\bibinfo {year} {2017})}\BibitemShut {NoStop}%
\bibitem [{\citenamefont {Hirata}(2021{\natexlab{b}})}]{Hirata_KL2021}%
  \BibitemOpen
  \bibfield  {author} {\bibinfo {author} {\bibfnamefont {S.}~\bibnamefont
  {Hirata}},\ }\bibfield  {title} {\enquote {\bibinfo {title} {Low-temperature
  breakdown of many-body perturbation theory for thermodynamics},}\ }\href@noop
  {} {\bibfield  {journal} {\bibinfo  {journal} {Phys. Rev. A}\ }\textbf
  {\bibinfo {volume} {103}},\ \bibinfo {pages} {012223} (\bibinfo {year}
  {2021}{\natexlab{b}})}\BibitemShut {NoStop}%
\bibitem [{\citenamefont {Hirata}(2022)}]{Hirata_KL2022}%
  \BibitemOpen
  \bibfield  {author} {\bibinfo {author} {\bibfnamefont {S.}~\bibnamefont
  {Hirata}},\ }\bibfield  {title} {\enquote {\bibinfo {title} {General solution
  to the {Kohn}--{Luttinger} nonconvergence problem},}\ }\href@noop {}
  {\bibfield  {journal} {\bibinfo  {journal} {Chem. Phys. Lett.}\ }\textbf
  {\bibinfo {volume} {800}},\ \bibinfo {pages} {139668} (\bibinfo {year}
  {2022})}\BibitemShut {NoStop}%
\bibitem [{\citenamefont {McQuarrie}(1975)}]{mcquarrie1975}%
  \BibitemOpen
  \bibfield  {author} {\bibinfo {author} {\bibfnamefont {D.}~\bibnamefont
  {McQuarrie}},\ }\href@noop {} {\emph {\bibinfo {title} {Statistical
  Mechanics}}},\ Chemistry Series\ (\bibinfo  {publisher} {Harper \& Row},\
  \bibinfo {year} {1975})\BibitemShut {NoStop}%
\bibitem [{\citenamefont {Yagi}\ \emph {et~al.}(2004)\citenamefont {Yagi},
  \citenamefont {Hirao}, \citenamefont {Taketsugu}, \citenamefont {Schmidt},\
  and\ \citenamefont {Gordon}}]{YagiQFF2004}%
  \BibitemOpen
  \bibfield  {author} {\bibinfo {author} {\bibfnamefont {K.}~\bibnamefont
  {Yagi}}, \bibinfo {author} {\bibfnamefont {K.}~\bibnamefont {Hirao}},
  \bibinfo {author} {\bibfnamefont {T.}~\bibnamefont {Taketsugu}}, \bibinfo
  {author} {\bibfnamefont {M.~W.}\ \bibnamefont {Schmidt}},\ and\ \bibinfo
  {author} {\bibfnamefont {M.~S.}\ \bibnamefont {Gordon}},\ }\bibfield  {title}
  {\enquote {\bibinfo {title} {{\it Ab initio} vibrational state calculations
  with a quartic force field: Applications to {H}$_2${CO}, {C}$_2${H}$_4$,
  {CH}$_3${OH}, {CH}$_3${CCH}, and {C}$_6${H}$_6$},}\ }\href@noop {} {\bibfield
   {journal} {\bibinfo  {journal} {J. Chem. Phys.}\ }\textbf {\bibinfo {volume}
  {121}},\ \bibinfo {pages} {1383--1389} (\bibinfo {year} {2004})}\BibitemShut
  {NoStop}%
\bibitem [{\citenamefont {Ashcroft}\ and\ \citenamefont
  {Mermin}(1976)}]{Ashcroft}%
  \BibitemOpen
  \bibfield  {author} {\bibinfo {author} {\bibfnamefont {N.~W.}\ \bibnamefont
  {Ashcroft}}\ and\ \bibinfo {author} {\bibfnamefont {N.~D.}\ \bibnamefont
  {Mermin}},\ }\href@noop {} {\emph {\bibinfo {title} {Solid State Physics}}}\
  (\bibinfo  {publisher} {Brooks/Cole, Belmont},\ \bibinfo {year}
  {1976})\BibitemShut {NoStop}%
\bibitem [{\citenamefont {Yagi}, \citenamefont {Hirata},\ and\ \citenamefont
  {Hirao}(2007)}]{YagiNMR2007}%
  \BibitemOpen
  \bibfield  {author} {\bibinfo {author} {\bibfnamefont {K.}~\bibnamefont
  {Yagi}}, \bibinfo {author} {\bibfnamefont {S.}~\bibnamefont {Hirata}},\ and\
  \bibinfo {author} {\bibfnamefont {K.}~\bibnamefont {Hirao}},\ }\bibfield
  {title} {\enquote {\bibinfo {title} {Multiresolution potential energy
  surfaces for vibrational state calculations},}\ }\href@noop {} {\bibfield
  {journal} {\bibinfo  {journal} {Theor. Chem. Acc.}\ }\textbf {\bibinfo
  {volume} {118}},\ \bibinfo {pages} {681--691} (\bibinfo {year}
  {2007})}\BibitemShut {NoStop}%
\bibitem [{\citenamefont {Gell-Mann}\ and\ \citenamefont
  {Low}(1951)}]{GellmannLow}%
  \BibitemOpen
  \bibfield  {author} {\bibinfo {author} {\bibfnamefont {M.}~\bibnamefont
  {Gell-Mann}}\ and\ \bibinfo {author} {\bibfnamefont {F.}~\bibnamefont
  {Low}},\ }\bibfield  {title} {\enquote {\bibinfo {title} {Bound states in
  quantum field theory},}\ }\href@noop {} {\bibfield  {journal} {\bibinfo
  {journal} {Phys. Rev.}\ }\textbf {\bibinfo {volume} {84}},\ \bibinfo {pages}
  {350--354} (\bibinfo {year} {1951})}\BibitemShut {NoStop}%
\bibitem [{\citenamefont {Brueckner}(1955)}]{brueckner}%
  \BibitemOpen
  \bibfield  {author} {\bibinfo {author} {\bibfnamefont {K.~A.}\ \bibnamefont
  {Brueckner}},\ }\bibfield  {title} {\enquote {\bibinfo {title} {Many-body
  problem for strongly interacting particles. {II}. {Linked} cluster
  expansion},}\ }\href@noop {} {\bibfield  {journal} {\bibinfo  {journal}
  {Phys. Rev.}\ }\textbf {\bibinfo {volume} {100}},\ \bibinfo {pages} {36--45}
  (\bibinfo {year} {1955})}\BibitemShut {NoStop}%
\bibitem [{\citenamefont {Goldstone}(1957)}]{goldstone}%
  \BibitemOpen
  \bibfield  {author} {\bibinfo {author} {\bibfnamefont {J.}~\bibnamefont
  {Goldstone}},\ }\bibfield  {title} {\enquote {\bibinfo {title} {Derivation of
  the {Brueckner} many-body theory},}\ }\href@noop {} {\bibfield  {journal}
  {\bibinfo  {journal} {Proc. Roy. Soc. A (London)}\ }\textbf {\bibinfo
  {volume} {239}},\ \bibinfo {pages} {267--279} (\bibinfo {year}
  {1957})}\BibitemShut {NoStop}%
\bibitem [{\citenamefont {Hugenholtz}(1957)}]{hugenholtz}%
  \BibitemOpen
  \bibfield  {author} {\bibinfo {author} {\bibfnamefont {N.~M.}\ \bibnamefont
  {Hugenholtz}},\ }\bibfield  {title} {\enquote {\bibinfo {title} {Perturbation
  theory of large quantum systems},}\ }\href@noop {} {\bibfield  {journal}
  {\bibinfo  {journal} {Physica}\ }\textbf {\bibinfo {volume} {23}},\ \bibinfo
  {pages} {481--532} (\bibinfo {year} {1957})}\BibitemShut {NoStop}%
\bibitem [{\citenamefont {Frantz}\ and\ \citenamefont {Mills}(1960)}]{Frantz}%
  \BibitemOpen
  \bibfield  {author} {\bibinfo {author} {\bibfnamefont {L.~M.}\ \bibnamefont
  {Frantz}}\ and\ \bibinfo {author} {\bibfnamefont {R.~L.}\ \bibnamefont
  {Mills}},\ }\bibfield  {title} {\enquote {\bibinfo {title} {Many-body basis
  for the optical model},}\ }\href@noop {} {\bibfield  {journal} {\bibinfo
  {journal} {Nucl. Phys.}\ }\textbf {\bibinfo {volume} {15}},\ \bibinfo {pages}
  {16--32} (\bibinfo {year} {1960})}\BibitemShut {NoStop}%
\bibitem [{\citenamefont {Manne}(1977)}]{Manne}%
  \BibitemOpen
  \bibfield  {author} {\bibinfo {author} {\bibfnamefont {R.}~\bibnamefont
  {Manne}},\ }\bibfield  {title} {\enquote {\bibinfo {title} {Linked-diagram
  expansion of ground state of a many-electron system: {A} time-independent
  derivation},}\ }\href@noop {} {\bibfield  {journal} {\bibinfo  {journal}
  {Int. J. Quantum Chem. Symp.}\ }\textbf {\bibinfo {volume} {11}},\ \bibinfo
  {pages} {175--192} (\bibinfo {year} {1977})}\BibitemShut {NoStop}%
\bibitem [{\citenamefont {Harris}, \citenamefont {Monkhorst},\ and\
  \citenamefont {Freeman}(1992)}]{Harris}%
  \BibitemOpen
  \bibfield  {author} {\bibinfo {author} {\bibfnamefont {F.~E.}\ \bibnamefont
  {Harris}}, \bibinfo {author} {\bibfnamefont {H.~J.}\ \bibnamefont
  {Monkhorst}},\ and\ \bibinfo {author} {\bibfnamefont {D.~L.}\ \bibnamefont
  {Freeman}},\ }\href@noop {} {\emph {\bibinfo {title} {Algebraic and
  Diagrammatic Methods in Many-Fermion Theory}}}\ (\bibinfo  {publisher}
  {Oxford University Press},\ \bibinfo {address} {Oxford},\ \bibinfo {year}
  {1992})\BibitemShut {NoStop}%
\bibitem [{\citenamefont {Dyson}(1993)}]{dyson_physicsworld}%
  \BibitemOpen
  \bibfield  {author} {\bibinfo {author} {\bibfnamefont {F.}~\bibnamefont
  {Dyson}},\ }\bibfield  {title} {\enquote {\bibinfo {title} {George {Green}
  and physics},}\ }\href@noop {} {\bibfield  {journal} {\bibinfo  {journal}
  {Physics World}\ }\textbf {\bibinfo {volume} {6}},\ \bibinfo {pages} {33--38}
  (\bibinfo {year} {1993})}\BibitemShut {NoStop}%
\bibitem [{\citenamefont {Mermin}(1963)}]{Mermin}%
  \BibitemOpen
  \bibfield  {author} {\bibinfo {author} {\bibfnamefont {N.~D.}\ \bibnamefont
  {Mermin}},\ }\bibfield  {title} {\enquote {\bibinfo {title} {Stability of the
  thermal {Hartree}--{Fock} approximation},}\ }\href {https://doi.org/Doi
  10.1016/0003-4916(63)90226-4} {\bibfield  {journal} {\bibinfo  {journal}
  {Ann. Phys.}\ }\textbf {\bibinfo {volume} {21}},\ \bibinfo {pages} {99--121}
  (\bibinfo {year} {1963})}\BibitemShut {NoStop}%
\bibitem [{\citenamefont {Pain}(2011)}]{Pain}%
  \BibitemOpen
  \bibfield  {author} {\bibinfo {author} {\bibfnamefont {J.~C.}\ \bibnamefont
  {Pain}},\ }\bibfield  {title} {\enquote {\bibinfo {title} {Koopmans' theorem
  in the statistical {Hartree}--{Fock} theory},}\ }\href {https://doi.org/Artn
  145001 10.1088/0953-4075/44/14/145001} {\bibfield  {journal} {\bibinfo
  {journal} {J. Phys. B. At. Mol. Opt.}\ }\textbf {\bibinfo {volume} {44}},\
  \bibinfo {pages} {145001} (\bibinfo {year} {2011})}\BibitemShut {NoStop}%
\bibitem [{\citenamefont {Ortiz}(2020)}]{OrtizDyson}%
  \BibitemOpen
  \bibfield  {author} {\bibinfo {author} {\bibfnamefont {J.~V.}\ \bibnamefont
  {Ortiz}},\ }\bibfield  {title} {\enquote {\bibinfo {title} {Dyson-orbital
  concepts for description of electrons in molecules},}\ }\href@noop {}
  {\bibfield  {journal} {\bibinfo  {journal} {J. Chem. Phys.}\ }\textbf
  {\bibinfo {volume} {153}},\ \bibinfo {pages} {070902} (\bibinfo {year}
  {2020})}\BibitemShut {NoStop}%
\bibitem [{\citenamefont {Knowles}(2016)}]{private2}%
  \BibitemOpen
  \bibfield  {author} {\bibinfo {author} {\bibfnamefont {P.~J.}\ \bibnamefont
  {Knowles}},\ }\href@noop {} {\enquote {\bibinfo {title} {private
  communications},}\ } (\bibinfo {year} {2016})\BibitemShut {NoStop}%
\bibitem [{\citenamefont {Knowles}\ \emph {et~al.}(1985)\citenamefont
  {Knowles}, \citenamefont {Somasundram}, \citenamefont {Handy},\ and\
  \citenamefont {Hirao}}]{Knowles}%
  \BibitemOpen
  \bibfield  {author} {\bibinfo {author} {\bibfnamefont {P.~J.}\ \bibnamefont
  {Knowles}}, \bibinfo {author} {\bibfnamefont {K.}~\bibnamefont
  {Somasundram}}, \bibinfo {author} {\bibfnamefont {N.~C.}\ \bibnamefont
  {Handy}},\ and\ \bibinfo {author} {\bibfnamefont {K.}~\bibnamefont {Hirao}},\
  }\bibfield  {title} {\enquote {\bibinfo {title} {The calculation of
  higher-order energies in the many-body perturbation theory series},}\
  }\href@noop {} {\bibfield  {journal} {\bibinfo  {journal} {Chem. Phys.
  Lett.}\ }\textbf {\bibinfo {volume} {113}},\ \bibinfo {pages} {8--12}
  (\bibinfo {year} {1985})}\BibitemShut {NoStop}%
\bibitem [{\citenamefont {Qin}(2021)}]{Qin_code2021}%
  \BibitemOpen
  \bibfield  {author} {\bibinfo {author} {\bibfnamefont {X.}~\bibnamefont
  {Qin}},\ }\href@noop {} {\enquote {\bibinfo {title} {P{yFTPT}},}\ }\bibinfo
  {howpublished} {\url{https://github.com/Terryqqy/PyFTPT}} (\bibinfo {year}
  {2021})\BibitemShut {NoStop}%
\bibitem [{\citenamefont {Olsen}\ \emph {et~al.}(1996)\citenamefont {Olsen},
  \citenamefont {Christiansen}, \citenamefont {Koch},\ and\ \citenamefont
  {Jorgensen}}]{Olsen}%
  \BibitemOpen
  \bibfield  {author} {\bibinfo {author} {\bibfnamefont {J.}~\bibnamefont
  {Olsen}}, \bibinfo {author} {\bibfnamefont {O.}~\bibnamefont {Christiansen}},
  \bibinfo {author} {\bibfnamefont {H.}~\bibnamefont {Koch}},\ and\ \bibinfo
  {author} {\bibfnamefont {P.}~\bibnamefont {Jorgensen}},\ }\bibfield  {title}
  {\enquote {\bibinfo {title} {Surprising cases of divergent behavior in
  {M}{\o}ller--{P}lesset perturbation theory},}\ }\href@noop {} {\bibfield
  {journal} {\bibinfo  {journal} {J. Chem. Phys.}\ }\textbf {\bibinfo {volume}
  {105}},\ \bibinfo {pages} {5082--5090} (\bibinfo {year} {1996})}\BibitemShut
  {NoStop}%
\bibitem [{\citenamefont {Hirata}\ and\ \citenamefont
  {Bartlett}(2000)}]{HirataCC}%
  \BibitemOpen
  \bibfield  {author} {\bibinfo {author} {\bibfnamefont {S.}~\bibnamefont
  {Hirata}}\ and\ \bibinfo {author} {\bibfnamefont {R.~J.}\ \bibnamefont
  {Bartlett}},\ }\bibfield  {title} {\enquote {\bibinfo {title} {High-order
  coupled-cluster calculations through connected octuple excitations},}\
  }\href@noop {} {\bibfield  {journal} {\bibinfo  {journal} {Chem. Phys.
  Lett.}\ }\textbf {\bibinfo {volume} {321}},\ \bibinfo {pages} {216--224}
  (\bibinfo {year} {2000})}\BibitemShut {NoStop}%
\end{thebibliography}

%aipnum4-2.bst 2019-01-14 (MD) hand-edited version of apsrev4-1.bst
%Control: key (0)
%Control: author (8) initials jnrlst
%Control: editor formatted (1) identically to author
%Control: production of article title (0) allowed
%Control: page (1) range
%Control: year (1) truncated
%Control: production of eprint (0) enabled
%

\end{document}